\documentclass[12pt]{article}
\usepackage{amsmath}
\usepackage{graphicx,psfrag,epsf}
\usepackage{enumerate}
\usepackage{natbib}
\usepackage{url} 
\usepackage{amsthm}

\usepackage{amsfonts}
\usepackage{multirow, booktabs}
\usepackage[table]{xcolor}
\usepackage{cleveref}
\usepackage[T1]{fontenc}
\usepackage[utf8]{inputenc}
\usepackage{authblk}
\usepackage[multiple]{footmisc}
\usepackage{blindtext,titlefoot}
\usepackage{sectsty}
\usepackage{bbm}
\usepackage{extarrows}
\usepackage{amsmath,graphicx,centernot}
\usepackage{mathtools}
\usepackage{tikz}
\usepackage{enumitem}
\usepackage{circuitikz}\usepackage{caption}
\usetikzlibrary{decorations.pathreplacing, calligraphy}

\usetikzlibrary{shapes,decorations,arrows,calc,arrows.meta,fit,positioning}
\tikzset{
    -Latex,auto,node distance =1 cm and 1 cm,semithick,
    state/.style ={ellipse, draw, minimum width = 0.7 cm},
    point/.style = {circle, draw, inner sep=0.04cm,fill,node contents={}},
    bidirected/.style={Latex-Latex,dashed},
    el/.style = {inner sep=2pt, align=left, sloped}
}

\newcommand{\iid}{\stackrel{iid}{\sim}}
\renewcommand{\arraystretch}{0.8}
\setlist[enumerate]{listparindent=0pt}

\usepackage{amsmath, amssymb}
\usepackage{graphicx,psfrag,epsf}
\usepackage{enumerate}
\usepackage{natbib}
\usepackage{url} 
\usepackage{amsthm}
\usepackage{subcaption}
\usepackage{xcolor}
\usepackage{graphicx}
\usepackage{caption}
\usepackage{sidecap}
\usepackage{multirow, booktabs}
\newcommand{\indep}{\rotatebox[origin=c]{90}{$\models$}}

\newtheorem{Lemma}{Lemma}
\newtheorem*{Lemma*}{Lemma}
\newtheorem{Definition}{Definition}
\newtheorem{Theorem}{Theorem}
\newtheorem*{Theorem*}{Theorem}

\newtheorem{Remark}{Remark}

\newtheorem{Assumption}{Assumption}

\title{\LARGE \bf {Confounding Heterogeneity in Multivariate Sensitivity Analysis for Observational Studies}}

\date{\today}

\author[1]{Anthony Frazier}
\author[2]{Siyu Heng}
\author[2]{Wen Zhou}

\affil[1]{Department of Statistics, Colorado State University, U.S.A.}
\affil[2]{Department of Biostatistics, New York University, U.S.A.}

\begin{document}

\sectionfont{\bfseries\large\sffamily}%
%

\subsectionfont{\bfseries\sffamily\normalsize}%
%


\def\spacingset#1{\renewcommand{\baselinestretch}%
{#1}\small\normalsize} \spacingset{1}

\begin{center}
    \Large \bf Confounding Heterogeneity in Multivariate Sensitivity Analysis for Observational Studies
\end{center}

\begin{center}
  \large $\text{Anthony Frazier}^{1}$, $\text{Siyu Heng}^{2}$, and $ \text{Wen Zhou}^{2}$
\end{center}

\begin{center}
   \large \textit{$^{1}$Department of Statistics, Colorado State University, U.S.A.}
\end{center}

\begin{center}
   \large \textit{$^{2}$Department of Biostatistics, School of Global Public Health, New York University, U.S.A.}
\end{center}


\begin{abstract}
In observational studies, unmeasured confounding may bias causal conclusions even after matching on measured confounders. Sensitivity analysis is therefore usually needed to assess the robustness of causal conclusions to unmeasured confounding. We consider multivariate sensitivity analysis that assesses the causal effect of a single treatment on multiple outcomes. We identify an inherent feature of such analyses, which we call ``confounding heterogeneity'': different outcomes may be affected by distinct unmeasured confounders, and a common unmeasured confounder may affect different outcomes to varying degrees. Building on matched observational studies, we formulate sensitivity analysis frameworks that accommodate confounding heterogeneity. We show that Type I error control for many existing sensitivity analysis methods is guaranteed under sensitivity analysis models and modes of inference that differ fundamentally from those used in the classic framework. We study asymptotic power using design sensitivity and Bahadur-Rosenbaum efficiency and examine finite-sample power through numerical studies. Our results show that the structure of unmeasured confounding affects sensitivity analysis, and that widely used single-parameter frameworks may mischaracterize the sensitivity of a causal conclusion to unmeasured confounding when confounding heterogeneity is considered. Simulations and a data application illustrate the practical importance of confounding heterogeneity in multivariate sensitivity analyses.
\end{abstract}


\spacingset{1.75} 
\vspace{-0.5cm}

\section{Introduction}

\vspace{-0.3cm}

In observational studies, unmeasured confounding remains a major concern even after careful adjustment for measured confounders. Sensitivity analyses are therefore routinely conducted to assess the robustness of causal conclusions to unmeasured confounding. In matched observational studies, arguably one of the most widely used observational study designs \citep{stuart2010matching, zubizarreta2023handbook}, the classic Rosenbaum $\Gamma$ sensitivity analysis framework specifies a single sensitivity parameter $\Gamma \geq 1$ that bounds how the odds of treatment assignment may differ between two units in a matched set due to unmeasured confounding \citep{rosenbaum2002observational,rosenbaum2020design}. The case $\Gamma=1$ corresponds to uniformly random treatment assignment within matched sets, whereas larger values of $\Gamma$ permit greater departures from random assignment. For each specified $\Gamma$, researchers calculate a worst-case $p$-value by maximizing the $p$-value over all within-set treatment assignment probabilities satisfying the sensitivity model  \citep{rosenbaum2002observational,rosenbaum2020design, fogarty2016sensitivity, kang2016full, ertefaie2018quantitative}. A causal conclusion is considered more robust to unmeasured confounding if the null hypothesis of no treatment effect can still be rejected at larger values of $\Gamma$, that is, if the worst-case $p$-value remains below the nominal significance level despite allowing for a greater magnitude of unmeasured confounding.

A natural complication arises because many matched observational studies assess the effects of a single treatment on multiple outcomes (\citealp{fogarty2016sensitivity, rosenbaum2016using, rosenbaum2020design, cohen2018multivariate, zhao2018cross, ye2022dimensions, bekerman2025planning, bekerman2026powerful}). A simple approach to sensitivity analysis in this setting is to apply a univariate procedure separately to each outcome and then use a Bonferroni correction to account for multiple comparisons. To improve power, several multivariate sensitivity analysis methods have been proposed, including the minimax method \citep{fogarty2016sensitivity}, the Scheff\'e projection method and its variants \citep{rosenbaum2016using, rosenbaum2020combining}, cross-screening \citep{zhao2018cross}, the chi-bar method \citep{cohen2018multivariate}, and sample splitting \citep{bekerman2025planning, bekerman2026powerful}.

These methods operate within the classic $\Gamma$ framework, which models how unmeasured confounding influences treatment assignment but does not explicitly model its associations with the outcomes. For a single outcome, the amplification framework of \citet{rosenbaum2009amplification} provides a two-parameter interpretation of the classic framework, with $\Lambda$ and $\Delta$ governing the associations of an unmeasured confounder with treatment assignment and the sign of the within-pair difference in potential outcomes under control, respectively. However, these frameworks leave a fundamental question unaddressed: \emph{how does the structure of unmeasured confounding across outcomes affect the robustness of a causal conclusion?} Different outcomes may be confounded by distinct unmeasured variables, and even a common unmeasured confounder may be associated with different outcomes to varying degrees. Neither the classic one-parameter $\Gamma$ framework nor the two-parameter amplification framework with a common outcome-related parameter explicitly parametrizes these differences. We formalize this overlooked dimension as \emph{confounding heterogeneity} and show that accounting for it can provide a more reliable and informative assessment of causal robustness. Our contributions are fourfold.

First, we distinguish two forms of confounding heterogeneity and develop corresponding sensitivity analysis frameworks for matched-pair studies. \emph{Type-I confounding heterogeneity} arises when different outcomes are subject to distinct unmeasured confounders, whereas \emph{Type-II confounding heterogeneity} arises when a common unmeasured confounder has different strengths of association with different outcomes. Both frameworks parameterize the associations of unmeasured confounding with treatment assignment and signs of within-pair differences in potential outcomes under control, allowing the outcome-related association strength to vary across outcomes. The Type-I framework additionally allows distinct unmeasured confounders to have different strengths of association with treatment assignment.

Second, we establish the statistical validity and computational feasibility of multivariate sensitivity analyses under the proposed frameworks. Specifically, we prove asymptotic Type~I error control for the minimax, Scheff\'e projection, and chi-bar methods adapted to these frameworks. We also develop an efficient computational strategy that combines asymptotic separability with local optimization to approximate the worst-case $p$-value and prove that the worst-case $p$-value can be approximated arbitrarily closely at large sample sizes. 

Third, we extend design sensitivity \citep{rosenbaum2004design} and Bahadur-Rosenbaum efficiency \citep{rosenbaum2015bahadur} to multi-parameter sensitivity analysis, providing theoretical tools for comparing analyses under different confounding structures. Design sensitivity characterizes the boundary separating sensitivity parameter configurations under which power converges to one from those under which it converges to zero. Within this consistency region, Bahadur-Rosenbaum efficiency provides a finer comparison through the exponential decay rates of worst-case $p$-values as the number of matched pairs increases.

Fourth, we demonstrate that \emph{assessments of causal robustness depend both on the magnitude of unmeasured confounding and its structure across outcomes}. Theoretical and numerical comparisons show that the classic $\Gamma$ framework may understate the sensitivity of a causal conclusion relative to frameworks that parameterize confounding heterogeneity. Moreover, even when the magnitude of confounding is held fixed under a common calibration, finite-sample power and Bahadur-Rosenbaum efficiency can vary across confounding structures. A single sensitivity parameter therefore cannot, in general, capture the consequences of confounding heterogeneity for multivariate sensitivity analysis. An empirical application illustrates how these distinctions affect assessments of robustness in practice.

Section~\ref{Sec: Notation} reviews the classic $\Gamma$ framework and existing multivariate sensitivity analysis methods. Section~\ref{Sec: CH Framework} extends the amplification framework \citep{rosenbaum2009amplification} to multiple outcomes and introduces the Type-I and Type-II frameworks. Section~\ref{Sec: Validity} establishes Type~I error control and computational feasibility under each framework, while Section~\ref{Sec: Power} develops the corresponding design sensitivities and Bahadur-Rosenbaum efficiencies. Section~\ref{Sec: Calculations} presents numerical comparisons of design sensitivity, Bahadur-Rosenbaum efficiency, and finite-sample power. Section~\ref{Sec: Application} compares the frameworks through a reanalysis of a matched observational study in \citet{rosenbaum2016using}. Section~\ref{Sec: Conclusion} concludes.

\vspace{-0.5cm}
\section{Notation and Background}
\label{Sec: Notation}
\vspace{-0.3cm}

\subsection{Matched Observational Studies with Multiple Outcomes}

We observe $\{(\mathbf{x}_{ij}, R_{ij1},\dots,R_{ijK}, Z_{ij})\}$ with measured confounders $\mathbf{x}_{ij}$, $K$ outcome variables ${R}_{ij1},\dots,{R}_{ijK}$, and a binary treatment $Z_{ij}$ for units $j=1,2$ in matched pairs $i=1,\dots I$. Let $Z_{ij}=1$ if unit $j$ in matched pair $i$ received treatment, otherwise let $Z_{ij}=0$. Matched pairs are formed by balancing measured confounders so that $\mathbf{x}_{i1}= \mathbf{x}_{i2}$ or $\mathbf{x}_{i1}\approx \mathbf{x}_{i2}$, but possibly failing to balance unmeasured confounders. Under the potential outcomes framework \citep{neyman1923application, rubin1974estimating} and the stable unit treatment value assumption (SUTVA), if unit $j$ in matched pair $i$ received treatment, we observe the potential outcomes $r_{Tij1},\dots,r_{TijK}$; otherwise, we observe the potential outcomes $r_{Cij1},\dots,r_{CijK}$. Therefore, the observed outcome for unit $j$ in matched pair $i$ is $R_{ijk}=Z_{ij}r_{Tijk}+(1-Z_{ij})r_{Cijk}$. Denote $Y_{ik}=R_{i1k}-R_{i2k}$ as the difference in the $k$th outcome for matched pair $i$, and denote $V_{i}=Z_{i1}-Z_{i2}$ as the difference in treatment indicators in matched pair $i$, so $V_{i} \in \{1, -1\}$. Denote $\mathbf{Z}=(Z_{11},Z_{12},\dots,Z_{I1}, Z_{I2})$ as the vector of treatment indicators, and denote $\mathbf{R}_{k} = (R_{11k},R_{12k},\dots,R_{I1k}, R_{I2k})$ as the vector of observed outcomes for outcome $k$.

Let $H_{0,k}$ denote Fisher's sharp null hypothesis of no treatment effect on the outcome $k$: $H_{0,k}: r_{Tijk} = r_{Cijk}$ for all $i,j$. In the classic multivariate sensitivity analysis setting, we test the overall null hypothesis of no treatment effect for all outcomes: 
\begin{equation}
\mathbf{H}_{0}: \bigwedge_{k=1}^{K} H_{0,k}.    
\end{equation}
Denote $Y_{Cik}=r_{Ci1k}-r_{Ci2k}$ as the difference in potential outcomes under control for outcome $k$ in matched pair $i$. Note that under $\mathbf{H}_{0}$, $Y_{ik}=Y_{Cik}$ for each outcome $k$ and matched pair $i$. Many commonly used test statistics in pair-matched studies take the form  $T_{k}(\mathbf{Z}, \mathbf{R}_{k})=\sum_{i=1}^{I}q_{ik}\cdot \operatorname{sign}(V_{i}Y_{ik}),$
where $q_{ik}$ is a non-negative function of $\{|Y_{ik}V_{i}|\}_{i=1,\dots,I}$ for each matched pair $i$ such that $Y_{ik}V_{i}=0$ implies $q_{ik}=0$, and $\operatorname{sign}(x)=1, 0$, or $-1$ as $x>0, x=0$, or $x<0$. For example, setting $q_{ik}=|Y_{ik}V_{i}|$ yields the permutation $t$-test. When $R_{ijk}\in \{0,1\}$, setting $q_{ik}=|Y_{ik}V_{i}|$ yields McNemar's test (\citealp{cox1989analysis}). If $\psi$ is an odd function (i.e., $\psi(-y)=-\psi(y)$), then $\sum_{i=1}^{I}\psi(V_{i}Y_{ik})=\sum_{i=1}^{I}q_{ik}\cdot \operatorname{sign}(V_{i}Y_{ik})$ with $q_{ik}=\psi(|Y_{ik}V_{i}|)$ and $\psi(y) \geq 0$ for all $y > 0$ yields a permutation $M$-test (\citealp{noether1973some}; \citealp{rosenbaum2007sensitivity}). Setting $q_{ik}=\mathbb{I}\{|Y_{ik}V_{i}| \neq 0\}$ for all matched pairs yields the sign test. Many test statistics are functions of the rank of $|Y_{ik}V_{i}|$ among $|Y_{1k}V_{1}|, \dots, |Y_{Ik}V_{I}|$, including Noether's quantile average test (\citealp{noether1973some}), Brown's combined quantile average test (\citealp{brown1981symmetric}), and the Wilcoxon signed-rank test. We provide the mild assumptions that $q_{ik}$ must satisfy in Appendix A.2.

\vspace{-0.3cm}

\subsection{The Classic Rosenbaum $\Gamma$ Sensitivity Analysis Framework}

Denote $\mathcal{F}_{k}=\{(r_{Tijk}, r_{Cijk}, \mathbf{x}_{ij}, u_{ij}):j=1, 2, i=1,\dots I\}$ as the set of potential outcomes for the $k$th outcome, measured confounders $\mathbf{x}_{ij}$ and the normalized unmeasured confounder $u_{ij}\in [0,1]$, and denote $\boldsymbol{\mathcal{F}} = \cup_{k=1}^{K} \mathcal{F}_{k}$ as the collection of all potential outcomes and confounders. Let $\pi_{ij}=P(Z_{ij}=1|\boldsymbol{\mathcal{F}})$ denote the probability that unit $j$ in matched pair $i$ receives treatment prior to matching. Denote $\mathcal{Z}$ as the set of all possible values of $\mathbf{Z}=(Z_{11},\dots, Z_{I2})$ assuming within-pair treatment assignment. That is, $\mathbf{Z}\in \mathcal{Z}$ if and only if $Z_{i1}+Z_{i2}=1$ for all $i$. The widely used Rosenbaum's $\Gamma$ sensitivity analysis framework, denoted as the ``classic framework'', bounds the ratio of odds of within-pair treatment assignment using $\Gamma\geq 1$:
\begin{equation}\label{bounds: odds ratio}
	\frac{1}{\Gamma}\leq \frac{\pi_{i1}(1-\pi_{i2})}{\pi_{i2}(1-\pi_{i1})}\leq \Gamma, \quad \text{for all $i$ and $\mathbf{u}\in \mathcal{U}:=[0, 1]^{2I}$.}
\end{equation}
The biased treatment assignment probability after matching under the constraint (\ref{bounds: odds ratio}) is
\begin{equation}\label{eqn: one parameter sensitivity model}
    P(\mathbf{Z}=\mathbf{z}|\boldsymbol{\mathcal{F}}, \mathcal{Z})=\prod_{i=1}^{I}\frac{\exp\{\gamma(z_{i1}u_{i1}+z_{i2}u_{i2})\}}{\exp(\gamma u_{i1})+\exp(\gamma u_{i2})},
\end{equation}
where $\gamma = \log(\Gamma)$. When $\Gamma=1$, (\ref{eqn: one parameter sensitivity model}) reduces to uniform treatment assignment probability $P(\mathbf{Z}=\mathbf{z}|\boldsymbol{\mathcal{F}}, \mathcal{Z})=2^{-I}$. As $\Gamma$ departs from $1$, (\ref{eqn: one parameter sensitivity model}) departs from a random treatment assignment mechanism. Under the null hypothesis $\mathbf{H}_0$ and (\ref{eqn: one parameter sensitivity model}), the one-sided $p$-value of the test statistic $T_{k}$ given the observed value $t$ in the presence of unmeasured confounding becomes
\begin{equation*}
   P(T_{k} \geq t |\boldsymbol{\mathcal{F}}, \mathcal{Z})=\sum_{\mathbf{z}\in \mathcal{Z}}\Big[\mathbb{I}\{ T_{k}(\mathbf{z}, \mathbf{R}_{k}) \geq t\}\cdot  \prod_{i=1}^{I}\frac{\exp\{\gamma(z_{i1}u_{i1}+z_{i2}u_{i2})\}}{\exp(\gamma u_{i1})+\exp(\gamma u_{i2})}\Big].
\end{equation*}
In a sensitivity analysis, researchers are interested in the worst-case $p$-value, which is the largest $p$-value over all possible arrangements of unmeasured confounders, reported by a test statistic $T$ given its observed value $t$ under constraint (\ref{eqn: one parameter sensitivity model}). The (one-sided) worst-case $p$-value reported by $T$ given the observed value $t$ under the sensitivity parameter $\Gamma$ is 
\begin{equation*}
   \max_{\mathbf{u}\in \mathcal{U}} P(T_{k} \geq t |\boldsymbol{\mathcal{F}}, \mathcal{Z})=\max_{\mathbf{u}\in \mathcal{U}}  \sum_{\mathbf{z}\in \mathcal{Z}}\Big[\mathbb{I}\{ T_{k}(\mathbf{z}, \mathbf{R}_{k}) \geq t\}\cdot  \prod_{i=1}^{I}\frac{\exp\{\gamma(z_{i1}u_{i1}+z_{i2}u_{i2})\}}{\exp(\gamma u_{i1})+\exp(\gamma u_{i2})}\Big].
\end{equation*}
Under constraint (\ref{bounds: odds ratio}), the worst-case $p$-value can be explicitly calculated:
\begin{equation*}
    \max_{\mathbf{u}\in \mathcal{U}} P(T_{k} \geq t |\boldsymbol{\mathcal{F}}, \mathcal{Z})= P(\overline{T}_{k, \Gamma} \geq t),
\end{equation*}
where $\overline{T}_{k, \Gamma}$ is the sum of $I$ independent random variables taking value $q_{ik}$ with probability $\Gamma/(1+\Gamma)$ and value $-q_{ik}$ with probability $1/(1+\Gamma)$ (\citealp{rosenbaum2002observational}).

\vspace{-0.3cm}

\subsection{Existing Methods for Multivariate Sensitivity Analysis}

When there are multiple outcomes ($K \geq 2$), calculating a worst-case $p$-value separately for each outcome and then applying the Bonferroni correction can be overly conservative \citep{fogarty2016sensitivity}. To improve power, several multivariate sensitivity analysis methods have been proposed. In particular, the minimax method \citep{fogarty2016sensitivity}, the Scheff\'e projection method \citep{rosenbaum2016using}, and the chi-bar method \citep{cohen2018multivariate} admit a common formulation based on standardized linear combinations of the outcome-specific test statistics $T_1,\ldots,T_K$. Specifically, let $\boldsymbol{\mu}_{\{\Gamma\}, \mathbf{u}}=(\mu_{\{\Gamma\}, 1,\mathbf{u}}, \dots, \mu_{\{\Gamma\}, K,\mathbf{u}})^{T}$ be the mean vector of $\mathbf{T}=(T_{1},\dots,T_{K})^{T}$ under $\mathbf{u}\in \mathcal{U}$ and $\Gamma\geq 1$. Let $\sigma_{\{\Gamma\}, k, k', \mathbf{u}}$ denote the covariance between $T_{k}$ and $T_{k'}$ given $\mathbf{u}\in \mathcal{U}$ and $\Gamma\geq 1$, and let $\Sigma_{\{\Gamma\}, \mathbf{u}}$ be the $K\times K$ covariance matrix with each component being $\sigma_{\{\Gamma\}, k, k', \mathbf{u}}$, where $k, k'=1,\dots,K$. Assume $\Sigma_{\{\Gamma\}, \mathbf{u}}$ is positive definite. Let $\boldsymbol{\kappa}=(\kappa_{1}, \dots, \kappa_{K})^{T}$ be a $K$-dimensional vector of constants. The minimax method \citep{fogarty2016sensitivity}, Scheff\'e projection method \citep{rosenbaum2016using}, and chi-bar method \citep{cohen2018multivariate} consider a test statistic of the form
\begin{equation*}\label{linear combination test stat}
    \min_{\mathbf{u}\in \mathcal{U}}\max_{\boldsymbol{\kappa} \in \mathcal{K}}\frac{\boldsymbol{\kappa}^{T}(\mathbf{T}-\boldsymbol{\mu}_{\{\Gamma\}, \mathbf{u}})}{\sqrt{\mathbf{\boldsymbol{\kappa}}^{T}\Sigma_{\{\Gamma\}, \mathbf{u}}\mathbf{\boldsymbol{\kappa}}}},
\end{equation*}
where $\mathcal{K} \subset \mathbb{R}^{K}$ characterizes the sensitivity analysis method used. For instance, the minimax method sets $\mathcal{K} = \{e_{1},\dots,e_{K}\}$, the standard basis vectors for $\mathbb{R}^{K}$. The Scheff\'e projection method sets $\mathcal{K} = \mathbb{R}^{K}\setminus \{\mathbf{0}\}$ and the chi-bar method sets $\mathcal{K} = \mathbb{R}_{+}^{K}\setminus \{\mathbf{0}\} := \{\boldsymbol{\kappa} \in \mathbb{R}^{K}:\kappa_{k} \geq 0, k=1,\dots,K, \sum_{k=1}^{K}\kappa_{k} > 0\}$. For the minimax, Scheff\'e projection, and chi-bar methods, we reject the global null hypothesis $\mathbf{H}_{0}$ if and only if
\begin{equation*}
    \min_{\mathbf{u}\in \mathcal{U}}\max_{\boldsymbol{\kappa} \in \mathcal{K}}\frac{\boldsymbol{\kappa}^{T}(\mathbf{T}-\boldsymbol{\mu}_{\{\Gamma\}, \mathbf{u}})}{\sqrt{\mathbf{\boldsymbol{\kappa}}^{T}\Sigma_{\{\Gamma\}, \mathbf{u}}\mathbf{\boldsymbol{\kappa}}}}\geq c_{\{\Gamma\}}(\alpha, \mathcal{K}),
\end{equation*}
where $c_{\{\Gamma\}}(\alpha, \mathcal{K})$ corresponds to the critical value that maintains asymptotic Type I error control at level $\alpha$ for the method characterized by $\mathcal{K}$. For the minimax method, $c_{\{\Gamma\}}(\alpha, \mathcal{K}) = \sqrt{\chi^{2}_{1, 1-2\alpha/K}}$, the square root of the $(1-2\alpha/K)$-quantile of a chi-square distribution with $1$ degree of freedom. For the Scheff\'e projection method, $c_{\{\Gamma\}}(\alpha, \mathcal{K}) = \sqrt{\chi^{2}_{K, 1-\alpha}}$, the square root of the $(1-\alpha)$-quantile of a chi-square distribution with $K$ degrees of freedom. Denote $\widetilde{\Sigma}_{\{\Gamma\}, \mathbf{u}, \rho}$ as the correlation matrix for $\mathbf{T}$ at a fixed $\mathbf{u} \in [0, 1]^{2I}$, and denote $\mathcal{S}_{\{\Gamma\}}$ as the set of all correlation matrices across unmeasured confounding values $\mathbf{u}$. For the chi-bar method, $c_{\{\Gamma\}}(\alpha, \mathcal{K}) = \sqrt{\sup_{\widetilde{\Sigma}_{\{\Gamma\}, \mathbf{u}, \rho} \in \mathcal{S}_{\{\Gamma\}}} \overline{\chi}^{2}_{1-\alpha}(\widetilde{\Sigma}_{\{\Gamma\}, \mathbf{u}, \rho}^{-1}, \mathbb{R}_{+}^{K}\setminus\{0\})}$, the square root of the largest $(1-\alpha)$-quantile of the chi-bar distribution over all feasible correlation matrices. Importantly, the critical value used in the chi-bar method depends on the sensitivity analysis framework and is calculated as the largest critical value over the unmeasured confounder. See \cite{cohen2018multivariate} for more details.

\vspace{-0.5cm}

\section{ Confounding-Heterogeneity-Aware Sensitivity Analysis}
\label{Sec: CH Framework}

\vspace{-0.3cm}

The methods reviewed in Section~\ref{Sec: Notation} combine evidence across outcomes within the classic $\Gamma$ framework. However, combining evidence across outcomes does not, by itself, address how unmeasured confounding relates to each outcome. The underlying sensitivity model constrains treatment assignment without explicitly parameterizing differences in the sources or strengths of unmeasured confounding across outcomes. To investigate the implications of this \emph{confounding heterogeneity}, we develop sensitivity analysis frameworks that explicitly model both treatment assignment and outcome-specific confounding relationships.

We begin by extending the amplification framework of \citet{rosenbaum2009amplification} to multiple outcomes. This extension provides a starting point for the Type-II framework, which allows a common unmeasured confounder to have different strengths of association with different outcomes, and the Type-I framework, which additionally allows distinct sources of unmeasured confounding. The multivariate amplification framework uses two sensitivity parameters, $\Lambda := \exp(\lambda)$ and $\Delta := \exp(\delta)$, to parameterize the associations of a common unmeasured confounder with treatment assignment and the signs of within-pair differences in potential outcomes under control, respectively, with a common $\Delta$ across outcomes. Denote $\mathcal{C} = \{(\mathbf{x}_{ij}, u_{ij})\}$ as the set of measured confounders and the unmeasured confounder for each unit $j$ and matched pair $i$. The amplification framework imposes the following treatment and outcome models:
\begin{align}
    &P(\mathbf{Z}=\mathbf{z}|\mathcal{C}, \mathcal{Z})=\prod_{i=1}^{I}\frac{\exp\{\lambda(z_{i1}u_{i1}+z_{i2}u_{i2})\}}{\exp(\lambda u_{i1})+\exp(\lambda u_{i2})}, \label{amplification model: treatment}  \\
    &P(Y_{Cik}\geq y|\mathcal{C}, \mathcal{Z})=\exp\{\delta (u_{i1}-u_{i2})\}P(Y_{Cik}\leq -y|\mathcal{C}, \mathcal{Z}) \hspace{3mm} \forall k=1,\dots,K, y>0. \label{amplification model: outcome}
\end{align}
The model for treatment assignment can be written in a symmetric form:
\begin{align*}
    &P(V_{i}=1 |\mathcal{C}, \mathcal{Z})=\exp\{\lambda (u_{i1}-u_{i2})\}P(V_{i}=-1|\mathcal{C}, \mathcal{Z}).
\end{align*}
Note that (\ref{amplification model: treatment}) is similar to (\ref{eqn: one parameter sensitivity model}), except $\mathcal{C}$ is conditioned on rather than $\boldsymbol{\mathcal{F}}$. $\lambda \geq 0$ affects the maximum amount by which unmeasured confounding causes departures from random treatment assignment, while $\delta \geq 0$ affects the maximum amount by which unmeasured confounding causes the distribution $Y_{Cik}$ to depart from a mean-zero symmetric distribution. \cite{rosenbaum2009amplification} note that, in the single outcome case, the worst-case $p$-value calculated under the classic framework with $\Gamma = \Gamma^{*}$ is equal to the worst-case $p$-value calculated under the amplification framework under the equality $\Gamma^{*} = (\Delta \Lambda + 1)/(\Delta + \Lambda )$. Moreover, if the above equality holds, then as $\Delta \to \infty$, $\Lambda \to \Gamma^{*}$. In the single outcome case, $\Gamma$ can be interpreted as bounding the odds of treatment assignment within pairs for an unmeasured confounder strongly correlated with $Y_{Ci}$. 

\begin{figure}[h]\label{fig: type I confounding heterogeneity}
    \centering
    \begin{subfigure}[b]{0.45\textwidth}
         \centering
         \begin{tikzpicture}
            \node[state] (1) {$Z$};
            \node[state] (2) [below =of 1, xshift=-1.7cm] {$U_{1}$};
            \node[state] (3) [below =of 1, xshift=1.7cm] {$U_{2}$};
            \node[state] (4) [below =of 2] {$R_{1}$};
            \node[state] (5) [below =of 3] {$R_{2}$};
        
            \path (2) edge node[left] {$\lambda_{1}$} (1);
            \path (3) edge node[right] {$\lambda_{2}$} (1);
            \path (1) edge node[el,above] {} (4);
            \path (1) edge node[el,above] {} (5);
            \path (2) edge node[left]{$\delta_{1}$} (4);
            \path (3) edge node[right]{$\delta_{2}$} (5);
        \end{tikzpicture}
        \caption{Type-I confounding heterogeneity}
    \end{subfigure}
        \begin{subfigure}[b]{0.45\textwidth}
         \centering
         \begin{tikzpicture}
            \node[state] (1) {$Z$};
            \node[state] (2) [below =of 1] {$U$};
            \node[state] (3) [below =of 2, xshift=-2cm] {$R_{1}$};
            \node[state] (4) [below =of 2, xshift=2cm] {$R_{2}$};
            
            \path (1) edge node[el,above] {} (3);
            \path (1) edge node[el,above] {} (4);
            \path (2) edge node[left]{$\lambda$} (1);
            \path (2) edge node[right]{$\delta_{1}$} (3);
            \path (2) edge node[left]{$\delta_{2}$} (4);
        \end{tikzpicture}
        \caption{Type-II confounding heterogeneity}
    \end{subfigure}
    \caption{Confounding heterogeneity with $K=2$ outcome variables}
    \label{fig: confounding heterogeneity}
\end{figure}
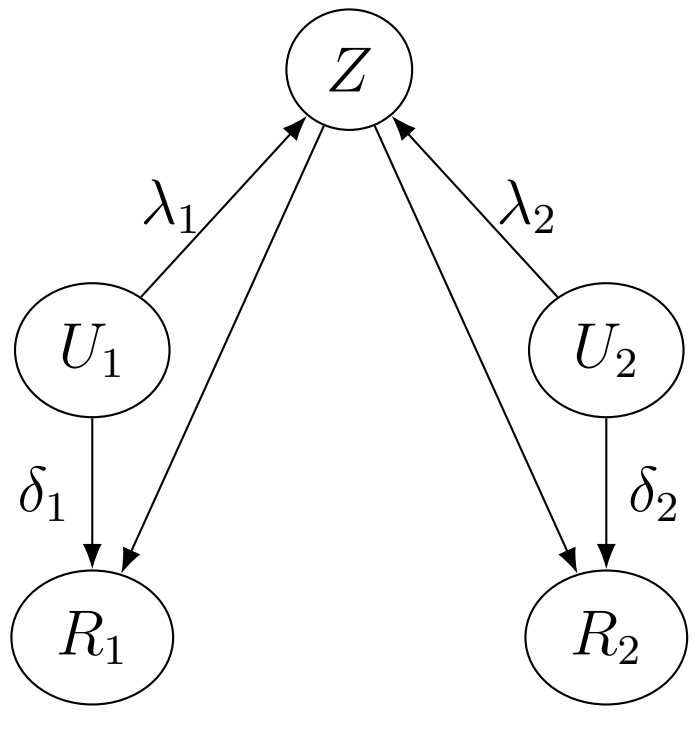
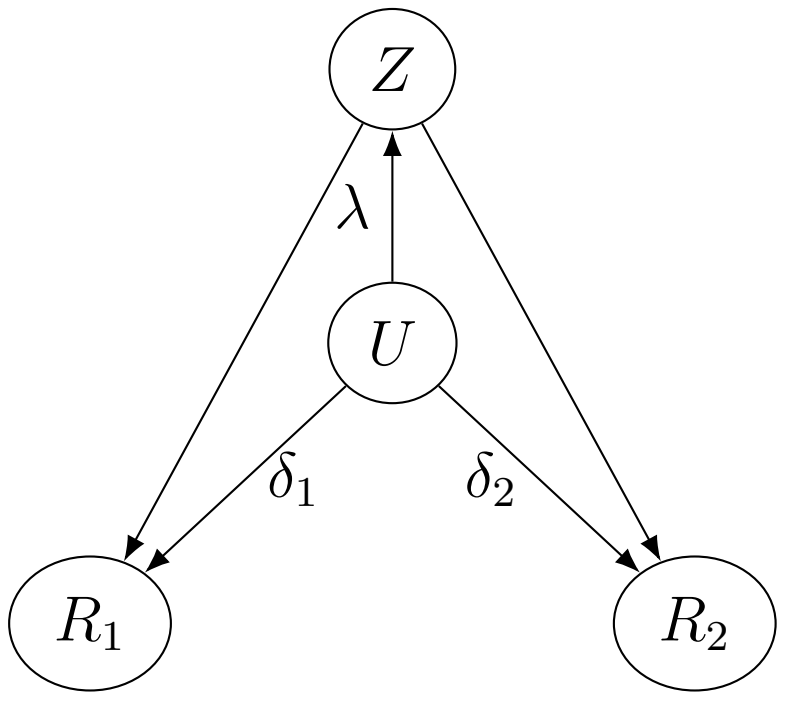

Now we develop the sensitivity analysis framework for Type-II confounding heterogeneity, in which we assume a single unmeasured confounder may have different strengths of association with different outcomes. Under Type-II confounding heterogeneity, one sensitivity parameter $\Lambda$ with $\Lambda = \exp\{\lambda\}$ quantifies the relationship between the unmeasured confounder and the treatment, while $K$ additional sensitivity parameters $\Delta_{1},\dots\Delta_{K}$ with $\Delta_{k}:=\exp\{\delta_{k}\}$ quantify varying strengths of relationship between the unmeasured confounder and outcomes, resulting in $K+1$ sensitivity parameters $\{\Lambda, \boldsymbol{\Delta}\}$, where $\boldsymbol{\Delta} = (\Delta_{1},\dots,\Delta_{K})$. Under the proposed Type-II confounding heterogeneity framework, the practitioner considers the following sensitivity model:
\begin{align}
& P(Z_{i1}=1,Z_{i2}=0|\mathcal{C}, \mathcal{Z}) = \exp\{\lambda(u_{i1}-u_{i2})\}P(Z_{i1}=0,Z_{i2}=1|\mathcal{C}, \mathcal{Z}), \label{eqn: Type-II CH treatment model} \\
& P(Y_{Cik} \geq y|\mathcal{C}, \mathcal{Z}) = \exp\{\delta_{k}(u_{i1}-u_{i2})\}P(Y_{Cik} \leq -y|\mathcal{C}, \mathcal{Z}) \hspace{5mm} \forall k=1,\dots,K, y>0.
\label{eqn: Type-II CH outcome model}
\end{align}
The Type-II confounding heterogeneity framework generalizes the amplification framework to the multivariate sensitivity analysis case. The worst-case $p$-value calculated from the amplification framework setting $\Lambda = \Lambda^{*}$ and $\Delta  = \Delta^{*}$ is equivalent to the worst-case $p$-value calculated under Type-II confounding heterogeneity, setting $\Lambda = \Lambda^{*}$ and $\Delta_{1},\dots \Delta_{K}=\Delta^{*}$. The right plot of Figure~\ref{fig: confounding heterogeneity} shows the directed acyclic graph corresponding to the Type-II confounding heterogeneity framework.

Next, we develop the sensitivity analysis framework for Type-I confounding heterogeneity, in which there are $K$ unmeasured confounding variables $\mathbf{u}_{1},\dots \mathbf{u}_{K}$ such that $\mathbf{u}_{k} \in [0, 1]^{2I}$. The $k$th unmeasured confounder $\mathbf{u}_{k}$ enters the $k$th outcome's sensitivity model only. Denote $\mathcal{C}_{k}=\{(\mathbf{x}_{ij}, u_{ijk}), i=1,\dots,I, j=1,2\}$ as the collection of measured confounders and the $k$th unmeasured confounder, and denote $\boldsymbol{\mathcal{C}} = \cup_{k=1}^{K}\mathcal{C}_{k}$ as the collection of all confounders. We assume a linear model for the log-odds of treatment assignment:
\begin{equation*}
\log\left\{\frac{P(Z_{ij}=1|\boldsymbol{\mathcal{C}})}{P(Z_{ij}=0|\boldsymbol{\mathcal{C}})}\right\} = g(\mathbf{x}_{ij}) + \sum_{k=1}^{K}\lambda_{k}u_{ijk},
\end{equation*}
where $g(\cdot)$ is some unknown function of $\mathbf{x}_{ij}$, and $\lambda_{1},\dots,\lambda_{K}\geq0$ quantify the strength of the relationship between the unmeasured confounders and treatment assignment. This extends the treatment model from \cite{Rosenbaum1987} to the case where more than one unmeasured confounder may affect treatment assignment. Under the Type-I confounding heterogeneity framework, a practitioner considers the following sensitivity model:
\begin{align}
& P(V_{i}=1|\boldsymbol{\mathcal{C}}, \mathcal{Z}) = \exp\left\{\sum_{k=1}^{K}\lambda_{k}(u_{i1k} - u_{i2k})\right\}P(V_{i}=-1|\boldsymbol{\mathcal{C}}, \mathcal{Z}), \label{eqn: Type-I CH treatment model} \\
& P(Y_{Cik} \geq y|\boldsymbol{\mathcal{C}}, \mathcal{Z}) = \exp\{\delta_{k}(u_{i1k}-u_{i2k})\}P(Y_{Cik} \leq -y|\boldsymbol{\mathcal{C}}, \mathcal{Z}) \hspace{5mm} \forall k=1,\dots,K, y>0. \label{eqn: Type-I CH outcome model}
\end{align}
Thus, there are $2K$ sensitivity parameters in the Type-I confounding heterogeneity framework $\{\boldsymbol{\Lambda, \boldsymbol{\Delta}}\}$, where $\boldsymbol{\Lambda} = (\Lambda_{1},\dots,\Lambda_{K})$. The Type-I confounding heterogeneity framework generalizes the Type-II confounding heterogeneity framework. Denote $\Lambda_{k} = \exp(\lambda_{k})$. The $p$-value calculated in the Type-II confounding heterogeneity framework setting $\Lambda = \Lambda^{*}$, $\boldsymbol{\Delta} = \boldsymbol{\Delta}^{*}$, and $\mathbf{u} = \mathbf{u}^{*}$ is equivalent to the $p$-value calculated under Type-I confounding heterogeneity when setting $\prod_{k=1}^{K}\Lambda_{k} = \Lambda^{*}$, setting $\Delta_{k} = \Delta_{k}^{*}$ for all $k=1,\dots K$, and setting $\mathbf{u}_{1}, \dots \mathbf{u}_{K} = \mathbf{u}^{*}$. The left plot in Figure~\ref{fig: confounding heterogeneity} shows the directed acyclic graph corresponding to the Type-I confounding heterogeneity framework when $K=2$.

Let $\boldsymbol{\mathcal{P}}:=\{\{\Gamma\}, \{\Lambda, \Delta\}, \{\Lambda_{1},\dots,\Lambda_{K}, \Delta_{1},\dots,\Delta_{K}\}, \{\Lambda, \Delta_{1},\dots,\Delta_{K}\}\}$ denote the collection of sensitivity parameterizations used for the classic, amplification, Type-I, and Type-II confounding heterogeneity frameworks. For any sensitivity analysis framework parameterized by $\mathcal{P} \in \boldsymbol{\mathcal{P}}$, denote $\mathcal{U}_{\mathcal{P}}$ as the support of the unmeasured confounder(s) so that $\mathcal{U}_{\{\Gamma\}} = \mathcal{U}_{\{\Lambda, \Delta\}} = \mathcal{U}_{\{\Lambda, \Delta_{1},\dots,\Delta_{K}\}} = [0, 1]^{2I}$ and $\mathcal{U}_{\{\Lambda_{1},\dots, \Lambda_{K}, \Delta_{1},\dots,\Delta_{K}\}} = [0, 1]^{2IK}$.
Denote the standardized deviate $d_{\mathcal{P},\kappa}(\mathbf{u}):=\kappa^{T}\left(\mathbf{T}-\boldsymbol{\mu}_{\mathcal{P},\mathbf{u}}\right)/\sqrt{\kappa^{T}\boldsymbol{\Sigma}_{\mathcal{P},\mathbf{u}}\kappa}$ for $\mathbf{u}\in\mathcal{U}_{\mathcal{P}}$. Denote $c_{\mathcal{P}}(\alpha, \mathcal{K})$ as the $100(1-\alpha)\%$ critical value used to compare the test statistic to when using the sensitivity analysis method characterized by $\mathcal{K}$, where the critical value may depend on the sensitivity analysis framework parameterized by $\mathcal{P}$. For any choice of sensitivity analysis framework parameterized by $\mathcal{P} \in \boldsymbol{\mathcal{P}}$ and sensitivity analysis method characterized by $\mathcal{K} \subset \mathbb{R}^{K}\setminus\{\mathbf{0}\}$, we reject the global null hypothesis $\mathbf{H}_{0}$ if and only if
\begin{equation}
\min _{\mathbf{u} \in \mathcal{U}_{\mathcal{P}}} \max _{\kappa \in \mathcal{K}} d_{\mathcal{P},\kappa}(\mathbf{u}) \geq c_{\mathcal{P}}(\alpha, \mathcal{K}).   
\end{equation}
Sections~\ref{Sec: Validity} and~\ref{Sec: Power} investigate the differences in statistical properties of multivariate sensitivity analysis methods across different sensitivity analysis frameworks.

\vspace{-0.5cm}
\section{Statistical Validity and Computational Feasibility of Multivariate Sensitivity Analysis Under Confounding Heterogeneity}
\label{Sec: Validity}
\vspace{-0.3cm}

\subsection{Statistical Validity}
Denote $\mathcal{A}_{k}:=(|Y_{1k}|,\dots,|Y_{Ik}|)$ as the vector of magnitudes of differences in the $k$th outcome across all matched pairs and $\boldsymbol{\mathcal{A}}=\{\mathcal{A}_{k}\}_{k=1}^{K}$ as the collection of $\mathcal{A}_{k}$ for all $K$ outcomes. Define the mapping $f:\boldsymbol{\mathcal{P}}\to\{\boldsymbol{\mathcal{F}}, \{\mathcal{C}, \boldsymbol{\mathcal{A}}\}, \{\boldsymbol{\mathcal{C}}, \boldsymbol{\mathcal{A}}\}\}$ such that $f(\{\Gamma\}) = \boldsymbol{\mathcal{F}}$, $f(\{\Lambda, \Delta\}) = f(\{\Lambda, \Delta_{1},\dots, \Delta_{K}\}) = \{\mathcal{C}, \boldsymbol{\mathcal{A}}\}$, and $f(\{\Lambda_{1},\dots, \Lambda_{K}, \Delta_{1},\dots,\Delta_{K}\}) =  \{\boldsymbol{\mathcal{C}}, \boldsymbol{\mathcal{A}}\}$. For a sensitivity analysis framework parameterized by $\mathcal{P} \in \boldsymbol{\mathcal{P}}$, the set $f(\mathcal{P})$ corresponds to the set of variables considered fixed when performing inference within that sensitivity analysis framework -- note that the amplification and Type-II confounding heterogeneity frameworks consider the same set of variables fixed. For all $\mathcal{P} \in \boldsymbol{\mathcal{P}}$, denote $\mu_{\mathcal{P}, k, \mathbf{u}}:=\mathbb{E}(T_{k}|f(\mathcal{P}), \mathcal{Z})$ as the expected value of the $k$th test statistic under the sensitivity analysis framework parameterized by $\mathcal{P}$. If $k \neq k'$, then the covariance between test statistics $T_{k}$ and $T_{k'}$ is denoted as $\sigma_{\mathcal{P},k,k',\mathbf{u}}:=\mathrm{Cov}(T_{k}, T_{k'}|f(\mathcal{P}),  \mathcal{Z})$, and if $k = k'$, the variance of the test statistic $T_{k}$ is denoted as $\sigma_{\mathcal{P},k,\mathbf{u}}^{2}:=\mathrm{Var}(T_{k}|f(\mathcal{P}),  \mathcal{Z})$. Write $\boldsymbol{\mu}_{\mathcal{P}, \mathbf{u}}=\left(\mu_{\mathcal{P}, 1, \mathbf{u}}, \ldots, \mu_{\mathcal{P}, K, \mathbf{u}}\right)^{T}$ and $\boldsymbol{\Sigma}_{\mathcal{P}, \mathbf{u}}$ denoting the covariance matrix of $\mathbf{T}$, containing $\sigma_{\mathcal{P},k,k',\mathbf{u}}$ and $\sigma_{\mathcal{P},k,\mathbf{u}}^{2}$ for all $k, k'=1,\dots,K$, and assume it to be positive definite. The closed-form expressions of the expected value and variance of the test statistic $T_{k}$ for each $\mathcal{P}$ are in Appendix B.1. Theorem~\ref{Theorem:Asymptotically Normal} states that for a fixed nonzero $\kappa \in \mathbb{R}^{K}$ the standardized linear combination of test statistics $(\boldsymbol{\kappa}^{T}(\mathbf{T}-\boldsymbol{\mu}_{\mathcal{P}, \mathbf{u}})) / \sqrt{\mathbf{\boldsymbol{\kappa}}^{T}\boldsymbol{\Sigma}_{\mathcal{P}, \mathbf{u}}\mathbf{\boldsymbol{\kappa}}}$ is asymptotically normally distributed.

\begin{Theorem}[Asymptotic Normality]
\label{Theorem:Asymptotically Normal}
Under the mild regularity conditions in Appendices A.1 and C.1, for any nonzero $\kappa \in \mathbb{R}^{K}$ and any sensitivity analysis parameterization $\mathcal{P} \in \boldsymbol{\mathcal{P}}$, under $\mathbf{H}_{0}$, as $I \to \infty$, we have:
\begin{equation*}
    \frac{\kappa^{T}\left(\mathbf{T}-\boldsymbol{\mu}_{\mathcal{P}, \mathbf{u}}\right)}{\sqrt{\kappa^{T} \boldsymbol{\Sigma}_{\mathcal{P}, \mathbf{u}}\kappa}} \overset{d}{\to} N(0, 1).
\end{equation*}
\end{Theorem}


The proofs of all theoretical results are in the supplementary materials. Let $\boldsymbol{\mathcal{K}} = \{\{e_{1},\dots,e_{K}\}, \mathbb{R}^{K}\setminus\{0\}, \mathbb{R}_{+}^{K}\setminus \{\mathbf{0}\}\}$ denote the set of linear combinations of test statistics considered by the minimax, Scheff\'e projection, and chi-bar methods, respectively. As a result of Theorem~\ref{Theorem:Asymptotically Normal}, the minimax, Scheff\'e projection, and chi-bar sensitivity analysis methods retain Type I error control across each sensitivity analysis framework.

\begin{Theorem}[Type I Error Control]
\label{Theorem:Type-I Error Controlled}
Assume the treatment $Z$ and response $\mathbf{R}$ follow the treatment and outcome models corresponding to the sensitivity analysis framework parameterized by $\mathcal{P}$. Under $\mathbf{H}_{0}$ and the regularity conditions in Appendices A.1 and C.1, for any $\mathcal{P} \in \boldsymbol{\mathcal{P}}$ and for any $\mathcal{K} \in \boldsymbol{\mathcal{K}} = \{\{e_{1},\dots,e_{K}\}, \mathbb{R}^{K}\setminus\{0\}, \mathbb{R}_{+}^{K}\setminus \{\mathbf{0}\}\}$, we have
$$
\limsup_{I\to\infty} P\left\{\min _{\mathbf{u} \in \mathcal{U}_{\mathcal{P}}} \max _{\kappa \in \mathcal{K}} d_{\mathcal{P},\kappa}(\mathbf{u}) \geq c_{\mathcal{P}}(\alpha, \mathcal{K}) |
f(\mathcal{P}), \mathcal{Z}\right\} \leq \alpha.
$$
\end{Theorem}

Type I error control is guaranteed under different modes of inference across frameworks, characterized by $f(\mathcal{P})$. Specifically, sensitivity analysis under the classic framework is guaranteed to maintain Type I error control under randomization-based inference, where all potential outcomes are considered fixed. In contrast, sensitivity analysis under the amplification and confounding heterogeneity frameworks retains Type I error control while the sign of $Y_{Cik}$ is treated as random. Table~\ref{tab: Sensitivity model summary} summarizes each sensitivity framework.

\begin{table}[ht!]
    \centering
    \small
    \resizebox{\linewidth}{!}{%
    \begin{tabular}{lllll}
    \toprule
    \multirow{2}{*}{Sensitivity model} & Treatment-related & Outcome-related & Unmeasured & Variables conditioned \\
    & parameters & parameters & confounding model(s) & on for inference \\
    \midrule 
    Classic & $\Gamma$ & No specified model & Equation~(\ref{eqn: one parameter sensitivity model}) & $\boldsymbol{\mathcal{F}}$, $\mathcal{Z}$ \\
    Amplification & $\Lambda$ & $\Delta$ & Equations~(\ref{amplification model: treatment}) and~(\ref{amplification model: outcome}) & $\boldsymbol{\mathcal{A}}$, $\mathcal{C}$, $\mathcal{Z}$ \\
    Type-II Confounding & \multirow{2}{*}{$\Lambda$} & \multirow{2}{*}{$\boldsymbol{\Delta} = (\Delta_{1},\dots,\Delta_{K})$} & \multirow{2}{*}{Equations~(\ref{eqn: Type-II CH treatment model}) and~(\ref{eqn: Type-II CH outcome model})} & \multirow{2}{*}{$\boldsymbol{\mathcal{A}}$, $\mathcal{C}$, $\mathcal{Z}$} \\
    Heterogeneity & & & & \\
    Type-I Confounding & \multirow{2}{*}{$\boldsymbol{\Lambda} = (\Lambda_{1},\dots,\Lambda_{K})$} & \multirow{2}{*}{$\boldsymbol{\Delta} = (\Delta_{1},\dots,\Delta_{K})$} & \multirow{2}{*}{Equations~(\ref{eqn: Type-I CH treatment model}) and~(\ref{eqn: Type-I CH outcome model})} & \multirow{2}{*}{$\boldsymbol{\mathcal{A}}$, $\boldsymbol{\mathcal{C}}$, $\mathcal{Z}$} \\
    Heterogeneity & & & & \\
    \bottomrule
    \end{tabular}}
    \caption{Summary of each sensitivity analysis framework.}
    \label{tab: Sensitivity model summary}
\end{table}

\vspace{-0.3cm}
\subsection{Computational Feasibility}

Under the amplification and confounding heterogeneity frameworks, the two-sided alternative version of the hypothesis test associated with sensitivity analyses admits a signomial programming formulation, for which direct global optimization can be computationally demanding \citep{boyd2007tutorial}. To facilitate computation, we adopt a similar strategy to that of the separable approximation algorithm in \citet{gastwirth2000asymptotic} to find an asymptotically equivalent and computationally efficient solution to the original optimization problem for any fixed nonzero $\kappa \in \mathbb{R}^{K}$. 

We approximate the worst-case standardized deviate by first minimizing the numerator of the standardized deviate \textit{separately} from the denominator, then, in the presence of ties, maximizing the denominator of the standardized deviate with respect to the unmeasured confounder $\mathbf{u}$. For the classic, amplification, and Type-II confounding heterogeneity frameworks, the separable approximation proceeds as follows: First, calculate the set of global minimizers of the numerator of the standardized deviate: $\mathcal{U}^{*} = \mathrm{argmin}_{\mathbf{u} \in [0, 1]^{2I}}\kappa^{T}\left(\mathbf{T}-\boldsymbol{\mu}_{\mathcal{P},\mathbf{u}}\right)$. Then, define $\widetilde{\mathbf{u}}^{*} = \mathrm{argmax}_{\mathbf{u} \in \mathcal{U}^{*}}\sqrt{\kappa^{T}\boldsymbol{\Sigma}_{\mathcal{P},\mathbf{u}}\kappa}$. Denote $\mathbf{u}^{*} = \mathrm{argmin}_{\mathbf{u} \in [0, 1]^{2I}}d_{\mathcal{P}, \kappa}(\mathbf{u})$. For a constant $x > 0$, denote $x_{I} = x \sqrt{\kappa^{T} \boldsymbol{\Sigma}_{\mathcal{P}, \mathbf{u}^{*}}\kappa} + \kappa^{T}\boldsymbol{\mu}_{\mathcal{P}, \mathbf{u}^{*}}$

\begin{Theorem}
\label{Theorem: Sufficiency of Local Optimization}
Fix any $\kappa \in \mathbb{R}^{K}_{+}$, a constant $x > 0$, and sensitivity parameter values under the classic, amplification, or Type-II confounding heterogeneity framework. Under the conditions in Appendix C.3 and under $\mathbf{H}_{0}$, for any $\epsilon > 0$ there exists an $I^{*}$ such that, for all $I \geq I^{*}$ we have $P(\kappa^{T}\mathbf{T} > x_{I} | f(\mathcal{P}), \mathcal{Z}, \mathbf{u} = \widetilde{\mathbf{u}}^{*}) \geq P(\kappa^{T}\mathbf{T} > x_{I} | f(\mathcal{P}), \mathcal{Z}, \mathbf{u} = \mathbf{u}^{*}) - \epsilon$.
\end{Theorem}

Theorem~\ref{Theorem: Sufficiency of Local Optimization} states that for a fixed cutoff $x$, the $p$-value from the separable approximation can be made arbitrarily close to the worst-case $p$-value with a sufficiently large sample size. Thus, the standardized deviate calculated by the separable approximation serves as a computationally feasible and close approximation to the worst-case standardized deviate. The signomial programming formulations, the proof of Theorem~\ref{Theorem: Sufficiency of Local Optimization}, and the comparison of the global minimum numerator values under the Type-I and Type-II confounding heterogeneity frameworks are in Appendix C.3.

\vspace{-0.5cm}

\section{Consistency and Efficiency of Multivariate Sensitivity Analysis Under Confounding Heterogeneity}
\label{Sec: Power}

\vspace{-0.3cm}

\subsection{Power of Sensitivity Analysis Across Frameworks}

Having established asymptotic Type~I error control, we next study the \emph{power of a sensitivity analysis}: the probability of rejecting the global null hypothesis $\mathbf{H}_0$ under a prespecified data-generating distribution with a nonzero treatment effect \citep{rosenbaum2020design}. Equivalently, power is the probability that the worst-case $p$-value falls below the significance level $\alpha$. It measures the ability to detect a treatment effect after accounting for the confounding permitted by the sensitivity model. For a fixed data-generating distribution and test statistic, comparing power across sensitivity analysis frameworks reveals how the assumed confounding structure affects the robustness of causal evidence.

Throughout this section, we focus on nonnegative linear combinations of outcome-specific test statistics, the class relevant to the minimax and chi-bar methods. For a fixed $\kappa \in \mathbb{R}_{+}^{K}\setminus\{0\}$, we study sensitivity analysis using $T_{\kappa} = I^{-1}\kappa^{T}\mathbf{T}
=
I^{-1}\sum_{i=1}^{I}\sum_{k=1}^{K}
\kappa_k q_{ik}\operatorname{sign}(Y_{ik}V_i)$. Denote $Y_{Tik} = r_{Ti1k} - r_{Ti2k}$ as the difference in potential outcomes under treatment for outcome $k$ in matched pair $i$. For the power calculations, suppose that
$(r_{Ti11},r_{Ci11},\dots,r_{Ti2K},r_{Ci2K}, V_i)\overset{\mathrm{iid}}{\sim}F$,
where $\mathbf{H}_0$ is false, and under $F$, we have $P\{\operatorname{sign}(Y_{Cik})=1\mid\mathcal{C},\mathcal{Z}\}
=
P\{\operatorname{sign}(Y_{Cik})=-1\mid\mathcal{C},\mathcal{Z}\}$ for $k=1,\ldots,K$, and $P(V_i=1\mid\mathcal{C},\mathcal{Z})
=
P(V_i=-1\mid\mathcal{C},\mathcal{Z})
=
1/2$. We hold $F$ fixed while varying the sensitivity analysis framework and its parameter values. To analyze power, we use a framework-specific bounding random variable $\overline{T}_{\mathcal{P},\kappa}$ whose upper-tail probability characterizes the worst-case $p$-value. For example, let $\overline{T}_{\{\Gamma\}, \kappa}$ be the average of $I$ independent random variables taking value $|\sum_{k=1}^{K}\kappa_{k}q_{ik}\operatorname{sign}(Y_{ik})|:=\eta_{\kappa i}$ with probability $\overline{p}_{\{\Gamma\}}:=\Gamma / (1+\Gamma)$ and value $-\eta_{\kappa i}$ with probability $ 1- \overline{p}_{\{\Gamma\}}$. Then, under the classic framework, for any fixed $x > 0$ we have
\begin{equation}
    \lim_{I\to\infty}\bigg|\max_{u}P(T_{\kappa} \geq x |\boldsymbol{\mathcal{F}}, \mathcal{Z}) - P(\overline{T}_{\{\Gamma\}, \kappa} \geq x|\boldsymbol{\mathcal{F}}, \mathcal{Z})\bigg| = 0,
\end{equation}
Denote $\phi_{\kappa i k}:=\kappa_{k}q_{ik}$. For sensitivity analysis parameterizations $\mathcal{P} \in \boldsymbol{\mathcal{P}}\setminus\{\{\Gamma\}\}$, define $\overline{T}_{\mathcal{P}, \kappa}=I^{-1}\sum_{i=1}^{I}\sum_{k=1}^K\phi_{\kappa ik}b_{ik}v_{i}$, where $b_{ik}, v_{i} \in \{1, -1\}$ for all $i=1,\dots,I$ and $k=1,\dots,K$, $P(b_{ik}v_{i} = 1) = \overline{p}_{\mathcal{P}, k}, P(b_{ik}v_{i} = -1) = 1 - \overline{p}_{\mathcal{P}, k}$, $\{b_{ik}:i=1,\dots,I, k=1,\dots,K\} \cup \{v_{i}:i=1,\dots,I\}$ form a mutually independent set, and $\overline{p}_{\mathcal{P}, k}$ is defined as:

\noindent (Amplification framework):
    \begin{align*}
    P(b_{ik} = 1) & = \frac{\Delta}{\Delta + 1}, ~P(v_{i} = 1) = \frac{\Lambda}{\Lambda + 1}\implies  
        \overline{p}_{\{\Lambda, \Delta\}, k}  = \frac{\Lambda \Delta + 1}{(\Lambda + 1)(\Delta + 1)}.
    \end{align*}
(Type-II confounding heterogeneity): 
    \begin{align*}
        P(b_{ik} = 1) & = \frac{\Delta_{k}}{\Delta_{k} + 1}, ~P(v_{i} = 1) = \frac{\Lambda}{\Lambda + 1}\implies  
        \overline{p}_{\{\Lambda, \boldsymbol{\Delta}\}, k}  = \frac{\Lambda \Delta_{k} + 1}{(\Lambda + 1)(\Delta_{k} + 1)}.
    \end{align*}
(Type-I confounding heterogeneity):
    \begin{align*}
        P(b_{ik} = 1) & = \frac{\Delta_{k}}{\Delta_{k} + 1}, ~P(v_{i} = 1) = \frac{(\prod_{\ell=1}^{K}\Lambda_{\ell})}{(\prod_{\ell=1}^{K}\Lambda_{\ell}) + 1}\implies  
        \overline{p}_{\{\boldsymbol{\Lambda}, \boldsymbol{\Delta}\}, k}  = \frac{(\prod_{\ell=1}^{K}\Lambda_{\ell}) \Delta_{k} + 1}{((\prod_{\ell=1}^{K}\Lambda_{\ell})  + 1)(\Delta_{k} + 1)}.
    \end{align*}
For the amplification framework, notice that $\overline{p}_{\{\Lambda, \Delta\}, 1}=\dots=\overline{p}_{\{\Lambda, \Delta\}, K}$, so denote $\overline{p}_{\{\Lambda, \Delta\}}$ as the probability $b_{ik}v_{i}$ equals $1$ under the amplification framework. Denote $\overline{\mu}_{I, \mathcal{P}, \kappa}$ and $\overline{\sigma}^{2}_{I, \mathcal{P}, \kappa}$ as the expected value and variance of $\overline{T}_{\mathcal{P}, \kappa}$ in a finite sample, then denote $\overline{\mu}_{\mathcal{P}, \kappa} = \lim_{I\to\infty}\overline{\mu}_{I, \mathcal{P}, \kappa}$ and $\overline{\sigma}^{2}_{\mathcal{P}, \kappa} =  \lim_{I\to\infty}I\cdot \overline{\sigma}^{2}_{I, \mathcal{P}, \kappa}$, where under the mild regularity conditions in Appendix A.2 both limits are well-defined. Lemma~\ref{Lem: Bounding Variable} establishes that $\overline{T}_{\mathcal{P}, \kappa}$ is asymptotically normal and is a valid bounding variable for $T_{\kappa}$.
\begin{Lemma}[Bounding Variable]
\label{Lem: Bounding Variable}
    Under the regularity conditions in Appendix A.2, for every sensitivity analysis parameterization $\mathcal{P} \in \boldsymbol{\mathcal{P}}$ we have $(\overline{T}_{\mathcal{P}, \kappa} -\overline{\mu}_{I, \mathcal{P}, \kappa})/\overline{\sigma}_{I, \mathcal{P}, \kappa}\overset{d}{\to}N(0, 1)$. Moreover, for any fixed $x > 0$ we have $\lim_{I\to\infty}\big|\max_{u}P(T_{\kappa} \geq x |f(\mathcal{P}), \mathcal{Z})  - P(\overline{T}_{\mathcal{P}, \kappa} \geq x|f(\mathcal{P}), \mathcal{Z})\big|=0$.
\end{Lemma}

The proof for Lemma~\ref{Lem: Bounding Variable} is in Appendix C.4. Denote $\mu_{F, \kappa}=\lim_{I\to\infty}\mathbb{E}_{F}(T_{\kappa})$ and $\sigma^{2}_{F, \kappa}=\lim_{I\to\infty}I\cdot \mathrm{Var}_{F}(T_{\kappa})$ as the limiting expected value and variance of $T_{\kappa}$ with respect to $F$. Denote $\Phi$ as the standard normal CDF and $\Phi^{-1}$ as the standard normal quantile function. Under each sensitivity analysis framework considered, $T_{\kappa}$ is asymptotically normally distributed as $I\to\infty$, so for a fixed test size $\alpha$, a large-sample approximation to the power of a sensitivity analysis conducted with $T_{\kappa}$ is calculated as: $P\left\{\frac{(T_{\kappa} - \overline{\mu}_{I, \mathcal{P}, \kappa})}{\overline{\sigma}_{I, \mathcal{P}, \kappa}} \geq \Phi^{-1}(1-\alpha)\right\} \approx \Phi\left(\frac{\sqrt{I}\cdot (\mu_{F, \kappa} - \overline{\mu}_{\mathcal{P}, \kappa}) - \Phi^{-1}(1-\alpha)\overline{\sigma}_{\mathcal{P}, \kappa}}{\sigma_{F, \kappa}}\right)$. 



\subsection{Generalization of Design Sensitivity to Multivariate Sensitivity Analysis}

Power describes the performance of a sensitivity analysis at a given sample size. To assess asymptotic robustness without specifying a particular number of matched pairs, we first ask whether the analysis is \emph{consistent}: under a fixed alternative distribution $F$ and fixed sensitivity parameter values, does its power converge to one as $I\to\infty$? In the classic $\Gamma$ framework, this property is characterized by \emph{design sensitivity}, a threshold $\widetilde{\Gamma}$ such that power converges to one for $\Gamma<\widetilde{\Gamma}$ and to zero for $\Gamma>\widetilde{\Gamma}$ \citep{rosenbaum2004design, rosenbaum2020design}. A larger design sensitivity indicates that the analysis remains consistent while allowing for stronger unmeasured confounding. Design sensitivity is a measure of asymptotic robustness dependent on the alternative distribution, the test statistic, and the sensitivity analysis framework. In the multi-parameter frameworks considered here, the scalar threshold becomes a \emph{design sensitivity boundary} in the space of sensitivity parameters. This boundary separates combinations of treatment- and outcome-related confounding strengths under which the analysis remains consistent from those under which its power converges to zero. Denote $d_{\mathcal{P}}$ as the number of parameters used in a sensitivity analysis parameterized by $\mathcal{P}$, so $d_{\{\Gamma\}}=1, d_{\{\Lambda, \Delta\}} = 2, d_{\{\Lambda, \boldsymbol{\Delta}\}} = K+1$, and $d_{\{\boldsymbol{\Lambda}, \boldsymbol{\Delta}\}} = 2K$. 

\begin{Definition}
Under mild regularity conditions (e.g., those specified in Appendix A.2), for any sensitivity analysis framework parameterized by $\mathcal{P} \in \boldsymbol{\mathcal{P}}$, there exists a set $\widetilde{D}_{\mathcal{P}, \kappa} \subset [1, \infty)^{d_{\mathcal{P}}}$ called the \textit{design sensitivity boundary} such that, when a sensitivity analysis is conducted with sensitivity parameter values below the design sensitivity boundary, the power of a sensitivity analysis goes to $1$ as $I \to \infty$, and if a sensitivity analysis is conducted with sensitivity parameter values above the design sensitivity boundary, the power of a sensitivity analysis goes to $0$ as $I \to \infty$. Specifically, denote $\overline{\mu}_{\mathcal{P}, \kappa}(\theta_{\mathcal{P}})$ as the limiting expected value of $\overline{T}_{\mathcal{P}, \kappa}$ evaluated at sensitivity parameter values $\theta_{\mathcal{P}} \in [1, \infty)^{d_{\mathcal{P}}}$. Then, $\widetilde{D}_{\mathcal{P}, \kappa} = \{\theta_{\mathcal{P}} \in [1, \infty)^{d_{\mathcal{P}}} : \overline{\mu}_{\mathcal{P}, \kappa}(\theta_{\mathcal{P}}) = \mu_{F, \kappa}\}$.

\end{Definition}

The power approximation in the preceding subsection makes this boundary intuitive. Because
$\bar{\sigma}_{\mathcal{P},\kappa}$ and $\sigma_{F,\kappa}$
are of constant order, a positive limiting difference between the alternative mean and the worst-case null mean yields power tending to one, whereas a negative limiting difference yields power tending to zero. The design sensitivity boundary $\widetilde{D}_{\mathcal{P}, \kappa}$ is therefore characterized by the equality $\bar{\mu}_{\mathcal{P},\kappa} = \mu_{F,\kappa}$. Our goal is to characterize this boundary under each framework and determine how confounding heterogeneity changes the combinations of sensitivity parameter values for which the analysis remains consistent.

To study design sensitivity, we restrict $\Gamma \geq 1$, and for the amplification and confounding heterogeneity frameworks, we set each $\Delta_{k} \geq 1$ and $\Lambda_{k} \geq 1$. For a fixed vector $\kappa \in \mathbb{R}_{+}^{K}\setminus \{\mathbf{0}\}$, denote $\eta_{\kappa} = \mathbb{E}(|\sum_{k=1}^{K}\kappa_{k}q_{ik}\operatorname{sign}(Y_{ik})|)$, $\phi_{\kappa k} = \mathbb{E}(\phi_{\kappa ik})$, $\phi_{\kappa i} = \sum_{k=1}^{K}\phi_{\kappa ik}$, and $\phi_{\kappa} = \mathbb{E}(\sum_{k=1}^{K}\phi_{\kappa ik})$, where each expectation is taken with respect to $F$. Theorem~\ref{Theorem:Design Sensitivity} describes the design sensitivity boundary of $T_{\kappa}$ under each sensitivity analysis framework.

\begin{Theorem}[Design Sensitivity Boundary]
\label{Theorem:Design Sensitivity}
Under the conditions in Appendix A.2,

\noindent (Classic framework): The design sensitivity $\widetilde{\Gamma}$ satisfies the equality
\begin{equation*}
\frac{\widetilde{\Gamma} - 1}{1 + \widetilde{\Gamma}}\eta_{\kappa} = \mu_{F, \kappa}.
\end{equation*}
(Amplification framework): $\widetilde{D}_{\{\Lambda, \Delta\}}$ is the set of all sensitivity parameter values $(\widetilde{\Lambda}, \widetilde{\Delta})$ satisfying the equality
\begin{equation*}
\frac{(\widetilde{\Lambda} - 1)}{(\widetilde{\Lambda} + 1)} \frac{(\widetilde{\Delta} - 1)}{(\widetilde{\Delta} + 1)}\phi_{\kappa} = \mu_{F, \kappa}.
\end{equation*}
(Type-II confounding heterogeneity): $\widetilde{D}_{\{\Lambda, \boldsymbol{\Delta}\}}$ is the set of all sensitivity parameter values $(\widetilde{\Lambda}, \widetilde{\Delta}_{1},\dots,\widetilde{\Delta}_{K})$ satisfying the equality 
\begin{equation*}
\frac{\widetilde{\Lambda} - 1}{\widetilde{\Lambda} + 1}\left(\sum_{k=1}^{K}\frac{(\widetilde{\Delta}_{k} - 1)}{(\widetilde{\Delta}_{k} + 1)}\phi_{\kappa k}\right) = \mu_{F, \kappa}.
\end{equation*}
(Type-I confounding heterogeneity): $\widetilde{D}_{\{\boldsymbol{\Lambda}, \boldsymbol{\Delta}\}}$ is the set of sensitivity parameter values 

\noindent $(\widetilde{\Lambda}_{1},\dots,\widetilde{\Lambda}_{K}, \widetilde{\Delta}_{1},\dots,\widetilde{\Delta}_{K})$ satisfying the equality
\begin{equation*}
\frac{(\prod_{\ell=1}^{K}\widetilde{\Lambda}_{\ell}) - 1}{(\prod_{\ell=1}^{K}\widetilde{\Lambda}_{\ell}) + 1}\left(\sum_{k=1}^{K}\frac{\widetilde{\Delta}_{k}-1}{\widetilde{\Delta}_{k}+1}\phi_{\kappa k}\right) = \mu_{F, \kappa}.
\end{equation*}
\end{Theorem}

Theorem~\ref{Theorem:Design Sensitivity} shows that even if sensitivity parameters $\widetilde{\Gamma}, \widetilde{\Lambda}, \widetilde{\Delta}$ satisfy the equality $(\widetilde{\Gamma} - 1)/(1 + \widetilde{\Gamma}) = ((\widetilde{\Lambda} - 1)(\widetilde{\Delta} - 1))/((1 + \widetilde{\Lambda})(1 + \widetilde{\Delta}))$, a sensitivity analysis conducted under the amplification framework will in general report a causal conclusion as less robust to unmeasured confounding compared to the classic framework, in large enough samples. This is because $\eta_{\kappa} \leq \phi_{\kappa}$, where equality is reached only if $\operatorname{sign}(Y_{ik}) = \operatorname{sign}(Y_{ik'})$ for all outcomes $k, k'$ in which $\kappa_{k}q_{ik}\operatorname{sign}(Y_{ik}) \neq 0$ and $\kappa_{k'}q_{ik'}\operatorname{sign}(Y_{ik'}) \neq 0$. If potential outcomes are viewed as fixed, as is under the classic framework, any amount of unmeasured confounding cannot unify the \textit{sign} of observed treatment effects across outcomes. In contrast, under the amplification framework, unmeasured confounding may unify the sign of observed scores across outcomes, thus requiring less overall magnitude of unmeasured confounding to overturn causal conclusions. Under both confounding heterogeneity frameworks, the design sensitivity boundary depends on outcome-specific expected score magnitudes ($\phi_{\kappa 1},\dots,\phi_{\kappa K}$), while the design sensitivity boundary for the classic and amplification frameworks only depends on aggregate measures of expected score ($\eta_{\kappa}$ or $\phi_{\kappa}$), highlighting the inability of the classic and amplification frameworks to account for heterogeneity across outcomes.

To facilitate interpretation and comparisons across frameworks, we summarize the design sensitivity boundaries through \emph{infimum treatment and outcome sensitivities}. These summaries characterize the smallest treatment- or outcome-related sensitivity magnitudes, under the criteria defined below, that reach the boundary at which the analysis loses consistency. The \textit{infimum outcome sensitivity} is denoted $\widetilde{\Delta}^{*}$ (amplification framework) and $(\widetilde{\Delta}^{*}_{1},\dots,\widetilde{\Delta}^{*}_{K})$ (confounding heterogeneity frameworks), while the \textit{infimum treatment sensitivity} is denoted $\widetilde{\Lambda}^{*}$ (amplification and Type-II confounding heterogeneity frameworks) and $(\widetilde{\Lambda}^{*}_{1},\dots,\widetilde{\Lambda}^{*}_{K})$ (Type-I confounding heterogeneity framework). The infimum treatment and outcome sensitivities are defined in the following way:
\begin{align*}
& \widetilde{\Delta}^{*} := \inf\{\Delta : (\Lambda, \Delta) \in \widetilde{D}_{\{\Lambda, \Delta\}}\}, \hspace{5mm} \widetilde{\Lambda}^{*}:= \inf\{\Lambda : (\Lambda, \Delta) \in \widetilde{D}_{\{\Lambda, \Delta\}}\}, \\
& (\widetilde{\Delta}^{*}_{1},\dots,\widetilde{\Delta}^{*}_{K}) := \mathrm{argmin}_{\Delta_{k} \geq 1}\sum_{k=1}^{K}\Delta_{k} \hspace{2mm} \text{subject to} \hspace{2mm} \sum_{k=1}^{K}\phi_{\kappa k}\frac{\Delta_{k} - 1}{\Delta_{k} + 1} \geq \mu_{F, \kappa}, \\
& (\widetilde{\Lambda}^{*}_{1},\dots,\widetilde{\Lambda}^{*}_{K}) := \mathrm{argmin}_{\Lambda_{k} \geq 1}\sum_{k=1}^{K}\Lambda_{k} \hspace{2mm} \text{subject to} \hspace{2mm}  \frac{\prod_{k=1}^{K}\Lambda_{k} - 1}{\prod_{k=1}^{K}\Lambda_{k} + 1}\phi_{\kappa} \geq   \mu_{F, \kappa}.
\end{align*}
Lemma~\ref{Lem: Infimum Outcome and Treatment Sensitivity} describes how the infimum outcome and treatment sensitivities are calculated. 

\begin{Lemma}[Infimum Outcome and Treatment Sensitivities]
\label{Lem: Infimum Outcome and Treatment Sensitivity}
Under the regularity conditions in Appendix A.2, the infimum outcome and treatment sensitivities are calculated as: (i) Infimum Single-Parameter Treatment and Outcome Sensitivities: $\widetilde{\Lambda}^{*} = \widetilde{\Delta}^{*} = (\phi_{\kappa} + \mu_{F, \kappa})/(\phi_{\kappa} - \mu_{F, \kappa})$; (ii) Infimum Treatment Sensitivity, Type-I Confounding Heterogeneity: For each $k=1,\dots K$, $\widetilde{\Lambda}^{*}_{k} = \left(\frac{\phi_{\kappa}+\mu_{F, \kappa}}{\phi_{\kappa}-\mu_{F, \kappa}}\right)^{1/K}$. (iii) Infimum Outcome Sensitivity, Type-I and Type-II Confounding Heterogeneity: For each $k=1,\dots K$, $\widetilde{\Delta}^{*}_{k} = \max\{1, \sqrt{2\nu\phi_{\kappa k}} - 1\}$, where $\nu>0$ satisfies: $\sum_{k: \nu \phi_{\kappa k}>2}\phi_{\kappa k}\left(1 -\sqrt{\frac{2}{\nu \phi_{\kappa k}}}\right) = \mu_{F, \kappa}$.
\end{Lemma}

Under both confounding heterogeneity frameworks, Lemma~2 shows that
$\widetilde{\Delta}_k^*>1$ precisely when
$\sqrt{2\nu\phi_{\kappa k}}>2$.
Thus, the allocation of outcome-related confounding that minimizes the sum of the sensitivity parameters depends on
$\phi_{\kappa k}=\kappa_k E_F(q_{ik})$,
which reflects the contribution of outcome $k$ to the joint statistic $T_\kappa$.
These outcome-specific components describe how confounding can be distributed across outcomes to reach the consistency boundary of the multivariate analysis. They therefore provide information that a single overall sensitivity threshold does not convey. Nevertheless, consistency alone does not distinguish sensitivity analyses whose powers both converge to one. A finer comparison requires examining their efficiency within the consistency region.

\vspace{-0.3cm}

\subsection{Bahadur-Rosenbaum Efficiency in Multivariate Sensitivity Analysis}

Design sensitivity identifies the sensitivity parameter values under which an analysis remains consistent, but it does not distinguish analyses that are both consistent. For example, consider two sensitivity analyses under the classic $\Gamma$ framework with design sensitivities $\widetilde{\Gamma}_1$ and $\widetilde{\Gamma}_2$. When
$\Gamma<\min\{\widetilde{\Gamma}_1,\widetilde{\Gamma}_2\}$,
both powers converge to one, even though their performance at practically relevant sample sizes may differ substantially. The same issue arises under multi-parameter frameworks when the sensitivity parameter configurations lie within the respective consistency regions of the analyses being compared.

Define $\widetilde{D}_{\mathcal{P}, \kappa}^{*}:=\{\theta_{\mathcal{P}} \in [1, \infty)^{d_{\mathcal{P}}}: \overline{\mu}_{\mathcal{P}, \kappa}(\theta_{\mathcal{P}}) < \mu_{F, \kappa}\}$ as the \textit{consistency region}, which describes the set of sensitivity parameter values for which a sensitivity analysis remains consistent. We adopt and generalize the approach and notation of \cite{rosenbaum2015bahadur} to derive the Bahadur-Rosenbaum slope for sensitivity analysis using the test statistic $T_{\kappa}$ under the classic, amplification, Type-I confounding heterogeneity, and Type-II confounding heterogeneity frameworks. Let $m_{I, \mathcal{P}, \kappa}(t):=\mathbb{E}(\exp\{I\cdot \overline{T}_{\mathcal{P}, \kappa}t\}|f(\mathcal{P}), \mathcal{Z})$ be the moment generating function of the bounding variable $I\cdot \overline{T}_{\mathcal{P}, \kappa}$ under sensitivity parameterization $\mathcal{P}$. Denote $\Psi_{I, \mathcal{P}, \kappa}^{(0)}(t) = \log m_{I, \mathcal{P}, \kappa}(t)$ as the cumulant generating function of $I\cdot \overline{T}_{\mathcal{P}, \kappa}$, $\Psi_{I, \mathcal{P}, \kappa}^{(d)}(t)$ as the $d$th derivative of the cumulant generating function, and $c^{(d)}_{\mathcal{P}, \kappa}(t)=\lim_{I\to\infty}I^{-1}\Psi_{I, \mathcal{P}, \kappa}^{(d)}(t)$. Theorem~\ref{Theorem:Bahadur-Rosenbaum slope} characterizes the exponential decay of the worst-case $p$-value and hence the Bahadur-Rosenbaum efficiency of the sensitivity analysis.

\begin{Theorem}[Bahadur-Rosenbaum slope] \label{Theorem:Bahadur-Rosenbaum slope}     Let $\{a_{1}, a_{2},\dots\}$ be a sequence of constants converging to $\mu_{F, \kappa}:=\mathbb{E}(\sum_{k=1}^{K}\kappa_{k}q_{ik}\operatorname{sign}(Y_{ik}V_{i}))$. Under the regularity conditions in Appendix A.2, for any sensitivity analysis done with parameterization $\mathcal{P} \in \boldsymbol{\mathcal{P}}$ and sensitivity parameter values in the consistency region $\widetilde{D}_{\mathcal{P}, \kappa}^{*}$, we have:     \begin{equation*}         -I^{-1}\log(P(\overline{T}_{\mathcal{P}, \kappa} \geq a_{I} |f(\mathcal{P}), \mathcal{Z})) \overset{P}{\to} \widetilde{\xi}_{\mathcal{P}}\mu_{F, \kappa} - c^{(0)}_{\mathcal{P}, \kappa}(\widetilde{\xi}_{\mathcal{P}}),     \end{equation*}     where $\widetilde{\xi}_{\mathcal{P}} = \xi$ is the unique solution to $\mu_{F, \kappa} = c^{(1)}_{\mathcal{P}, \kappa}(\xi)$. The Bahadur-Rosenbaum slope is $2\{ \widetilde{\xi}_{\mathcal{P}}\mu_{F,\kappa} -c_{\mathcal{P},\kappa}^{(0)}(\widetilde{\xi}_{\mathcal{P}}) \}$. \end{Theorem}

A larger positive Bahadur-Rosenbaum slope corresponds to faster exponential decay of the worst-case $p$-value as the number of matched pairs increases. This comparison remains informative even when the limiting power is one for every analysis being compared. Thus, design sensitivity characterizes the boundary of consistency, whereas Bahadur-Rosenbaum efficiency distinguishes performance within the consistency region.

To compare these slopes across sensitivity analysis frameworks, we place their parameter values on a common calibration. Under the amplification and confounding heterogeneity frameworks, efficiency can depend on how confounding is allocated between treatment assignment and the outcomes, as well as across outcomes. We therefore compare Bahadur-Rosenbaum efficiencies under a common overall magnitude of confounding and a specified relative scaling of treatment- and outcome-related sensitivity parameters. For $a>0$, we restrict the sensitivity parameter values to $B_{\Gamma,a}$, defined by $\mathcal{B}_{\Gamma, a} = \left\{\{\Lambda, \Delta, \boldsymbol{\Lambda}, \boldsymbol{\Delta}\}: \Delta = a\Lambda, \overline{p}_{\{\Gamma\}} = \overline{p}_{\{\Lambda, \Delta\}} = K^{-1}\sum_{k=1}^{K}\overline{p}_{\{\Lambda, \boldsymbol{\Delta}\}, k}   = K^{-1}\sum_{k=1}^{K}\overline{p}_{\{\boldsymbol{\Lambda}, \boldsymbol{\Delta}\}, k}\right\}$. At $\Gamma = 1$, we define $\mathcal{B}_{\Gamma, a} = \left\{\{\Lambda, \Delta, \boldsymbol{\Lambda}, \boldsymbol{\Delta}\}: \Delta = a\Lambda, \Delta_{1},\dots,\Delta_{K} = \Delta, \prod_{k=1}^{K}\Lambda_{k}=\Lambda\right\}$, representing the limit of the set $\mathcal{B}_{\Gamma, a}$ as $\Gamma \to 1^{+}$. $\mathcal{B}_{\Gamma, a}$ generalizes the relationship between the sensitivity parameters of the classic and amplification frameworks to include both confounding heterogeneity frameworks \citep{rosenbaum2009amplification}. If $\{\Lambda, \Delta, \boldsymbol{\Lambda}, \boldsymbol{\Delta}\} \in \mathcal{B}_{\Gamma, a}$, the average of the probability parameters for each bounding variable $\overline{T}_{\mathcal{P}, \kappa}$ will be equal to each other, making the magnitude of unmeasured confounding comparable across frameworks. Denote $b_{\{\Gamma\}, \kappa}(\Gamma), b_{\{\Lambda, \Delta\}, \kappa}(\Lambda, \Delta), b_{\{\Lambda, \boldsymbol{\Delta}\}, \kappa}(\Lambda, \boldsymbol{\Delta}),$ and $b_{\{\boldsymbol{\Lambda}, \boldsymbol{\Delta}\}, \kappa}(\boldsymbol{\Lambda}, \boldsymbol{\Delta})$ as the Bahadur-Rosenbaum slope under the classic, amplification, Type-I and Type-II confounding heterogeneity frameworks. Denote $\overline{p}_{\Delta, k} = \Delta_{k} / (1 + \Delta_{k})$, where the restriction $\Delta_{k} \geq 1$ implies $\overline{p}_{\Delta, k} \in [1/2, 1]$, and $\overline{p}_{\Delta, k}=1$ is understood as a limiting value of $\overline{p}_{\Delta, k}$ as $\Delta_{k} \to \infty$. Last, denote the \textit{infimum Bahadur-Rosenbaum slope} as $b^{-}_{\kappa}(\Gamma, a)$ and the \textit{supremum Bahadur-Rosenbaum slope} as $b^{+}_{\kappa}(\Gamma, a)$, where
\begin{align*}
b^{-}_{\kappa}(\Gamma, a) & = \inf_{\overline{p}_{\Delta, 1},\dots,\overline{p}_{\Delta, K} \in [1/2, 1]}\{b_{\{\Lambda, \boldsymbol{\Delta}\}, \kappa}(\Lambda, \boldsymbol{\Delta}) : \{\Lambda, \Delta, \boldsymbol{\Lambda}, \boldsymbol{\Delta}\} \in \mathcal{B}_{\Gamma, a}\}, \\
b^{+}_{\kappa}(\Gamma, a) & = \sup_{\overline{p}_{\Delta, 1},\dots,\overline{p}_{\Delta, K} \in [1/2, 1]}\{b_{\{\Lambda, \boldsymbol{\Delta}\}, \kappa}(\Lambda, \boldsymbol{\Delta}) : \{\Lambda, \Delta, \boldsymbol{\Lambda}, \boldsymbol{\Delta}\} \in \mathcal{B}_{\Gamma, a}\}.
\end{align*}
At a fixed overall magnitude of confounding described by $\Gamma$ and relative scaling of treatment- and outcome-related sensitivity parameters $a$, the infimum Bahadur-Rosenbaum slope describes the efficiency of sensitivity analysis under the worst-case configuration of confounding heterogeneity, while the supremum Bahadur-Rosenbaum slope describes the efficiency under the most favorable scenario of confounding heterogeneity. Theorem~\ref{Theorem:Bahadur-Rosenbaum slope relationship} states a necessary initial condition under which the best-case Bahadur-Rosenbaum efficiency under confounding heterogeneity dominates the efficiency under the classic framework.



\begin{Theorem}
\label{Theorem:Bahadur-Rosenbaum slope relationship}
Assume the regularity conditions in Appendix A.2 hold, and suppose the sensitivity parameter values $\{\Lambda, \Delta, \boldsymbol{\Lambda}, \boldsymbol{\Delta}\} \in \mathcal{B}_{\Gamma, a}$. Then, with $a = 1$,
\begin{equation*}
\frac{\partial}{\partial \Gamma} b^{+}_{\kappa}(\Gamma, a)\bigg|_{\Gamma \to 1^{+}} \geq \frac{\partial}{\partial \Gamma}b_{\{\Gamma\}, \kappa}(\Gamma)\bigg|_{\Gamma \to 1^{+}}
\end{equation*}
only if the following inequality holds:
\begin{equation*}
\mathbb{E}_{F}\big(\tanh(\eta_{\kappa i}\widetilde{\xi}_{\{\Gamma\}})\big) \geq K\min_{1\leq k\leq K} \mathbb{E}_{F}\big(\tanh(\phi_{\kappa ik}\widetilde{\xi}_{\{\Lambda, \Delta\}})\big).
\end{equation*}
\end{Theorem}
The hyperbolic tangent function in Theorem~\ref{Theorem:Bahadur-Rosenbaum slope relationship} arises from the partial derivative of the Bahadur-Rosenbaum slope with respect to $\Gamma$. While the necessary condition in Theorem~\ref{Theorem:Bahadur-Rosenbaum slope relationship} specifically describes the case when $\Gamma$ approaches $1$ and $a=1$, it still serves as a useful description of under what conditions the efficiency of a sensitivity analysis under confounding heterogeneity may decay slower than the efficiency of a sensitivity analysis under the classic framework with increasing sensitivity parameter values. The condition is guaranteed to hold if $\kappa \in \{e_{1},\dots e_{K}\}$ and, in general, is easier to satisfy if $\mathbb{E}(\phi_{\kappa i k})$ is close to zero for at least one outcome $k$. The necessary condition is generally more difficult to satisfy if many outcomes are being tested, or if $\mathbb{E}_{F}(\phi_{\kappa i})$ is large under the data-generating distribution $F$.

\vspace{-0.5cm}

\section{Simulation Studies}
\label{Sec: Calculations}

\vspace{-0.3cm}


\subsection{Simulation Setting}
\label{Sec: simulation setting}
We consider an example with two outcomes $(Y_{i1}V_{i}, Y_{i2}V_{i}) \iid N(\boldsymbol{\tau}, \Sigma)$, with diagonal entries $\Sigma_{11} = \Sigma_{22} = 1$ and off-diagonal entries $\Sigma_{12} = \Sigma_{21} = \rho$. For the design sensitivity calculations, we consider two cases: (i) Case 1: $(\tau_{1}, \tau_{2})=(0.25, 0.25)$ and (ii) Case 2: $(\tau_{1}, \tau_{2})=(0.1, 0.5)$, while we focus on Case 2 for the Bahadur-Rosenbaum efficiency and finite-sample power calculations. We set $q_{ik}=|Y_{ik}V_{i}|$, corresponding to the permutation t-test. The design sensitivity and Bahadur-Rosenbaum efficiency are obtained via the newly derived formulas in Section~\ref{Sec: Power}, and finite-sample power is simulated using $1000$ datasets with $I=300$ matched pairs. For both finite-sample power and Bahadur-Rosenbaum efficiency, we restrict $\{\Lambda, \Delta, \boldsymbol{\Lambda}, \boldsymbol{\Delta}\} \in \mathcal{B}_{\Gamma, a}$ for $a = 1/3, 1$, and $3$, representing settings where the outcome-related sensitivity parameters are weighted less than, equal to, and greater than the treatment-related sensitivity parameters. 

\vspace{-0.3cm}

\subsection{Design Sensitivity}

For both cases described in Section~\ref{Sec: simulation setting}, we calculate the infimum design sensitivities $\widetilde{\Lambda}^{*}, \widetilde{\Delta}^{*}, (\widetilde{\Lambda}^{*}_{1},\dots \widetilde{\Lambda}^{*}_{K})$ and $(\widetilde{\Delta}^{*}_{1},\dots \widetilde{\Delta}^{*}_{K})$, as well as $\widetilde{\Gamma}_{\rho}$, which denotes the design sensitivity $\widetilde{\Gamma}$ for the classic framework when the correlation between $Y_{Ci1}V_{i}$ and $Y_{Ci2}V_{i}$ equals $\rho$ for $\rho = -0.5, 0, 0.5$. Note that the correlation between $Y_{Ci1}$ and $Y_{Ci2}$ does not affect the infimum design sensitivities for the amplification and confounding heterogeneity frameworks. Design sensitivities for this example are shown in Table~\ref{tab: design sensitivity table}. Additional simulations and calculations for rank-based test statistics and heavy-tailed outcomes are in Appendix D.1.

\begin{table}[ht]
\footnotesize
\centering
\resizebox{\linewidth}{!}{%
\begin{tabular}[t]{c|ccccccc|ccccccc}
\toprule
& \multicolumn{7}{c|}{Case 1: $(\tau_1, \tau_2) = (0.25, 0.25)$} & \multicolumn{7}{c}{Case 2: $(\tau_1, \tau_2) = (0.1, 0.5)$} \\
\midrule
$(\kappa_{1}, \kappa_{2})$ & $\widetilde{\Gamma}_{-0.5}$ & $\widetilde{\Gamma}_{0}$ & $\widetilde{\Gamma}_{0.5}$ & $\widetilde{\Lambda}^{*}$, $\widetilde{\Delta}^{*}$ & $\widetilde{\Lambda}^{*}_{1}$, $\widetilde{\Lambda}^{*}_{2}$ & $\widetilde{\Delta}^{*}_{1}$ & $\widetilde{\Delta}^{*}_{2}$ & $\widetilde{\Gamma}_{-0.5}$ & $\widetilde{\Gamma}_{0}$ & $\widetilde{\Gamma}_{0.5}$ & $\widetilde{\Lambda}^{*}$, $\widetilde{\Delta}^{*}$ & $\widetilde{\Lambda}^{*}_{1}$, $\widetilde{\Lambda}^{*}_{2}$ & $\widetilde{\Delta}^{*}_{1}$ & $\widetilde{\Delta}^{*}_{2}$ \\
\hline 
(1, 0) & 1.87 & 1.87 & 1.87 & 1.87 & 1.37 & 1.87 & 1.00 & 1.28 &  1.28 & 1.28 & 1.28 & 1.13 & 1.28 & 1.00\\
(0.75, 0.25) & 2.59 & 2.21 & 2.01 & 1.87 & 1.37 & 2.36 & 1.00 & 2.14 & 1.89 & 1.75 & 1.64  & 1.28 & 2.00 & 1.00 \\
(0.5, 0.5) & 3.53 & 2.43 & 2.06 & 1.87 & 1.37 & 1.87 & 1.87 & 4.56 & 2.91 & 2.39 & 2.09  & 1.45 & 2.01 & 2.18 \\
(0.25, 0.75) & 2.59 & 2.21 & 2.01 & 1.87 & 1.37 & 1.00 & 2.36 & 4.61 & 3.58 & 3.06 & 2.69 & 1.64 & 1.40 & 3.40 \\
(0, 1) & 1.87 & 1.87 & 1.87 & 1.87 & 1.37 & 1.00 & 1.87 & 3.53 & 3.53 & 3.53 & 3.53 & 1.88 & 1.00 & 3.53 \\
\bottomrule
\end{tabular}}
\caption{Infimum design sensitivity values across sensitivity analysis frameworks when $Y_{i1}V_{i} \sim N(\tau_1, 1), Y_{i2}V_i \sim N(\tau_2, 1)$ and $q_{ik}$ is chosen to be $|Y_{ik}V_{i}|$, corresponding to the permutation t-test. We consider the two cases: Case 1: $(\tau_1, \tau_2) = (0.25, 0.25)$, and Case 2: $(\tau_1, \tau_2) = (0.1, 0.5)$.}
\label{tab: design sensitivity table}
\end{table}

\begin{figure}[ht] 
    \centering
    \makebox[\textwidth][c]{%
    \begin{tikzpicture}

        \node[] at (-6, 0) {\includegraphics[width=0.38\textwidth]{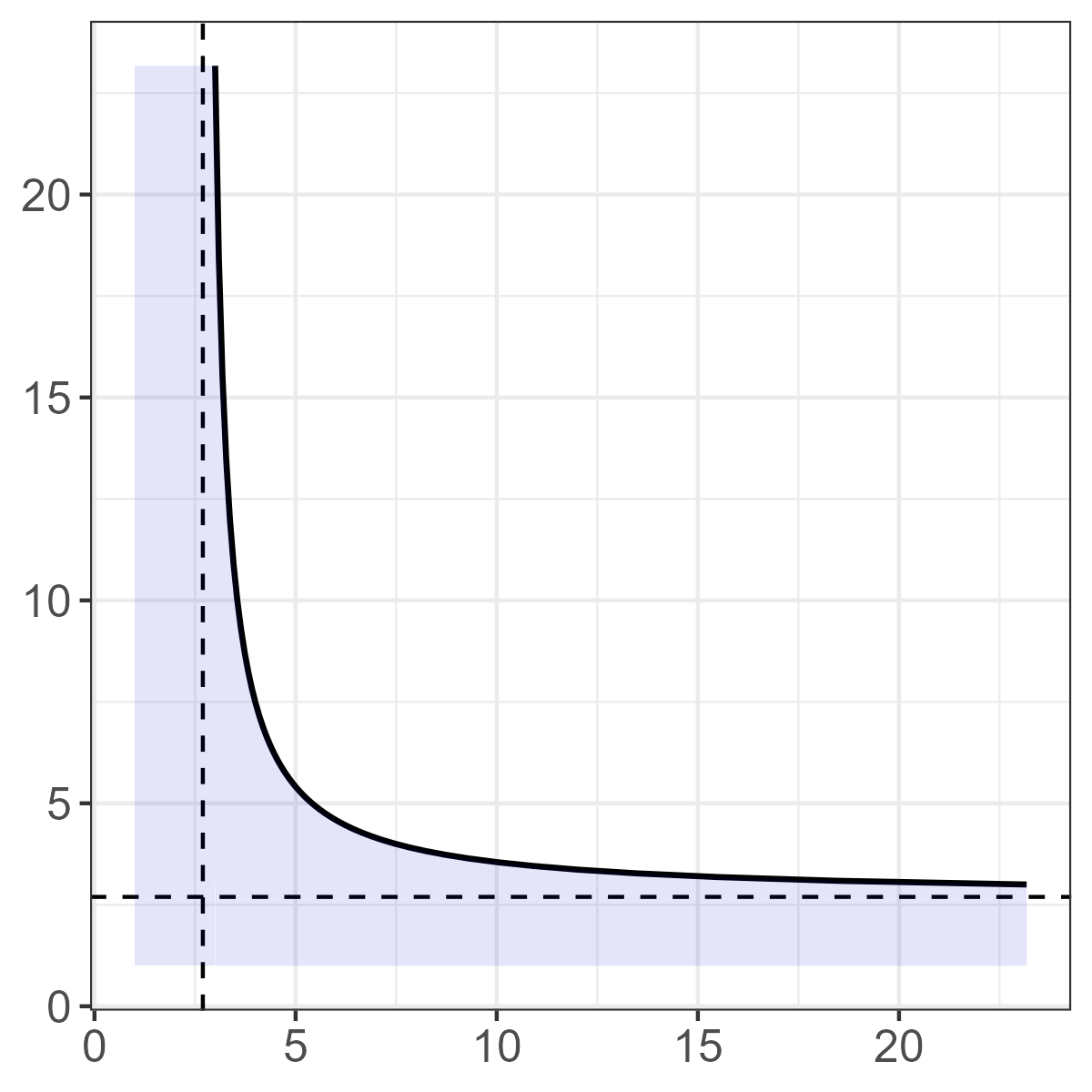}};
        \node[] at (0, 0) {\includegraphics[width=0.38\textwidth]{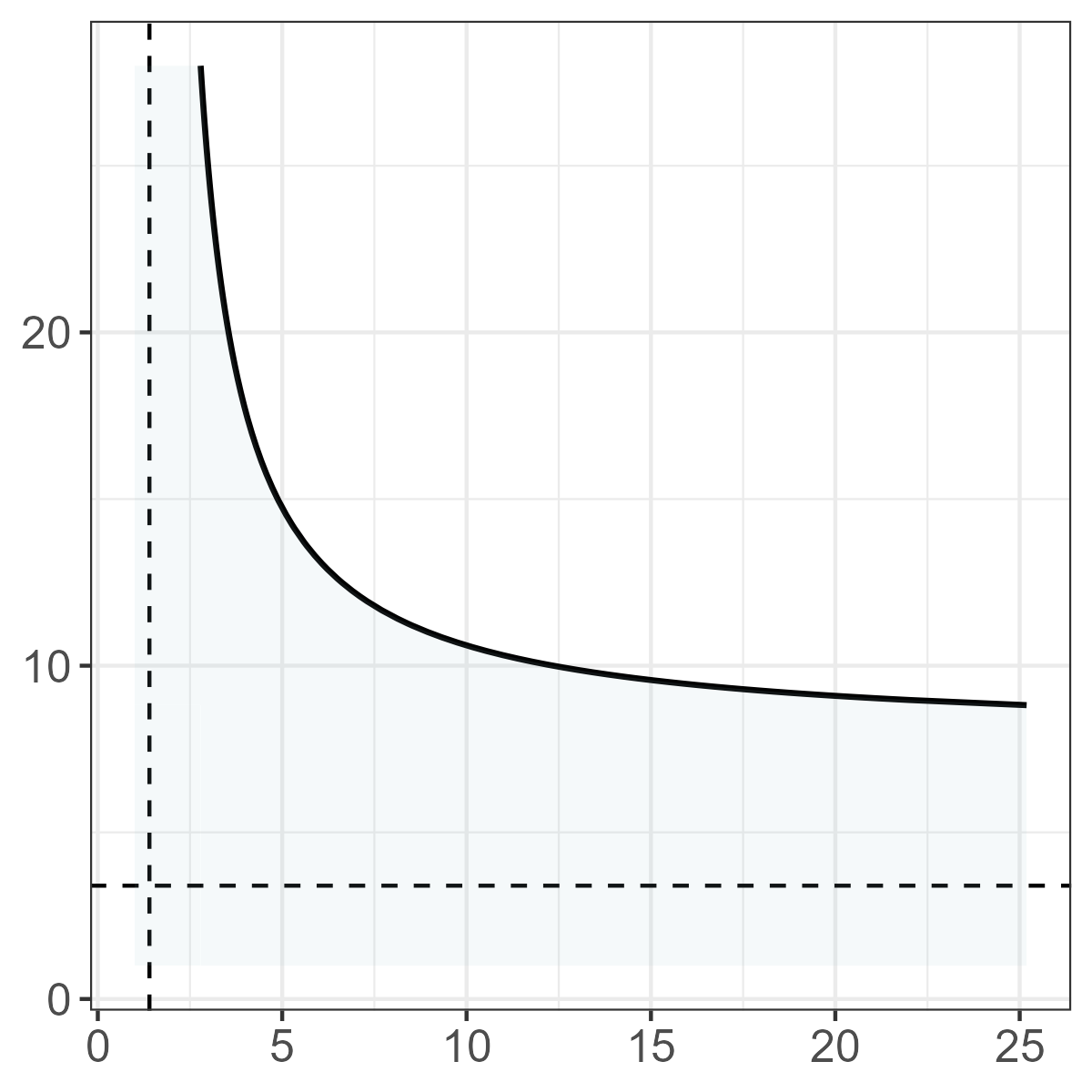}};
        \node[] at (6, 0) {\includegraphics[width=0.38\textwidth]{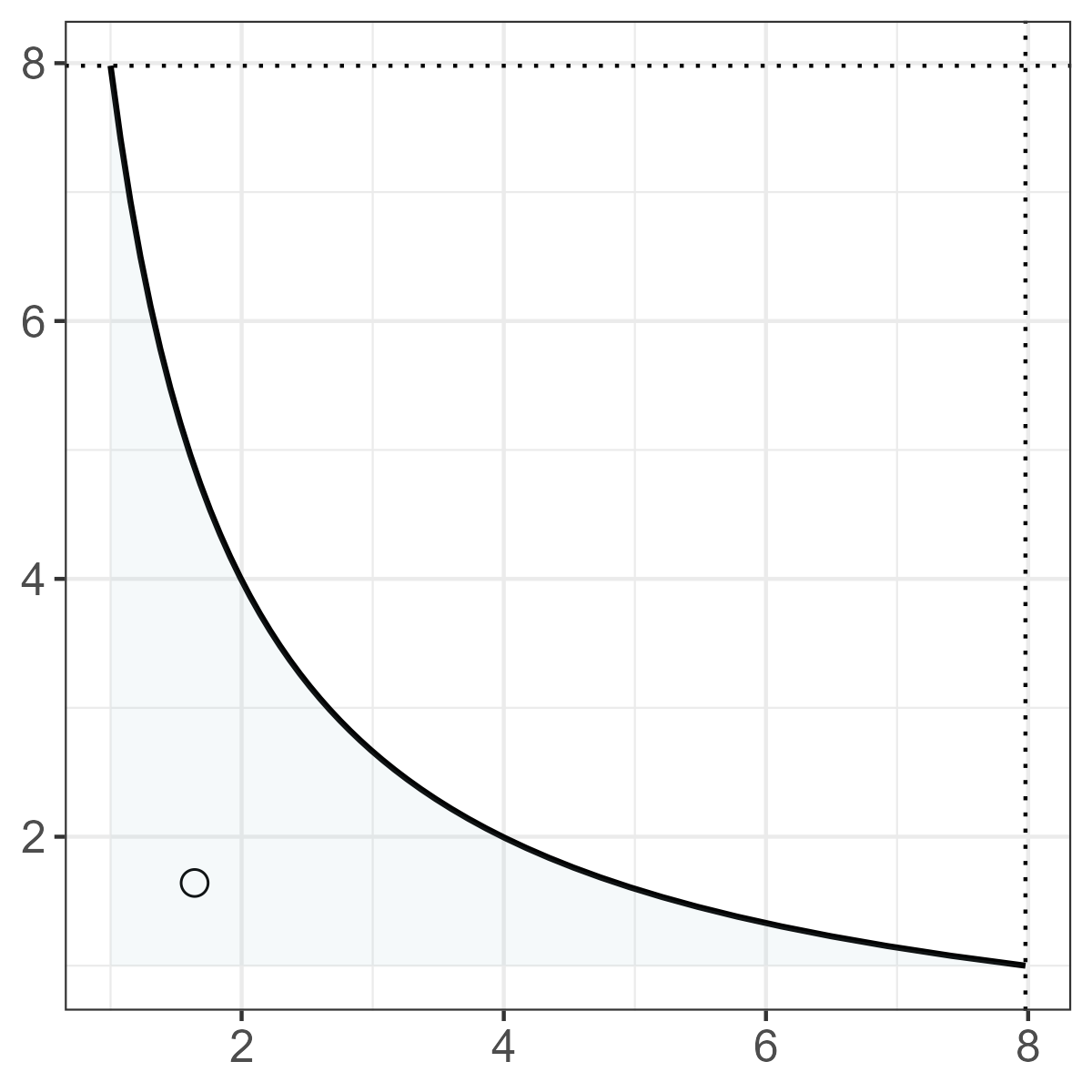}};

        \node[font = \footnotesize] at (-6, 3.35) {(a) Amplification};
        \node[font = \footnotesize, align = center] at (0 , 3.35) {(b) Type-II CH, $\Lambda = 3.5$};
        \node[font = \footnotesize, align = center] at (6, 3.35) {(c) Type-I CH, $\boldsymbol{\Delta} = (2, 5)$};

        \node[font = \footnotesize] at (-6.7, 0.95) {$\widetilde{D}^{*}_{\{\Lambda, \Delta\}}$};
        \draw[] (-7.35, 0.95) -> (-8.17, 0.95);
        \draw[] (-7.95, -1.99) circle [radius = 2mm];
        \draw[] (-7.15, -1.3) -> (-7.73, -1.73);
        \node[font = \footnotesize] at (-6.5, -1.2) {$(\widetilde{\Lambda}^{*}, \widetilde{\Delta}^{*})$};
        \node[font = \footnotesize] at (-8.5, 3.20) {$\Lambda$};
        \node[font = \footnotesize] at (-3.2, -2.85) {$\Delta$};

        \node[font = \footnotesize, align = center] at (-0.45, 0.95) {$\widetilde{D}^{*}_{\{\Lambda, \Delta_{1}, \Delta_{2}\}}$};
        \draw[] (-1.35, 0.95) -> (-2.05, 0.95);
        \draw[] (-2.25, -1.93) circle [radius = 2mm];
        \draw[] (-1.5, -1.3) -> (-2.05, -1.7);
        \node[font = \footnotesize] at (-0.5, -1.2) {$(3.5, \widetilde{\Delta}^{*}_{1}, \widetilde{\Delta}^{*}_{2})$};
        \node[font = \footnotesize] at (-2.5, 3.2) {$\Delta_{2}$};
        \node[fill = white, font = \footnotesize] at (2.8, -2.85) {$\Delta_{1}$};

        \node[font = \footnotesize,  align = center] at (5.45, 0.95) {$\widetilde{D}^{*}_{\{\Lambda_{1}, \Lambda_{2}, \Delta_{1}, \Delta_{2}\}}$};
        \draw[] (4.3, 0.95) -> (3.6, 0.95);
        \draw[] (5.1, -1) -> (4.15, -1.65);
        \node[font = \footnotesize] at (6.1, -0.9) {$(\widetilde{\Lambda}^{*}_{1}, \widetilde{\Lambda}^{*}_{2}, 2, 5)$};
        \draw[] (4.01, -1.9) circle [radius = 2mm];
        \node[font = \footnotesize] at (3.5, 3.2) {$\Lambda_{1}$};
        \node[fill = white, font = \footnotesize] at (8.8, -2.85) {$\Lambda_{2}$};
                
    \end{tikzpicture}
    }
    \caption{Visualizing the design sensitivity boundary (solid lines) for Case 2 with $\kappa = (0.25, 0.75)$ for the amplification (Figure a), Type-II confounding heterogeneity (Figure b), and Type-I confounding heterogeneity (Figure c) frameworks. Figure b shows a cross-section of the design sensitivity boundary and consistency region setting $\Lambda = 3.5$, and Figure c shows a cross-section of the design sensitivity boundary and consistency region when $\Delta_{1} = 2$ and $\Delta_{2} = 5$. Note that $(7.98, 2, 5) \in \widetilde{D}_{(\Lambda, \Delta_{1}, \Delta_{2})}$.}
    \label{fig: Design Sensitivity Figure}
\end{figure}
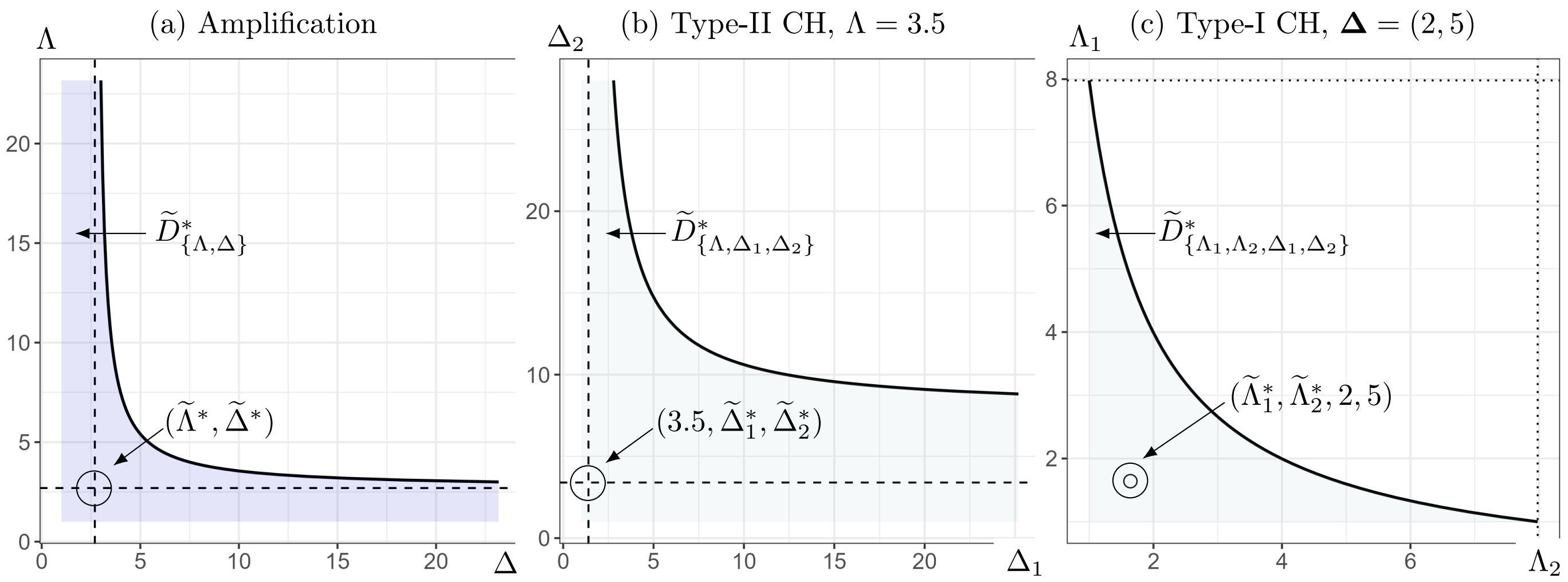

Figure~\ref{fig: Design Sensitivity Figure} visualizes the design sensitivity for the amplification and confounding heterogeneity frameworks for the case where $Y_{i1}V_{i} \sim N(0.1, 1), Y_{i2}V_{i} \sim N(0.5, 1)$, and $\kappa = (0.25, 0.75)$. Cross-sections of the design sensitivity are shown for the confounding heterogeneity frameworks, as the design sensitivity for Type-I and Type-II confounding heterogeneity is a surface in four- and three-dimensional parameter spaces. Some results shown in Table~\ref{tab: design sensitivity table} agree with the existing sensitivity analysis literature. For instance, the infimum design sensitivity for the amplification framework equals the design sensitivity for the classic framework in the univariate testing case, setting $\kappa = (1, 0)$ or $\kappa = (0, 1)$ \citep{rosenbaum2009amplification}. Additionally, in the multivariate testing case $\kappa \notin \{(1, 0), (0, 1)\}$, positive correlation between potential outcomes under control decreases design sensitivity, while negative correlation increases design sensitivity \citep{rosenbaum2016using, cohen2018multivariate}. 

There are several new insights gained from Table~\ref{tab: design sensitivity table}. First, in the multivariate testing case, if outcomes are not perfectly positively correlated $(\rho \neq 1)$, then the design sensitivity for the classic framework will be no less than the infimum treatment and outcome sensitivity for the amplification framework. As $I \to \infty$, for certain sensitivity parameter values satisfying the relation $\Gamma = (\Lambda \Delta + 1)/(\Lambda + \Delta)$, the worst-case $p$-value under the amplification framework may tend to one while the worst-case $p$-value under the classic framework may tend to zero. Second, the average infimum outcome sensitivity for the Type-I and Type-II confounding heterogeneity frameworks is no greater than the infimum outcome sensitivity for the amplification framework, suggesting that if unmeasured confounding strength is allowed to differ across outcomes, less overall confounding strength is required to overturn causal conclusions. Third, if unmeasured confounding strength is allowed to differ across outcomes, not all outcomes with positive treatment effects need be affected by unmeasured confounding to overturn causal conclusions. For example, in Case 2 letting $\kappa = (0.75, 0.25)$, the second outcome is expected to contribute a positive treatment effect of $0.25 \cdot 0.5 = 0.125$, yet if the unmeasured confounder(s) exhibit a strong relationship with the treatment ($\Lambda \text{ or } \Lambda_{1}, \Lambda_{2} \to \infty$) and has some relationship with the first outcome ($\Delta_{1} > 1$), no confounding effect on the second outcome may be required to overturn causal conclusions. Last, the average infimum treatment sensitivity for the Type-I confounding heterogeneity framework is less than the infimum treatment sensitivity for the amplification and Type-II confounding heterogeneity frameworks, suggesting that if confounding is allowed to come from multiple sources, each source of confounding can exhibit a weaker relationship with treatment while still being able to overturn causal conclusions. 

\vspace{-0.3cm}

\subsection{Finite-sample Power}

\begin{figure}[h] 
    \centering
    \makebox[\textwidth][c]{%
    \begin{tikzpicture}

        \node[] at (-3.75, 2.25) {\includegraphics[width=0.4\textwidth]{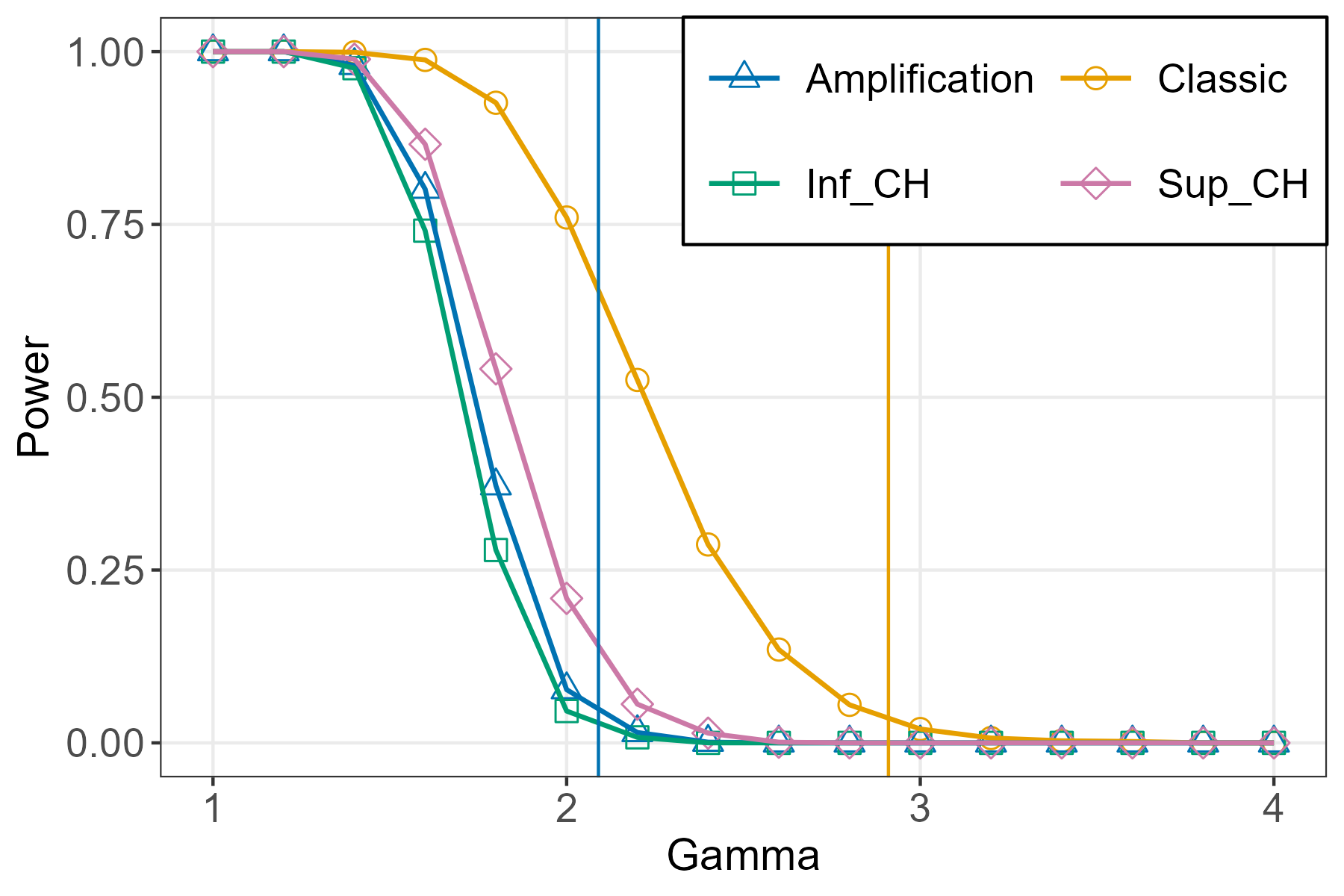}};
        \node[] at (3.25, 2.25) {\includegraphics[width=0.4\textwidth]{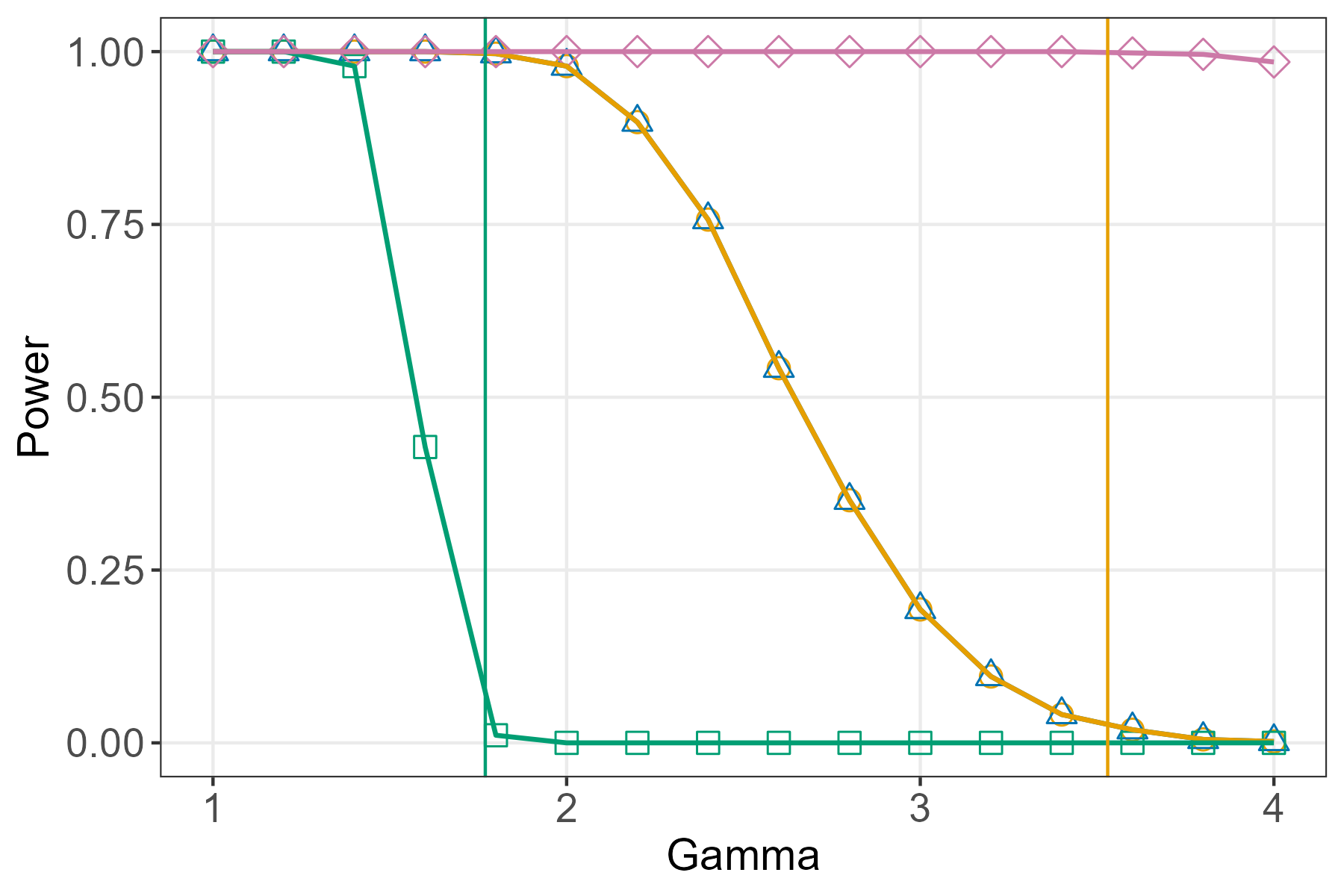}};
        \node[] at (-3.75, -2.25) {\includegraphics[width=0.4\textwidth]{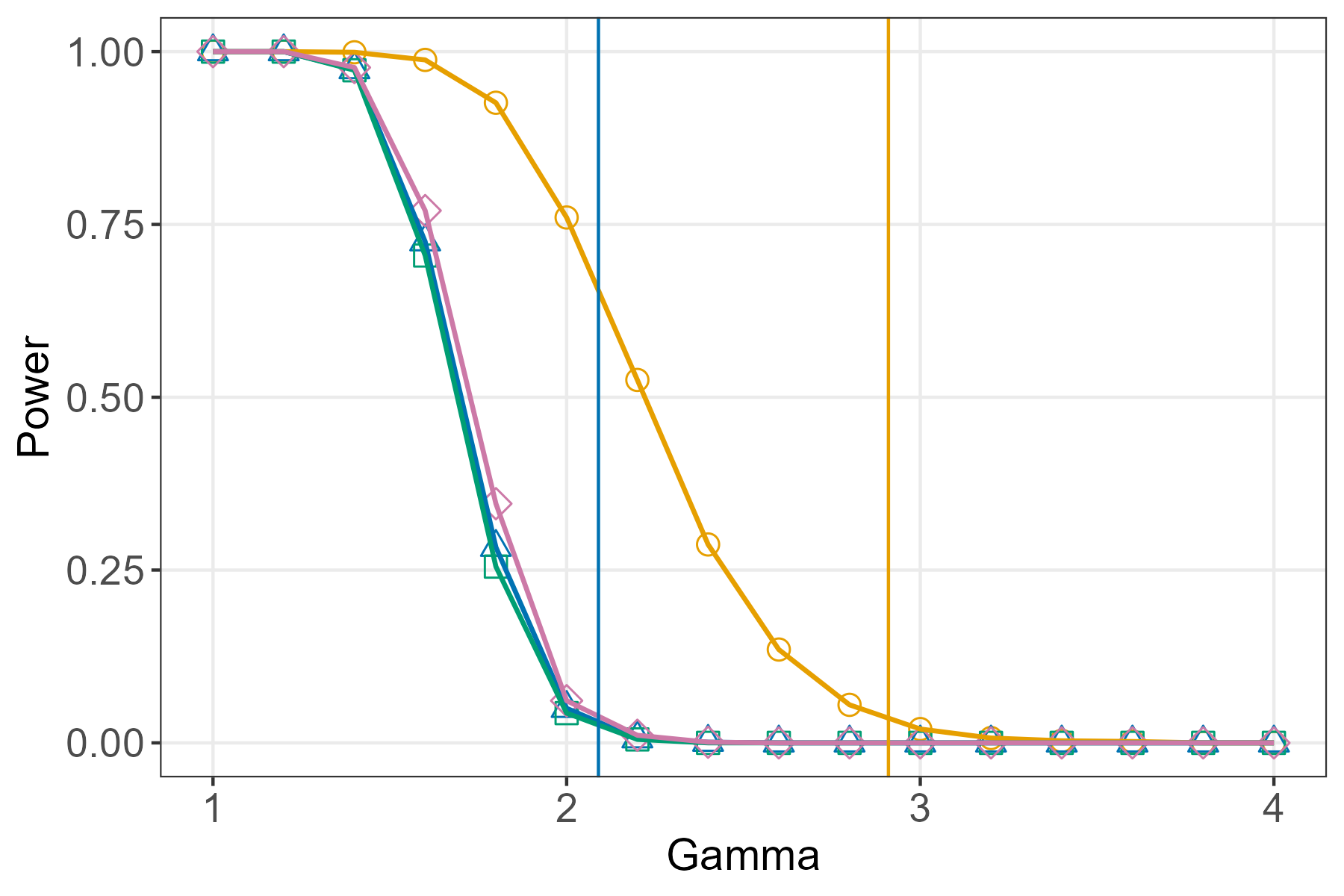}};
        \node[] at (3.25, -2.25) {\includegraphics[width=0.4\textwidth]{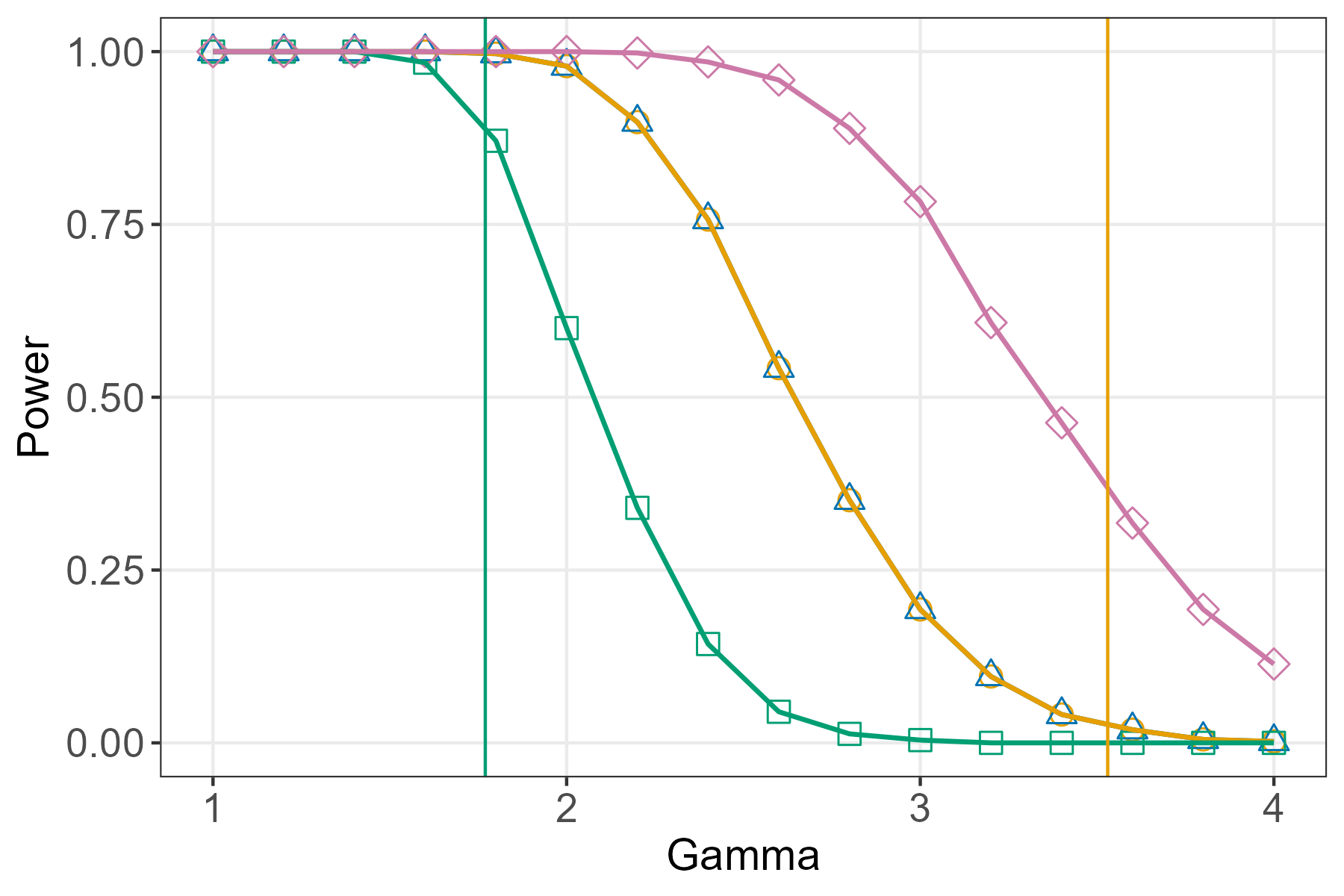}};

        \node[] at (-3.5, 4.75) {$\kappa = (0.5, 0.5)$};
        \node[] at (3.5, 4.75) {$\kappa = (0, 1)$};
        \node[rotate=90] at (-7.5, 2.25) {$a = 1/3$};
        \node[rotate=90] at (-7.5, -2.25) {$a = 3$};
                
    \end{tikzpicture}
    }
    \caption{The simulated power (rejection rate of the global null) over $1000$ simulated datasets of $I=300$ matched pairs, where $\kappa = (0.5, 0.5)$ (left) and $(0, 1)$ (right) with $a=1/3, 3$. The vertical line for the classic framework (orange) corresponds to $\widetilde{\Gamma}_{0}$, the vertical line for the amplification framework (blue) corresponds to the value of $\Gamma$ satisfying $\Gamma = \widetilde{\Delta}^{*}$, and the vertical line for the worst-case confounding heterogeneity scenario (green) corresponds to the value of $\Gamma$ satisfying $\Gamma/(\Gamma + 1) = 0.5(\widetilde{\Delta}^{*}_{1}/(\widetilde{\Delta}^{*}_{1} + 1) + \widetilde{\Delta}^{*}_{2}/(\widetilde{\Delta}^{*}_{2} + 1))$. Note that the vertical line calculations are independent of the choice of $a$. Additionally, note that for $\kappa = (0.5, 0.5)$, the vertical lines for the amplification and infimum confounding heterogeneity frameworks overlap, and for $\kappa = (0, 1)$, the vertical lines for the classic and amplification frameworks overlap. The formulas for each vertical line are justified in Appendix B.5.}
    \label{fig: finite-sample power}
\end{figure}
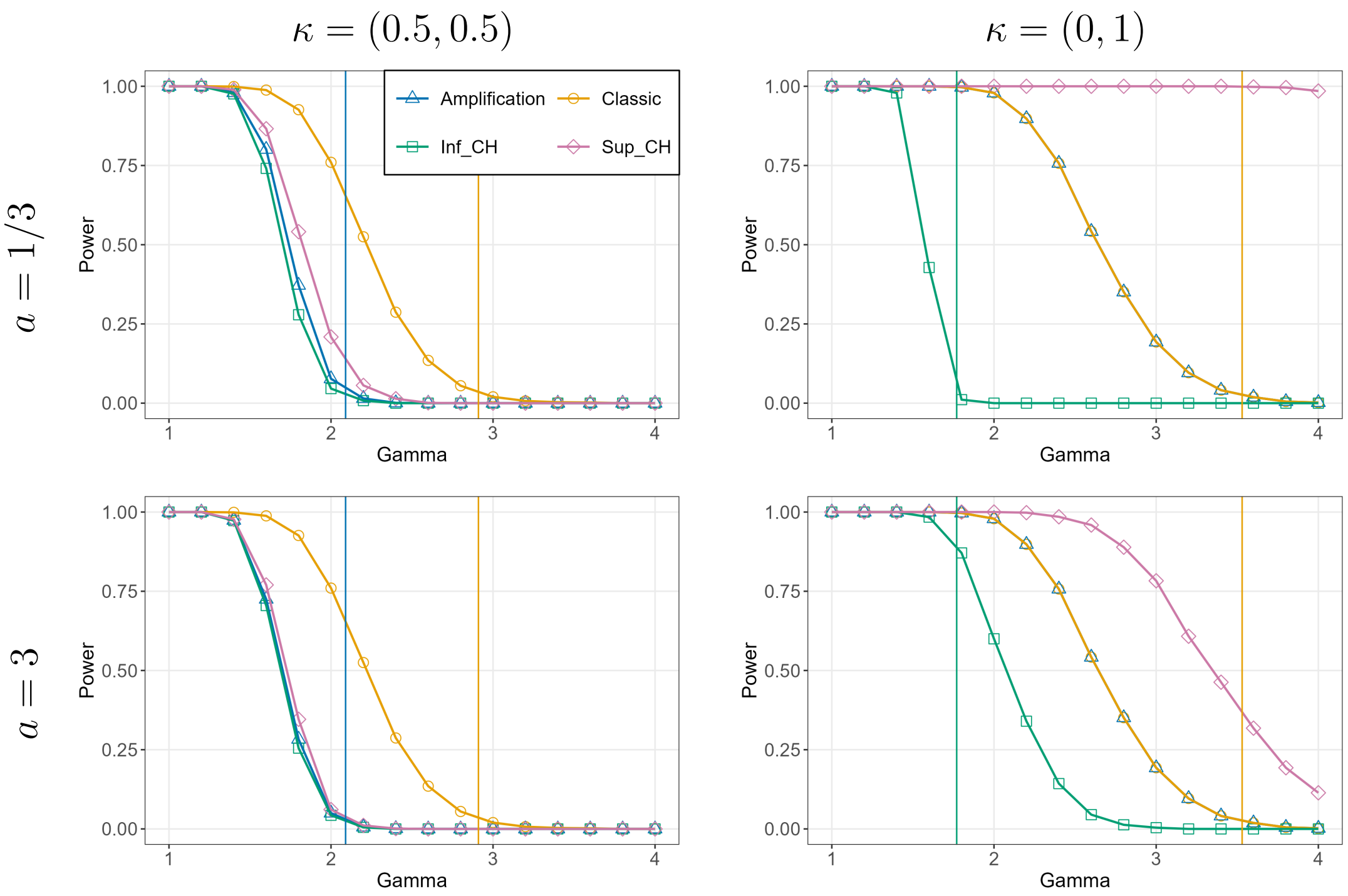

Under Case 2, considering $\kappa = (0.5, 0.5), (0.25, 0.75), (0, 1)$ and with $\rho = 0$, the simulated power of a sensitivity analysis is calculated under the classic and amplification frameworks, as well as under the most favorable configuration of confounding heterogeneity (``Sup\_CH'') and the least favorable configuration of confounding heterogeneity (``Inf\_CH''). Figure~\ref{fig: finite-sample power} presents finite-sample power results when $a = 1/3, 3$ and $\kappa = (0.5, 0.5), (0, 1)$. Additional simulation results for $a = 1$ and $\kappa = (0.25, 0.75)$ are in Appendix D.2. We simulate $1000$ datasets with $I = 300$ matched pairs and use a significance level of $\alpha = 0.05$.

For a fixed choice of test statistic and strength of unmeasured confounding quantified by $\Gamma$, a test statistic's finite-sample power may change based on the structure of unmeasured confounding considered. When $\kappa$ is chosen to heavily weight one outcome over the other, such as $\kappa = (0.25, 0.75)$ or $\kappa = (0, 1)$, a sensitivity analysis under the most favorable configuration of confounding heterogeneity may have higher finite-sample power compared to sensitivity analysis under the classic framework. At the same time, under the least favorable case of confounding heterogeneity, the finite-sample power of a sensitivity analysis may be much worse compared to the classic framework. 

\subsection{Bahadur-Rosenbaum Slope}

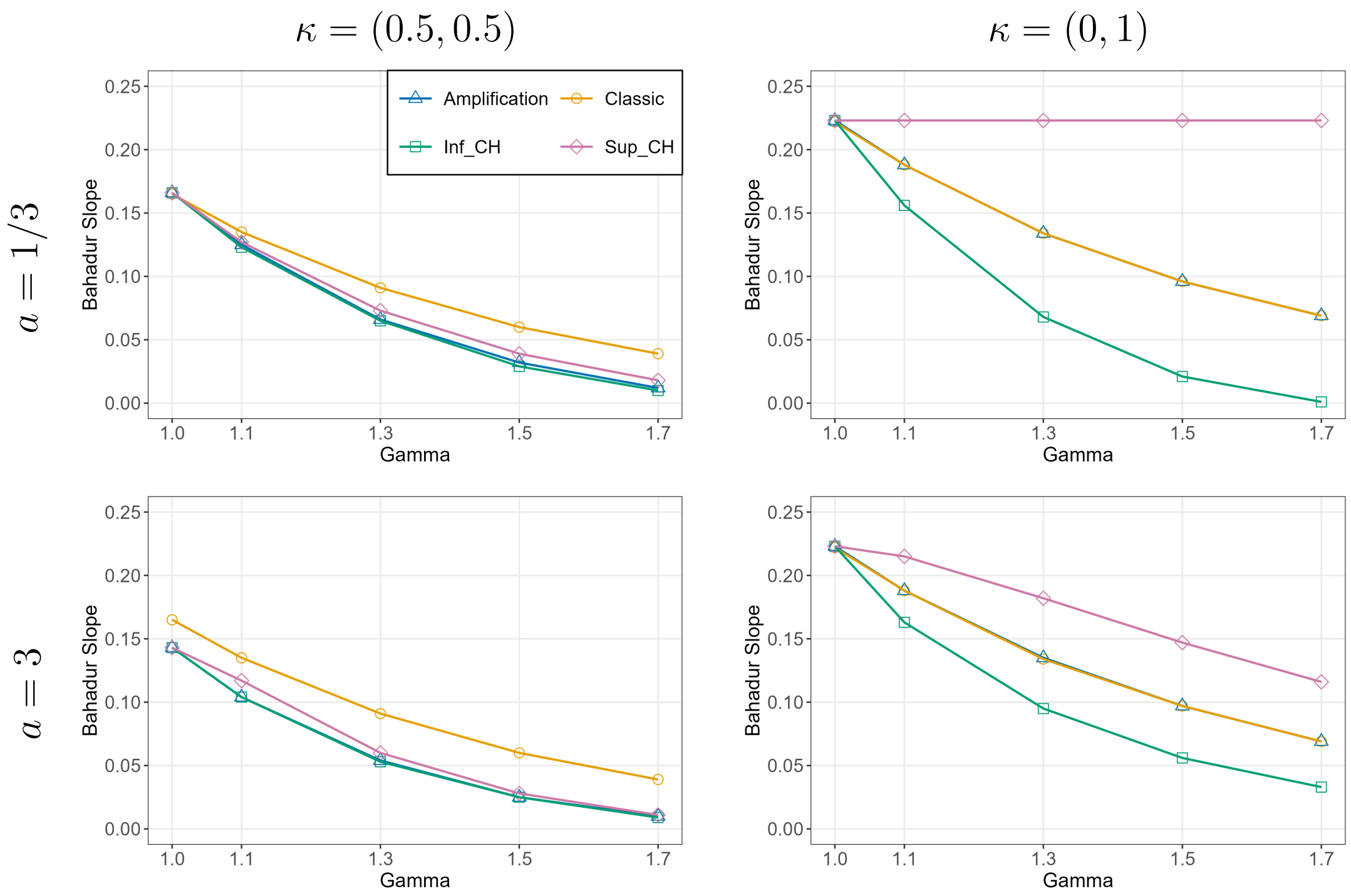
\begin{figure}[h] 
    \centering
    \makebox[\textwidth][c]{%
    \begin{tikzpicture}

        \node[] at (-3.75, 2.25) {\includegraphics[width=0.4\textwidth]{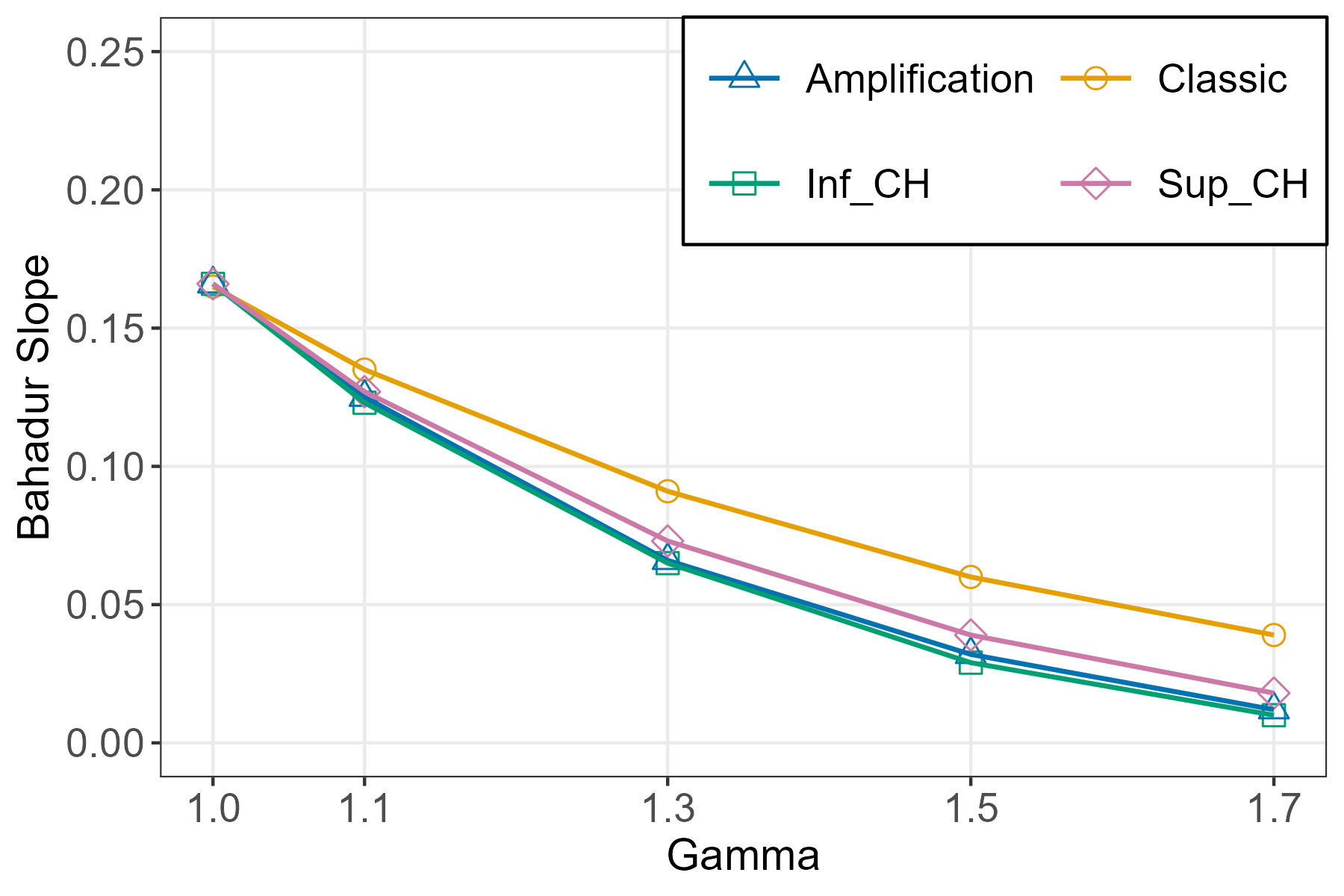}};
        \node[] at (3.25, 2.25) {\includegraphics[width=0.4\textwidth]{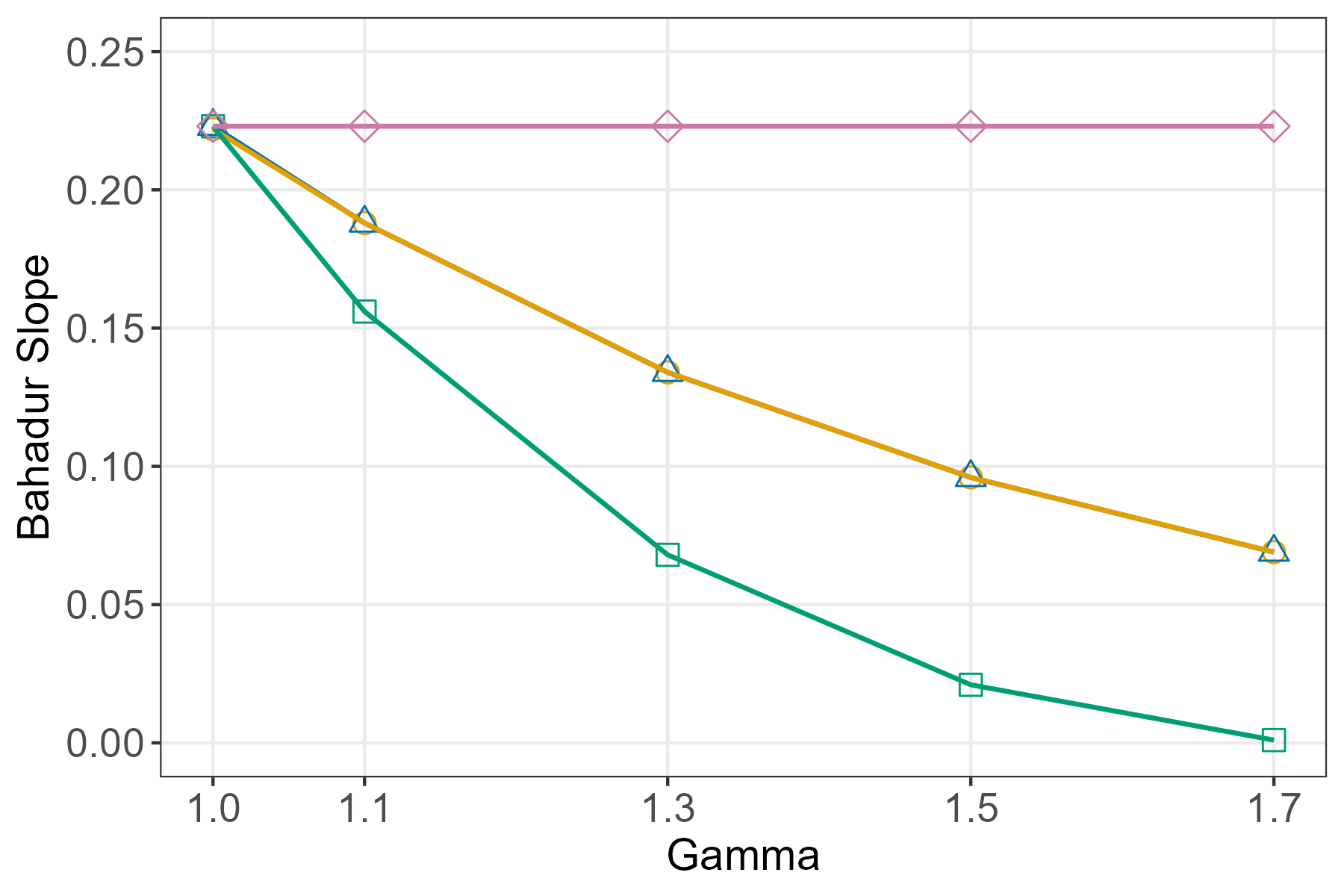}};
        \node[] at (-3.75, -2.25) {\includegraphics[width=0.4\textwidth]{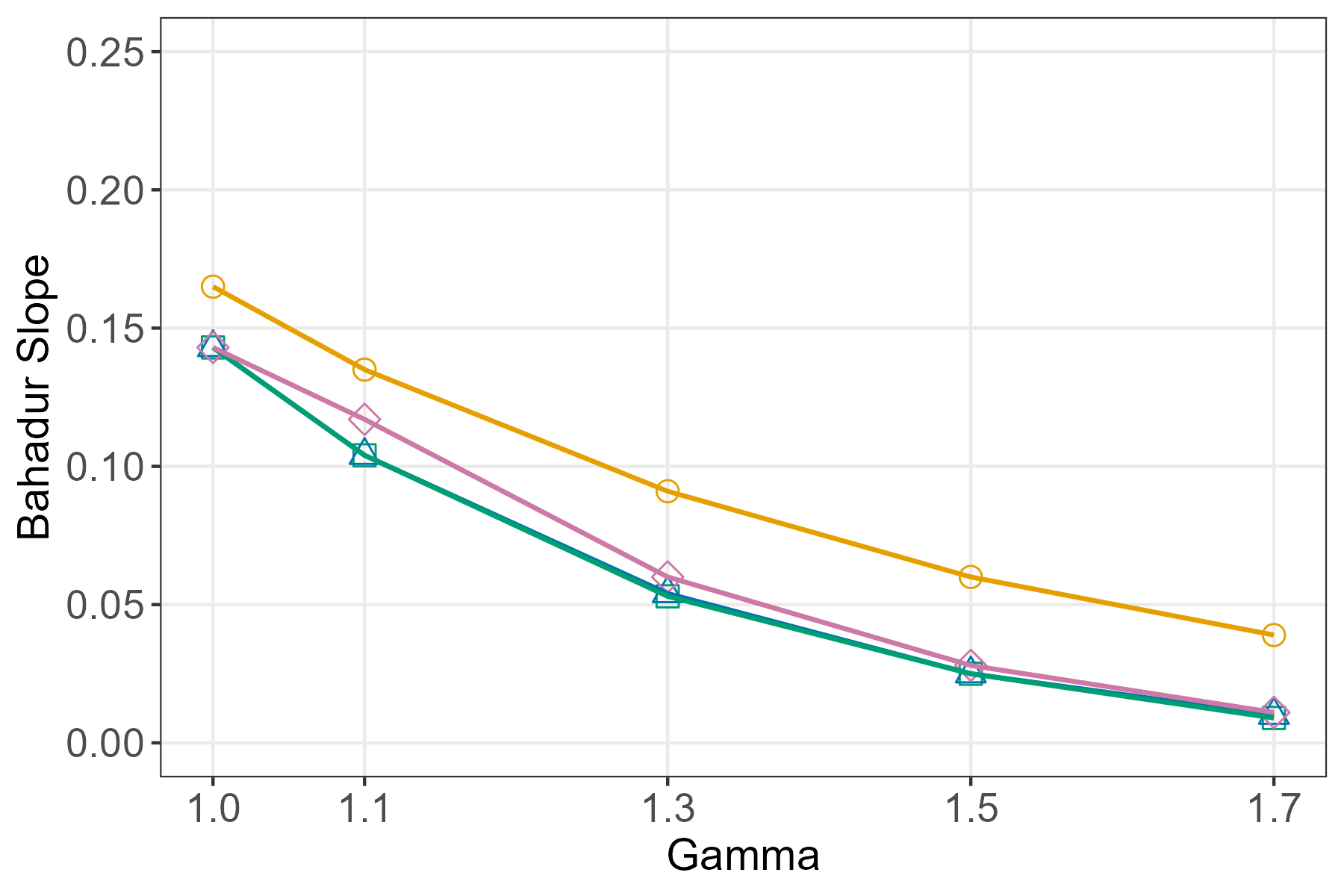}};
        \node[] at (3.25, -2.25) {\includegraphics[width=0.4\textwidth]{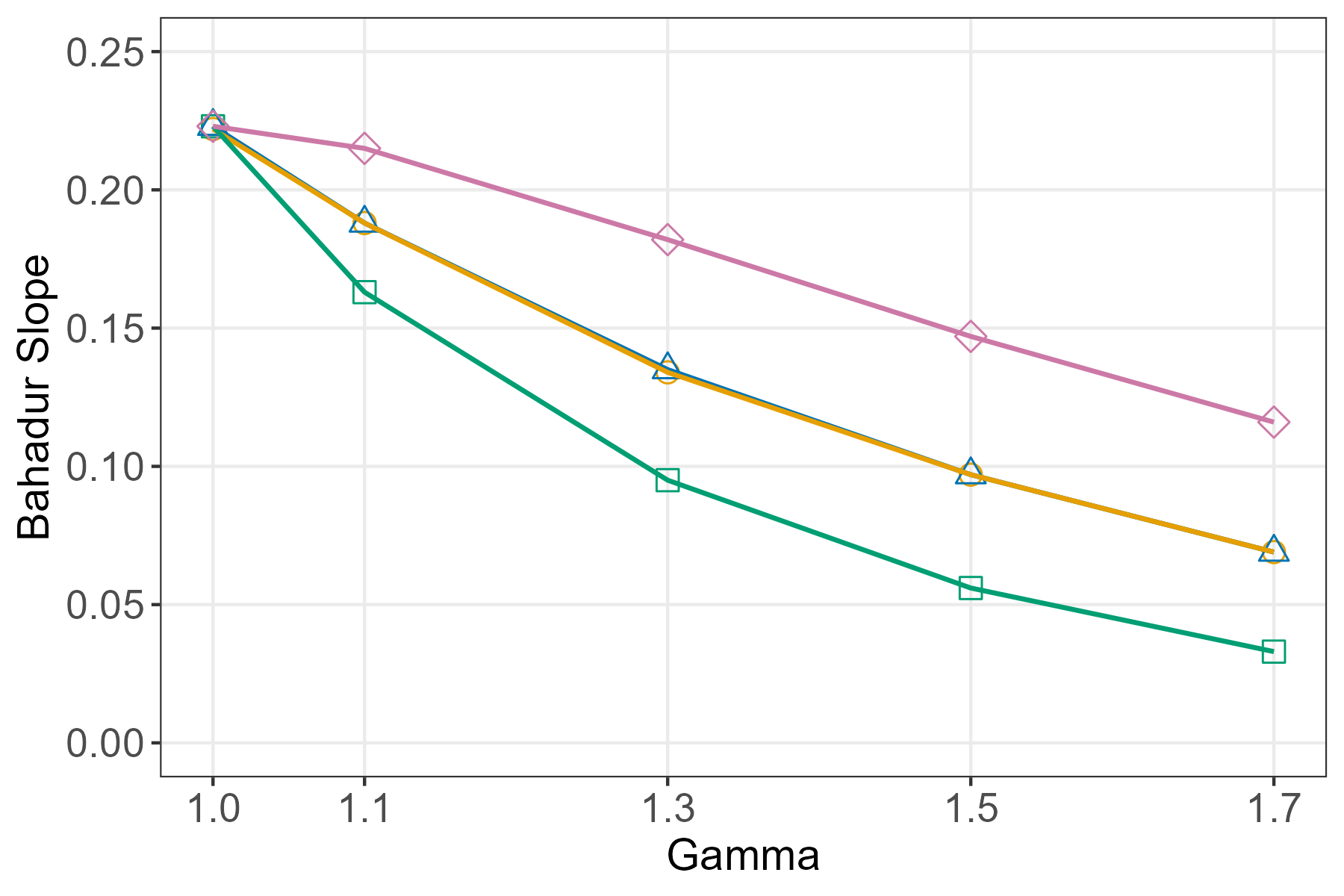}};

        \node[] at (-3.5, 4.75) {$\kappa = (0.5, 0.5)$};
        \node[] at (3.5, 4.75) {$\kappa = (0, 1)$};
        \node[rotate=90] at (-7.5, 2.25) {$a = 1/3$};
        \node[rotate=90] at (-7.5, -2.25) {$a = 3$};
                
    \end{tikzpicture}
    }
    \caption{The Bahadur-Rosenbaum slope when $Y_{i1}V_{i} \sim N(0.1, 1), Y_{i2}V_i \sim N(0.5, 1)$, $q_{ik} = |Y_{ik}V_{i}|$, $a = 1/3$ (top row) and $a=3$ (bottom row), and $\kappa = (0.5, 0.5)$ (left column) and $\kappa = (0, 1)$ (right column).}
    \label{fig: Bahadur Slope figure}
\end{figure}

Under Case 2, with $\rho = 0$, we estimate the Bahadur-Rosenbaum slope under the classic and amplification frameworks, as well as the infimum and supremum Bahadur-Rosenbaum slopes for $\Gamma = 1, 1.1, 1.3, 1.5, 1.7$ via Monte Carlo estimation using $I=1{,}000{,}000$ matched pairs. Figure~\ref{fig: Bahadur Slope figure} displays the Bahadur-Rosenbaum slopes across frameworks for increasing $\Gamma$ when $a = 1/3, 3$ and $\kappa = (0.5, 0.5), (0, 1)$. Additional simulation results for $a = 1$ and $\kappa = (0.25, 0.75)$ are in Appendix D.3.

As with finite-sample power, for a fixed choice of test statistic and strength of confounding quantified by $\Gamma$, a test statistic has a range of possible Bahadur-Rosenbaum efficiency under the confounding heterogeneity framework. Additionally, the choice of $\kappa$ affects whether the efficiency of a sensitivity analysis under confounding heterogeneity can dominate the efficiency under the classic framework or not. Both the finite-sample power and Bahadur-Rosenbaum efficiency calculations present one unified message: the structure of unmeasured confounding influences the power of a sensitivity analysis, not just the magnitude of unmeasured confounding.

\vspace{-0.5cm}

\section{Application: Smoking and Periodontal Disease}
\label{Sec: Application}
\vspace{-0.3cm}

Smoking is an established risk factor for periodontal disease, previously estimated to account for up to 42\% of periodontal disease cases \citep{tomar2000smoking}. To assess the claim that smoking causes periodontal disease, \cite{rosenbaum2016using} formed $441$ matched pairs of smoking and non-smoking individuals among the participants in the 2011-2012 NHANES survey; see \cite{rosenbaum2016using} for details regarding confounder balance and formation of matched pairs. The treatment $Z$ is a binary indicator of smoking status, where $Z=1$ indicates daily smokers and $Z=0$ indicates never smokers. Loss of attachment and pocket depth are measured in six locations on each tooth, where a pocket depth or loss of attachment of at least $4$ mm is considered evidence of periodontal disease. Each tooth provides a score of 0-6, and the two outcome variables are the summed score for the 14 lower teeth (outcome 1) and the 14 upper teeth (outcome 2). Under the assumption of no unmeasured confounding, there is sufficient evidence to reject the global null hypothesis and conclude evidence of a causal effect on the lower and/or upper teeth, so a sensitivity analysis is needed to assess whether this causal evidence is insensitive to unmeasured confounding. 

\cite{rosenbaum2016using} and \cite{cohen2018multivariate} use the Scheff\'e projection and chi-bar methods to conduct sensitivity analyses for this study, both under the classic framework. By conducting sensitivity analysis under the classic framework, both analyses fail to describe the impact of unmeasured confounding that exhibits confounding heterogeneity. An unmeasured confounder exhibiting confounding heterogeneity between upper and lower teeth may reasonably exist in this case. For example, individuals who regularly use chewing tobacco are more likely to also be regular smokers \citep{felicione2021cigarette, tam2015systematic}, and this association persists across common demographic and socioeconomic groups in the United States \citep{mcclave2011smokers}, meaning the unmeasured confounder of chewing tobacco usage may be imbalanced within pairs, even after matching on demographic and socioeconomic factors. Additionally, chewing tobacco products are commonly used near the upper gums or beneath the upper lip, which has been linked to adverse oral health outcomes, particularly in the upper teeth \citep{quadri2024smokeless}. Thus, an unmeasured confounder like use of chewing tobacco is expected to exhibit confounding heterogeneity between lower and upper teeth. We reanalyze the smoking and periodontal disease matched observational study by conducting sensitivity analyses under the classic, amplification, and confounding heterogeneity frameworks to assess to what extent the causal conclusion that smoking causes periodontal disease is insensitive to unmeasured confounding that exhibits confounding heterogeneity, as well as how the reported sensitivity to unmeasured confounding differs between sensitivity analysis frameworks. We conduct sensitivity analyses across frameworks, keeping the sensitivity analysis method (chi-bar method) and test statistic (setting $q_{ik} = |Y_{ik}|$) fixed to highlight differences in the robustness of evidence against the global null to unmeasured confounding.

The sensitivity value \citep{zhao2019sensitivityvalue} is the largest value of the sensitivity parameter such that a sensitivity analysis conducted with sensitivity parameter values below it is still able to reject the global null hypothesis, given some significance level $\alpha$. For sensitivity parameter values $\{\Lambda, \Delta, \boldsymbol{\Lambda}, \boldsymbol{\Delta}\} \in \mathcal{B}_{\Gamma, a}$ with $a = 1/3, 1$, and $3$, we calculate the empirical sensitivity values with significance level $\alpha = 0.05$ for the classic and amplification frameworks, as well as the sensitivity value corresponding to the allocation of confounding heterogeneity maximizing the worst-case $p$-value (``Infimum CH'') and the sensitivity value corresponding to the allocation of confounding heterogeneity minimizing the worst-case $p$-value (``Supremum CH''), reported in Table~\ref{tab: sensitivity values}.

\begin{table}[ht]
    \centering
    \begin{tabular}{c|c|ccc|ccc|ccc}
        \toprule
         & Classic & \multicolumn{3}{c}{Amplification} & \multicolumn{3}{c}{Infimum CH} & \multicolumn{3}{c}{Supremum CH} \\
         a & - & 1/3 & 1 & 3 & 1/3 & 1 & 3 & 1/3 & 1 & 3 \\
         \midrule
        $\Gamma$ & 2.76 & 2.52 & 2.52 & 2.52 & 2.38 & 2.35 & 2.30 & 8.89 & 5.92 & 3.29\\
        $\Lambda$ & -- & 9.77 & 4.83 & 3.26 & 9.19 & 4.48 & 2.95 & 35.48 & 11.75 & 4.31 \\
        $\Delta_{1}$ & -- & 3.26 & 4.83 & 9.77 & 2.50 & 3.31 & 5.11 & $\infty$ & $\infty$ & $\infty$ \\
        $\Delta_{2}$ & -- & 3.26 & 4.83 & 9.77 & 3.85 & 6.51 & 24.49 & 5.41 & 5.38 & 5.96 \\
        $\kappa_{1}$ & 0 & 0.10 & 0 & 0 & 0.55 & 0.54 & 0.54 & 0 & 0 & 0 \\
        $\kappa_{2}$ & 1 & 0.90 & 1 & 1 & 0.45 & 0.46 & 0.46 & 1 & 1 & 1 \\
        \hline 
    \end{tabular}
    \caption{Sensitivity values across sensitivity analysis frameworks. $\kappa$ is normalized so $\kappa_{1} + \kappa_{2} = 1$. 
    }
    \label{tab: sensitivity values}
\end{table}

From Table~\ref{tab: sensitivity values}, if potential outcomes are viewed as fixed (classic framework), the causal conclusion that smoking causes periodontal disease is insensitive to an unmeasured confounder that, within each matched pair, increases the odds of daily smoking status $2.76$-fold. If potential outcomes are viewed as random but still assuming no confounding heterogeneity (amplification framework), the causal conclusion is insensitive to an unmeasured confounder that, within each matched pair, increases the odds of daily smoking status $3.26$-fold and the odds of greater loss of attachment and pocket depth by $9.77$-fold in both upper and lower teeth in the absence of daily smoking, as one example. Under the worst-case scenario of confounding heterogeneity tested, the causal conclusion is insensitive to an unmeasured confounder that, within each matched pair, increases the odds of daily smoking status by $2.95$-fold, the odds of greater loss of attachment and pocket depth by $5.11$-fold for lower teeth in the absence of daily smoking, and the odds of greater loss of attachment and pocket depth by $24.49$-fold for upper teeth in the absence of daily smoking. In the best-case scenario of confounding heterogeneity tested, the causal conclusion is insensitive to an unmeasured confounder that, within each matched pair, increases the odds of daily smoking status about $35$-fold, essentially determines the odds of greater loss of attachment and pocket depth for lower teeth in the absence of daily smoking, and increases the odds of greater loss of attachment and pocket depth by $5.41$-fold for upper teeth in the absence of daily smoking.

Across all sensitivity analysis frameworks, we find that the causal conclusion that smoking causes periodontal disease is insensitive to a moderate amount of unmeasured confounding. However, the claim of insensitivity to unmeasured confounding under the classic framework does not explore the impact of confounding heterogeneity, weakening the evidence that the causal conclusion is insensitive to unmeasured confounding. In contrast, using the proposed confounding-heterogeneity-aware sensitivity analysis frameworks, we find that the causal conclusion remains insensitive to a moderate amount of unmeasured confounding even after accounting for confounding heterogeneity, strengthening the claim of insensitivity to unmeasured confounding.

\vspace{-0.5cm}

\section{Conclusion and Discussion}
\label{Sec: Conclusion}
\vspace{-0.3cm}

This work formalizes confounding heterogeneity in multivariate sensitivity analysis and develops the Type-I and Type-II frameworks for matched observational studies. We establish asymptotic Type~I error control for several existing multivariate methods under these frameworks and extend design sensitivity and Bahadur-Rosenbaum efficiency to characterize consistency and efficiency under different confounding structures. 
Our results show that causal robustness depends on both the magnitude and the structure of unmeasured confounding. The classic $\Gamma$ framework may understate sensitivity relative to frameworks that explicitly parameterize confounding heterogeneity. Even when overall confounding magnitude is held fixed under a common calibration, finite-sample power and asymptotic efficiency can vary across confounding structures. For practitioners, the proposed frameworks provide a way to incorporate subject-matter knowledge about plausible unmeasured confounders and assess which outcome-specific confounding configurations a causal conclusion can withstand. Future research could extend these frameworks beyond matched designs and investigate the implications of confounding heterogeneity for other approaches to addressing unmeasured confounding, including negative controls. More broadly, our results motivate a shift from asking only \emph{how much unmeasured confounding would be needed to overturn a causal conclusion} to also asking \emph{how that confounding would need to be structured across outcomes}.

\vspace{-0.5cm}

\section*{Supplementary Materials}

\vspace{-0.3cm}

The online supplementary materials contain detailed proofs, additional simulation studies, and further discussions and remarks.

\vspace{-0.5cm}

\section*{Declaration of Generative AI Use}

\vspace{-0.3cm}
The authors used AI to assist with proofreading for grammatical, typographical, and notational inconsistencies, checking the logical consistency of mathematical proofs, and improving their rigor. The authors confirm the originality and integrity of the manuscript and take full responsibility for the proposed methods and theoretical and numerical results. 

\newpage

\renewcommand*{\thesection}{\Alph{section}}
\renewcommand{\thefigure}{S\arabic{figure}}
\renewcommand{\thetable}{S\arabic{table}}
\renewcommand\theAssumption{S\arabic{Assumption}}
\renewcommand\theLemma{S\arabic{Lemma}}
\renewcommand\theRemark{S\arabic{Remark}}
\setcounter{section}{0}

\begin{center}
    {\Large \bf Supplementary Materials for ``Confounding Heterogeneity in Multivariate Sensitivity Analysis for Observational Studies''}
\end{center}

\begin{abstract}
The Supplementary Materials contain proofs for theoretical results and additional simulation results. Specifically, Section A describes the set of causal assumptions used for the theoretical results under each sensitivity analysis framework, as well as additional assumptions for theoretical analysis of the power of a sensitivity analysis. Section B states and proves supplementary lemmas used to prove the main theoretical results, as well as details regarding computation. Section C formally proves the main theoretical results. Section D presents additional simulation results for power, design sensitivity, and Bahadur-Rosenbaum efficiency. Section E presents an additional analysis of the smoking and periodontal disease application. 
\end{abstract}

\section{Assumptions}

\vspace{-0.3cm}

\subsection{Causal assumptions under each sensitivity analysis framework}

In the classic, amplification, and Type-II confounding heterogeneity sensitivity analysis frameworks, we model unmeasured confounding as coming from a single source $\mathbf{u}$. In order for these sensitivity analysis frameworks to be valid, we assume there is no additional source of confounding not captured by $\mathbf{u}$ and independence holds across matched pairs conditioned on measured confounders and the single unmeasured confounder $\mathbf{u}$, as well as consistency:

\begin{Assumption}[Causal assumptions for the classic framework]
\label{Assumption: classic}
\noindent
\begin{enumerate}
    \item (No additional confounders): $(r_{Ti11}, r_{Ci11}, \dots, r_{Ti1K}, r_{Ci1K},\dots,r_{Ti21}, r_{Ci21}, \dots, r_{Ti2K}, r_{Ci2K})$ $\indep (Z_{i1},Z_{i2}) |(\mathbf{x}_{i1}, \mathbf{x}_{i2}, u_{i1}, u_{i2})$ for all $i=1,\dots, I$.
    \item (Positivity): $0<P(Z_{ij}=1|\mathbf{x}_{ij}, u_{ij})<1$, for all $\mathbf{x}_{ij}, u_{ij}$.
    \item (Independence across matched pairs): $\{(Y_{i1},\dots,Y_{iK}, V_{i})\}_{i=1}^{I}$ form a mutually independent set conditioned on $\mathbf{x}_{ij}, u_{ij}$ for all $j=1,2,i=1,\dots,I$.
    \item (Consistency): $R_{ijk} = Z_{ij}r_{Tijk} + (1-Z_{ij})r_{Cijk}$.
\end{enumerate}
\end{Assumption}

In the amplification and Type-II confounding heterogeneity frameworks, we additionally require that outcome differences under control are mutually independent within each pair, conditioned on measured and unmeasured confounders. 

\begin{Assumption}[Causal assumptions for amplification and Type-II confounding heterogeneity frameworks]
\label{Assumption: Amp and Type-II CH}
\noindent
\begin{enumerate}
    \item (No additional confounders): $(r_{Ti11}, r_{Ci11}, \dots, r_{Ti1K}, r_{Ci1K},\dots,r_{Ti21}, r_{Ci21}, \dots, r_{Ti2K}, r_{Ci2K})$ $\indep (Z_{i1},Z_{i2}) |(\mathbf{x}_{i1}, \mathbf{x}_{i2}, u_{i1}, u_{i2})$ for all $i=1,\dots, I$. Additionally, $\{Y_{Cik}\}_{k=1}^{K}$ form a mutually independent set conditioned on $\mathbf{x}_{ij}, u_{ij}$ for all $j=1,2,i=1,\dots,I$.
    \item (Positivity): $0<P(Z_{ij}=1|\mathbf{x}_{ij}, u_{ij})<1$, for all $\mathbf{x}_{ij}, u_{ij}$.
    \item (Independence across matched pairs): $\{(Y_{i1},\dots,Y_{iK}, V_{i})\}_{i=1}^{I}$ form a mutually independent set conditioned on $\mathbf{x}_{ij}, u_{ij}$ for all $j=1,2,i=1,\dots,I$.
    \item (Consistency): $R_{ijk} = Z_{ij}r_{Tijk} + (1-Z_{ij})r_{Cijk}$.
\end{enumerate}
\end{Assumption}

In the Type-I confounding heterogeneity framework, we assume there are $K$ different unmeasured confounders. We must assume the same causal assumptions hold, conditioned on measured confounders and the $K$ unmeasured confounders. 

\begin{Assumption}[Causal assumptions for Type-I confounding heterogeneity framework]
\label{Assumption: Type-I CH}
\noindent
\begin{enumerate}
    \item (No additional confounders): $(r_{Ti11}, r_{Ci11}, \dots, r_{Ti1K}, r_{Ci1K},\dots,r_{Ti21}, r_{Ci21}, \dots, r_{Ti2K}, r_{Ci2K})$ $\indep (Z_{i1},Z_{i2}) |(\mathbf{x}_{i1}, \mathbf{x}_{i2}, u_{i11},\dots,u_{i1K}, u_{i21},\dots,u_{i2K})$ for all $i=1,\dots, I$. Additionally, $\{Y_{Cik}\}_{k=1}^{K}$ form a mutually independent set conditioned on $\mathbf{x}_{ij}, u_{ij1},\dots,u_{ijK}$ for all $j=1,2,i=1,\dots,I$.
    \item (Positivity) $0<P(Z_{ij}=1|\mathbf{x}_{ij}, u_{ij1},\dots,u_{ijK})<1$, for all $\mathbf{x}_{ij}, u_{ij1},\dots,u_{ijK}$.
    \item (Independence across matched pairs): $\{(Y_{i1},\dots,Y_{iK}, V_{i})\}_{i=1}^{I}$ form a mutually independent set conditioned on $\mathbf{x}_{ij}, u_{ij1},\dots,u_{ijK}$ for all $j=1,2,i=1,\dots,I$.
    \item (Consistency): $R_{ijk} = Z_{ij}r_{Tijk} + (1-Z_{ij})r_{Cijk}$.
\end{enumerate}
\end{Assumption}

\vspace{-0.3cm}

\subsection{Additional assumptions for theoretical analysis of power}

\begin{Assumption}[Assumptions for results related to power]
\label{Assumption: Power analysis}
\noindent
\begin{enumerate}
    \item (Constraints on $\kappa$ and sensitivity parameter values): $\kappa \in \mathbb{R}^{K}_{+}\setminus\{0\}$, and 
    
    $\Gamma, \Lambda, \Delta, \Lambda_{1},\dots,\Lambda_{K}, \Delta_{1},\dots,\Delta_{K} \geq 1$.
    \item (Sampling assumption in the favorable scenario) 
    \\ $(r_{Ti11}, r_{Ci11}, \dots, r_{Ti1K}, r_{Ci1K},\dots,r_{Ti21}, r_{Ci21}, \dots, r_{Ti2K}, r_{Ci2K},V_i)\overset{\mathrm{iid}}{\sim}F$, where $\mu_{F, \kappa} > 0$ for any $\kappa \in \mathbb{R}^{K}_{+} \setminus \{0\}$.
    \item (Positive probability of negative observed treatment effect): For a fixed choice of test statistic $q_{ik}$, we have $P_{F}(\sum_{k=1}^{K} \kappa_{k}q_{ik}\operatorname{sign}(Y_{ik}V_{i}) < 0) > 0$ for all pairs $i$, and for any fixed $\kappa \in \mathbb{R}^{K}_{+}\setminus\{0\}$.
    \item (Bounded fourth moments): $\mathbb{E}_{F}((\kappa_{k}q_{ik})^{4})<\infty$ for each outcome $k$. 
    \item (Fourth moments for the cumulant generating function): For fixed $\kappa \in \mathbb{R}^{K}_{+} \setminus \{0\}$, and any finite $T$ and $|t| \leq T$, we have $|\Psi_{I, \mathcal{P}, \kappa}^{(0)}(t)| \leq T\sum_{i=1}^{I}L_{i}, \hspace{2mm} |\Psi_{I, \mathcal{P}, \kappa}^{(1)}(t)| \leq \sum_{i=1}^{I}L_{i},$ $0 \leq \Psi_{I, \mathcal{P}, \kappa}^{(2)}(t) \leq \sum_{i=1}^{I}L_{i}^{2}$, $|\Psi_{I, \mathcal{P}, \kappa}^{(3)}(t)| \leq \sum_{i=1}^{I}8L_{i}^{3}$, and $|\Psi_{I, \mathcal{P}, \kappa}^{(4)}(t)| \leq \sum_{i=1}^{I}64L_{i}^{4}$, where $\mathbb{E}_{F}(L_{i}^{4}) < \infty$ for each $i=1,\dots,I$.
    Additionally, we have $c^{(d)}_{\mathcal{P}, \kappa}(t)$.
    \item (Non-degeneracy): For fixed $\kappa \in \mathbb{R}^{K}_{+} \setminus \{0\}$, we have $\overline{\sigma}^{2}_{I, \mathcal{P}, \mathcal{\kappa}} > 0, \overline{\sigma}^{2}_{ \mathcal{P}, \mathcal{\kappa}} > 0,$ and $\sigma^{2}_{F, \kappa} > 0$.
\end{enumerate}
\end{Assumption}

\begin{Assumption}[$q_{ik}$ is asymptotically a function of $|Y_{ik}V_{i}|$] 
\label{A: qik condition}
    Denote $q_{ik}:= q_{i}(\{|Y_{ik}V_{i}|\}_{i=1}^{I})$ for some function $q_{i}$ satisfying $q_{i}(0)=0$. There exists a function $\widetilde{q}_{i}:\mathbb{R}\mapsto\mathbb{R}$ such that, as $I\to\infty$, we have
    \begin{enumerate}
        \item $q_{i}(\{|Y_{ik}V_{i}|\}_{i=1}^{I}) \overset{a.s.}{\to} \widetilde{q}_{i}(|Y_{ik}V_{i}|)$, 
        \item $I^{-1}\sum_{i=1}^{I}\sum_{k=1}^{K}|q_{i}(\{|Y_{ik}V_{i}|\}_{i=1}^{I}) - \widetilde{q}_{i}(|Y_{ik}V_{i}|)|^{4} \overset{a.s}{\to} 0$.
        \item $\lim_{I\to\infty}I^{-1/2}\sum_{i=1}^{I}\sum_{k=1}^{K}|q_{i}(\{|Y_{ik}V_{i}|\}_{i=1}^{I}) - \widetilde{q}_{i}(|Y_{ik}V_{i}|)| \overset{P_{F}}{\to} 0$.     
    \end{enumerate}
\end{Assumption}

\vspace{-0.5cm}

\section{Supplementary Lemmas}

\begin{figure}[h]\label{fig: type I confounding heterogeneity}
     \centering
     \begin{tikzpicture}


    \node[draw, minimum height=0.7cm] at (-1.9, 5) {Assumption~\ref{Assumption: CLT}};
    \draw[-Stealth] (-0.3, 5) -- (0, 3.85);
    \node[draw, minimum height=0.7cm] at (5.13, 5) {Assumption~\ref{Assumption: Separable approximation}};
    \draw[-Stealth] (6.8, 5) -- (7.38, 3.85);


    \node[draw, minimum height=0.7cm] at (1.25, 3.75) {Theorem 1};
    \draw[-Stealth] (2.50, 3.75) -- (3.50, 3.75);
    \node[draw, minimum height=0.7cm] at (4.75, 3.75) {Theorem 2};
    \draw[-Stealth] (6.05, 3.75) -- (7.38, 3.75);
    \node[draw, minimum height=0.7cm] at (8.65, 3.75) {Theorem 3};

    \node[draw, minimum height=0.7cm, align=center] at (-2.27, 2.5) {Lemma~\ref{Lem: Exp_var T_k, classic} \\[-8pt]
    Lemma~\ref{Lem: Exp_var T_k, amp & Type-II CH}};
    \draw[-Stealth] (-1.05, 2.5) -- (0, 3.75);
    \draw[-Stealth] (-1.05, 2.5) -- (3.5, 2.5);
    \draw[-Stealth] (-1.05, 2.5) -- (0, 1.35);
    \node[draw, minimum height=0.7cm, align=center] at (4.86, 2.5) {Lemma~\ref{Lem: Sep approx, classic} \\[-8pt]
    Lemma~\ref{Lem: Sep approx, amp}};    
    \draw[-Stealth] (6.25, 2.5) -- (7.38, 3.65);

    \node[draw, minimum height=0.7cm, align=center] at (-2.27, 1.25) {Lemma~\ref{Lem: Exp_var bounding stat, classic} \\[-8pt]
    Lemma~\ref{Lem: Exp_var bounding stat, amp and CH}};
    \draw[-Stealth] (-1.05, 1.25) -- (0, 1.25);
    \node[draw, minimum height=0.7cm] at (1.35, 1.25) {Lemma~\ref{Lem: bounding variable verification}};
    \draw[-Stealth] (2.65, 1.25) -- (3.5, 1.25);
    \draw[-Stealth] (2.65, 1.25) -- (3.5, 0);
    \node[draw, minimum height=0.7cm] at (4.63, 1.25) {Lemma 1};

    \draw[arrows=-] (-3.4, 0.52) -- (9.5, 0.52);

    \node[draw, minimum height=0.7cm] at (4.79, 0) {Theorem 4};
    \draw[-Stealth] (6.08, 0) -- (7.38, 0);
    \node[draw, minimum height=0.7cm] at (8.5, 0) {Lemma 2};

    \node[draw, minimum height=0.7cm] at (-2.22, -1.25) {Section B.3};
    \draw[-Stealth] (-0.9,-1.25) -- (0, -1.25);
    \node[draw, minimum height=0.7cm] at (1.25, -1.25) {Theorem 5};

    \node[draw, minimum height=0.7cm] at (-2.27, -2.5) {Lemma~\ref{Lem: Bahadur-Rosenbaum slope convex in Gamma}};
    \draw[-Stealth] (-1.05,-2.5) -- (-0.05, -3.65);

    \node[draw, minimum height=0.7cm] at (-2.27, -3.75) {Lemma~\ref{Lem: Bahadur-Rosenbaum slope CH strictly decreasing in Delta}};
    \draw[-Stealth] (-1.05,-3.75) -- (-0.05, -3.75);
    \node[draw, minimum height=0.7cm] at (1.25, -3.75) {Theorem 6};


    \end{tikzpicture}
    \caption{Map of theoretical results. Each result uses the causal Assumptions~\ref{Assumption: classic} through~\ref{Assumption: Type-I CH} appropriate to the framework considered, while only the theoretical results below the horizontal line additionally require Assumptions~\ref{Assumption: Power analysis} and~\ref{A: qik condition} to hold.}
    \label{fig: theory map}
\end{figure}
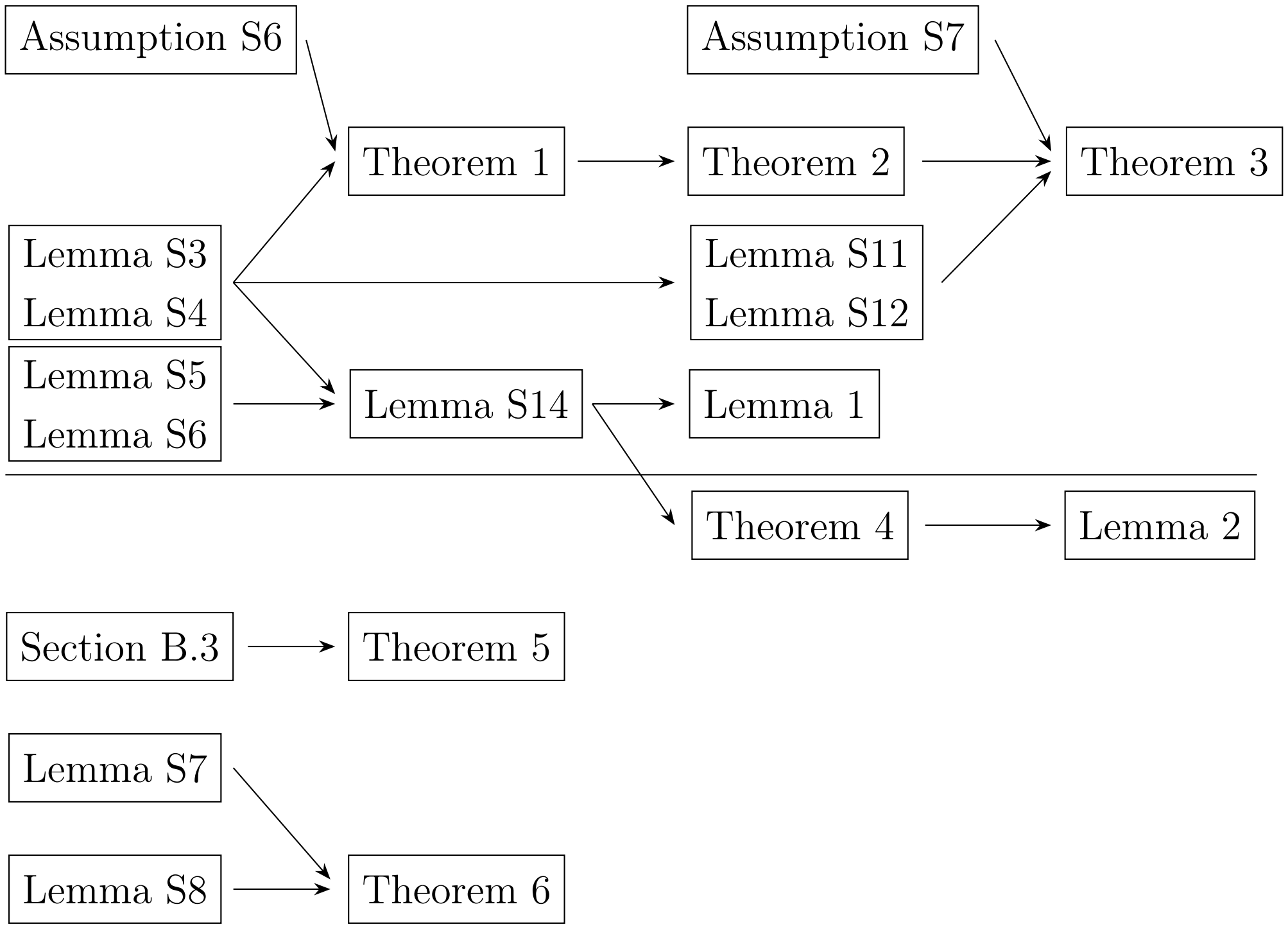

\vspace{-0.3cm}

\subsection{Expected Value and Variance of $T_{k}$ Under Different Sensitivity Analysis Frameworks}

\begin{Lemma}[Expected value and variance of $T_{k}$ under $\mathbf{H}_{0}$, classic framework]
\label{Lem: Exp_var T_k, classic}
    Under Rosenbaum's $\Gamma$ framework and Assumption~\ref{Assumption: classic}, and assuming the global null hypothesis $\mathbf{H}_{0}$ holds, the expected value of the test statistic $T_{k}$ is
    \begin{equation*}
    \mu_{\{\Gamma\}, k, \mathbf{u}}:=\mathbb{E}(T_{k}|\boldsymbol{\mathcal{F}}, \mathcal{Z}) = \sum_{i=1}^{I}q_{ik}\operatorname{sign}(Y_{Cik})\frac{\exp\{\gamma(u_{i1}-u_{i2})\} - 1}{1+\exp\{\gamma(u_{i1}-u_{i2})\}}.
    \end{equation*}
    and the covariance of test statistics $T_{k}, T_{k'}$ under Rosenbaum's $\Gamma$ sensitivity model is given by
    \begin{equation*}
    \mathrm{Cov}_{\{\Gamma\}, k, k', \mathbf{u}}(T_{k}, T_{k'}|\boldsymbol{\mathcal{F}}, \mathcal{Z}) = \sum_{i=1}^{I}q_{ik}q_{ik'}\operatorname{sign}(Y_{Cik})\operatorname{sign}(Y_{Cik'})\left(1-\left(\frac{\exp\{\gamma(u_{i1}-u_{i2})\} - 1}{1+\exp\{\gamma(u_{i1}-u_{i2})\}}\right)^{2}\right),
    \end{equation*}
    where $\gamma = \log(\Gamma)$.
\end{Lemma}
\begin{proof}
First, by Assumption~\ref{Assumption: classic}, $q_{ik}$ is fixed conditioned on $\boldsymbol{\mathcal{F}}, \mathcal{Z}$. So, the expectation of $T_{k}$ can be rewritten as:
\begin{align*}
\mathbb{E}(T_{k}|\boldsymbol{\mathcal{F}}, \mathcal{Z}) & = \sum_{i=1}^{I}\mathbb{E}(q_{ik}\operatorname{sign}(V_{i}Y_{Cik})|\boldsymbol{\mathcal{F}}, \mathcal{Z})  = \sum_{i=1}^{I}q_{ik}\mathbb{E}(\operatorname{sign}(V_{i}Y_{Cik})|\boldsymbol{\mathcal{F}}, \mathcal{Z}).
\end{align*} \allowdisplaybreaks
Now, note:
\begin{align*}
\mathbb{E}(\operatorname{sign}(V_{i}Y_{Cik})|\boldsymbol{\mathcal{F}}, \mathcal{Z}) & = P(\operatorname{sign}(V_{i}Y_{Cik}) = 1|\boldsymbol{\mathcal{F}}, \mathcal{Z}) - P(\operatorname{sign}(V_{i}Y_{Cik}) = -1|\boldsymbol{\mathcal{F}}, \mathcal{Z}) \\
& = \big[P(\{\operatorname{sign}(V_{i}) = 1 \}\cap\{\operatorname{sign}(Y_{Cik}) = 1 \}|\boldsymbol{\mathcal{F}}, \mathcal{Z})\} \\
& + P(\{\operatorname{sign}(V_{i}) = -1 \}\cap\{\operatorname{sign}(Y_{Cik}) = -1 \}|\boldsymbol{\mathcal{F}}, \mathcal{Z})\}  \big] \\
& -  \big[P(\{\operatorname{sign}(V_{i}) = 1 \}\cap\{\operatorname{sign}(Y_{Cik}) = -1 \}|\boldsymbol{\mathcal{F}}, \mathcal{Z})\} \\
& + P(\{\operatorname{sign}(V_{i}) = -1 \}\cap\{\operatorname{sign}(Y_{Cik}) = 1 \}|\boldsymbol{\mathcal{F}}, \mathcal{Z})\}  \big] \\
& = \mathbb{I}\{\operatorname{sign}(Y_{Cik}) = 1\}[P(\operatorname{sign}(V_{i}) = 1|\boldsymbol{\mathcal{F}}, \mathcal{Z}) - P(\operatorname{sign}(V_{i}) = -1|\boldsymbol{\mathcal{F}}, \mathcal{Z})] \\
& + \mathbb{I}\{\operatorname{sign}(Y_{Cik}) = -1\}[P(\operatorname{sign}(V_{i}) = -1|\boldsymbol{\mathcal{F}}, \mathcal{Z}) - P(\operatorname{sign}(V_{i}) = 1|\boldsymbol{\mathcal{F}}, \mathcal{Z})] \\
& = \{\mathbb{I}\{\operatorname{sign}(Y_{Cik}) = 1\}-\mathbb{I}\{\operatorname{sign}(Y_{Cik}) = -1\}\}\frac{\exp\{\gamma(u_{i1}-u_{i2})\} - 1}{1+\exp\{\gamma(u_{i1}-u_{i2})\}} \\
& = \operatorname{sign}(Y_{Cik})\frac{\exp\{\gamma(u_{i1}-u_{i2})\} - 1}{1+\exp\{\gamma(u_{i1}-u_{i2})\}}.
\end{align*}
The first equality follows from the law of total probability, the third equality follows from $\operatorname{sign}(Y_{Cik})$ being in the sigma algebra generated by $\boldsymbol{\mathcal{F}}$, the fourth equality leverages Rosenbaum's $\Gamma$ sensitivity model, and the last equality uses the fact that $\operatorname{sign}(x)\equiv\{\mathbb{I}\{\operatorname{sign}(x) = 1\} - \mathbb{I}\{\operatorname{sign}(x) = -1\} \}$. Now, note:
\begin{align*}
\mathbb{E}(q_{ik}q_{ik'}\operatorname{sign}(V_{i}Y_{Cik})\operatorname{sign}(V_{i}Y_{Cik'})|\boldsymbol{\mathcal{F}}, \mathcal{Z}) & = \mathbb{E}(q_{ik}q_{ik'}\operatorname{sign}(Y_{Cik})\operatorname{sign}(Y_{Cik'})|\boldsymbol{\mathcal{F}}, \mathcal{Z}) \\
& = q_{ik}q_{ik'}\operatorname{sign}(Y_{Cik})\operatorname{sign}(Y_{Cik'}).
\end{align*}
The first equality uses the identity $\operatorname{sign}(ax)\operatorname{sign}(ay)=\operatorname{sign}(xy)$, which holds since $V_{i} \in \{-1, 1\}$. By independence of matched pairs, we have
\begin{align*}
\mathrm{Cov}_{\{\Gamma\}, k, k', \mathbf{u}}(T_{k}, T_{k'}|\boldsymbol{\mathcal{F}}, \mathcal{Z}) & = \sum_{i=1}^{I}\mathrm{Cov}\{q_{ik}\operatorname{sign}(Y_{Cik}V_{i}), q_{ik'}\operatorname{sign}(Y_{Cik'}V_{i})|\boldsymbol{\mathcal{F}}, \mathcal{Z}\} \\
& = \sum_{i=1}^{I} q_{ik}q_{ik'}\operatorname{sign}(Y_{Cik})\operatorname{sign}(Y_{Cik'})\mathrm{Var}(V_{i} | \boldsymbol{\mathcal{F}}, \mathcal{Z}) \\
& = \sum_{i=1}^{I}q_{ik}q_{ik'}\operatorname{sign}(Y_{Cik})\operatorname{sign}(Y_{Cik'})\left(1-\left(\frac{\exp\{\gamma(u_{i1}-u_{i2})\} - 1}{1+\exp\{\gamma(u_{i1}-u_{i2})\}}\right)^{2}\right).
\end{align*}
\end{proof}

\begin{Lemma}[Expected value and variance of $T_{k}$ under $\mathbf{H}_{0}$, amplification and confounding heterogeneity frameworks]
\label{Lem: Exp_var T_k, amp & Type-II CH}
    Under the Type-II confounding heterogeneity sensitivity analysis framework and Assumption~\ref{Assumption: Amp and Type-II CH}, and assuming the global null hypothesis $\mathbf{H}_{0}$ holds, the expected value of the test statistic $T_{k}$ is:
    \begin{equation*}
    \mu_{\{\Lambda, \boldsymbol{\Delta}\}, k, \mathbf{u}}:=E(T_{k}|\mathcal{C}, \boldsymbol{\mathcal{A}}, \mathcal{Z})=\sum_{i=1}^{I}q_{ik}\cdot \frac{[\exp\{\lambda(u_{i1}-u_{i2}) \} - 1][\exp\{\delta_{k}(u_{i1}-u_{i2})\} - 1]}{[1+\exp\{\lambda(u_{i1}-u_{i2})\}][1+\exp\{\delta_{k}(u_{i1}-u_{i2})\}]}
    \end{equation*}
    The variance of the test statistic $T_{k}$ is:
    \begin{align*}
     \sigma_{\{\Lambda, \boldsymbol{\Delta}\},k,\mathbf{u}}^{2}:=&\mathrm{Var}(T_{k}|\mathcal{C}, \boldsymbol{\mathcal{A}}, \mathcal{Z})\\
     =& \sum_{i=1}^I q_{ik}^2\bigg(1 -  \bigg(\frac{[\exp\{\lambda(u_{i1}-u_{i2})\}-1]}{[1+\exp\{\lambda(u_{i1}-u_{i2})\}]}\frac{[\exp\{\delta_{k}(u_{i1}-u_{i2})\}-1]}{[1+\exp\{\delta_{k}(u_{i1}-u_{i2})\}]}\bigg)^{2}\bigg).
    \end{align*}
    If $k \neq k'$, the covariance of the test statistics $T_{k}, T_{k'}$ is:
    \begin{align*}
    & \sigma_{\{\Lambda, \boldsymbol{\Delta}\},k,k',\mathbf{u}} :=\mathrm{Cov}(T_{k}, T_{k'}|\mathcal{C}, \boldsymbol{\mathcal{A}}, \mathcal{Z})\\
    =&\sum_{i=1}^I q_{ik}q_{ik'}\frac{[\exp\{\delta_{k}(u_{i1}-u_{i2})\}-1]}{[1+\exp\{\delta_{k}(u_{i1}-u_{i2})\}]}\frac{[\exp\{\delta_{k'}(u_{i1}-u_{i2})\}-1]}{[1+\exp\{\delta_{k'}(u_{i1}-u_{i2})\}]}\bigg(1-\left(\frac{\exp\{\lambda(u_{i1}-u_{i2})\}-1}{1+\exp\{\lambda(u_{i1}-u_{i2})\}}\right)^{2}\bigg),
    \end{align*}
where $\lambda = \log(\Lambda), \delta_{k} = \log(\Delta_{k})$. The expected value and variance of $T_{k}$ under the amplification framework are obtained by setting $\Delta_{1},\dots,\Delta_{K} = \Delta$, and the expected value and variance of $T_{k}$ under Type-I confounding heterogeneity are obtained by replacing $\lambda(u_{i1} - u_{i2})$ with $\sum_{\ell = 1}^{K}\lambda_{\ell}(u_{i1\ell} - u_{i2\ell})$ and $\delta_{k}(u_{i1} - u_{i2})$ with $\delta_{k}(u_{i1k} - u_{i2k})$.
\end{Lemma}
\begin{proof}
Let $\mathbf{Y}_{Ck} := (Y_{C1k},\dots,Y_{CIk})$ and $\mathbf{S}_{k}:=(\operatorname{sign}(Y_{C1k}),...,\operatorname{sign}(Y_{CIk}))$. Recall that under the global null hypothesis, $Y_{ik} = Y_{Cik}$, so define $\mathcal{A}_{Ck}=(|Y_{C1k}|,\dots,|Y_{CIk}|)$ and $\boldsymbol{\mathcal{A}}_{C} = \cup_{k=1}^{K}\mathcal{A}_{Ck}$. By Assumption~\ref{Assumption: Amp and Type-II CH}, under the global null, we have $\mathbf{Y}_{Ck} \indep \mathbf{Z}|\mathcal{C}$. Then, by the outcome models under the amplification and confounding heterogeneity frameworks, for any $a > 0$, for all outcomes $k=1,\dots,K$, we have
\begin{equation*}
P(\operatorname{sign}(Y_{Cik}) = 1 | \mathcal{C}, \mathcal{Z}, |Y_{Cik}| = a) = \frac{\exp\{\delta_{k}(u_{i1} - u_{i2})\}}{1 + \exp\{\delta_{k}(u_{i1} - u_{i2})\}}.
\end{equation*}
Therefore $\mathbf{S}_{k} \indep \mathcal{A}_{Ck}|\mathcal{C}$ whenever $|Y_{Cik}| \neq 0$. That is, the sign of $Y_{Cik}$ is independent of its magnitude, conditioned on $\mathcal{C}$. Combining this fact and Assumption~\ref{Assumption: Amp and Type-II CH}, we have $\mathbf{S}_{k} \indep \mathbf{Z}|(\mathcal{C}, \mathcal{A}_{Ck}, \mathcal{Z})$ whenever $|Y_{Cik}| \neq 0$. Then, by Assumption~\ref{Assumption: Amp and Type-II CH}, we have $\mathbf{S}_{k} \indep \mathbf{Z}|(\mathcal{C}, \boldsymbol{\mathcal{A}}_{C}, \mathcal{Z})$ whenever $|Y_{Cik}| \neq 0$. Thus, 
\begin{align*}
\mathbb{E}(\operatorname{sign}(V_{i}Y_{Cik})|(\mathcal{C}, \boldsymbol{\mathcal{A}}_{C}, \mathcal{Z})) & = \mathbb{I}\{|Y_{Cik}| \neq 0\} \cdot \{\{P(\operatorname{sign}(V_{i})=1|(\mathcal{C}, \boldsymbol{\mathcal{A}}_{C}, \mathcal{Z}))P(\operatorname{sign}(Y_{Cik})=1|(\mathcal{C}, \boldsymbol{\mathcal{A}}_{C}, \mathcal{Z})) \\
& + P(\operatorname{sign}(V_{i})=-1|(\mathcal{C}, \boldsymbol{\mathcal{A}}_{C}, \mathcal{Z}))P(\operatorname{sign}(Y_{Cik})=-1|(\mathcal{C}, \boldsymbol{\mathcal{A}}_{C}, \mathcal{Z}))\} \\
& - \{P(\operatorname{sign}(V_{i})=1|(\mathcal{C}, \boldsymbol{\mathcal{A}}_{C}, \mathcal{Z}))P(\operatorname{sign}(Y_{Cik})=-1|(\mathcal{C}, \boldsymbol{\mathcal{A}}_{C}, \mathcal{Z})) \\
& + P(\operatorname{sign}(V_{i})=-1|(\mathcal{C}, \boldsymbol{\mathcal{A}}_{C}, \mathcal{Z}))P(\operatorname{sign}(Y_{Cik})=1|(\mathcal{C}, \boldsymbol{\mathcal{A}}_{C}, \mathcal{Z}))\}\} \\
& + \mathbb{I}\{|Y_{Cik}| = 0\}\cdot \mathbb{E}(\operatorname{sign}(V_{i}Y_{Cik})|(\mathcal{C}, \boldsymbol{\mathcal{A}}_{C}, \mathcal{Z})) \\
& = \{P(\operatorname{sign}(V_{i})=1|(\mathcal{C}, \boldsymbol{\mathcal{A}}_{C}, \mathcal{Z})) - P(\operatorname{sign}(V_{i})=-1|(\mathcal{C}, \boldsymbol{\mathcal{A}}_{C}, \mathcal{Z}))\} \\
& \cdot \{P(\operatorname{sign}(Y_{Cik})=1|(\mathcal{C}, \boldsymbol{\mathcal{A}}_{C}, \mathcal{Z})) - P(\operatorname{sign}(Y_{Cik})=-1|(\mathcal{C}, \boldsymbol{\mathcal{A}}_{C}, \mathcal{Z}))\}.
\end{align*}
The first equality follows from $\mathbf{S}_{k} \indep \mathbf{Z}|(\mathcal{C}, \boldsymbol{\mathcal{A}}_{C}, \mathcal{Z})$ whenever $|Y_{Cik}| \neq 0$, the second equality follows after simplification. Now, note:
\begin{align*}
&~ P(\operatorname{sign}(V_{i})=1|(\mathcal{C}, \boldsymbol{\mathcal{A}}_{C}, \mathcal{Z})) - P(\operatorname{sign}(V_{i})=-1|(\mathcal{C}, \boldsymbol{\mathcal{A}}_{C}, \mathcal{Z})) \\
 =&~ P(\operatorname{sign}(V_{i})=1|\mathcal{C}, \mathcal{Z}) - P(\operatorname{sign}(V_{i})=-1|\mathcal{C}, \mathcal{Z}) = \frac{(\exp\{\lambda(u_{i1}-u_{i2})\}-1)}{(1+\exp\{\lambda(u_{i1}-u_{i2})\})},
\end{align*}
where the first equality follows from $\mathbf{Y}_{Ck} \indep \mathbf{Z}|\mathcal{C} \implies |\mathbf{Y}_{Ck}| \indep \mathbf{Z}|\mathcal{C}$, and the second equality follows from the sensitivity model on $\mathbf{Z}$. Additionally, we have
\begin{align*}
&~ P(\operatorname{sign}(Y_{Cik})=1|(\mathcal{C}, \boldsymbol{\mathcal{A}}_{C}, \mathcal{Z})) - P(\operatorname{sign}(Y_{Cik})=-1|(\mathcal{C}, \boldsymbol{\mathcal{A}}_{C}, \mathcal{Z})) \\
 =&~ P(\operatorname{sign}(Y_{Cik})=1|\mathcal{C}, \mathcal{Z}) - P(\operatorname{sign}(Y_{Cik})=-1|\mathcal{C}, \mathcal{Z}) \\
= &~ \frac{(\exp\{\delta_{k}(u_{i1}-u_{i2})\} - 1)}{(1+\exp\{\delta_{k}(u_{i1}-u_{i2})\})},
\end{align*}
where the first equality follows from $\mathbf{S}_{k} \indep \boldsymbol{\mathcal{A}}_{C}|\mathcal{C}$ whenever $|Y_{Cik}| \neq 0$ and the second equality uses the sensitivity model on $\operatorname{sign}(Y_{Cik})$. Next, we prove the forms for the variance and covariance of the test statistic. Note \allowdisplaybreaks
\begin{align*}
\mathrm{Cov}(T_{k}, T_{k'}|\mathcal{C}, \boldsymbol{\mathcal{A}}_{C}, \mathcal{Z}) & = \sum_{i=1}^{I}\mathrm{Cov}\left(q_{ik} \operatorname{sign}(V_{i}Y_{Cik}), q_{ik'}\operatorname{sign}(V_{i}Y_{Cik'})|\mathcal{C}, \boldsymbol{\mathcal{A}}_{C}, \mathcal{Z}\right) \\
& = \sum_{i=1}^{I}\big\{\mathbb{E}(q_{ik} \operatorname{sign}(V_{i}Y_{Cik})q_{ik'}\operatorname{sign}(V_{i}Y_{Cik'})|\mathcal{C}, \boldsymbol{\mathcal{A}}_{C}, \mathcal{Z}) \\
& - \mathbb{E}(q_{ik}\operatorname{sign}(V_{i}Y_{Cik})|\mathcal{C}, \boldsymbol{\mathcal{A}}_{C}, \mathcal{Z})\big\}\mathbb{E}(q_{ik'}\operatorname{sign}(V_{i}Y_{Cik'})|\mathcal{C}, \boldsymbol{\mathcal{A}}_{C}, \mathcal{Z})\big\} \\
& = \sum_{i=1}^{I}q_{ik}q_{ik'}\mathbb{E}(\operatorname{sign}(V_{i}Y_{Cik})\operatorname{sign}(V_{i}Y_{Cik'})|\mathcal{C}, \boldsymbol{\mathcal{A}}_{C}, \mathcal{Z}) \\
& - q_{ik}q_{ik'}\frac{[\exp\{\delta_{k}(u_{i1}-u_{i2})\}-1][\exp\{\delta_{k'}(u_{i1}-u_{i2})\}-1]}{[1+\exp\{\delta_{k}(u_{i1}-u_{i2})\}][1+\exp\{\delta_{k'}(u_{i1}-u_{i2})\}]} \\
& \cdot \left(\frac{\exp\{\lambda(u_{i1}-u_{i2})\}-1}{1+\exp\{\lambda(u_{i1}-u_{i2})\}}\right)^{2}.
\end{align*}
The first equality follows from independence of matched pairs described in Assumption~\ref{Assumption: Amp and Type-II CH}, the second equality uses the definition of covariance, and the third equality uses the fact that $q_{ik}, q_{ik'}$ are functions of $\boldsymbol{\mathcal{A}}_{C}$. Now, if $k \neq k'$, note:
\begin{align*}
\mathbb{E}(\operatorname{sign}(V_{i}Y_{Cik})\operatorname{sign}(V_{i}Y_{Cik'})|\mathcal{C}, \boldsymbol{\mathcal{A}}_{C}, \mathcal{Z}) 
& = \mathbb{E}(\operatorname{sign}(V_{i})^{2} \operatorname{sign}(Y_{Cik}Y_{Cik'})|\mathcal{C}, \boldsymbol{\mathcal{A}}_{C}, \mathcal{Z}) \\
& = \mathbb{E}(\operatorname{sign}(Y_{Cik})|\mathcal{C}, \boldsymbol{\mathcal{A}}_{C}, \mathcal{Z})\mathbb{E}(\operatorname{sign}(Y_{Cik'})|\mathcal{C}, \boldsymbol{\mathcal{A}}_{C}, \mathcal{Z}) \\
& = \frac{[\exp\{\delta_{k'}(u_{i1}-u_{i2})\} - 1][\exp\{\delta_{k}(u_{i1}-u_{i2})\}-1]}{[1+\exp\{\delta_{k'}(u_{i1}-u_{i2})\}][1+\exp\{\delta_{k}(u_{i1}-u_{i2})\}]}.
\end{align*}
The first and second equalities follow from Assumption~\ref{Assumption: Amp and Type-II CH}, and the last equality uses the sensitivity model for $Y_{Cik}, Y_{Cik'}$. If $k = k'$, then
\begin{align*} &q_{ik}q_{ik'}\mathbb{E}(\operatorname{sign}(V_{i}Y_{Cik})\operatorname{sign}(V_{i}Y_{Cik'})|\mathcal{C}, \boldsymbol{\mathcal{A}}_{C}, \mathcal{Z})  \\
 =&~ q_{ik}^2\mathbb{E}(\operatorname{sign}(V_{i}Y_{Cik})^{2}|\mathcal{C}, \boldsymbol{\mathcal{A}}_{C}, \mathcal{Z}) \\
 =& ~(\mathbb{I}\{q_{ik} \neq 0\} + \mathbb{I}\{q_{ik} = 0\})q_{ik}^2\mathbb{E}(\operatorname{sign}(Y_{Cik})^{2}|\mathcal{C}, \boldsymbol{\mathcal{A}}_{C}, \mathcal{Z}) \\
  =&~ q_{ik}^2\mathbb{E}(\mathbb{I}\{q_{ik} \neq 0\} \cdot \operatorname{sign}(Y_{Cik})^{2}|\mathcal{C}, \boldsymbol{\mathcal{A}}_{C}, \mathcal{Z}) + \mathbb{I}\{q_{ik} = 0\}(0)^2(0) = q_{ik}^2(1) = q_{ik}^2.
\end{align*}
Under the amplification framework (without confounding heterogeneity), simply replace each $\delta_{k}, \Delta_{k}$ with a common $\delta, \Delta$ to describe the expected value and variance of $T_{k}$. Under Type-I confounding heterogeneity, simply replace each $\lambda(u_{i1} - u_{i2})$ with $\sum_{\ell = 1}^{K}\lambda_{\ell}(u_{i1\ell} - u_{i2\ell})$, each $\delta_{k}(u_{i1}-u_{i2})$ with $\delta_{k}(u_{i1k}-u_{i2k})$, assume Assumption~\ref{Assumption: Type-I CH} holds, and replace $\mathcal{C}$ with $\boldsymbol{\mathcal{C}}$ in all conditioning sets.
\end{proof}

\vspace{-0.3cm}

\subsection{Expected Value and Variance of $\overline{T}_{\kappa}$ Under Different Sensitivity Analysis Frameworks}

\begin{Lemma}[Expected value and variance of $\overline{T}_{\kappa}$ classic framework]
\label{Lem: Exp_var bounding stat, classic}
Recall $|\sum_{k=1}^{K}\kappa_{k} q_{ik}\operatorname{sign}(Y_{ik})|:=\eta_{\kappa i}$. Then,
\begin{align*}
    \mathbb{E}(\overline{T}_{\{\Gamma\}, \kappa}|\boldsymbol{\mathcal{F}}, \mathcal{Z}) &:= \overline{\mu}_{I, \{\Gamma\}, \kappa} = \frac{\Gamma - 1}{1 + \Gamma}I^{-1}\sum_{i=1}^{I}\eta_{\kappa i}, \\
    \mathrm{Var}(\overline{T}_{\{\Gamma\}, \kappa}|\boldsymbol{\mathcal{F}}, \mathcal{Z}) &:= \overline{\sigma}_{I, \{\Gamma\}, \kappa}^{2} = \frac{4\Gamma}{(1+\Gamma)^2}I^{-2}\sum_{i=1}^{I}\eta_{\kappa i}^{2},
\end{align*}
\end{Lemma}
\begin{proof}
Recall that $\overline{T}_{\{\Gamma\}, \kappa} = I^{-1}\sum_{i=1}^{I}b_{i}$ where $b_{i}$ takes value $\eta_{\kappa i}$ with probability $\Gamma / (1+\Gamma)$ and value $-\eta_{\kappa i}$ with probability $1 / (1+\Gamma)$, and $b_{i} \indep b_{i'}$ for all $i \neq i'$. Thus, 
\begin{align*}
& \mathbb{E}(\overline{T}_{\{\Gamma\}, \kappa}|\boldsymbol{\mathcal{F}}, \mathcal{Z}) = I^{-1}\sum_{i=1}^{I}\mathbb{E}(b_{i}|\boldsymbol{\mathcal{F}}, \mathcal{Z}) = I^{-1}\sum_{i=1}^{I}\frac{\Gamma}{1+\Gamma}\eta_{\kappa i} - \frac{1}{1+\Gamma}\eta_{\kappa i} = \frac{\Gamma - 1}{1 + \Gamma}I^{-1}\sum_{i=1}^{I}\eta_{\kappa i}, \\
& \mathrm{Var}(\overline{T}_{\{\Gamma\}, \kappa}|\boldsymbol{\mathcal{F}}, \mathcal{Z}) = I^{-2}\sum_{i=1}^{I}\mathrm{Var}(b_{i}|\boldsymbol{\mathcal{F}}, \mathcal{Z}) = I^{-2}\sum_{i=1}^{I} 4\frac{\Gamma}{1+\Gamma}\cdot \frac{1}{1+\Gamma}\eta_{\kappa, i}^{2}  = \frac{4\Gamma}{(1+\Gamma)^2}I^{-2}\sum_{i=1}^{I}\eta_{\kappa i}^{2}.
\end{align*}
\end{proof}

\begin{Lemma}[Expected value and variance of $\overline{T}_{\kappa}$ amplification and confounding heterogeneity frameworks]
\label{Lem: Exp_var bounding stat, amp and CH}
Recall $\kappa_{k} q_{ik}:=\phi_{\kappa i k}$. We prove the form of the expected value and variance of $\overline{T}_{\kappa}$ under the Type-I confounding heterogeneity framework, as the form under Type-II confounding heterogeneity and the amplification framework follow as special cases. Recall that $\overline{T}_{\{\boldsymbol{\Lambda}, \boldsymbol{\Delta}\}, \kappa}=I^{-1}\sum_{i=1}^{I}\sum_{k=1}^K\phi_{\kappa_{ik}}b_{ik}v_{i}$, where $\{b_{ik}\}_{i=1,\dots,I, k=1,\dots,K}$ are mutually independent, $v_{i} \indep v_{i'}$ for all $i \neq i'$, $b_{ik} \indep v_{i}$ for all $i$ and $k$, and
    \begin{align*}
        P(b_{ik} = 1) & = \frac{\Delta_{k}}{\Delta_{k} + 1}, P(v_{i} = 1) = \frac{(\prod_{\ell=1}^{K}\Lambda_{\ell})}{(\prod_{\ell=1}^{K}\Lambda_{\ell}) + 1},\implies \overline{p}_{\{\boldsymbol{\Lambda}, \boldsymbol{\Delta}\}, k} = \frac{(\prod_{\ell=1}^{K}\Lambda_{\ell}) \Delta_{k} + 1}{((\prod_{\ell=1}^{K}\Lambda_{\ell})  + 1)(\Delta_{k} + 1)}.
    \end{align*}
Then, we have
\begin{align*}
    \mathbb{E}(\overline{T}_{\{\boldsymbol{\Lambda}, \boldsymbol{\Delta}\}, \kappa}|\boldsymbol{\mathcal{C}}, \boldsymbol{\mathcal{A}}, \mathcal{Z}) := \overline{\mu}_{I, \{\boldsymbol{\Lambda}, \boldsymbol{\Delta}\}, \kappa} & = I^{-1}\frac{((\prod_{\ell=1}^{K}\Lambda_{\ell}) - 1)}{((\prod_{\ell=1}^{K}\Lambda_{\ell}) + 1)}\sum_{i=1}^{I}\sum_{k=1}^{K}\frac{(\Delta_{k} - 1)}{(\Delta_{k} + 1)}\phi_{\kappa ik}, \\
    \mathrm{Var}(\overline{T}_{\{\boldsymbol{\Lambda}, \boldsymbol{\Delta}\}, \kappa}|\boldsymbol{\mathcal{C}}, \boldsymbol{\mathcal{A}}, \mathcal{Z}) := \overline{\sigma}_{I, \{\boldsymbol{\Lambda}, \boldsymbol{\Delta}\}, \kappa}^{2} & = I^{-2}\sum_{i=1}^{I} \bigg\{ \sum_{k=1}^{K}\frac{4((\prod_{\ell=1}^{K}\Lambda_{\ell}) + \Delta_{k})((\prod_{\ell=1}^{K}\Lambda_{\ell}) \Delta_{k} + 1)}{(((\prod_{\ell=1}^{K}\Lambda_{\ell}) + 1)(\Delta_{k} + 1))^{2}}\phi_{\kappa ik}^{2} \\
    & + \sum_{k \neq k'}^{K} \phi_{\kappa i k}\phi_{\kappa i k'} \bigg(\frac{(\Delta_{k} - 1)(\Delta_{k'} - 1)}{(\Delta_{k} + 1)(\Delta_{k'} + 1)}\bigg)\bigg(1 - \bigg(\frac{(\prod_{\ell=1}^{K}\Lambda_{\ell}) - 1}{(\prod_{\ell=1}^{K}\Lambda_{\ell}) + 1}\bigg)^2\bigg)\bigg\}.
\end{align*}
\end{Lemma}
\allowdisplaybreaks
\begin{proof} First, we show the expected value and variance of $\overline{T}_{\{\boldsymbol{\Lambda}, \boldsymbol{\Delta}\}, \kappa}$ in the Type-I confounding heterogeneity framework:
\begin{align*}
& \mathbb{E}(\overline{T}_{\{\boldsymbol{\Lambda}, \boldsymbol{\Delta}\}, \kappa}|\boldsymbol{\mathcal{C}}, \boldsymbol{\mathcal{A}}, \mathcal{Z}) = I^{-1}\sum_{i=1}^{I}\sum_{k=1}^{K} \overline{p}_{\{\boldsymbol{\Lambda}, \boldsymbol{\Delta}\}, k}\phi_{\kappa ik} - (1-\overline{p}_{\{\boldsymbol{\Lambda}, \boldsymbol{\Delta}\}, k})\phi_{\kappa ik} \\
& = I^{-1}\sum_{i=1}^{I}\sum_{k=1}^{K} (2\overline{p}_{\{\boldsymbol{\Lambda}, \boldsymbol{\Delta}\}, k}-1)\phi_{\kappa ik}  = I^{-1}\frac{((\prod_{\ell=1}^{K}\Lambda_{\ell}) - 1)}{((\prod_{\ell=1}^{K}\Lambda_{\ell}) + 1)}\sum_{i=1}^{I}\sum_{k=1}^{K}\frac{(\Delta_{k} - 1)}{(\Delta_{k} + 1)}\phi_{\kappa ik}, \\
& \mathrm{Var}(\overline{T}_{\{\boldsymbol{\Lambda}, \boldsymbol{\Delta}\}, \kappa}|\boldsymbol{\mathcal{C}}, \boldsymbol{\mathcal{A}}, \mathcal{Z}) = I^{-2}\sum_{i=1}^{I}\bigg\{\sum_{k=1}^{K} 4\overline{p}_{\{\boldsymbol{\Lambda}, \boldsymbol{\Delta}\}, k}(1-\overline{p}_{\{\boldsymbol{\Lambda}, \boldsymbol{\Delta}\}, k})\phi_{\kappa ik}^{2} \\
& + \sum_{k \neq k'}^{K} \phi_{\kappa i k}\phi_{\kappa i k'} \bigg(\frac{(\Delta_{k} - 1)(\Delta_{k'} - 1)}{(\Delta_{k} + 1)(\Delta_{k'} + 1)}\bigg)\bigg(1 - \bigg(\frac{(\prod_{\ell=1}^{K}\Lambda_{\ell}) - 1}{(\prod_{\ell=1}^{K}\Lambda_{\ell}) + 1}\bigg)^2\bigg)\bigg\} \\
& = I^{-2}\sum_{i=1}^{I} \bigg\{ \sum_{k=1}^{K}\phi_{\kappa ik}^{2}\frac{4((\prod_{\ell=1}^{K}\Lambda_{\ell}) + \Delta_{k})((\prod_{\ell=1}^{K}\Lambda_{\ell}) \Delta_{k} + 1)}{(((\prod_{\ell=1}^{K}\Lambda_{\ell}) + 1)(\Delta_{k} + 1))^{2}} \\
& + \sum_{k \neq k'}^{K} \phi_{\kappa i k}\phi_{\kappa i k'} \bigg(\frac{(\Delta_{k} - 1)(\Delta_{k'} - 1)}{(\Delta_{k} + 1)(\Delta_{k'} + 1)}\bigg)\bigg(1 - \bigg(\frac{(\prod_{\ell=1}^{K}\Lambda_{\ell}) - 1}{(\prod_{\ell=1}^{K}\Lambda_{\ell}) + 1}\bigg)^2\bigg)\bigg\}.
\end{align*}
For Type-II confounding heterogeneity, $\mathbb{E}(\overline{T}_{\{\Lambda, \boldsymbol{\Delta}\}, \kappa}|\mathcal{C}, \boldsymbol{\mathcal{A}}, \mathcal{Z}) := \overline{\mu}_{I, \{\Lambda, \boldsymbol{\Delta}\}, \kappa}$ which has the same expression as $\overline{\mu}_{I, \{\boldsymbol{\Lambda}, \boldsymbol{\Delta}\}, \kappa}$ when $\Lambda = \prod_{k=1}^{K}\Lambda_{k}$. Further, under the amplification framework, $\mathbb{E}(\overline{T}_{\{\Lambda, \Delta\}, \kappa}|\mathcal{C}, \boldsymbol{\mathcal{A}}, \mathcal{Z}) := \overline{\mu}_{I, \{\Lambda, \Delta\}, \kappa}$ which has the same expression as $\overline{\mu}_{\{\Lambda, \boldsymbol{\Delta}\}, \kappa}$ when $\Delta_{1}=\dots={\Delta}_{K} = \Delta$.
\end{proof}

\vspace{-0.3cm}

\subsection{Moment Generating Function and Cumulant Generating Function for $\overline{T}_{\kappa}$ Under Different Sensitivity Analysis Frameworks}

\allowdisplaybreaks
We derive the moment generating function, the cumulant generating function, and the first three derivatives of the cumulant generating function for $\overline{T}_{\mathcal{P}, \kappa}$ for each sensitivity analysis framework parameterized by $\mathcal{P} \in \boldsymbol{\mathcal{P}}$. Recall that $\overline{p}_{\{\Gamma\}} = \Gamma / (\Gamma + 1)$. First, for the classic framework:
\begin{align*}
m_{I, \{\Gamma\}, \kappa}(t) & = \mathbb{E}(\exp\{I\cdot \overline{T}_{\{\Gamma\}, \kappa} t\}|\boldsymbol{\mathcal{F}}, \mathcal{Z}) = \prod_{i=1}^{I} \bigg\{\overline{p}_{\{\Gamma\}}\exp\{\eta_{\kappa i}t\} + (1 - \overline{p}_{\{\Gamma\}})\exp\{-\eta_{\kappa i}t\}\bigg\}, \\
\Psi^{(0)}_{I, \{\Gamma\}, \kappa}(t)& = \sum_{i=1}^{I}\log (\overline{p}_{\{\Gamma\}}\exp\{\eta_{\kappa i}t\} + (1 - \overline{p}_{\{\Gamma\}})\exp\{-\eta_{\kappa i}t\}), \\
\Psi^{(1)}_{I, \{\Gamma\}, \kappa}(t) & = \sum_{i=1}^{I} \eta_{\kappa i} \frac{\overline{p}_{\{\Gamma\}} \exp\{\eta_{\kappa i}t\} - (1 - \overline{p}_{\{\Gamma\}})\exp\{-\eta_{\kappa i}t\}}{\overline{p}_{\{\Gamma\}}\exp\{\eta_{\kappa i}t\} + (1-\overline{p}_{\{\Gamma\}}) \exp\{-\eta_{\kappa i}t\}}, \\
\Psi^{(2)}_{I, \{\Gamma\}, \kappa}(t) & = \sum_{i=1}^{I} \eta_{\kappa i}^{2} \frac{4\overline{p}_{\{\Gamma\}}(1-\overline{p}_{\{\Gamma\}})}{\{\overline{p}_{\{\Gamma\}}\exp\{\eta_{\kappa i}t\} + (1-\overline{p}_{\{\Gamma\}})\exp\{-\eta_{\kappa i}t\}\}^{2}}, \\
\Psi^{(3)}_{I, \{\Gamma\}, \kappa}(t) & =-\sum_{i=1}^{I} \eta_{\kappa i}^{3} \frac{8\overline{p}_{\{\Gamma\}}(1 - \overline{p}_{\{\Gamma\}})\{\overline{p}_{\{\Gamma\}} \exp\{\eta_{\kappa i}t\} - (1-\overline{p}_{\{\Gamma\}})\exp\{-\eta_{\kappa i}t\}\}}{\{\overline{p}_{\{\Gamma\}}\exp\{\eta_{\kappa i}t\} + (1-\overline{p}_{\{\Gamma\}})\exp\{-\eta_{\kappa i}t\}\}^{3}}.
\end{align*}
Denote $C_{F, \eta, \kappa}(\cdot)$ as the cumulative distribution function of $\eta_{\kappa i}$ under Assumption~\ref{Assumption: Power analysis}. Then, $c^{(0)}_{\{\Gamma\}, \kappa}(t), c^{(1)}_{\{\Gamma\}, \kappa}(t), c^{(2)}_{\{\Gamma\}, \kappa}(t)$ are written as:
\begin{align*}
c^{(0)}_{\{\Gamma\}, \kappa}(t) & = \lim_{I\to \infty}I^{-1}\Psi_{I, \{\Gamma\}, \kappa}^{(0)}(t) = \int_{\mathbb{R}}\log(\overline{p}_{\{\Gamma\}}\exp\{\eta_{\kappa i} t\} + (1 - \overline{p}_{\{\Gamma\}})\exp\{-\eta_{\kappa i} t\})dC_{F, \eta, \kappa}(\eta_{\kappa i}), \\
c^{(1)}_{\{\Gamma\}, \kappa}(t) & = \lim_{I\to \infty}I^{-1}\Psi_{I, \{\Gamma\}, \kappa}^{(1)}(t) = \int_{\mathbb{R}} \eta_{\kappa i} \frac{\overline{p}_{\{\Gamma\}} \exp\{\eta_{\kappa i} t\} - (1 - \overline{p}_{\{\Gamma\}})\exp\{-\eta_{\kappa i} t\}}{\overline{p}_{\{\Gamma\}}\exp\{\eta_{\kappa i} t\} + (1-\overline{p}_{\{\Gamma\}}) \exp\{-\eta_{\kappa i} t\}}dC_{F, \eta, \kappa}(\eta_{\kappa i}), \\
c^{(2)}_{\{\Gamma\}, \kappa}(t) & = \lim_{I\to \infty}I^{-1}\Psi_{I, \{\Gamma\}, \kappa}^{(2)}(t) = \int_{\mathbb{R}} \eta_{\kappa i}^{2} \frac{4\overline{p}_{\{\Gamma\}}(1-\overline{p}_{\{\Gamma\}})}{\{\overline{p}_{\{\Gamma\}}\exp\{\eta_{\kappa i} t\} + (1-\overline{p}_{\{\Gamma\}})\exp\{-\eta_{\kappa i} t\}\}^{2}}dC_{F, \eta, \kappa}(\eta_{\kappa i}).
\end{align*}

For the amplification and Type-II confounding heterogeneity frameworks, recall that $\overline{T}_{\mathcal{P}, \kappa}=I^{-1}\sum_{i=1}^{I}\sum_{k=1}^K\phi_{\kappa ik}b_{ik}v_{i}$, where $b_{ik} \indep b_{i'k'}$ for all $(i, k) \neq (i', k')$, $v_{i} \indep v_{i'}$ for all $i \neq i'$, $b_{ik} \indep v_{i}$ for all $i$ and $k$, and
    \begin{align*}
        P(b_{ik} = 1) & = \frac{\Delta_{k}}{\Delta_{k} + 1}, P(v_{i} = 1) = \frac{\Lambda}{\Lambda + 1},\implies \overline{p}_{\{\Lambda, \boldsymbol{\Delta}\}, k} = \frac{\Lambda \Delta_{k} + 1}{(\Lambda + 1)(\Delta_{k} + 1)},
    \end{align*}
where under the amplification framework, $\Delta_{1}=\dots=\Delta_{K}=\Delta$. Denote $\mathbf{v} = (v_{1},\dots,v_{I}).$ We derive the moment generating function, the cumulant generating function, and the first three derivatives of the cumulant generating function for the Type-II confounding heterogeneity framework, noting that the derivations also apply to the amplification framework by setting $\Delta_{1}=\dots =\Delta_{K}=\Delta$. 
\begin{align*}
m_{I, \{\Lambda, \boldsymbol{\Delta}\}, \kappa}(t) & = \mathbb{E}(\exp\{I\cdot \overline{T}_{\{\Lambda, \boldsymbol{\Delta}\}, \kappa} t\}|f(\mathcal{P}), \mathcal{Z}) \\
& = \mathbb{E}_{\mathbf{v}}(\mathbb{E}(\exp\{I\cdot \overline{T}_{\{\Lambda, \boldsymbol{\Delta}\}, \kappa} t\}|f(\mathcal{P}), \mathcal{Z}, \mathbf{v}))\\
& = \prod_{i=1}^{I}\bigg\{\frac{\Lambda}{\Lambda + 1}\prod_{k=1}^{K} \bigg\{\frac{\Delta_{k}}{\Delta_{k}+1}\exp\{\phi_{\kappa ik}t\} + \frac{1}{\Delta_{k}+1}\exp\{-\phi_{\kappa ik}t\}\bigg\} \\
& + \frac{1}{1+\Lambda}\prod_{k=1}^{K} \bigg\{\frac{\Delta_{k}}{\Delta_{k}+1}\exp\{-\phi_{\kappa ik}t\} + \frac{1}{\Delta_{k}+1}\exp\{\phi_{\kappa ik}t\}\bigg\}\bigg\}.
\end{align*}
Now, define the following functions:
\begin{align*}
M_{i}(t) & = \sum_{k=1}^{K}\log\bigg\{\frac{\Delta_{k}\exp\{\phi_{\kappa i k}t\} + \exp\{-\phi_{\kappa i k}t\}}{\Delta_{k} + 1}\bigg\}, \\
h_{i}(t) & = M_{i}(-t) - M_{i}(t) - \log(\Lambda), \\
Q_{i}(t) & = \text{logistic}(h_{i}(t)) = \frac{\exp\{h_{i}(t)\}}{1 + \exp\{h_{i}(t)\}}.
\end{align*}
Denote $M_{i}^{(d)}(t)$ as the $d$th derivative of $M_{i}(t)$ with respect to $t$. The first three derivatives of $M_{i}(t)$ are shown below:
\begin{align*}
M_{i}^{(1)}(t) & = \sum_{k=1}^{K}\phi_{\kappa ik} \frac{\Delta_{k}\exp\{\phi_{\kappa ik}t\} - \exp\{-\phi_{\kappa ik}t\}}{\Delta_{k}\exp\{\phi_{\kappa ik}t\} + \exp\{-\phi_{\kappa ik}t\}}, \\
M_{i}^{(2)}(t) & = \sum_{k=1}^{K}\phi_{\kappa ik}^{2} \frac{4\Delta_{k}}{(\Delta_{k}\exp\{\phi_{\kappa ik}t\} + \exp\{-\phi_{\kappa ik}t\})^{2}}, \\
M_{i}^{(3)}(t) & = \sum_{k=1}^{K}-\phi_{\kappa ik}^{3} \frac{8\Delta_{k}(\Delta_{k}\exp\{\phi_{\kappa ik}t\} - \exp\{-\phi_{\kappa ik}t\})}{(\Delta_{k}\exp\{\phi_{\kappa ik}t\} + \exp\{-\phi_{\kappa ik}t\})^{3}}. 
\end{align*}
Back to the moment generating function $m_{I, \{\Lambda, \boldsymbol{\Delta}\}, \kappa}(t)$, we have:
\begin{align*}
m_{I, \{\Lambda, \boldsymbol{\Delta}\}, \kappa}(t) & = \prod_{i=1}^{I}\bigg\{\prod_{k=1}^{K}\frac{\Delta_{k}\exp\{\phi_{\kappa i k}t\} + \exp\{-\phi_{\kappa i k}t\}}{\Delta_{k} + 1}\bigg\} \\
& \cdot \bigg\{\frac{\Lambda}{\Lambda + 1} + \frac{1}{\Lambda + 1}\prod_{k=1}^{K} \frac{\Delta_{k}\exp\{-\phi_{\kappa i k}t\} + \exp\{\phi_{\kappa i k}t\}}{\Delta_{k}\exp\{\phi_{\kappa i k}t\} + \exp\{-\phi_{\kappa i k}t\}}\bigg\} \\
& = \prod_{i=1}^{I}\exp\{M_{i}(t)\}\cdot \frac{\Lambda + \exp\{M_{i}(-t)-M_{i}(t)\}}{\Lambda + 1} \implies \\
\Psi^{(0)}_{I, \{\Lambda, \boldsymbol{\Delta}\}, \kappa}(t)& = \sum_{i=1}^{I}M_{i}(t) + \log\bigg\{\frac{\Lambda + \exp\{M_{i}(-t) - M_{i}(t)\}}{\Lambda + 1}\bigg\}, \\
\Psi^{(1)}_{I, \{\Lambda, \boldsymbol{\Delta}\}, \kappa}(t)& = \sum_{i=1}^{I}M_{i}^{(1)}(t) - Q_{i}(t)(M_{i}^{(1)}(-t) + M_{i}^{(1)}(t)),  \\
\Psi^{(2)}_{I, \{\Lambda, \boldsymbol{\Delta}\}, \kappa}(t)& = \sum_{i=1}^{I}M_{i}^{(2)}(t) + Q_{i}(t)(M_{i}^{(2)}(-t) - M_{i}^{(2)}(t))  \\
& + Q_{i}(t)(1-Q_{i}(t))(M_{i}^{(1)}(-t) + M_{i}^{(1)}(t))^{2}, \\
\Psi^{(3)}_{I, \{\Lambda, \boldsymbol{\Delta}\}, \kappa}(t)& = \sum_{i=1}^{I}M_{i}^{(3)}(t) - Q_{i}(t)(M_{i}^{(3)}(-t) + M_{i}^{(3)}(t)) \\
& - 3Q_{i}(t)(1-Q_{i}(t))(M_{i}^{(1)}(-t) + M_{i}^{(1)}(t))(M_{i}^{(2)}(-t) - M_{i}^{(2)}(t)) \\
& - Q_{i}(t)(1-Q_{i}(t))(1-2Q_{i}(t))(M_{i}^{(1)}(-t) + M_{i}^{(1)}(t))^{3}.
\end{align*}

Denote $C_{F, \phi, \kappa}(\cdot)$ as the joint cumulative distribution function of $(\phi_{\kappa i 1},\dots \phi_{\kappa i K})$ under Assumption~\ref{Assumption: Power analysis}, and assume each $q_{ik}$ follows Assumption~\ref{A: qik condition}. Then, $c^{(0)}_{\{\Lambda, \boldsymbol{\Delta}\}, \kappa}(t), c^{(1)}_{\{\Lambda, \boldsymbol{\Delta}\}, \kappa}(t), c^{(2)}_{\{\Lambda, \boldsymbol{\Delta}\}, \kappa}(t)$ are written as:
\begin{align*}
& c^{(0)}_{\{\Lambda, \boldsymbol{\Delta}\}, \kappa}(t) = \lim_{I\to \infty}I^{-1}\Psi_{I, \{\Lambda, \boldsymbol{\Delta}\}, \kappa}^{(0)}(t) = \int_{\mathbb{R}^{K}}M_{i}(t) + \log\bigg\{\frac{\Lambda + \exp\{M_{i}(-t) - M_{i}(t)\}}{\Lambda + 1}\bigg\}dC_{F, \phi, \kappa}(\phi_{\kappa i1},\dots \phi_{\kappa iK}), \\
& c^{(1)}_{\{\Lambda, \boldsymbol{\Delta}\}, \kappa}(t) = \lim_{I\to \infty}I^{-1}\Psi_{I, \{\Lambda, \boldsymbol{\Delta}\}, \kappa}^{(1)}(t) = \int_{\mathbb{R}^{K}} M_{i}^{(1)}(t) - Q_{i}(t)(M_{i}^{(1)}(-t) + M_{i}^{(1)}(t))dC_{F, \phi, \kappa}(\phi_{\kappa i1},\dots \phi_{\kappa iK}),\\
& c^{(2)}_{\{\Lambda, \boldsymbol{\Delta}\}, \kappa}(t) = \lim_{I\to \infty}I^{-1}\Psi_{I, \{\Lambda, \boldsymbol{\Delta}\}, \kappa}^{(2)}(t) \\
& = \int_{\mathbb{R}^{K}}M_{i}^{(2)}(t) + Q_{i}(t)(M_{i}^{(2)}(-t) - M_{i}^{(2)}(t))  + Q_{i}(t)(1-Q_{i}(t))(M_{i}^{(1)}(-t) + M_{i}^{(1)}(t))^{2}dC_{F, \phi, \kappa}(\phi_{\kappa i1},\dots \phi_{\kappa iK}).
\end{align*}

For the Type-I confounding heterogeneity frameworks, the definitions of $m_{I, \{\boldsymbol{\Lambda}, \boldsymbol{\Delta}\}, \kappa}(t),$ $M_{i}(t),$ $Q_{i}(t),$ $M_{i}^{(d)}(t),$ $\Psi^{(d)}_{I, \{\boldsymbol{\Lambda}, \boldsymbol{\Delta}\}, \kappa}(t)$ and $c^{(0)}_{\{\boldsymbol{\Lambda}, \boldsymbol{\Delta}\}, \kappa}(t)$ are equivalent to the definitions of $m_{I, \{\Lambda, \boldsymbol{\Delta}\}, \kappa}(t),$ $M_{i}(t),$ $Q_{i}(t),$ $M_{i}^{(d)}(t),$ $\Psi^{(d)}_{I, \{\Lambda, \boldsymbol{\Delta}\}, \kappa}(t)$ and $c^{(0)}_{\{\Lambda, \boldsymbol{\Delta}\}, \kappa}(t)$ by replacing $\Lambda$ with $\prod_{k=1}^{K} \Lambda_{k}$. 

\vspace{-0.3cm}

\subsection{Additional Lemmas for Bahadur-Rosenbaum slope Results}

Denote $\Psi^{(d)}_{I, \{\Gamma\}, \kappa}(\Gamma, t)$ and $c^{(d)}_{\{\Gamma\}, \kappa}(\Gamma, t)$ as the functions $\Psi^{(d)}_{I, \{\Gamma\}, \kappa}(t)$ and $c^{(d)}_{\{\Gamma\}, \kappa}(t)$ evaluated at sensitivity parameter value $\Gamma \geq 1$. Then, define $M_{i}^{(d)}(\Gamma, a, t)$, $Q_{i}(\Gamma, a, t)$, $h_{i}(\Gamma, a, t)$, $\Psi^{(d)}_{I, \{\Lambda, \boldsymbol{\Delta}\}, \kappa}(\Gamma, a, t)$ and $c^{(d)}_{\{\Lambda, \boldsymbol{\Delta}\}, \kappa}(\Gamma, a, t)$ as the functions $M_{i}^{(d)}(t)$, $Q_{i}(t)$, $h_{i}(t)$ $\Psi^{(d)}_{I, \{\Lambda, \boldsymbol{\Delta}\}, \kappa}(t)$ and $c^{(d)}_{\{\Lambda, \boldsymbol{\Delta}\}, \kappa}(t)$ evaluated at sensitivity parameter values $\{\Lambda, \Delta, \boldsymbol{\Lambda}, \boldsymbol{\Delta}\} \in \mathcal{B}_{\Gamma, a}$. Additionally, denote $\widetilde{\xi}_{\{\Gamma\}, \kappa}(\Gamma) = \xi$ as the unique solution to $\mu_{F, \kappa} = c^{(1)}_{\{\Gamma\}, \kappa}(\Gamma, \xi)$. For sensitivity analysis parameterizations $\mathcal{P} \in \boldsymbol{\mathcal{P}}\setminus\{\{\Gamma\}\}$, denote $\widetilde{\xi}_{\mathcal{P}, \kappa}(\Gamma, a) = \xi$ as the unique solution to $\mu_{F, \kappa} = c^{(1)}_{\mathcal{P}, \kappa}(\Gamma, a, \xi)$.

\begin{Lemma}[Bahadur-Rosenbaum slope is convex in $\Gamma$]
Assume $\eta_{\kappa i}>0$. Then, under the regularity conditions in Assumption~\ref{Assumption: Power analysis}, for any fixed $a > 0$, $b_{\{\Gamma\}, \kappa}(\Gamma)$ is strictly decreasing and convex in $\Gamma$ for $\Gamma \in [1, \widetilde{\Gamma})$. 
\label{Lem: Bahadur-Rosenbaum slope convex in Gamma}
\end{Lemma}
\begin{proof}
We first prove the desired result for the classic framework. For any $\Gamma \geq 1$, $t > 0$ and $\omega \geq 0$ we have:
\begin{equation*}
\frac{d}{dt} \log (\overline{p}_{\{\Gamma\}} \exp \{\omega t\} + (1-\overline{p}_{\{\Gamma\}})\exp\{-\omega t\}) = \omega  \frac{\overline{p}_{\{\Gamma\}} \exp \{\omega t\} - (1-\overline{p}_{\{\Gamma\}})\exp\{-\omega t\}}{\overline{p}_{\{\Gamma\}} \exp \{\omega t\} + (1-\overline{p}_{\{\Gamma\}})\exp\{-\omega t\}} \leq \omega.
\end{equation*}
Thus, by the Dominated Convergence Theorem, we have $d/dt(c^{(0)}_{\{\Gamma\}, \kappa }(\Gamma, t)) = c^{(1)}_{\{\Gamma\}, \kappa }(\Gamma, t)$. Since $c^{(0)}_{\{\Gamma\}, \kappa }(\Gamma, 0) = 0$, we have
\begin{equation*}
\{\widetilde{\xi}_{\{\Gamma\}, \kappa}(\Gamma) \mu_{F,  \kappa} - c^{(0)}_{\{\Gamma\}, \kappa }(\Gamma, \widetilde{\xi}_{\{\Gamma\}, \kappa}(\Gamma))\} = \int_{0}^{\widetilde{\xi}_{\{\Gamma\}, \kappa}(\Gamma)} \mu_{F,  \kappa} - c^{(1)}_{\{\Gamma\}, \kappa}(\Gamma, t) dt.
\end{equation*}
Taking the partial derivative of $b_{\{\Gamma\}, \kappa}(\Gamma)$ with respect to $\Gamma$, note:
\begin{align*}
& \frac{\partial}{\partial \Gamma}b_{\{\Gamma\}, \kappa}(\Gamma) = \frac{\partial}{\partial \Gamma}2\bigg\{\int_{0}^{\widetilde{\xi}_{\{\Gamma\}, \kappa}(\Gamma)} \mu_{F,  \kappa} - c^{(1)}_{\{\Gamma\}, \kappa}(\Gamma, t) dt\bigg\} \\
& = 2\bigg\{\int_{0}^{\widetilde{\xi}_{\{\Gamma\}, \kappa}(\Gamma)} -\frac{\partial}{\partial \Gamma}c^{(1)}_{\{\Gamma\}, \kappa}(\Gamma, t) dt + (\mu_{F,  \kappa} - c^{(1)}_{\{\Gamma\}, \kappa}(\Gamma, \widetilde{\xi}_{\{\Gamma\}, \kappa}(\Gamma)))\frac{\partial}{\partial \Gamma}\widetilde{\xi}_{\{\Gamma\}, \kappa}(\Gamma)\bigg\} \\
& = 2\int_{0}^{\widetilde{\xi}_{\{\Gamma\}, \kappa}(\Gamma)} -\frac{\partial}{\partial \Gamma}c^{(1)}_{\{\Gamma\}, \kappa}(\Gamma, t) dt.
\end{align*}
The second equality uses the Leibniz integral rule, and the third equality follows from the definition of $\widetilde{\xi}_{\{\Gamma\}, \kappa}(\Gamma)$. Now, note that for any $\Gamma \geq 1$ and $t \geq 0$, the following inequality holds:
\begin{equation*}
0\leq\eta_{\kappa i}\frac{2}{(\Gamma \exp\{\eta_{\kappa i}t\} + \exp\{-\eta_{\kappa i}t\})^{2}} \leq \eta_{\kappa i}/2.
\end{equation*}
Thus, using the Dominated Convergence Theorem, we have:
\begin{equation*}
\frac{\partial}{\partial \Gamma}c^{(1)}_{\{\Gamma\}, \kappa}(\Gamma, t) = \int_{\mathbb{R}}\eta_{\kappa i}\frac{2}{(\Gamma \exp\{\eta_{\kappa i}t\} + \exp\{-\eta_{\kappa i}t\})^{2}}dC_{F, \eta, \kappa}(\eta_{\kappa i}).
\end{equation*}
Now, looking at the second derivative of $b_{\{\Gamma\}, \kappa}(\Gamma)$ with respect to $\Gamma$, note:
\begin{align*}
\frac{\partial^{2}}{\partial \Gamma^{2}}b_{\{\Gamma\}, \kappa}(\Gamma) & = -2\frac{\partial}{\partial \Gamma} \int_{0}^{\widetilde{\xi}_{\{\Gamma\}, \kappa}(\Gamma)} \frac{\partial}{\partial \Gamma}c^{(1)}_{\{\Gamma\}, \kappa}(\Gamma, t) dt \\
& = -2\bigg\{\int_{0}^{\widetilde{\xi}_{\{\Gamma\}, \kappa}(\Gamma)} \frac{\partial^{2}}{\partial \Gamma^{2}}c^{(1)}_{\{\Gamma\}, \kappa}(\Gamma, t) dt + \frac{\partial}{\partial \Gamma}c^{(1)}_{\{\Gamma\}, \kappa}(\Gamma, t)\bigg|_{t = \widetilde{\xi}_{\{\Gamma\}, \kappa}(\Gamma)}\cdot \frac{\partial}{\partial \Gamma}\widetilde{\xi}_{\{\Gamma\}, \kappa}(\Gamma)\bigg\},
\end{align*}
where the second equality uses the Leibniz integral rule. Now, for any $\Gamma \geq 1$ and $t \geq 0$, the following inequality holds:
\begin{equation*}
\eta_{\kappa i}^{2} \frac{4\overline{p}_{\{\Gamma\}}(1-\overline{p}_{\{\Gamma\}})}{\{\overline{p}_{\{\Gamma\}}\exp\{\eta_{\kappa i}t\} + (1-\overline{p}_{\{\Gamma\}})\exp\{-\eta_{\kappa i}t\}\}^{2}} \leq \eta_{\kappa i}^{2}.
\end{equation*}
Thus, by the Dominated Convergence Theorem, we have $(d/dt)c^{(1)}_{\{\Gamma\}, \kappa}(\Gamma, t) = c^{(2)}_{\{\Gamma\}, \kappa}(\Gamma, t)$. By the Implicit Function Theorem, we have
\begin{equation*}
\frac{\partial}{\partial \Gamma}\widetilde{\xi}_{\{\Gamma\}, \kappa}(\Gamma) = -\frac{\frac{\partial}{\partial \Gamma}c^{(1)}_{\{\Gamma\}, \kappa}(\Gamma, t)\bigg|_{t = \widetilde{\xi}_{\{\Gamma\}, \kappa}(\Gamma)}}{c^{(2)}_{\{\Gamma\}, \kappa}(\Gamma, \widetilde{\xi}_{\{\Gamma\}, \kappa}(\Gamma))}.
\end{equation*}
Therefore, we have
\begin{equation*}
- \frac{\partial}{\partial \Gamma}c^{(1)}_{\{\Gamma\}, \kappa}(\Gamma, t)\bigg|_{t = \widetilde{\xi}_{\{\Gamma\}, \kappa}(\Gamma)}\cdot \frac{\partial}{\partial \Gamma}\widetilde{\xi}_{\{\Gamma\}, \kappa}(\Gamma) = \frac{\left(\frac{\partial}{\partial \Gamma}c^{(1)}_{\{\Gamma\}, \kappa}(\Gamma, t)\bigg|_{t = \widetilde{\xi}_{\{\Gamma\}, \kappa}(\Gamma)}\right)^{2}}{c^{(2)}_{\{\Gamma\}, \kappa}(\Gamma, \widetilde{\xi}_{\{\Gamma\}, \kappa}(\Gamma))} > 0
\end{equation*}
since $c^{(2)}_{\{\Gamma\}, \kappa}(\Gamma, t) > 0$ for all $t > 0$. Now, note that for any $\Gamma \geq 1$ and $t \geq 0$, the following inequality holds:
\begin{equation*}
0<\eta_{\kappa i}\frac{4\exp\{\eta_{\kappa i}t\}}{(\Gamma \exp\{\eta_{\kappa i}t\} + \exp\{-\eta_{\kappa i}t\})^{3}} < \eta_{\kappa i}.
\end{equation*}
Thus, by the Dominated Convergence Theorem, we have:
\begin{equation*}
-\frac{\partial^{2}}{\partial \Gamma^{2}}c^{(1)}_{\{\Gamma\}, \kappa}(\Gamma, t) = \int_{\mathbb{R}}\eta_{\kappa i}\frac{4\exp\{\eta_{\kappa i}t\}}{(\Gamma \exp\{\eta_{\kappa i}t\} + \exp\{-\eta_{\kappa i}t\})^{3}}dC_{F, \eta, \kappa}(\eta_{\kappa i}),
\end{equation*}
which is positive for all $t > 0$. Therefore, $(\partial^{2}/\partial \Gamma^{2})b_{\{\Gamma\}, \kappa}(\Gamma) > 0$, proving the desired result.
\end{proof}

Define $M_{i}^{(d)}(\Lambda, \boldsymbol{\Delta}, t)$, $Q_{i}(\Lambda, \boldsymbol{\Delta}, t)$, $\Psi^{(d)}_{I, \{\Lambda, \boldsymbol{\Delta}\}, \kappa}(\Lambda, \boldsymbol{\Delta}, t)$ and $c^{(d)}_{\{\Lambda, \boldsymbol{\Delta}\}, \kappa}(\Lambda, \boldsymbol{\Delta}, t)$ as the functions $M_{i}^{(d)}(t)$, $Q_{i}(t)$, $\Psi^{(d)}_{I, \{\Lambda, \boldsymbol{\Delta}\}, \kappa}(t)$ and $c^{(d)}_{\{\Lambda, \boldsymbol{\Delta}\}, \kappa}(t)$ evaluated at sensitivity parameter values $\Lambda, \boldsymbol{\Delta}$. Additionally, denote $\widetilde{\xi}_{\{\Lambda, \boldsymbol{\Delta}\}, \kappa}(\Lambda, \boldsymbol{\Delta}) = \xi$ as the unique solution to $\mu_{F, \kappa} = c^{(1)}_{\{\Lambda, \boldsymbol{\Delta}\}, \kappa}(\Lambda, \boldsymbol{\Delta}, \xi)$. 


\begin{Lemma}[Bahadur-Rosenbaum slope under confounding heterogeneity is non-increasing in each component of $\boldsymbol{\Delta}$ below the design sensitivity] 
Assume the regularity conditions in Assumption~\ref{Assumption: Power analysis}, fix $\Lambda \geq 1$ and restrict $\Lambda, \Delta_{1},\dots\Delta_{K}$ to satisfy
\begin{equation*}
\frac{\Lambda - 1}{\Lambda + 1}\left(\sum_{k=1}^{K}\frac{(\Delta_{k} - 1)}{(\Delta_{k} + 1)}\phi_{\kappa k}\right) < \mu_{F, \kappa}.
\end{equation*}
Then, $b_{\{\Lambda, \boldsymbol{\Delta}\}, \kappa}(\Lambda,  \boldsymbol{\Delta})$ is non-increasing in each component of $\boldsymbol{\Delta}$.
\label{Lem: Bahadur-Rosenbaum slope CH strictly decreasing in Delta}
\end{Lemma}
\begin{proof}
It is sufficient to show that the moment generating function is non-decreasing in $\Delta_{k}$. Recall the moment generating function:
\begin{align*}
m_{I, \{\Lambda, \boldsymbol{\Delta}\}, \kappa}(t) & = \prod_{i=1}^{I}\bigg\{\frac{\Lambda}{\Lambda + 1}\prod_{k=1}^{K} \bigg\{\frac{\Delta_{k}}{\Delta_{k}+1}\exp\{\phi_{\kappa ik}t\} + \frac{1}{\Delta_{k}+1}\exp\{-\phi_{\kappa ik}t\}\bigg\} \\
& + \frac{1}{1+\Lambda}\prod_{k=1}^{K} \bigg\{\frac{\Delta_{k}}{\Delta_{k}+1}\exp\{-\phi_{\kappa ik}t\} + \frac{1}{\Delta_{k}+1}\exp\{\phi_{\kappa ik}t\}\bigg\}\bigg\}.
\end{align*}
Taking the derivative with respect to $\Delta_{k}$, note
\begin{align*}
\frac{\partial}{\partial \Delta_{k}}m_{I, \{\Lambda, \boldsymbol{\Delta}\}, \kappa}(t) & = \frac{\exp\{t\phi_{\kappa ik}\} - \exp\{-t\phi_{\kappa ik}\}}{(\Lambda + 1)(\Delta_{k} + 1)^{2}}\bigg\{\Lambda \prod_{k' \neq k}\frac{\Delta_{k'}\exp\{t\phi_{\kappa ik}\} + \exp\{-t\phi_{\kappa ik}\}}{\Delta_{k'} + 1} \\
& - \prod_{k' \neq k}\frac{\Delta_{k'}\exp\{-t\phi_{\kappa ik}\} + \exp\{t\phi_{\kappa ik}\}}{\Delta_{k'} + 1}\bigg\} \geq 0.
\end{align*}
The inequality follows since $\phi_{\kappa ik} \geq 0$, $\Delta_{1,\dots,K}, \Lambda \geq 1$, implying 
\begin{equation*}
\Lambda \prod_{k' \neq k}\frac{\Delta_{k'}\exp\{t\phi_{\kappa ik}\} + \exp\{-t\phi_{\kappa ik}\}}{\Delta_{k'} + 1} \geq \prod_{k' \neq k}\frac{\Delta_{k'}\exp\{-t\phi_{\kappa ik}\} + \exp\{t\phi_{\kappa ik}\}}{\Delta_{k'} + 1}.
\end{equation*}
Thus, $m_{I, \{\Lambda, \boldsymbol{\Delta}\}, \kappa}(t)$ is non-decreasing in $\Delta_{k}$. Therefore, the log transform of $m_{I, \{\Lambda, \boldsymbol{\Delta}\}, \kappa}(t)$ is non-decreasing in $\Delta_{k}$, therefore making $b_{\{\Lambda, \boldsymbol{\Delta}\}, \kappa}(\Lambda,  \boldsymbol{\Delta})$ non-increasing in each component of $\boldsymbol{\Delta}$.
\end{proof}

\vspace{-0.3cm}

\subsection{Relationship Between Infimum Design Sensitivity and Bahadur-Rosenbaum slope}

\begin{Lemma}[Relationship between infimum outcome sensitivity and Bahadur-Rosenbaum slope, amplification framework]
    $\lim_{\Lambda \to \infty} b_{\{\Lambda, \Delta\}, \kappa}(\Lambda, \Delta) = 0$ when $\Delta = \widetilde{\Delta}^{*}$.
\end{Lemma}
\begin{proof}
    Note that as $\Lambda \to \infty$, $(\Lambda, \widetilde{\Delta}^{*})$ approach the design sensitivity $\widetilde{D}_{\{\Lambda, \Delta\}}$. Thus, $\lim_{\Lambda\to\infty}c_{\{\Lambda, \Delta\}, \kappa}^{(1)}(\Lambda, \widetilde{\Delta}^{*}, 0) = (\widetilde{\Delta}^{*} - 1)/(\widetilde{\Delta}^{*} + 1)\phi_{\kappa} = \mu_{F, \kappa}$ by definition of $\widetilde{\Delta}^{*}$, which implies $\lim_{\Lambda\to\infty}\widetilde{\xi}_{\{\Lambda, \Delta\}, \kappa}(\Lambda, \widetilde{\Delta}^{*}) = 0$. Thus, $\lim_{\Lambda \to \infty} b_{\{\Lambda, \Delta\}, \kappa}(\Lambda, \widetilde{\Delta}^{*}) = 0.$
\end{proof}

\begin{Lemma}[Relationship between infimum outcome sensitivity and Bahadur-Rosenbaum slope, confounding heterogeneity framework]
    $\lim_{\Lambda \to \infty} b_{\{\Lambda, \boldsymbol{\Delta}\}, \kappa}(\Lambda, \boldsymbol{\Delta}) = 0$ when $\boldsymbol{\Delta} = \widetilde{\Delta}^{*}_{1},\dots,\widetilde{\Delta}^{*}_{K}$.
\end{Lemma}
\begin{proof}
    As $\Lambda \to \infty$, $(\Lambda, \widetilde{\Delta}^{*}_{1},\dots, \widetilde{\Delta}^{*}_{K})$ approach the design sensitivity $\widetilde{D}_{\{\Lambda, \boldsymbol{\Delta}\}}$. Thus, 
    
    $\lim_{\Lambda\to\infty}c_{\{\Lambda, \boldsymbol{\Delta}\}, \kappa}^{(1)}(\Lambda, \widetilde{\Delta}^{*}_{1},\dots,\widetilde{\Delta}^{*}_{K}, 0) = \sum_{k=1}^{K}(\widetilde{\Delta}^{*}_{k} - 1)/(\widetilde{\Delta}^{*}_{k} + 1)\phi_{\kappa k} = \mu_{F, \kappa}$ by definition of $\widetilde{\Delta}^{*}$, implying $\lim_{\Lambda\to\infty}\widetilde{\xi}_{\{\Lambda, \boldsymbol{\Delta}\}, \kappa}(\Lambda, \widetilde{\Delta}^{*}_{1},\dots,\widetilde{\Delta}^{*}_{K}) = 0$. Thus, $\lim_{\Lambda \to \infty} b_{\{\Lambda, \boldsymbol{\Delta}\}, \kappa}(\Lambda, \widetilde{\Delta}^{*}_{1},\dots,\widetilde{\Delta}^{*}_{K}) = 0.$
\end{proof}

\section{Proofs of Main Theoretical Results}

\vspace{-0.3cm}

\subsection{Theorem 1}

Denote $\mathbf{T}_{i} = (q_{i1}\operatorname{sign}(Y_{Ci1}V_{i}),\dots,q_{iK}\operatorname{sign}(Y_{CiK}V_{i}))$ as the $i$th matched pair's contribution to the test statistic $\mathbf{T}$. Denote $\boldsymbol{\mu}_{i, \mathcal{P}, \mathbf{u}} = (\mathbb{E}(T_{i1}|f(\mathcal{P}), \mathcal{Z}),\dots,\mathbb{E}(T_{iK}|f(\mathcal{P}), \mathcal{Z}))$ as the expected value of $\mathbf{T}_{i}$ in the sensitivity analysis framework parameterized by $\mathcal{P}$. We assume the following regularity conditions hold:
\begin{Assumption}[Regularity conditions for asymptotic normality]
\label{Assumption: CLT}
\begin{enumerate}
\noindent
    \item Either Assumption~\ref{Assumption: classic}, Assumption~\ref{Assumption: Amp and Type-II CH} or Assumption~\ref{Assumption: Type-I CH} holds, depending on the sensitivity analysis parameterization $\mathcal{P}$.
    \item As $I \to \infty$, $I^{-1}\boldsymbol{\Sigma}_{\mathcal{P}, \mathbf{u}}$ has a positive definite limit $V_{\mathcal{P}}$.
    \item For any fixed nonzero vector $\kappa \in \mathbb{R}^{K}$, there exists an $\epsilon>0$ such that Lyapunov's condition is satisfied: $$\lim\limits_{I\to\infty}\frac{\sum_{i=1}^{I}\mathbb{E}(|\kappa^{T}(\mathbf{T}_{i} - \boldsymbol{\mu}_{i, \mathcal{P}, \mathbf{u}})|^{2+\epsilon}|f(\mathcal{P}), \mathcal{Z})}{( \kappa^{T}\boldsymbol{\Sigma}_{\mathcal{P}, \mathbf{u}}\kappa)^{1+\epsilon/2}} = 0.$$
\end{enumerate}
\end{Assumption}

First, we will show that the following standardized deviate is asymptotically normal:
\begin{equation*}
\frac{\kappa^{T}(\mathbf{T} - \boldsymbol{\mu}_{\mathcal{P}, \mathbf{u}})}{\sqrt{I\cdot \kappa^{T} V_{\mathcal{P}} \kappa}} \to N(0, 1).
\end{equation*}
\begin{proof}
By condition (i), the set
\begin{equation*}
G_{I^{*}} = \{\kappa^{T}(V_{\mathcal{P}})^{-1/2}\mathbf{T}_{1},...,\kappa^{T}(V_{\mathcal{P}})^{-1/2}\mathbf{T}_{I^{*}}\}
\end{equation*}
is a set of independent random variables. Note that the sequence $\{G_{1}, G_{2},...\}$ forms a triangular array of random variables. By setting $\kappa=\kappa^{T}(V_{\mathcal{P}})^{-1/2}$ in condition (iii) and applying the Lyapunov central limit theorem (\cite{Billingsley1995-BILMAP}, Theorem 27.3) to $\{G_{1}, G_{2},...\}$ yields:
\begin{equation*}
\frac{\kappa^{T}(V_{\mathcal{P}})^{-1/2}(\mathbf{T} - \boldsymbol{\mu}_{\mathcal{P}, \mathbf{u}})}{\sqrt{I\cdot \kappa^{T}(V_{\mathcal{P}})^{-1/2}(I^{-1}\boldsymbol{\Sigma}_{\mathcal{P}, \mathbf{u}})(V_{\mathcal{P}})^{-1/2}\kappa}} \to N(0, 1).
\end{equation*}
This implies
\begin{align*}
\frac{\kappa^{T}(V_{\mathcal{P}})^{-1/2}(\mathbf{T} - \boldsymbol{\mu}_{\mathcal{P}, \mathbf{u}})}{\sqrt{I\cdot \kappa^{T}\kappa}} & = \frac{\kappa^{T}(V_{\mathcal{P}})^{-1/2}(\mathbf{T} - \boldsymbol{\mu}_{\mathcal{P}, \mathbf{u}})}{\sqrt{I\cdot \kappa^{T}(V_{\mathcal{P}})^{-1/2}(I^{-1}\boldsymbol{\Sigma}_{\mathcal{P}, \mathbf{u}})(V_{\mathcal{P}})^{-1/2}\kappa}} \\
& \cdot \frac{\sqrt{\kappa^{T}(V_{\mathcal{P}})^{-1/2}(I^{-1}\boldsymbol{\Sigma}_{\mathcal{P}, \mathbf{u}})(V_{\mathcal{P}})^{-1/2}\kappa}}{\sqrt{\kappa^{T}\kappa}} \\
& \to N(0, 1) \cdot 1 \hspace{5mm} \text{by Slutsky's theorem and condition (ii)}.
\end{align*}
This implies $\kappa^{T}(\mathbf{T} - \boldsymbol{\mu}_{\mathcal{P}, \mathbf{u}})/\sqrt{I\cdot \kappa^{T} V_{\mathcal{P}} \kappa} \overset{d}{\to} N(0, 1)$, which implies
\begin{align*}
\frac{\kappa^{T}(\mathbf{T} - \boldsymbol{\mu}_{\mathcal{P}, \mathbf{u}})}{\sqrt{I\cdot \kappa^{T} V_{\mathcal{P}} \kappa}} \frac{\sqrt{I\cdot \kappa^{T} V_{\mathcal{P}} \kappa}}{\sqrt{I\cdot \kappa^{T} I^{-1}\boldsymbol{\Sigma}_{\mathcal{P}, \mathbf{u}} \kappa}} \overset{d}{\to} N(0, 1) \cdot 1.
\end{align*}
Therefore, we have 
\begin{equation*}
    \frac{\kappa^{T}(\mathbf{T} - \boldsymbol{\mu}_{\mathcal{P}, \mathbf{u}})}{\sqrt{\kappa^{T} \boldsymbol{\Sigma}_{\mathcal{P}, \mathbf{u}} \kappa}} \overset{d}{\to} N(0, 1).
\end{equation*}
\end{proof}

\subsection{Theorem 2}

\vspace{-0.3cm}
\subsubsection{Minimax Method}

\begin{proof}
First, as a result of Theorem 1, as $I\to\infty$ and under the global null $\mathbf{H}_{0}$, for any sensitivity analysis parameterization $P \in \mathcal{P}$ and any outcome $k=1,\dots K$, for some $\mathbf{u}^{*} \in \mathcal{U}$, we have $(T_{k} - \mu_{\mathcal{P}, k, \mathbf{u}^{*}})^{2} /(\sigma_{\mathcal{P} k,  \mathbf{u}^{*}})^{2} \to \chi^{2}_{1}$. Letting $\mathcal{K} = \{e_{1},\dots e_{K}\}$, for a one-sided test, by the union bound, we have
\begin{align*}
& \limsup_{I\to\infty}\operatorname{Pr}\left\{\max _{\kappa \in \mathcal{K}} \frac{\kappa^{T}\left(\mathbf{T}-\boldsymbol{\mu}_{\mathcal{P}, \mathbf{u}^{*}}\right)}{\sqrt{\kappa^{T} \boldsymbol{\Sigma}_{\mathcal{P}, \mathbf{u}^{*}}\kappa}} \geq \sqrt{\chi^{2}_{1-2\alpha/K}} |
f(\mathcal{P}), \mathcal{Z}\right\} \\
& \leq \limsup_{I\to\infty}\operatorname{Pr} \left\{\bigcup_{k=1}^{K} \{ \frac{\kappa^{T}\left(\mathbf{T}-\boldsymbol{\mu}_{\mathcal{P}, \mathbf{u}^{*}}\right)}{\sqrt{\kappa^{T} \boldsymbol{\Sigma}_{\mathcal{P}, \mathbf{u}^{*}}\kappa}} \geq \sqrt{\chi^{2}_{1-2\alpha/K}} |
f(\mathcal{P}), \mathcal{Z}\right\} \\
& \leq \limsup_{I\to\infty}\sum_{k=1}^{K} \operatorname{Pr}\left\{ \frac{\kappa^{T}\left(\mathbf{T}-\boldsymbol{\mu}_{\mathcal{P}, \mathbf{u}^{*}}\right)}{\sqrt{\kappa^{T} \boldsymbol{\Sigma}_{\mathcal{P}, \mathbf{u}^{*}}\kappa}} \geq \sqrt{\chi^{2}_{1-2\alpha/K}} |
f(\mathcal{P}), \mathcal{Z}\right\}  = K \cdot \frac{\alpha}{K} = \alpha.
\end{align*}
Thus, we have
\begin{equation*}
\limsup_{I\to\infty}\operatorname{Pr}\left\{\min_{\mathbf{u} \in \mathcal{U}_{\mathcal{P}}}\max _{\kappa \in \mathcal{K}} \frac{\kappa^{T}\left(\mathbf{T}-\boldsymbol{\mu}_{\mathcal{P}, \mathbf{u}}\right)}{\sqrt{\kappa^{T} \boldsymbol{\Sigma}_{\mathcal{P}, \mathbf{u}}\kappa}} \geq \sqrt{\chi^{2}_{1-2\alpha/K}} |
f(\mathcal{P}), \mathcal{Z}\right\} \leq \alpha. \\
\end{equation*}
\end{proof}

\vspace{-0.3cm}

\subsubsection{Scheff\'e Projections}

\begin{proof}
First, as a result of Theorem 1, as $I\to\infty$ and under the global null $\mathbf{H}_{0}$, for any sensitivity analysis parameterization $P \in \mathcal{P}$, we have $(\mathbf{T} - \boldsymbol{\mu}_{\mathcal{P}, \mathbf{u}^{*}})^{T} (\boldsymbol{\Sigma}_{\mathcal{P}, \mathbf{u}^{*}})^{-1}(\mathbf{T} - \boldsymbol{\mu}_{\mathcal{P}, \mathbf{u}^{*}}) \overset{d}{\to} \chi^{2}_{K}$ for some $\mathbf{u}^{*} \in \mathcal{U}$. Thus, as $I\to\infty$, we have $P((\mathbf{T} - \boldsymbol{\mu}_{\mathcal{P}, \mathbf{u}^{*}})^{T} (\boldsymbol{\Sigma}_{\mathcal{P}, \mathbf{u}^{*}})^{-1}(\mathbf{T} - \boldsymbol{\mu}_{\mathcal{P}, \mathbf{u}^{*}}) \geq \chi^{2}_{K, 1-\alpha}|f(\mathcal{P}), \mathcal{Z}) \leq \alpha$. Since $\boldsymbol{\Sigma}_{\mathcal{P}, \mathbf{u}}$ is assumed to be positive definite, the following result on maxima of quadratic forms holds for any $\mathbf{u} \in \mathcal{U}_{\mathcal{P}}$ \citep{rao1973linear}.
\begin{equation*}
\sqrt{(\mathbf{T} - \boldsymbol{\mu}_{\mathcal{P}, \mathbf{u}})^{T} (\boldsymbol{\Sigma}_{\mathcal{P}, \mathbf{u}})^{-1}(\mathbf{T} - \boldsymbol{\mu}_{\mathcal{P}, \mathbf{u}})} = \max_{\kappa\neq \mathbf{0}}\frac{|\kappa^{T}(\mathbf{T} - \boldsymbol{\mu}_{\mathcal{P}, \mathbf{u}})|}{\sqrt{\kappa^{T}\boldsymbol{\Sigma}_{\mathcal{P}, \mathbf{u}}\kappa}} = \max_{\kappa\neq \mathbf{0}}\frac{\kappa^{T}(\mathbf{T} - \boldsymbol{\mu}_{\mathcal{P}, \mathbf{u}})}{\sqrt{\kappa^{T}\boldsymbol{\Sigma}_{\mathcal{P}, \mathbf{u}}\kappa}}.
\end{equation*}
Since the above equality holds for any $\mathbf{u} \in \mathcal{U}_{\mathcal{P}}$ we have:
\begin{align*}
  \sqrt{(\mathbf{T} - \boldsymbol{\mu}_{P, \mathbf{u}^{*}})^{T} (\boldsymbol{\Sigma}_{P, \mathbf{u}^{*}})^{-1}(\mathbf{T} - \boldsymbol{\mu}_{P, \mathbf{u}^{*}})}   \geq &\min_{\mathbf{u} \in \mathcal{U}_{\mathcal{P}}} \sqrt{(\mathbf{T} - \boldsymbol{\mu}_{\mathcal{P}, \mathbf{u}})^{T} (\boldsymbol{\Sigma}_{\mathcal{P}, \mathbf{u}})^{-1}(\mathbf{T} - \boldsymbol{\mu}_{\mathcal{P}, \mathbf{u}})} \\
 = &\min_{\mathbf{u} \in \mathcal{U}_{\mathcal{P}}}\max_{\kappa\neq \mathbf{0}}\frac{\kappa^{T}(\mathbf{T} - \boldsymbol{\mu}_{\mathcal{P}, \mathbf{u}})}{\sqrt{\kappa^{T}\boldsymbol{\Sigma}_{\mathcal{P}, \mathbf{u}}\kappa}}.
\end{align*}
Thus, under the global null hypothesis, we have:
$$
\limsup_{I\to\infty}\operatorname{Pr}\left\{\min _{\mathbf{u} \in \mathcal{U}_{\mathcal{P}}} \max _{\kappa \neq \mathbf{0}} \frac{\kappa^{T}\left(\mathbf{T}-\boldsymbol{\mu}_{\mathcal{P}, \mathbf{u}}\right)}{\sqrt{\kappa^{T} \boldsymbol{\Sigma}_{\mathcal{P}, \mathbf{u}} \kappa}} \geq \sqrt{\chi^{2}_{K, 1-\alpha}} |
f(\mathcal{P}), \mathcal{Z}\right\} \leq \alpha.
$$
\end{proof}

\vspace{-0.3cm}

\subsubsection{Chi-bar Test}

\begin{proof}

Assume $I^{-1}\Sigma_{\mathcal{P}, \mathbf{u}}$ has a positive definite limit $V_{\mathcal{P}}$. From Theorem 1 of~\cite{cohen2018multivariate}, letting $\mathcal{K} = \mathbb{R}_{+}^{K}\setminus\{0\}:= \{\boldsymbol{\kappa} \in \mathbb{R}^{K}:\kappa_{k} \geq 0, k=1,\dots,K, \sum_{k=1}^{K}\kappa_{k} > 0\}$ and assuming the global null hypothesis is true, for some $\mathbf{u}^{*} \in \mathcal{U}$, we have
\begin{equation*}
\left\{\max\left(0, \max _{\kappa \in \mathcal{K}} \frac{\kappa^{T}\left(\mathbf{T}-\boldsymbol{\mu}_{\mathcal{P}, \mathbf{u}^{*}}\right)}{\sqrt{\kappa^{T} \boldsymbol{\Sigma}_{\mathcal{P}, \mathbf{u}^{*}}\kappa}}\right)\right\}^{2} \overset{d}{\to} \overline{\chi}^{2}((V_{\mathcal{P}})^{-1}, \mathbb{R}_{+}^{K}\setminus\{0\}).
\end{equation*}

Let $\widetilde{\Sigma}_{\mathcal{P}, \mathbf{u}, \rho}$ denote the correlation matrix for $\mathbf{T}$ at a fixed $\mathbf{u} \in \mathcal{U}_{\mathcal{P}}$. Then, denote $\mathcal{S}_{\mathcal{P}} = \{\widetilde{\Sigma}_{\mathcal{P}, \mathbf{u}, \rho}:\mathbf{u} \in \mathcal{U}_{\mathcal{P}}\}$ as the set of all correlation matrices across unmeasured confounding values. Following \cite{cohen2018multivariate}, a large-sample approximation for an upper bound of the critical value of the worst-case deviate is calculated as the largest $1-\alpha$ quantile of $C_{\mathcal{P}}^{*}$, where $C_{\mathcal{P}}^{*} = \max_{\widetilde{\Sigma}_{\mathcal{P}, \mathbf{u}, \rho} \in \mathcal{S}_{\mathcal{P}}} \overline{\chi}^{2}(\widetilde{\Sigma}_{\mathcal{P}, \mathbf{u}, \rho}^{-1}, \mathbb{R}_{+}^{K}\setminus\{0\})$. 
\begin{equation*}
\limsup_{I\to\infty}\operatorname{Pr}\left\{\min _{\mathbf{u} \in \mathcal{U}_{\mathcal{P}}} \max _{\kappa \in \mathbb{R}^{K}_{+}\setminus\{0\}} \frac{\kappa^{T}\left(\mathbf{T}-\boldsymbol{\mu}_{\mathcal{P}, \mathbf{u}}\right)}{\sqrt{\kappa^{T} \boldsymbol{\Sigma}_{\mathcal{P}, \mathbf{u}} \kappa}} \geq \sqrt{\sup_{\widetilde{\Sigma}_{\mathcal{P}, \mathbf{u}, \rho} \in \mathcal{S}_{\mathcal{P}}} \overline{\chi}^{2}_{1-\alpha}(\widetilde{\Sigma}_{\mathcal{P}, \mathbf{u}, \rho}^{-1}, \mathbb{R}_{+}^{K}\setminus\{0\})} |
f(\mathcal{P}), \mathcal{Z}\right\} \leq \alpha.
\end{equation*}
See \cite{cohen2018multivariate} for more details.
\end{proof}

\subsection{Theorem 3}

Denote $\mathcal{B}_{K}:=\{\kappa \in \mathbb{R}^{K} : \sum_{k=1}^{K}\kappa^{2} = 1\}$ as the unit sphere in $\mathbb{R}^{K}$. Under any sensitivity analysis framework $P$ with parameterization in 

\noindent $\mathcal{P}:=\{\Gamma, (\Lambda, \Delta), (\Lambda_{1},...,\Lambda_{K}, \Delta_{1},...,\Delta_{K}), (\Lambda, \Delta_{1},...,\Delta_{K})\}$ and method characterized by maximizing the standardized deviate over linear combinations $\mathcal{K}$ in 

\noindent $\boldsymbol{\mathcal{K}} = \{\{e_{1},\dots,e_{K}\}, \mathbb{R}^{K}\setminus\{0\}, \mathbb{R}_{+}^{K}\setminus\{0\}\}$, a two-sided hypothesis test rejects the global null hypothesis if:
\begin{align*}
\min _{\mathbf{u} \in \mathcal{U}_{\mathcal{P}}} \max _{\kappa \in \mathcal{K} \cap \mathcal{B}_{K}} \kappa^{T}\left(\mathbf{T}-\boldsymbol{\mu}_{\mathcal{P}, \mathbf{u}}\right)\left(\mathbf{T}-\boldsymbol{\mu}_{\mathcal{P}, \mathbf{u}}\right)^{T}\kappa - c_{\mathcal{P}}(\alpha, \mathcal{K})^{2}\kappa^{T} \boldsymbol{\Sigma}_{\mathcal{P}, \mathbf{u}}\kappa \geq 0.
\end{align*}

$\kappa$ can be restricted to $\mathcal{K} \cap \mathcal{B}_{K}$ without loss of generality, as $d_{\mathcal{P}, \kappa}(\mathbf{u}) = d_{\mathcal{P}, \widetilde{\kappa}}(\mathbf{u})$ with $\widetilde{\kappa} = \kappa / ||\kappa||$. The above formulation is an equivalent hypothesis testing procedure to the one described in the main text, given a fixed $\alpha > 0$ and critical value $c_{\mathcal{P}}(\alpha, \mathcal{K})$ calibrated to maintain asymptotic Type-1 error control at level $\alpha$ against a two-sided alternative. While minimizing the above inequality with respect to $\mathbf{u}$ in general does not find the values of $\mathbf{u}$ minimizing the standardized deviate, a hypothesis test conducted by verifying the inequality above leads to an equivalent decision to the hypothesis test conducted with the standardized deviate, given a fixed cutoff $c_{\mathcal{P}}(\alpha, \mathcal{K})$. We derive the signomial geometric program associated with minimizing the left-hand side of the above inequality with respect to $\mathbf{u}$.

\vspace{-0.3cm}

\subsubsection*{Signomial Geometric Program for Amplification / Type-II Confounding Heterogeneity}

For a fixed $\kappa \in \mathbb{R}^{K}$, we derive the optimization problem that minimizes 

\noindent $\kappa^{T}\left(\mathbf{T}-\boldsymbol{\mu}_{\mathcal{P}, \mathbf{u}}\right)\left(\mathbf{T}-\boldsymbol{\mu}_{\mathcal{P}, \mathbf{u}}\right)^{T}\kappa - c_{\mathcal{P}}(\alpha, \mathcal{K})^{2}\kappa^{T} \boldsymbol{\Sigma}_{\mathcal{P}, \mathbf{u}}\kappa$ with respect to $\mathbf{u} \in [0, 1]^{2I}$ for the Type-II confounding heterogeneity framework and the amplification framework. Define $\omega_{i}:=\exp\{u_{i1}-u_{i2}\}$ and $\nu_{ik}:=\left((1+\omega_{i}^{\lambda})(1+\omega_{i}^{\delta_{k}})\right)^{-1}$. We rewrite the function 

\noindent $\kappa^{T}\left(\mathbf{T}-\boldsymbol{\mu}_{\mathcal{P}, \mathbf{u}}\right)\left(\mathbf{T}-\boldsymbol{\mu}_{\mathcal{P}, \mathbf{u}}\right)^{T}\kappa - c_{\mathcal{P}}(\alpha, \mathcal{K})^{2}\kappa^{T} \boldsymbol{\Sigma}_{\mathcal{P}, \mathbf{u}}\kappa$ as:
\begin{align*}
g(\boldsymbol{\omega}, \boldsymbol{\nu}) & = \sum_{k=1}^{K}\sum_{k'=1}^{K}\kappa_{k}\kappa_{k'}\sum_{i=1}^{I}\bigg\{\sum_{i'=1}^{I}q_{ik}q_{i'k'}\operatorname{sign}(Y_{ik}V_{i})\operatorname{sign}(Y_{i'k'}V_{i'}) \\
& - 2q_{ik}q_{i'k'}\operatorname{sign}(Y_{ik}V_{i}) \nu_{i'k'}(\omega_{i'}^{\lambda}-1)(\omega_{i'}^{\delta_{k'}}-1) \\
& + q_{ik}q_{i'k'}\nu_{ik}\nu_{i'k'}(\omega_{i}^{\lambda}-1)(\omega_{i}^{\delta_{k}}-1)(\omega_{i'}^{\lambda}-1)(\omega_{i'}^{\delta_{k'}}-1)\bigg\}  \\
& - c_{\mathcal{P}}(\alpha, \mathcal{K})^{2}\bigg\{\mathbb{I}\{k \neq k'\}4q_{ik}q_{ik'}\nu_{ik}\nu_{ik'}\omega_{i}^{\lambda}(\omega_{i}^{\delta_{k}}-1)(\omega_{i}^{\delta_{k'}}-1) \\
& + \mathbb{I}\{k = k'\}q_{ik}^{2}(1-(\nu_{ik}(\omega_{i}^{\lambda}-1)(\omega_{i}^{\delta_{k}}-1))^{2})\bigg\}.
\end{align*}
where $\boldsymbol{\omega}:=(\omega_{1},\dots \omega_{I})$ and $\boldsymbol{\nu}:=(\nu_{11}\dots\nu_{IK})$. We find the set of $\mathbf{u}$ solving the fixed-cutoff quadratic threshold problem by first solving the following optimization problem
\begin{align*}
\underset{\boldsymbol{\omega}, \boldsymbol{\nu}}{\mathrm{minimize}} & \hspace{5mm} g(\boldsymbol{\omega}, \boldsymbol{\nu}) \\
\mathrm{\text{subject to}} & \hspace{5mm} e^{-1} \leq \omega_i \leq e \hspace{2mm} \forall i, \hspace{3mm} \nu_{ik}(1+\omega_{i}^{\lambda})(1+\omega_{i}^{\delta_{k}}) = 1 \hspace{2mm} \forall i, k.
\end{align*}
By taking log-transformations of each $\omega_{i}$, we find the set of $u_{i1}-u_{i2}$ that minimizes the squared deviate. The above optimization problem applies to sensitivity analysis under the amplification framework by setting $\delta_{1}=\dots =\delta_{K}$.

\vspace{-0.3cm}

\subsubsection*{Signomial Geometric Program for Type-I Confounding Heterogeneity}

Under the Type-I confounding heterogeneity framework, we seek to minimize 

\noindent $\kappa^{T}\left(\mathbf{T}-\boldsymbol{\mu}_{\{\boldsymbol{\Lambda}, \boldsymbol{\Delta}\}, \mathbf{u}}\right)\left(\mathbf{T}-\boldsymbol{\mu}_{\{\boldsymbol{\Lambda}, \boldsymbol{\Delta}\}, \mathbf{u}}\right)^{T}\kappa - c_{\mathcal{P}}(\alpha, \mathcal{K})^{2}\kappa^{T} \boldsymbol{\Sigma}_{\{\boldsymbol{\Lambda}, \boldsymbol{\Delta}\}, \mathbf{u}}\kappa$ with respect to $\mathbf{u}_{1},\dots \mathbf{u}_{K} \in [0, 1]^{2I} \times \dots \times [0, 1]^{2I}$.
Define $\omega_{ik} = \exp\{u_{i1k} - u_{i2k}\}$ and $\nu_{ik} = \left((1+\prod_{\ell=1}^{K}\omega_{i\ell}^{\lambda_{\ell}})(1+\omega_{ik}^{\delta_{k}})\right)^{-1}$. We rewrite the function 

\noindent $\kappa^{T}\left(\mathbf{T}-\boldsymbol{\mu}_{\{\boldsymbol{\Lambda}, \boldsymbol{\Delta}\}, \mathbf{u}}\right)\left(\mathbf{T}-\boldsymbol{\mu}_{\{\boldsymbol{\Lambda}, \boldsymbol{\Delta}\}, \mathbf{u}}\right)^{T}\kappa - c_{\mathcal{P}}(\alpha, \mathcal{K})^{2}\kappa^{T} \boldsymbol{\Sigma}_{\{\boldsymbol{\Lambda}, \boldsymbol{\Delta}\}, \mathbf{u}}\kappa$ as:
\begin{align*}
g(\boldsymbol{\omega}, \boldsymbol{\nu}) & = \sum_{k=1}^{K}\sum_{k'=1}^{K}\kappa_{k}\kappa_{k'}\sum_{i=1}^{I}\bigg\{\sum_{i'=1}^{I}q_{ik}q_{i'k'}\operatorname{sign}(Y_{ik}V_{i})\operatorname{sign}(Y_{i'k'}V_{i'}) \\
& - 2q_{ik}q_{i'k'}\operatorname{sign}(Y_{ik}V_{i}) \nu_{i'k'}(\prod_{\ell=1}^{K}\omega_{i'\ell}^{\lambda_{\ell}}-1)(\omega_{i'k'}^{\delta_{k'}}-1) \\
& + q_{ik}q_{i'k'}\nu_{ik}\nu_{i'k'}(\prod_{\ell=1}^{K}\omega_{i\ell}^{\lambda_{\ell}}-1)(\omega_{ik}^{\delta_{k}}-1)(\prod_{\ell=1}^{K}\omega_{i'\ell}^{\lambda_{\ell}}-1)(\omega_{i'k'}^{\delta_{k'}}-1)\bigg\}  \\
& - c_{\mathcal{P}}(\alpha, \mathcal{K})^{2}\bigg\{\mathbb{I}\{k \neq k'\}4q_{ik}q_{ik'}\nu_{ik}\nu_{ik'}(\prod_{\ell=1}^{K}\omega_{i\ell}^{\lambda_{\ell}})(\omega_{ik}^{\delta_{k}}-1)(\omega_{ik'}^{\delta_{k'}}-1) \\
& + \mathbb{I}\{k = k'\}q_{ik}^{2}(1-\big(\nu_{ik}(\prod_{\ell=1}^{K}\omega_{i\ell}^{\lambda_{\ell}}-1)(\omega_{ik}^{\delta_{k}}-1)\big)^{2})\bigg\},
\end{align*}
where $\boldsymbol{\omega}:=(\omega_{11},\dots \omega_{IK})$ and $\boldsymbol{\nu}:=(\nu_{11}\dots\nu_{IK})$. We find the set of $\mathbf{u}$ solving the fixed-cutoff quadratic threshold problem by solving the optimization problem:
\begin{align*}
\underset{\boldsymbol{\omega}, \boldsymbol{\nu}}{\mathrm{minimize}} & \hspace{5mm} g(\boldsymbol{\omega}, \boldsymbol{\nu}) \\
\mathrm{\text{subject to}} & \hspace{5mm} e^{-1} \leq \omega_{ik} \leq e \hspace{2mm} \forall i, k, \hspace{3mm} \nu_{ik}\left((1+\prod_{\ell=1}^{K}\omega_{i\ell}^{\lambda_{\ell}})(1+\omega_{ik}^{\delta_{k}})\right) = 1 \hspace{2mm} \forall i, k.
\end{align*}
Again, by taking log-transformations of each $\omega_{ik}$ we obtain the set of $u_{i1k}-u_{i2k}$ that minimize the squared deviate under the Type-I confounding heterogeneity framework.

\subsubsection{Proof of Theorem 3}

First, recall that $\mathcal{U}^{*} = \mathrm{argmin}_{\mathbf{u} \in [0, 1]^{2I}}\kappa^{T}\left(\mathbf{T}-\boldsymbol{\mu}_{\mathcal{P},\mathbf{u}}\right)$ denotes the set of unmeasured confounding values minimizing the numerator of the standardized deviate $d_{\mathcal{P},\kappa}(\mathbf{u})$. Then, $\widetilde{\mathbf{u}}^{*}$ is defined as $\widetilde{\mathbf{u}}^{*} = \mathrm{argmax}_{\mathbf{u} \in \mathcal{U}^{*}}\sqrt{\kappa^{T}\boldsymbol{\Sigma}_{\mathcal{P},\mathbf{u}}\kappa}$, the set of unmeasured confounding values that minimize the numerator and maximize the denominator of the standardized deviate. First we show that for each sensitivity analysis framework $\widetilde{u}^{*}_{ij} \in \{0, 1\}$ for all $j=1,2,i=1,\dots I$. 

\begin{Lemma}[Separable Approximation, Classic Framework]
\label{Lem: Sep approx, classic}
    Under the classic sensitivity analysis framework, $\widetilde{u}^{*}_{i1}-\widetilde{u}^{*}_{i2}=1$ for all $i$ with $\sum_{k=1}^{K}\kappa_{k}q_{ik}\operatorname{sign}(Y_{ik})>0$ and $\widetilde{u}^{*}_{i1}-\widetilde{u}^{*}_{i2}=-1$ otherwise.
\end{Lemma}
\begin{proof}
    Under the treatment assignment model for the classic sensitivity analysis framework, for any $\mathbf{u} \in [0, 1]^{2I}$ we have
\begin{align*}
    & \kappa^{T}\boldsymbol{\mu}_{\{\Gamma\}, \mathbf{u}} = \sum_{i=1}^{I}\sum_{k=1}^{K}\kappa_{k}q_{ik}\operatorname{sign}(Y_{ik})\frac{\exp\{\gamma(u_{i1}-u_{i2})\} - 1}{1 + \exp\{\gamma(u_{i1}-u_{i2})\}}.
\end{align*}
Note that $\mathrm{argmin}_{\mathbf{u}}\kappa^{T}\left(\mathbf{T}-\boldsymbol{\mu}_{\{\Gamma\}, \mathbf{u}}\right) = \mathrm{argmax}_{\mathbf{u}}\kappa^{T}\boldsymbol{\mu}_{\{\Gamma\}, \mathbf{u}}$. Then, since $(\exp\{\gamma(u_{i1}-u_{i2})\} - 1)/(1 + \exp\{\gamma(u_{i1}-u_{i2})\})$ is monotonically increasing in $u_{i1}-u_{i2}$, $\kappa^{T}\boldsymbol{\mu}_{\{\Gamma\}, \mathbf{u}}$ is maximized when, for all matched pairs, $\widetilde{u}^{*}_{i1}-\widetilde{u}^{*}_{i2} = 1$ when $\sum_{k=1}^{K}\kappa_{k}q_{ik}\operatorname{sign}(Y_{ik})>0$ and $\widetilde{u}^{*}_{i1}-\widetilde{u}^{*}_{i2} = -1$ otherwise. If $\sum_{k=1}^{K}\kappa_{k}q_{ik}\operatorname{sign}(Y_{ik}) = 0$, note that the expected value and variance of $\kappa^{T}\mathbf{T}$ is unchanged regardless of the values assigned to $\widetilde{u}^{*}_{i1}-\widetilde{u}^{*}_{i2} \in [-1, 1]$. Thus, $\widetilde{u}^{*}_{i1}-\widetilde{u}^{*}_{i2}$ can be assigned as $-1$ arbitrarily.
\end{proof}

\begin{Lemma}[Separable Approximation, Amplification and Type-II Confounding Heterogeneity Framework]
\label{Lem: Sep approx, amp}
    For $\kappa \in \mathbb{R}^{K}_{+}$, under the Type-II confounding heterogeneity sensitivity analysis framework $\widetilde{u}^{*}_{i1} - \widetilde{u}^{*}_{i2} \in \{0, 1\}$ for all $i$. 
\end{Lemma}
\begin{proof}
    Under the treatment assignment model for the Type-II confounding heterogeneity sensitivity analysis framework, for any $\mathbf{u} \in [0, 1]^{2I}$ we have
\begin{align*}
    \kappa^{T}\boldsymbol{\mu}_{\{\Lambda, \boldsymbol{\Delta}\}, \mathbf{u}} & = \sum_{i=1}^{I}\sum_{k=1}^{K}\kappa_{k}q_{ik}\frac{[\exp\{\lambda(u_{i1}-u_{i2})\}-1]}{[1+\exp\{\lambda(u_{i1}-u_{i2})\}]}\frac{[\exp\{\delta_{k}(u_{i1}-u_{i2})\}-1]}{[1+\exp\{\delta_{k}(u_{i1}-u_{i2})\}]}.
\end{align*}
Then, $\mathrm{argmin}_{\mathbf{u}}\kappa^{T}\left(\mathbf{T}-\boldsymbol{\mu}_{\{\Lambda, \boldsymbol{\Delta}\}, \mathbf{u}}\right) = \mathrm{argmax}_{\mathbf{u}}\kappa^{T}\boldsymbol{\mu}_{\{\Lambda, \boldsymbol{\Delta}\}, \mathbf{u}}$. Now, note that 
\begin{equation*}
\frac{[\exp\{\lambda(u_{i1}-u_{i2})\}-1]}{[1+\exp\{\lambda(u_{i1}-u_{i2})\}]}\frac{[\exp\{\delta_{k}(u_{i1}-u_{i2})\}-1]}{[1+\exp\{\delta_{k}(u_{i1}-u_{i2})\}]}
\end{equation*}
is maximized only when $u_{i1}-u_{i2}=1$ or $u_{i1}-u_{i2}=-1$ and is minimized whenever $u_{i1}=u_{i2}$. Then, based on the variance of $T_{k}$ derived in Lemma~\ref{Lem: Exp_var T_k, amp & Type-II CH}, notice that the covariance of $T_{k}$ and $T_{k'}$ is equal in the cases when $u_{i1}-u_{i2}=1$ or $u_{i1}-u_{i2}=-1$. Additionally, the covariance of $T_{k}$ and $T_{k'}$ is equal across all cases when $u_{i1}=u_{i2}$. Thus, $\kappa^{T}\boldsymbol{\mu}_{\{\Lambda, \boldsymbol{\Delta}\}, \mathbf{u}}$ is maximized when $\widetilde{u}^{*}_{i1}-\widetilde{u}^{*}_{i2}= 1$ for all $i$ such that $\sum_{k=1}^{K}\kappa_{k}q_{ik} > 0$ and $\widetilde{u}^{*}_{i1}-\widetilde{u}^{*}_{i2} = 0$ for all $i$ such that $\sum_{k=1}^{K}\kappa_{k}q_{ik} < 0$. Additionally, $\boldsymbol{\widetilde{u}}^{*}$ maximizes $\sqrt{\kappa^{T}\boldsymbol{\Sigma}_{\mathcal{P},\mathbf{u}}\kappa}$. If $\sum_{k=1}^{K}\kappa_{k}q_{ik} = 0$ for a matched pair $i$, both the expected value and variance of $\kappa^{T}\mathbf{T}$ is unchanged setting $\widetilde{u}^{*}_{i1}-\widetilde{u}^{*}_{i2} \in [-1, 1]$, so $\widetilde{u}^{*}_{i1}-\widetilde{u}^{*}_{i2}$ can be made equal to $1$ arbitrarily. Therefore, $\widetilde{u}^{*}_{i1} - \widetilde{u}^{*}_{i2} \in \{0, 1\}$ for all $i$. The above statement and proof generalize to the amplification framework by setting $\Delta_{1}=\dots=\Delta_{K} = \Delta$.
\end{proof}

Now we proceed with proving Theorem 3. Denote $$\kappa^{T}\sigma^{2}_{\mathcal{P}, i, \mathbf{u}}\kappa = \sum_{k, k'=1}^{K}\kappa_{k}\kappa_{k'}\mathrm{Cov}(q_{ik}\operatorname{sign}(Y_{ik}V_{i}), q_{ik'}\operatorname{sign}(Y_{ik'}V_{i})| f(\mathcal{P}), \mathcal{Z})$$ so that $\sum_{i=1}^{I}\kappa^{T}\sigma^{2}_{\mathcal{P}, i, \mathbf{u}}\kappa = \kappa^{T}\Sigma_{\mathcal{P}, \mathbf{u}}\kappa$. To prove Theorem 3, we require the following regularity conditions:
\begin{Assumption}[Regularity Conditions, Separable Approximation]
\label{Assumption: Separable approximation}
\begin{enumerate}
    \item (No matched pairs with no contribution) Assume $\sum_{k=1}^{K}q_{ik} > 0$ for all matched pairs $i$
    \item (Uniformly bounded variance) Assume there exists a sequence of constants $\nu_{I, -}^{2}, \nu_{I, +}^{2}$ such that $0 < \nu_{I, -}^{2} \leq \nu_{I, +}^{2} < \infty$ for all $I$, $\min_{i, \mathbf{u} \in [0, 1]^{2I}}\kappa^{T}\sigma^{2}_{\mathcal{P}, i, \mathbf{u}}\kappa \geq \nu_{I, -}^{2}$ and $\max_{i, \mathbf{u} \in [0, 1]^{2I}}\kappa^{T}\sigma^{2}_{\mathcal{P}, i, \mathbf{u}}\kappa \leq \nu_{I, +}^{2}$ for all $I$, and $(\nu_{I, +}^{2}-\nu_{I, -}^{2})/\nu_{I, -} = o(\sqrt{I})$.
    \item (Non-trivial separation) Let $\delta = \lim_{I\to\infty}\min_{i}\{\kappa^{T}\mu_{\mathcal{P}, i, \widetilde{\mathbf{u}}^{*}} -\kappa^{T} \mu_{\mathcal{P}, i, \mathbf{u}^{*}} :\kappa^{T}\mu_{\mathcal{P}, i, \widetilde{\mathbf{u}}^{*}} > \kappa^{T}\mu_{\mathcal{P}, i, \mathbf{u}^{*}}\}$ and assume $\delta > 0$.
\end{enumerate}
\end{Assumption}
Now, denote $\mathbf{u}^{*} = \mathrm{argmin}_{\mathbf{u} \in \mathcal{U}_{\mathcal{P}}} d_{\mathcal{P}, \kappa}(\mathbf{u})$. Fix an $\epsilon > 0$ and constant $x > 0$ and let $x_{I} = x \sqrt{\kappa^{T} \boldsymbol{\Sigma}_{\mathcal{P}, \mathbf{u}^{*}}\kappa} + \kappa^{T}\boldsymbol{\mu}_{\mathcal{P}, \mathbf{u}^{*}}$. It is sufficient for us to show there exists an $I^{*}$ such that for all $I \geq I^{*}$ we have
\begin{equation*}
P\bigg(\kappa^{T}\mathbf{T} > x_{I} |f(\mathcal{P}), \mathcal{Z}, \mathbf{u} = \widetilde{\mathbf{u}}^{*}\bigg) \geq P\bigg(\kappa^{T}\mathbf{T} > x_{I} |f(\mathcal{P}), \mathcal{Z}, \mathbf{u} = \mathbf{u}^{*}\bigg) - \epsilon.
\end{equation*}


From the asymptotic normality result proven in Theorem 1, there exists an $I_{1}$ such that
\begin{align*}
P\bigg(\frac{\kappa^{T}\left(\mathbf{T}-\boldsymbol{\mu}_{\mathcal{P}, \widetilde{\mathbf{u}}^{*}}\right)}{\sqrt{\kappa^{T} \boldsymbol{\Sigma}_{\mathcal{P}, \widetilde{\mathbf{u}}^{*}}\kappa}} > x |f(\mathcal{P}), \mathcal{Z}, \mathbf{u} = \widetilde{\mathbf{u}}^{*} \bigg) & \geq 1 - \Phi(x) - \epsilon / 2, \\
P\bigg(\kappa^{T}\mathbf{T} > x_{I} |f(\mathcal{P}), \mathcal{Z},  \mathbf{u} = \mathbf{u}^{*}\bigg) & \leq 1 - \Phi(x) + \epsilon / 2.
\end{align*}
Then, note:
\begin{align*}
& P\bigg(\kappa^{T}\mathbf{T} > x_{I} |f(\mathcal{P}), \mathcal{Z}, \mathbf{u} = \widetilde{\mathbf{u}}^{*} \bigg) \\
& = P\bigg(\frac{\kappa^{T}\left(\mathbf{T}-\boldsymbol{\mu}_{\mathcal{P}, \widetilde{\mathbf{u}}^{*}}\right)}{\sqrt{\kappa^{T} \boldsymbol{\Sigma}_{\mathcal{P}, \widetilde{\mathbf{u}}^{*}}\kappa}} > \frac{\kappa^{T}(\boldsymbol{\mu}_{\mathcal{P}, \mathbf{u}^{*}} -\boldsymbol{\mu}_{\mathcal{P}, \widetilde{\mathbf{u}}^{*}}) + x\sqrt{\kappa^{T} \boldsymbol{\Sigma}_{\mathcal{P}, \mathbf{u}}^{*}}\kappa}{\sqrt{\kappa^{T} \boldsymbol{\Sigma}_{\mathcal{P}, \widetilde{\mathbf{u}}^{*}}\kappa}} |f(\mathcal{P}), \mathcal{Z}, \mathbf{u} = \widetilde{\mathbf{u}}^{*}\bigg).
\end{align*}
Now, if there existed an $I_{2}$ such that
\begin{equation*}
x \geq \frac{\kappa^{T}(\boldsymbol{\mu}_{\mathcal{P}, \mathbf{u}^{*}} -\boldsymbol{\mu}_{\mathcal{P}, \widetilde{\mathbf{u}}^{*}}) + x\sqrt{\kappa^{T} \boldsymbol{\Sigma}_{\mathcal{P}, \mathbf{u}^{*}}}\kappa}{\sqrt{\kappa^{T} \boldsymbol{\Sigma}_{\mathcal{P}, \widetilde{\mathbf{u}}^{*}}\kappa}},
\end{equation*}
Then for all $I \geq \max\{I_{1},I_{2}\}$ we would have 
\begin{align*}
P\bigg(\kappa^{T}\mathbf{T} > x_{I} |f(\mathcal{P}), \mathcal{Z},  \mathbf{u} = \mathbf{u}^{*}\bigg) & \leq 1 - \Phi(x) + \epsilon / 2, \\
P\bigg(\kappa^{T}\mathbf{T} > x_{I} |f(\mathcal{P}), \mathcal{Z}, \mathbf{u} = \widetilde{\mathbf{u}}^{*}\bigg) & \geq 1  - \Phi(x) - \epsilon / 2 \implies \\
P\bigg(\kappa^{T}\mathbf{T} > x_{I} |f(\mathcal{P}), \mathcal{Z}, \mathbf{u} = \widetilde{\mathbf{u}}\bigg) & \geq P\bigg(\kappa^{T}\mathbf{T} > x_{I} |f(\mathcal{P}), \mathcal{Z}, \mathbf{u} = \mathbf{u}^{*}\bigg) - \epsilon.
\end{align*}
Thus, it is sufficient to prove that for a sufficiently large $I$ the following inequality holds:
\begin{equation*}
\frac{1}{I}\bigg(\kappa^{T}\boldsymbol{\mu}_{\mathcal{P}, \widetilde{\mathbf{u}}^{*}} - \kappa^{T}\boldsymbol{\mu}_{\mathcal{P}, \mathbf{u}^{*}}\bigg) \geq \frac{x}{\sqrt{I}} \bigg(\sqrt{\frac{1}{I}\kappa^{T} \boldsymbol{\Sigma}_{\mathcal{P}, \mathbf{u}^{*}}\kappa} - \sqrt{\frac{1}{I}\kappa^{T} \boldsymbol{\Sigma}_{\mathcal{P}, \widetilde{\mathbf{u}}^{*}}\kappa}\bigg).
\end{equation*}
Let $A_{I} = \{i \in \{1,\dots, I\}:\kappa^{T}\mu_{\mathcal{P}, i, \widetilde{\mathbf{u}}^{*}} > \kappa^{T}\mu_{\mathcal{P}, i, \mathbf{u}^{*}}\}$, and denote $\pi_{I} = |A_{I}|/I$. Then, we have
\begin{equation*}
\frac{1}{I}\bigg(\kappa^{T}\boldsymbol{\mu}_{\mathcal{P}, \widetilde{\mathbf{u}}^{*}} - \kappa^{T}\boldsymbol{\mu}_{\mathcal{P}, \mathbf{u}^{*}}\bigg) \geq \delta \pi_{I}.
\end{equation*}
Now, denote $\psi_{I}^{*} = (I - |A_{I}|)^{-1}\sum_{i \notin A_{I}} \kappa^{T}\sigma^{2}_{\mathcal{P}, i, \mathbf{u}^{*}}\kappa$ and $\widetilde{\psi}_{I}^{*} = (I - |A_{I}|)^{-1}\sum_{i \notin A_{I}} \kappa^{T}\sigma^{2}_{\mathcal{P}, i, \widetilde{\mathbf{u}}^{*}}\kappa$ if $|A_{I}| < I$ and $0$ if $|A_{I}| = I$. By Condition~\ref{Assumption: Separable approximation}, we have $I^{-1}\sum_{i\in A_{I}}\kappa^{T}\sigma^{2}_{\mathcal{P}, i, \mathbf{u}^{*}}\kappa \leq \pi_{I}\nu_{I, +}^{2}$ and $I^{-1}\sum_{i\in A_{I}}\kappa^{T}\sigma^{2}_{\mathcal{P}, i, \widetilde{\mathbf{u}}^{*}}\kappa \geq \pi_{I}\nu_{I, -}^{2}$. Therefore, 
\begin{align*}
& \frac{x}{\sqrt{I}} \bigg(\sqrt{\frac{1}{I}\kappa^{T} \boldsymbol{\Sigma}_{\mathcal{P}, \mathbf{u}^{*}}\kappa} - \sqrt{\frac{1}{I}\kappa^{T} \boldsymbol{\Sigma}_{\mathcal{P}, \widetilde{\mathbf{u}}^{*}}\kappa}\bigg) \leq \frac{x}{\sqrt{I}} \bigg(\sqrt{\pi_{I}\nu_{I, +}^{2} + (1-\pi_{I})\psi_{I}^{*}} - \sqrt{\pi_{I}\nu_{I, -}^{2} + (1-\pi_{I})\widetilde{\psi}_{I}^{*}}\bigg) \\
& \leq \frac{x}{\sqrt{I}} \frac{\pi_{I}}{2\sqrt{\pi_{I}\nu_{I, -}^{2} + (1-\pi_{I})\widetilde{\psi}_{I}^{*}}}(\nu_{I, +}^{2} - \nu_{I, -}^{2}) \leq \frac{x}{\sqrt{I}}\frac{\pi_{I}}{2\nu_{I, -}}(\nu_{I, +}^{2} - \nu_{I, -}^{2}) = o(1).
\end{align*}
Thus, there exists an $I_{2}$ such that 
\begin{equation*}
\delta \pi_{I_{2}} \geq \frac{k}{\sqrt{I_{2}}}\frac{\pi_{I_{2}}}{2\nu_{I, -}}(\nu_{I, +}^{2} - \nu_{I, -}^{2}),
\end{equation*}
which would imply that
\begin{equation*}
x \geq \frac{\kappa^{T}(\boldsymbol{\mu}_{\mathcal{P}, \mathbf{u}^{*}} -\boldsymbol{\mu}_{\mathcal{P}, \widetilde{\mathbf{u}}^{*}}) + x\sqrt{\kappa^{T} \boldsymbol{\Sigma}_{\mathcal{P}, \mathbf{u}}^{*}}\kappa}{\sqrt{\kappa^{T} \boldsymbol{\Sigma}_{\mathcal{P}, \widetilde{\mathbf{u}}^{*}}\kappa}},
\end{equation*}
Setting $\widetilde{I} = \max\{I_{1}, I_{2}\}$ proves the following inequality holds:
\begin{equation*}
P\bigg(\kappa^{T}\mathbf{T} > x_{I} |f(\mathcal{P}), \mathcal{Z}, \mathbf{u} = \widetilde{\mathbf{u}}^{*}\bigg) \geq P\bigg(\kappa^{T}\mathbf{T} > x_{I} |f(\mathcal{P}), \mathcal{Z}, \mathbf{u} = \mathbf{u}^{*}\bigg) - \epsilon.
\end{equation*}


\vspace{-0.3cm}

\subsubsection{Global Minimum Equivalence Under Type-I and Type-II Confounding Heterogeneity}

Denote $\mathcal{U}^{\text{Type-I}} = \mathrm{argmin}_{\mathbf{u} \in \mathcal{U}_{\{\boldsymbol{\Lambda}, \boldsymbol{\Delta}\}}}\kappa^{T}\left(\mathbf{T}-\boldsymbol{\mu}_{\{\boldsymbol{\Lambda}, \boldsymbol{\Delta}\},\mathbf{u}}\right)$. Then, denote $\widetilde{\mathbf{u}}^{\text{Type-I}} \in \mathcal{U}^{\text{Type-I}}$ as the set of unmeasured confounding values satisfying $\widetilde{\mathbf{u}}^{\text{Type-I}} = \mathrm{argmax}_{\mathbf{u} \in \mathcal{U}^{\text{Type-I}}}\sqrt{\kappa^{T}\Sigma_{\{\boldsymbol{\Lambda}, \boldsymbol{\Delta}\}, \mathbf{u}}\kappa}$.

\begin{Lemma}
\label{Lem: Minimum equiv for CH}
Fix $\kappa \in \mathbb{R}^{K}_{+} \setminus\{0\}$. Then, under the Type-I confounding heterogeneity framework, one can choose $\widetilde{u}^{\text{Type-I}}_{i1k}-\widetilde{u}^{\text{Type-I}}_{i2k} = 1$ for all matched pairs $i$ and outcomes $k$. Further, if $\Lambda = \prod_{k=1}^{K}\Lambda_{k}$, then $\kappa^{T}\left(\mathbf{T}-\boldsymbol{\mu}_{\{\boldsymbol{\Lambda}, \boldsymbol{\Delta}\},\widetilde{\mathbf{u}}^{\text{Type-I}}}\right) = \kappa^{T}\left(\mathbf{T}-\boldsymbol{\mu}_{\{\Lambda,\boldsymbol{\Delta}\}, \widetilde{\mathbf{u}}^{*}}\right)$ and $\sqrt{\kappa^{T}\Sigma_{\{\boldsymbol{\Lambda}, \boldsymbol{\Delta}\}, \widetilde{\mathbf{u}}^{\text{Type-I}}}\kappa} = \sqrt{\kappa^{T}\Sigma_{\{\Lambda, \boldsymbol{\Delta}\}, \widetilde{\mathbf{u}}^{*}}\kappa}$.
\end{Lemma}
\begin{proof}
As a result of Lemma~\ref{Lem: Exp_var T_k, amp & Type-II CH}, the expected value of $\kappa^{T}\mathbf{T}$ takes the form
\begin{equation*}
\kappa^{T}\boldsymbol{\mu}_{\{\boldsymbol{\Lambda}, \boldsymbol{\Delta}\}, \mathbf{u}}= \sum_{i=1}^{I}\sum_{k=1}^{K}\kappa_{k}q_{ik}\frac{\exp\{\sum_{\ell=1}^{K}\lambda_{\ell}(u_{i1\ell} - u_{i2\ell})\} - 1}{1 + \exp\{\sum_{\ell=1}^{K}\lambda_{\ell}(u_{i1\ell} - u_{i2\ell})\}}\frac{\exp\{\delta_{k} (u_{i1k} - u_{i2k})\} - 1}{1 + \exp\{\delta_{k} (u_{i1k} - u_{i2k})\}}
\end{equation*}
Note that by assuming $\kappa \in \mathbb{R}^{K}_{+} \setminus\{0\}$, $\mathbb{E}(\kappa^{T}\mathbf{T} | f(\{\boldsymbol{\Lambda}, \boldsymbol{\Delta}\})$ is maximized with respect to $\mathbf{u}$ when $u_{i1k}-u_{i2k} = 1$ for each matched pair $i$ and each outcome $k$. Therefore, $\widetilde{u}^{\text{Type-I}}_{i1k}-\widetilde{u}^{\text{Type-I}}_{i2k} = 1$ for all matched pairs $i$ and outcomes $k$. Now, recall the expected value of $\kappa^{T}\mathbf{T}$ under the Type-II confounding heterogeneity framework:
\begin{align*}
    \kappa^{T}\boldsymbol{\mu}_{\{\Lambda, \boldsymbol{\Delta}\}, \mathbf{u}} & = \sum_{i=1}^{I}\sum_{k=1}^{K}\kappa_{k}q_{ik}\frac{[\exp\{\lambda(u_{i1}-u_{i2})\}-1]}{[1+\exp\{\lambda(u_{i1}-u_{i2})\}]}\frac{[\exp\{\delta_{k}(u_{i1}-u_{i2})\}-1]}{[1+\exp\{\delta_{k}(u_{i1}-u_{i2})\}]}.
\end{align*}
If $\Lambda = \prod_{k=1}^{K}\Lambda_{k}$, then $\kappa^{T}\boldsymbol{\mu}_{\{\boldsymbol{\Lambda}, \boldsymbol{\Delta}\}, \widetilde{\mathbf{u}}^{\text{Type-I}}} = \kappa^{T}\boldsymbol{\mu}_{\{\Lambda, \boldsymbol{\Delta}\}, \widetilde{\mathbf{u}}^{*}}$, so $\kappa^{T}\left(\mathbf{T}-\boldsymbol{\mu}_{\{\boldsymbol{\Lambda}, \boldsymbol{\Delta}\},\widetilde{\mathbf{u}}^{\text{Type-I}}}\right) = \kappa^{T}\left(\mathbf{T}-\boldsymbol{\mu}_{\{\Lambda,\boldsymbol{\Delta}\}, \widetilde{\mathbf{u}}^{*}}\right)$. Then, note that if $u_{i1}-u_{i2} = 1$ for all matched pairs under the Type-II confounding heterogeneity framework and $u_{i1k}-u_{i2k} = 1$ for all matched pairs and all outcomes, then $\sqrt{\kappa^{T}\Sigma_{\{\boldsymbol{\Lambda}, \boldsymbol{\Delta}\}, \widetilde{\mathbf{u}}^{\text{Type-I}}}\kappa} = \sqrt{\kappa^{T}\Sigma_{\{\Lambda, \boldsymbol{\Delta}\}, \widetilde{\mathbf{u}}^{*}}\kappa}$.
\end{proof}

\vspace{-0.3cm}

\subsection{Lemma 1}

We first prove the following lemma:

\begin{Lemma}
\label{Lem: bounding variable verification}
For any sensitivity analysis parameterization $\mathcal{P} \in \boldsymbol{\mathcal{P}}$, we have $\max_{\mathbf{u}}\mathbb{E}(T_{\kappa}|f(\mathcal{P}), \mathcal{Z})=\overline{\mu}_{I, \mathcal{P}, \kappa}$ and $\mathrm{Var}(T_{\kappa}|f(\mathcal{P}), \mathcal{Z}, \mathbf{u} = \widetilde{\mathbf{u}})=\overline{\sigma}^{2}_{I, \mathcal{P}, \kappa}$, where $\widetilde{\mathbf{u}}=\mathrm{argmax}_{\mathbf{u}}\mathbb{E}(T_{\kappa}|f(\mathcal{P}), \mathcal{Z})$, and in the presence of ties, $\widetilde{\mathbf{u}}$ satisfies $\widetilde{\mathbf{u}} = \mathrm{argmax}_{\mathbf{u}}\mathrm{Var}(T_{\kappa}|f(\mathcal{P}), \mathcal{Z})$.
\end{Lemma}

\begin{proof}
(Classic Framework) Note:
\begin{align*}
    \max_{\mathbf{u}}\mathbb{E}(T_{\kappa}|\boldsymbol{\mathcal{F}}, \mathcal{Z}) & = \max_{\mathbf{u}}I^{-1}\sum_{i=1}^{I}\sum_{k=1}^{K}\kappa_{k}q_{ik}\operatorname{sign}(Y_{ik})\frac{\exp\{\gamma(u_{i1}-u_{i2})\}-1}{1+\exp\{\gamma(u_{i1}-u_{i2})\}} \\
    & = I^{-1} \sum_{i=1}^{I}\bigg\{\mathbb{I}\{\operatorname{sign}\left(\sum_{k=1}^{K}\kappa_{k}q_{ik}\operatorname{sign}(Y_{ik})\right) = 1\}\frac{\Gamma - 1}{\Gamma + 1} \\
    & + \mathbb{I}\{\operatorname{sign}\left(\sum_{k=1}^{K}\kappa_{k}q_{ik}\operatorname{sign}(Y_{ik})\right) = -1\}\frac{1-\Gamma}{\Gamma + 1} \bigg\}\bigg(\sum_{k=1}^{K}\kappa_{k}q_{ik}\operatorname{sign}(Y_{ik})\bigg) \\
    & = I^{-1} \frac{\Gamma - 1}{\Gamma + 1}\sum_{i=1}^{I}\bigg|\sum_{k=1}^{K}\kappa_{k}q_{ik}\operatorname{sign}(Y_{ik})\bigg| = \overline{\mu}_{I, \{\Gamma\}, \kappa}.
\end{align*}

In the second equality, the worst-case expected value is maximized by placing the largest possible weight on matched pairs satisfying $\mathbb{I}\{\operatorname{sign}\left(\sum_{k=1}^{K}\kappa_{k}q_{ik}\operatorname{sign}(Y_{ik})\right) = 1\}$ and placing the smallest weight on matched pairs satisfying $\mathbb{I}\{\operatorname{sign}\left(\sum_{k=1}^{K}\kappa_{k}q_{ik}\operatorname{sign}(Y_{ik})\right) = -1\}$. Let $\widetilde{\mathbf{u}} = \underset{\mathbf{u}}{\mathrm{argmax}}\mathbb{E}(T_{\kappa}|\boldsymbol{\mathcal{F}}, \mathcal{Z})$. Then, note:
\begin{align*}
    \mathrm{Var}(T_{\kappa}|\boldsymbol{\mathcal{F}}, \mathcal{Z}) & = I^{-2} \sum_{i=1}^{I} \left(1 - \bigg(\frac{\exp\{\gamma(u_{i1}-u_{i2})\}-1}{1+\exp\{\gamma(u_{i1}-u_{i2})\}}\bigg)^{2}\right) \left(\sum_{k=1}^{K}\kappa_{k}q_{ik}\operatorname{sign}(Y_{ik})\right)^2.
\end{align*}
Then, setting $\mathbf{u} = \widetilde{\mathbf{u}}$, we have
\begin{align*}
    \mathrm{Var}(T_{\kappa}|\boldsymbol{\mathcal{F}}, \mathcal{Z}) & = \frac{4\Gamma}{(\Gamma + 1)^{2}}I^{-2}\sum_{i=1}^{I}\eta^{2}_{\kappa i} = \overline{\sigma}_{I, \{\Gamma\}, \kappa}^{2}.
\end{align*}

(Amplification and Type-II confounding heterogeneity Frameworks) Because the amplification framework is obtained from the Type-II confounding heterogeneity framework by setting $\Delta_{1}=\dots=\Delta_{K} = \Delta$, we prove the lemma under Type-II confounding heterogeneity. Note:
\begin{align*}
\max_{\mathbf{u}}\mathbb{E}(T_{\kappa} |\mathcal{C}, \boldsymbol{\mathcal{A}}, \mathcal{Z}) & = \max_{\mathbf{u}}I^{-1}\sum_{i=1}^{I}\sum_{k=1}^{K}\kappa_{k}q_{ik}\frac{\exp\{\lambda(u_{i1}-u_{i2})\}-1}{1+\exp\{\lambda(u_{i1}-u_{i2})\}}\frac{\exp\{\delta_{k}(u_{i1}-u_{i2})\}-1}{1+\exp\{\delta_{k}(u_{i1}-u_{i2})\}} \\
& = I^{-1}\sum_{i=1}^{I}\bigg\{\sum_{k=1}^{K}\kappa_{k}q_{ik}\frac{\Lambda-1}{\Lambda+1}\frac{\Delta_{k}-1}{\Delta_{k}+1}\bigg\} \\
& = I^{-1}\frac{\Lambda-1}{\Lambda+1}\sum_{i=1}^{I}\sum_{k=1}^{K}\frac{\Delta_{k}-1}{\Delta_{k}+1}\kappa_{k}q_{ik} = \overline{\mu}_{I, \{\Lambda, \boldsymbol{\Delta}\}, \kappa}.
\end{align*}
Let $\widetilde{\mathbf{u}} = \underset{\mathbf{u}}{\mathrm{argmax}}\mathbb{E}(T_{\kappa}|\mathcal{C}, \boldsymbol{\mathcal{A}}, \mathcal{Z})$. Under Type-II confounding heterogeneity, note:
\begin{align*}
\mathrm{Var}(T_{\kappa}|\mathcal{C}, \boldsymbol{\mathcal{A}}, \mathcal{Z}, \mathbf{u} = \widetilde{\mathbf{u}}) & = I^{-2}\sum_{i=1}^{I}\bigg\{\sum_{k=1}^{K}\phi_{\kappa ik}^{2} \frac{4(\Lambda + \Delta_{k})(\Lambda \Delta_{k} + 1)}{((\Lambda + 1)(\Delta_{k} + 1))^{2}} \\
& + \sum_{k \neq k'}^{K} \phi_{\kappa i k}\phi_{\kappa i k'} \bigg(\frac{(\Delta_{k} - 1)(\Delta_{k'} - 1)}{(\Delta_{k} + 1)(\Delta_{k'} + 1)}\bigg)\bigg(1 - \bigg(\frac{\Lambda - 1}{\Lambda + 1}\bigg)^2\bigg)\bigg\} = \overline{\sigma}_{I, \{\Lambda, \boldsymbol{\Delta}\}, \kappa}^{2}.
\end{align*}
(Type-I Confounding Heterogeneity Framework) note:
\begin{align*}
\max_{\mathbf{u}}\mathbb{E}(T_{\kappa} |\boldsymbol{\mathcal{C}}, \boldsymbol{\mathcal{A}}, \mathcal{Z}) & = \max_{\mathbf{u}}I^{-1}\sum_{i=1}^{I}\sum_{k=1}^{K}\kappa_{k}q_{ik}\frac{\exp\{\sum_{\ell = 1}^{K}\lambda_{\ell}(u_{i1\ell}-u_{i2\ell})\}-1}{1+\exp\{\sum_{\ell = 1}^{K}\lambda_{\ell}(u_{i1\ell}-u_{i2\ell})\}}\frac{\exp\{\delta_{k}(u_{i1k}-u_{i2k})\}-1}{1+\exp\{\delta_{k}(u_{i1k}-u_{i2k})\}} \\
& = I^{-1}\sum_{i=1}^{I}\bigg(\frac{(\prod_{\ell=1}^{K}\Lambda_{\ell})-1}{(\prod_{\ell=1}^{K}\Lambda_{\ell})+1}\bigg)\sum_{k=1}^{K}\kappa_{k}q_{ik}\frac{\Delta_{k}-1}{\Delta_{k}+1} \\
& = I^{-1}\frac{(\prod_{\ell=1}^{K}\Lambda_{\ell})-1}{(\prod_{\ell=1}^{K}\Lambda_{\ell})+1}\sum_{i=1}^{I}\sum_{k=1}^{K}\frac{\Delta_{k}-1}{\Delta_{k}+1}\kappa_{k}q_{ik} = \overline{\mu}_{I, \{\boldsymbol{\Lambda}, \boldsymbol{\Delta}\}, \kappa}.
\end{align*}
Let $\widetilde{\mathbf{u}} = \underset{\mathbf{u}}{\mathrm{argmax}}\mathbb{E}(T_{\kappa}|\boldsymbol{\mathcal{C}}, \boldsymbol{\mathcal{A}}, \mathcal{Z})$. Under Type-I confounding heterogeneity, note:
\begin{align*}
& \mathrm{Var}(T_{\kappa}|\boldsymbol{\mathcal{C}}, \boldsymbol{\mathcal{A}}, \mathcal{Z}, \mathbf{u} = \widetilde{\mathbf{u}}) = I^{-2}\sum_{i=1}^{I} \bigg\{ \sum_{k=1}^{K}\phi_{\kappa ik}^{2}\frac{4((\prod_{\ell=1}^{K}\Lambda_{\ell}) + \Delta_{k})((\prod_{\ell=1}^{K}\Lambda_{\ell}) \Delta_{k} + 1)}{(((\prod_{\ell=1}^{K}\Lambda_{\ell}) + 1)(\Delta_{k} + 1))^{2}} \\
& + \sum_{k \neq k'}^{K} \phi_{\kappa i k}\phi_{\kappa i k'} \bigg(\frac{(\Delta_{k} - 1)(\Delta_{k'} - 1)}{(\Delta_{k} + 1)(\Delta_{k'} + 1)}\bigg)\bigg(1 - \bigg(\frac{(\prod_{\ell=1}^{K}\Lambda_{\ell}) - 1}{(\prod_{\ell=1}^{K}\Lambda_{\ell}) + 1}\bigg)^2\bigg)\bigg\} = \overline{\sigma}_{I, \{\boldsymbol{\Lambda}, \boldsymbol{\Delta}\}, \kappa}^{2}.
\end{align*}
\end{proof}
Now, if Lindeberg-type conditions are satisfied, $(\overline{T}_{\mathcal{P}, \kappa} -\overline{\mu}_{I, \mathcal{P}, \kappa})/\overline{\sigma}_{I, \mathcal{P}, \kappa}\overset{d}{\to}N(0, 1)$. For example, under the classic sensitivity analysis framework, if $\max_{i}\eta^{2}_{\kappa i} / \sum_{i=1}^{I}\eta^{2}_{\kappa i} \to 0$, the Lindeberg condition is satisfied. For the amplification, Type-II confounding heterogeneity, and Type-I confounding heterogeneity frameworks, if $(\max_{i}\sum_{k=1}^{K}\phi^{2}_{\kappa ik}) / (\sum_{i=1}^{I}\sum_{k=1}^{K}\phi^{2}_{\kappa ik})\to 0$, the Lindeberg condition will be satisfied. Now, for any fixed $x > 0$, note
\begin{align*}
& \lim_{I\to\infty}\bigg|\max_{u}P(T_{\kappa} \geq x |f(\mathcal{P}), \mathcal{Z})  - P(\overline{T}_{\mathcal{P}, \kappa} \geq x|f(\mathcal{P}), \mathcal{Z})\bigg| \\
& \leq \lim_{I\to\infty}\bigg\{\bigg|\max_{u}P(T_{\kappa} \geq x |f(\mathcal{P}), \mathcal{Z}) - \bigg(1 - \Phi \bigg(\min_{\mathbf{u}} \frac{x - \mathbb{E}(T_{\kappa} |f(\mathcal{P}), \mathcal{Z})}{\sqrt{\mathrm{Var}(T_{\kappa} |f(\mathcal{P}), \mathcal{Z})}} \bigg)\bigg)\bigg|   \\
& + \bigg|\bigg(1 - \Phi \bigg(\min_{\mathbf{u}} \frac{x - \mathbb{E}(T_{\kappa} |f(\mathcal{P}), \mathcal{Z})}{\sqrt{\mathrm{Var}(T_{\kappa} |f(\mathcal{P}), \mathcal{Z})}} \bigg)\bigg) - \bigg(1 - \Phi \bigg(\frac{x - \max_{\mathbf{u}}\mathbb{E}(T_{\kappa} |f(\mathcal{P}), \mathcal{Z})}{\sqrt{\mathrm{Var}(T_{\kappa} |f(\mathcal{P}), \mathcal{Z}, \mathbf{u} = \widetilde{\mathbf{u}})}}\bigg)\bigg)\bigg| \\
& + \bigg| \bigg(1 - \Phi \bigg(\frac{x - \max_{\mathbf{u}}\mathbb{E}(T_{\kappa} |f(\mathcal{P}), \mathcal{Z})}{\sqrt{\mathrm{Var}(T_{\kappa} |f(\mathcal{P}), \mathcal{Z}, \mathbf{u} = \widetilde{\mathbf{u}})}}\bigg)\bigg)- P(\overline{T}_{\mathcal{P}, \kappa} \geq x|f(\mathcal{P}), \mathcal{Z})\bigg|\bigg\}.
\end{align*}
First, as a result of the asymptotic normality of $T_{k}$ proven in Theorem 1 we have
\begin{equation*}
\lim_{I\to\infty}\bigg|\max_{u}P(T_{\kappa} \geq x |f(\mathcal{P}), \mathcal{Z}) - \bigg(1 - \Phi \bigg(\min_{\mathbf{u}} \frac{x - \mathbb{E}(T_{\kappa} |f(\mathcal{P}), \mathcal{Z})}{\sqrt{\mathrm{Var}(T_{\kappa} |f(\mathcal{P}), \mathcal{Z})}} \bigg)\bigg)\bigg| = 0.
\end{equation*}
Then, by the separable approximation proven in Theorem 3, we have
\begin{equation*}
\lim_{I\to\infty}\bigg|\bigg(1 - \Phi \bigg(\min_{\mathbf{u}} \frac{x - \mathbb{E}(T_{\kappa} |f(\mathcal{P}), \mathcal{Z})}{\sqrt{\mathrm{Var}(T_{\kappa} |f(\mathcal{P}), \mathcal{Z})}} \bigg)\bigg) - \bigg(1 - \Phi \bigg(\frac{x - \max_{\mathbf{u}}\mathbb{E}(T_{\kappa} |f(\mathcal{P}), \mathcal{Z})}{\sqrt{\mathrm{Var}(T_{\kappa} |f(\mathcal{P}), \mathcal{Z}, \mathbf{u} = \widetilde{\mathbf{u}})}}\bigg)\bigg)\bigg| = 0.
\end{equation*}
Then, by the asymptotic normality of $\overline{T}_{\mathcal{P}, \kappa}$ and Lemma~\ref{Lem: bounding variable verification} we have
\begin{equation*}
\lim_{I\to\infty}\bigg| \bigg(1 - \Phi \bigg(\frac{x - \max_{\mathbf{u}}\mathbb{E}(T_{\kappa} |f(\mathcal{P}), \mathcal{Z})}{\sqrt{\mathrm{Var}(T_{\kappa} |f(\mathcal{P}), \mathcal{Z}, \mathbf{u} = \widetilde{\mathbf{u}})}}\bigg)\bigg)- P(\overline{T}_{\mathcal{P}, \kappa} \geq x|f(\mathcal{P}), \mathcal{Z})\bigg| = 0.
\end{equation*}
Therefore, we have $\lim_{I\to\infty}\big|\max_{u}P(T_{\kappa} \geq x |f(\mathcal{P}), \mathcal{Z})  - P(\overline{T}_{\mathcal{P}, \kappa} \geq x|f(\mathcal{P}), \mathcal{Z})\big| = 0$, which is the desired result.

\vspace{-0.3cm}

\subsection{Theorem 4}

\begin{proof}
Recall that $\lim_{I\to\infty}\Phi\left((\sqrt{I}(\mu_{I, F, \kappa} - \overline{\mu}_{I, \mathcal{P}, \kappa}) - \Phi^{-1}(1-\alpha)\overline{\sigma}_{I, \mathcal{P}, \kappa})/\sigma_{I, F, \kappa}\right) = 1$ if and only if $\mu_{F, \kappa}>\overline{\mu}_{ \mathcal{P}, \kappa}$ and $\lim_{I\to\infty}\Phi\left((\sqrt{I}(\mu_{I, F, \kappa} - \overline{\mu}_{I, \mathcal{P}, \kappa}) - \Phi^{-1}(1-\alpha)\overline{\sigma}_{I, \mathcal{P}, \kappa})/\sigma_{I, F, \kappa}\right) = 0$ if and only if $\mu_{F, \kappa}<\overline{\mu}_{\mathcal{P}, \kappa}$. Thus, the design sensitivity is the set of sensitivity parameter values $\widetilde{D}_{\mathcal{P}, \kappa}$ such that $\overline{\mu}_{\mathcal{P}, \kappa} = \mu_{F, \kappa}$ for any sensitivity parameter values in $\widetilde{D}_{\mathcal{P}, \kappa}$.

Let Assumptions~\ref{Assumption: Power analysis} and~\ref{A: qik condition} hold. Recall that under Assumption~\ref{A: qik condition}, there exists a function $\widetilde{q}$ such that as $I \to \infty$, we have $q_{ik} \overset{a.s.}{\to} \widetilde{q}(|Y_{ik}V_{i}|)$. Using a Law of Large Numbers argument, one can show the following convergence results:
\begin{align*}
I^{-1}\sum_{i=1}^{I}\sum_{k=1}^{K}\kappa_{k}q_{ik}\operatorname{sign}(Y_{ik}V_{i}) & \overset{p}{\to} \mathbb{E}(\sum_{k=1}^{K}\kappa_{k}q_{ik}\operatorname{sign}(Y_{ik}V_{i})) = \mu_{F,  \kappa}, \\
I^{-1}\sum_{i=1}^{I}\bigg|\sum_{k=1}^{K}\kappa_{k}q_{ik}\operatorname{sign}(Y_{ik})\bigg| & \overset{p}{\to} \mathbb{E}\left(\bigg|\sum_{k=1}^{K}\kappa_{k}q_{ik}\operatorname{sign}(Y_{ik})\bigg|\right) =\eta_{\kappa}, \\
I^{-1}\sum_{i=1}^{I}\kappa_{k}q_{ik} & \overset{p}{\to} \mathbb{E}\left(\kappa_{k}q_{ik}\right) =\phi_{\kappa k}, \\
I^{-1}\sum_{i=1}^{I}\sum_{k=1}^{K}\kappa_{k}q_{ik} & \overset{p}{\to} \mathbb{E}\left(\sum_{k=1}^{K}\kappa_{k}q_{ik}\right) = \phi_{\kappa}.
\end{align*}
Then, by Lemma~\ref{Lem: bounding variable verification}, we have:
\begin{align*}
\lim_{I\to\infty}\max_{\mathbf{u}}\mu_{I, \{\Gamma\}, \kappa, \mathbf{u}}& = \frac{\Gamma-1}{\Gamma + 1}\eta_{\kappa}, \\
\lim_{I\to\infty}\max_{\mathbf{u}}\mu_{I, \{\Lambda, \Delta\}, \kappa, \mathbf{u}} & = \frac{(\Lambda - 1)(\Delta - 1)}{(\Lambda + 1)(\Delta + 1)} \phi_{\kappa}, \\
\lim_{I\to\infty}\max_{\mathbf{u}}\mu_{I, \{\Lambda, \boldsymbol{\Delta}\}, \kappa, \mathbf{u}} & = \frac{(\Lambda - 1)}{(\Lambda + 1)} \sum_{k=1}^{K}\frac{(\Delta_{k} - 1)}{(\Delta_{k} + 1)}\phi_{\kappa k}, \\
\lim_{I\to\infty}\max_{\mathbf{u}}\mu_{I, \{\boldsymbol{\Lambda}, \boldsymbol{\Delta}\}, \kappa, \mathbf{u}} & = \frac{(\prod_{\ell = 1}^{K}\Lambda_{\ell} - 1)}{(\prod_{\ell = 1}^{K}\Lambda_{\ell} + 1)} \sum_{k=1}^{K}\frac{(\Delta_{k} - 1)}{(\Delta_{k} + 1)}\phi_{\kappa k},
\end{align*}
By setting $\lim_{I\to\infty}\max_{\mathbf{u}}\mu_{I, \mathcal{P}, \kappa, \mathbf{u}} = \mu_{F,  \kappa}$ for each sensitivity analysis framework parameterized by $\mathcal{P}$, we achieve the desired result. 
\end{proof}

\vspace{-0.3cm}

\subsection{Lemma 2}

\begin{proof}
First, under the amplification framework, for any finite $\widetilde{\Lambda} 
\geq 1$, we have $(\widetilde{\Lambda} - 1)/(\widetilde{\Lambda} + 1) < 1$, and as $\widetilde{\Lambda}\to\infty$, we have $(\widetilde{\Lambda} - 1)/(\widetilde{\Lambda} + 1) \to 1$. Thus, the infimum outcome sensitivity $\widetilde{\Delta}^{*}$ satisfies $(\widetilde{\Delta}^{*} - 1)/(\widetilde{\Delta}^{*} + 1)\phi_{\kappa} = \mu_{F,  \kappa}$, which implies $\widetilde{\Delta}^{*} = (\phi_{\kappa} + \mu_{F,  \kappa})/(\phi_{\kappa} - \mu_{F,  \kappa}).$ Similarly, for any finite $\widetilde{\Delta} 
\geq 1$, we have $(\widetilde{\Delta} - 1)/(\widetilde{\Delta} + 1) < 1$, and as $\widetilde{\Delta}\to\infty$, we have $(\widetilde{\Delta} - 1)/(\widetilde{\Delta} + 1) \to 1$. Using a similar argument for showing the infimum outcome sensitivity, one can show that the infimum treatment sensitivity for the amplification and Type-II confounding heterogeneity frameworks must satisfy $\widetilde{\Lambda}^{*} = (\phi_{\kappa} + \mu_{F,  \kappa})/(\phi_{\kappa} - \mu_{F,  \kappa}).$

Next, for the Type-I confounding heterogeneity framework, since $\Lambda_{k} \geq 1$ for all $k=1\dots K$, we also have $(\prod_{\ell=1}^{K}\Lambda_{\ell} - 1)/(\prod_{\ell=1}^{K}\Lambda_{\ell} + 1) < 1$ for a fixed $\Lambda_{1},\dots \Lambda_{K}$, and as each $\Lambda_{k} \to \infty$, we have $(\prod_{\ell=1}^{K}\Lambda_{\ell} - 1)/(\prod_{\ell=1}^{K}\Lambda_{\ell} + 1) \to 1$. Thus, in both the Type-I and Type-II confounding heterogeneity framework, the infimum outcome sensitivity $\widetilde{\boldsymbol{\Delta}}^{*}$ is defined by the minimization problem:
\begin{align*}
    \widetilde{\boldsymbol{\Delta}}^{*} & = \mathrm{argmin} \sum_{k=1}^{K} \widetilde{\Delta}_{k} \\
    \text{subject to } & \sum_{k=1}^{K}\frac{(\widetilde{\Delta}_{k} - 1)}{(\widetilde{\Delta}_{k} + 1)}\phi_{\kappa k} = \mu_{F,  \kappa}, \hspace{3mm} \widetilde{\Delta}_{k} \geq 1 \hspace{2mm} \forall k=1\dots K.
\end{align*}
We solve this minimization problem using the method of Lagrange multipliers. Define the Lagrangian
\begin{equation*}
\mathcal{L}(\widetilde{\boldsymbol{\Delta}}, \nu) = \sum_{k=1}^{K} \widetilde{\Delta}_{k} - \nu \left(\sum_{k=1}^{K}\frac{(\widetilde{\Delta}_{k} - 1)}{(\widetilde{\Delta}_{k} + 1)}\phi_{\kappa k} - \mu_{F,  \kappa}\right).
\end{equation*}
First, note that the partial derivative of $\mathcal{L}(\widetilde{\boldsymbol{\Delta}}, \nu)$ with respect to $\widetilde{\Delta}_{k}$ is $\partial \mathcal{L}/\partial \widetilde{\Delta}_{k} = 1 - 2\nu \phi_{\kappa k}/(\widetilde{\Delta}_{k} + 1)^{2}$. Setting $\partial \mathcal{L}/\partial \widetilde{\Delta}_{k} = 0$ and solving for $\widetilde{\Delta}_{k}$, we have $\widetilde{\Delta}_{k} = \pm \sqrt{2\nu \phi_{\kappa k}}-1$. Since $\widetilde{\Delta}_{k}$ must be positive, the negative square-root solution is discarded so that $\widetilde{\Delta}_{k} = \sqrt{2\nu \phi_{\kappa k}}-1$. Then, enforcing the constraint $\widetilde{\Delta}_{k} \geq 1$ for all $k$, the solution for the minimum outcome sensitivity takes the form $\widetilde{\Delta}_{k}^{*} = \max\{1, \sqrt{2\nu \phi_{\kappa k}}-1\}$
for a certain choice of $\nu$. Enforcing the equality constraint and plugging in the solution for $\widetilde{\Delta}_{k}^{*}$, we find that $\nu$ must satisfy:
\begin{align*}
& \sum_{k=1}^{K}\phi_{\kappa k}\left(\frac{\widetilde{\Delta}^{*}_{k} - 1}{\widetilde{\Delta}^{*}_{k} + 1}\right) = \mu_{F,  \kappa} \implies \sum_{k: \nu \phi_{\kappa k}>2}\phi_{\kappa k}\left(\frac{\widetilde{\Delta}^{*}_{k} - 1}{\widetilde{\Delta}^{*}_{k} + 1}\right) = \mu_{F,  \kappa} \implies \\
& \sum_{k: \nu \phi_{\kappa k}>2}\phi_{\kappa k}\left(1 -\sqrt{\frac{2}{\nu \phi_{\kappa k}}}\right) = \mu_{F,  \kappa}.
\end{align*}
In the second equation, if $\nu \phi_{\kappa k}\leq 2$ that implies $\widetilde{\Delta}^{*}_{k} = 1$, which implies $(\widetilde{\Delta}^{*}_{k} - 1)/(\widetilde{\Delta}^{*}_{k} + 1) = 0$. The third equation follows from replacing $\widetilde{\Delta}^{*}_{k}$ with $\sqrt{2\nu \phi_{\kappa k}}-1$.
\end{proof}

\begin{Remark}[An algorithm for determining $\nu$]

Denote $A(\nu) := \{k: \nu\phi_{\kappa k} > 2\}$ as the ``active set'' of outcome indices $k$ such that $\widetilde{\Delta}^{*}_{k} > 1$. For a given active set $A(\nu)$, the condition for $\nu$ admits a closed form:
\begin{align*}
\sum_{k \in A(\nu)}\phi_{\kappa k}\left(1 -\sqrt{\frac{2}{\nu \phi_{\kappa k}}}\right) & = \mu_{F,  \kappa} \implies \nu = 2\left(\frac{\sum_{k \in A(\nu)}\sqrt{\phi_{\kappa k}}}{\sum_{k \in A(\nu)}\phi_{\kappa k} - \mu_{F,  \kappa}}\right)^{2}.
\end{align*}
We propose the following algorithm for finding $\nu$.
\begin{enumerate}
    \item Define $\phi_{\kappa (j)}$ as the $j$th order statistic from the collection $\{\phi_{\kappa 1},\dots ,\phi_{\kappa K}\}$ such that $\phi_{\kappa (1)} \geq \dots \geq \phi_{\kappa (K)}$. 
    \item Starting with $m=1$, if $\sum_{k=1}^{m}\phi_{\kappa (k)} > \mu_{F,  \kappa}$, compute the candidate value for $\nu$, denoted $\nu_{m}$:
    \begin{equation*}
        \nu_{m} = 2\left(\frac{\sum_{k = 1}^{m}\sqrt{\phi_{\kappa(k)}}}{\sum_{k = 1}^{m}\phi_{\kappa (k)} - \mu_{F,  \kappa}}\right)^{2}.
    \end{equation*}
    Otherwise, set $\nu_{m} = 0.$
    \item Check the candidate set satisfies both a) $\nu_{m}\phi_{\kappa (m)} > 2$ and b) either $m=K$ or $\nu_{m}\phi_{\kappa(m + 1)} \leq 2$. If both a) and b) hold, return $\nu = \nu_{m}$. Otherwise, increment $m$ by $1$ and return to step 2.
\end{enumerate}
\end{Remark}

Now we derive the infimum treatment sensitivity for Type-I confounding heterogeneity.
\begin{proof}
Note that for any finite $\widetilde{\Delta}_{1},\dots \widetilde{\Delta}_{K} \geq 1$, we have $\sum_{k=1}^{K}(\widetilde{\Delta}_{k}-1)/(\widetilde{\Delta}_{k} + 1)\phi_{\kappa k} < \phi_{\kappa}$, and as $\widetilde{\Delta}_{k} \to \infty$ for each $k=1,\dots K$, we have $\sum_{k=1}^{K}(\widetilde{\Delta}_{k}-1)/(\widetilde{\Delta}_{k} + 1)\phi_{\kappa k} \to \phi_{\kappa}$. Thus, the infimum treatment sensitivity for Type-I confounding heterogeneity $\widetilde{\boldsymbol{\Lambda}}^{*}$ is defined by the minimization problem:
\begin{align*}
    \widetilde{\boldsymbol{\Lambda}}^{*} & = \mathrm{argmin} \sum_{k=1}^{K} \widetilde{\Lambda}_{k} \\
    \text{subject to } & \frac{(\prod_{\ell=1}^{K}\widetilde{\Lambda}_{\ell} - 1)}{(\prod_{\ell=1}^{K}\widetilde{\Lambda}_{\ell} + 1)}\phi_{\kappa} = \mu_{F,  \kappa}, \hspace{3mm} \widetilde{\Lambda}_{k} \geq 1 \hspace{2mm} \forall k=1\dots K.
\end{align*}
By solving for $\prod_{k=1}^{K} \widetilde{\Lambda}_{k}$ on one side of the equation and taking logs on both sides, the equality constraint can be rewritten as:
\begin{equation*}
    \sum_{k=1}^{K} \log \widetilde{\Lambda}_{k} = \log \left(\frac{\phi_{\kappa} + \mu_{F,  \kappa}}{\phi_{\kappa} - \mu_{F,  \kappa}}\right).
\end{equation*}
Define the Lagrangian
\begin{equation*}
\mathcal{L}(\widetilde{\boldsymbol{\Lambda}}, \nu) = \sum_{k=1}^{K} \widetilde{\Lambda}_{k} + \nu \left(\sum_{k=1}^{K} \log \widetilde{\Lambda}_{k} - \log \left(\frac{\phi_{\kappa} + \mu_{F,  \kappa}}{\phi_{\kappa} - \mu_{F,  \kappa}}\right)\right).
\end{equation*}
The partial derivative of $\mathcal{L}(\widetilde{\boldsymbol{\Lambda}}, \nu)$ with respect to $\widetilde{\Lambda}_{k}$ for each $k=1,\dots K$ is $\partial \mathcal{L}/\partial \widetilde{\Lambda}_{k} = 1 + \nu/\widetilde{\Lambda}_{k}$. Setting the partial derivative of the Lagrangian to zero, we have $\widetilde{\Lambda}_{k} = -\nu$ for all $k=1,\dots K$, which implies $\widetilde{\Lambda}_{1}=\dots \widetilde{\Lambda}_{K}$. Using the equality constraint of the minimization problem, the infimum treatment sensitivity $\widetilde{\boldsymbol{\Lambda}}^{*}_{k}$ for each $k=1,\dots K$ is found as:
\begin{align*}
\sum_{k=1}^{K} \log \widetilde{\Lambda}_{k}^{*} & = \log \left(\frac{\phi_{\kappa} + \mu_{F,  \kappa}}{\phi_{\kappa} - \mu_{F,  \kappa}}\right) \implies \widetilde{\Lambda}_{k}^{*} = \left(\frac{\phi_{\kappa} + \mu_{F,  \kappa}}{\phi_{\kappa} - \mu_{F,  \kappa}}\right)^{1/K}.
\end{align*}
\end{proof}

\vspace{-0.3cm}

\subsection{Theorem 5}

\begin{proof}
We leverage the weaker version of Sievers' theorem described in \cite{plachky1971theorem}. First, we show that for any sensitivity analysis parameterization $\mathcal{P}$ with sensitivity parameter values greater than or equal to one, for all $t \in [0, T), T<\infty$, the four conditions for Sievers' theorem hold: $(i): m_{I, \mathcal{P}, \kappa}(t) < \infty$, $(ii): c^{(d)}_{\mathcal{P}, \kappa}(t) < \infty$ for $d=0, 1, 2$, $(iii): c^{(2)}_{\mathcal{P}, \kappa}(t) > 0$, and $(iv): \Psi^{(3)}_{I, \mathcal{P}, \kappa}(t)<\infty$. 

\noindent (i): Recall that under the classic framework, $\overline{T}_{\{\Gamma\}, \kappa}$ is the average of $I$ independent random variables taking value $|\sum_{k=1}^{K}\kappa_{k}q_{ik}\operatorname{sign}(Y_{ik})|:=\eta_{\kappa i}$ with probability $\overline{p}_{\{\Gamma\}}:=\Gamma / (1+\Gamma)$ and value $-\eta_{\kappa i}$ with probability $ 1- \overline{p}_{\{\Gamma\}}$. Since $\eta_{\kappa i}$ is treated as fixed in the definition of $m_{I, \{\Gamma\}, \kappa}(t)$, this implies $m_{I, \{\Gamma\}, \kappa}(t) < \infty$. Similarly, recall that for the amplification and confounding heterogeneity frameworks, $I\cdot \overline{T}_{\mathcal{P}, \kappa} = \sum_{i=1}^{I}\sum_{k=1}^{K} \phi_{\kappa ik} b_{ik}v_{i}$, where $b_{ik} \indep b_{i'k'}$ for all $(i, k) \neq (i', k')$, $v_{i} \indep v_{i'}$ for all $i \neq i'$, $b_{ik} \indep v_{i}$ for all $i$ and $k$, and $P(b_{ik}v_{i} = 1) = \overline{p}_{\mathcal{P}, k}, P(b_{ik}v_{i} = -1) = 1 - \overline{p}_{\mathcal{P}, k}$. Since $b_{ik}, v_{i}$ are bounded and $\phi_{\kappa i k}$ are treated as fixed in the definition of $m_{I, \mathcal{P}, \kappa}(t)$ for $\mathcal{P} \in \boldsymbol{\mathcal{P}}\setminus\{\{\Gamma\}\}$, this implies $m_{I, \mathcal{P}, \kappa}(t) < \infty$ for all $\mathcal{P} \in \boldsymbol{\mathcal{P}}\setminus\{\{\Gamma\}\}$. 

\noindent (ii): First, note that for any $p \in [1/2, 1]$ and $t \geq 0$, the following inequalities hold:
\begin{align*}
& \log\{p\exp\{\eta_{\kappa i}t\} + (1-p)\exp\{-\eta_{\kappa i}t\}\} \leq \eta_{\kappa i}t, \\
& \eta_{\kappa i} \frac{p \exp\{\eta_{\kappa i}t\} - (1 - p)\exp\{-\eta_{\kappa i}t\}}{p\exp\{\eta_{\kappa i}t\} + (1-p) \exp\{-\eta_{\kappa i}t\}} \leq \eta_{\kappa i}, \\
& \eta_{\kappa i}^{2} \frac{4p(1-p)}{\{p\exp\{\eta_{\kappa i}t\} + (1-p)\exp\{-\eta_{\kappa i}t\}\}^{2}} \leq \eta_{\kappa i}^{2}.
\end{align*}
Thus, under Assumption~\ref{Assumption: Power analysis}, $c^{(d)}_{\{\Gamma\}, \kappa}(t) < \infty$ for $d=0, 1, 2$. Recall the notation $\phi_{\kappa i } = \sum_{k=1}^{K}\phi_{\kappa i k}$. For any $\Lambda, \Delta_{1},\dots ,\Delta_{K} \geq 1$, the following inequalities hold:
\begin{align*}
& \sum_{k=1}^{K}\log\bigg\{\frac{\Delta_{k}\exp\{\phi_{\kappa i k}t\} + \exp\{-\phi_{\kappa i k}t\}}{\Delta_{k} + 1}\bigg\}< \phi_{\kappa i}t, \\
& \lim_{\boldsymbol{\Delta} \to \infty}\sum_{k=1}^{K}\log\bigg\{\frac{\Delta_{k}\exp\{\phi_{\kappa i k}t\} + \exp\{-\phi_{\kappa i k}t\}}{\Delta_{k} + 1}\bigg\}  = \phi_{\kappa i}t, \\
& 0 \leq \frac{\exp\{M_{i}(-t) - M_{i}(t)\}}{\Lambda + \exp\{M_{i}(-t) - M_{i}(t)\}} \leq 1, \\
& \sum_{k=1}^{K}\phi_{\kappa ik} \frac{\Delta_{k}\exp\{\phi_{\kappa ik}t\} - \exp\{-\phi_{\kappa ik}t\}}{\Delta_{k}\exp\{\phi_{\kappa ik}t\} + \exp\{-\phi_{\kappa ik}t\}} < \phi_{\kappa i}, \\
& \lim_{\boldsymbol{\Delta} \to \infty} \sum_{k=1}^{K}\phi_{\kappa ik} \frac{\Delta_{k}\exp\{\phi_{\kappa ik}t\} - \exp\{-\phi_{\kappa ik}t\}}{\Delta_{k}\exp\{\phi_{\kappa ik}t\} + \exp\{-\phi_{\kappa ik}t\}} = \phi_{\kappa i}, \\
& \sum_{k=1}^{K}\phi_{\kappa ik}^{2} \frac{4\Delta_{k}}{(\Delta_{k}\exp\{\phi_{\kappa ik}t\} + \exp\{-\phi_{\kappa ik}t\})^{2}} \leq \phi_{\kappa i}^{2}.
\end{align*}
Thus, under Assumption~\ref{Assumption: Power analysis}, $c^{(d)}_{\mathcal{P}, \kappa}(t) < \infty$ for $d=0, 1, 2$ under the amplification and Type-II confounding heterogeneity sensitivity analysis frameworks. The same result holds under the Type-I sensitivity analysis framework setting $\Lambda = \prod_{k=1}^{K}\Lambda_{k}$. 

\noindent (iii): Under Assumption~\ref{Assumption: Power analysis}, $0 <\mathbb{E}_{F}(\eta_{\kappa i}^{2}) < \infty$ and $0 <\mathbb{E}_{F}(\phi_{\kappa i}^{2}) < \infty$. Thus, for $p \in [1/2, 1)$, $\boldsymbol{\Delta} \geq 1$ and $t \geq 0$ the following inequalities hold:
\begin{align*}
0 & < \eta_{\kappa i}^{2} \frac{4p(1-p)}{\{p\exp\{\eta_{\kappa i}t\} + (1-p)\exp\{-\eta_{\kappa i}t\}\}^{2}}, \\
0 & < \sum_{k=1}^{K}\phi_{\kappa ik}^{2} \frac{4\Delta_{k}}{(\Delta_{k}\exp\{\phi_{\kappa ik}t\} + \exp\{-\phi_{\kappa ik}t\})^{2}}.
\end{align*}
Thus, $c^{(2)}_{\mathcal{P}, \kappa}(t) > 0$ for each sensitivity analysis parameterization $\mathcal{P}$.

\noindent (iv): For $p \in [1/2, 1]$, $\boldsymbol{\Delta} \geq 1$ and $t \geq 0$, note that $\Psi^{(3)}_{I, \{\Gamma\}, \kappa}(t) > -\sum_{i=1}^{I}\eta_{\kappa i}^{3}$ and $M_{i}^{(3)}(t) > -\sum_{k=1}^{K}\phi_{\kappa i k}^{3}$. Thus, under Assumption~\ref{Assumption: Power analysis}, considering $\eta_{\kappa i}$ and $\phi_{\kappa i k}$ fixed for each $i, k$, we have $|\Psi^{(3)}_{I, \mathcal{P}, \kappa}(t)| < \infty$. 

Next, we show that for any sensitivity analysis parameterization $\mathcal{P}$, there exists a solution $\widetilde{\xi} = \xi$ to $\mu_{F,  \kappa} = c^{(1)}_{\mathcal{P}, \kappa}(\xi)$. It suffices to show $c^{(1)}_{\mathcal{P}, \kappa}(0) < \mu_{F,  \kappa} < \lim_{t\to \infty}c^{(1)}_{\mathcal{P}, \kappa}(t)$. First, for the classic framework, we have:
\begin{align*}
c^{(1)}_{\{\Gamma\}, \kappa}(0) & = \int_{\mathbb{R}} \omega \frac{\Gamma \exp\{\omega (0)\} - \exp\{-\omega (0)\}}{\Gamma \exp\{\omega (0)\} + \exp\{-\omega (0)\}}dC_{F, \eta, \kappa}(\omega) = \frac{\Gamma -1}{\Gamma+1}\eta_{\kappa}, \\
\lim_{t\to \infty}c^{(1)}_{\{\Gamma\}, \kappa}(t) & = \int_{\mathbb{R}}\lim_{t\to \infty}\omega \frac{\Gamma \exp\{\omega t\} - \exp\{-\omega t\}}{\Gamma \exp\{\omega t\} + \exp\{-\omega t\}}dC_{F, \eta, \kappa}(\omega) = \int_{\mathbb{R}}\omega dC_{F, \eta, \kappa}(\omega) = \eta_{\kappa}.
\end{align*}
For any sensitivity parameter $\Gamma < \widetilde{\Gamma}$ we have $c^{(1)}_{\{\Gamma\}, \kappa}(0) < \mu_{F,  \kappa}$, and under Assumption~\ref{Assumption: Power analysis}, we have $\mu_{F,  \kappa} < \eta_{\kappa}$. Next, under the Type-II confounding heterogeneity framework, we have:
\begin{align*}
M_{i}(0) & = \sum_{k=1}^{K}\log\bigg\{\frac{\Delta_{k}\exp\{\phi_{\kappa i k}(0)\} + \exp\{-\phi_{\kappa i k}(0)\}}{\Delta_{k} + 1}\bigg\} = 0, \\
M_{i}^{(1)}(0) & = \sum_{k=1}^{K}\phi_{\kappa ik}\frac{\Delta_{k} \exp\{\phi_{\kappa ik}(0)\} - \exp\{-\phi_{\kappa ik}(0)\}}{\Delta_{k} \exp\{\phi_{\kappa ik}(0)\} + \exp\{-\phi_{\kappa ik}(0)\}}  = \sum_{k=1}^{K}\frac{\Delta_{k} - 1}{\Delta_{k} + 1}\phi_{\kappa ik}, \\
Q_{i}(0) & = \frac{\exp\{M_{i}(0) - M_{i}(0)\}}{\Lambda + \exp\{M_{i}(0) - M_{i}(0)\}} = \frac{1}{\Lambda + 1} \implies \\
c_{\{\Lambda, \Delta\}}^{(1)}(0) & = \int_{\mathbb{R}^{K}}\bigg\{1 - \frac{2}{\Lambda + 1}\bigg\}\sum_{k=1}^{K}\frac{\Delta_{k} - 1}{\Delta_{k} + 1}\phi_{\kappa ik}dC_{F, \phi, \kappa}(\phi_{\kappa i1},\dots \phi_{\kappa iK}) = \frac{\Lambda - 1}{\Lambda + 1}\sum_{k=1}^{K}\frac{\Delta_{k} - 1}{\Delta_{k} + 1}\phi_{\kappa k}. 
\end{align*}
Similarly, one can show that under the amplification and Type-I confounding heterogeneity frameworks, we have:
\begin{align*}
c_{\{\Lambda, \Delta\}}^{(1)}(0) & = \frac{\Lambda - 1}{\Lambda + 1}\frac{\Delta - 1}{\Delta + 1}\phi_{\kappa}, \hspace{5mm} c_{\{\boldsymbol{\Lambda}, \boldsymbol{\Delta}\}}^{(1)}(0)  = \frac{\prod_{\ell=1}^{K}\Lambda_{\ell} - 1}{\prod_{\ell=1}^{K}\Lambda_{\ell} + 1} \sum_{k=1}^{K}\frac{\Delta_{k} - 1}{\Delta_{k} + 1}\phi_{\kappa k}.
\end{align*}
Then, under the amplification framework, note the following limits:
\begin{align*}
\lim_{t\to\infty}M_{i}(t) & = \lim_{t\to\infty}\sum_{k=1}^{K}\log\bigg\{\frac{\Delta\exp\{\phi_{\kappa i k}t\} + \exp\{-\phi_{\kappa i k}t\}}{\Delta + 1}\bigg\}  = \infty, \\
\lim_{t\to\infty}M_{i}^{(1)}(t) & = \lim_{t\to\infty}\sum_{k=1}^{K}\phi_{\kappa ik}\frac{\Delta \exp\{\phi_{\kappa ik}t\} - \exp\{-\phi_{\kappa ik}t\}}{\Delta \exp\{\phi_{\kappa ik}t\} + \exp\{-\phi_{\kappa ik}t\}}   = \phi_{\kappa i}, \\
\lim_{t\to\infty}M_{i}^{(1)}(-t) & = \lim_{t\to\infty}\sum_{k=1}^{K}\phi_{\kappa ik}\frac{\Delta \exp\{-\phi_{\kappa ik}t\} - \exp\{\phi_{\kappa ik}t\}}{\Delta \exp\{-\phi_{\kappa ik}t\} + \exp\{\phi_{\kappa ik}t\}}   = -\phi_{\kappa i}, \\
\lim_{t\to\infty}Q_{i}(t) & = \lim_{t\to\infty}\frac{\exp\{M_{i}(-t) - M_{i}(t)\}}{\Lambda + \exp\{M_{i}(-t) - M_{i}(t)\}} = \frac{1}{1+\Lambda \prod_{k: \phi_{\kappa k} \neq 0}\Delta_{k}} \implies \\
\lim_{t\to\infty}c_{\{\Lambda, \Delta\}}^{(1)}(t) & = \int_{\mathbb{R}^{K}}\phi_{\kappa i} - \frac{1}{1+\Lambda \prod_{k: \phi_{\kappa ik} \neq 0}\Delta_{k}}(\phi_{\kappa i} - \phi_{\kappa i}) dC_{F, \phi, \kappa}(\phi_{\kappa i1},\dots \phi_{\kappa iK}) = \phi_{\kappa}.
\end{align*}
Then, one can show $\lim_{t\to\infty}c_{\{\Lambda, \boldsymbol{\Delta}\}}^{(1)}(t) = \lim_{t\to\infty}c_{\{\boldsymbol{\Lambda}, \boldsymbol{\Delta}\}}^{(1)}(t) = \phi_{\kappa}$ as well. Since $\eta_{\kappa} \leq \phi_{\kappa}$ for all $\kappa \in \mathbb{R}^{K}_{+}$ we immediately have $\mu_{F,  \kappa} < \phi_{\kappa}$.  Let $\{a_{1}, a_{2},\dots\}$ be a sequence of constants converging to $\mu_{F, \kappa}$. Then, evaluated at any sensitivity parameter values $p \in \widetilde{D}_{\mathcal{P}, \kappa}^{*}$, $c^{(1)}_{\mathcal{P}, \kappa}(0) < \mu_{F,  \kappa}$. Thus, by the theorem in \cite{plachky1971theorem}, we have:
\begin{equation*}
    -I^{-1}\log (P(\overline{T}_{\mathcal{P}, \kappa} \geq a_{I})) \overset{P}{\to} \widetilde{\xi}\mu_{F,  \kappa} - c^{(0)}_{\mathcal{P}, \kappa}(\widetilde{\xi}),
\end{equation*}
where $\widetilde{\xi} = \xi$ is the unique solution to $\mu_{F,  \kappa} = c^{(1)}_{\mathcal{P}, \kappa}(\xi)$. The desired result follows from Theorem 2 of \cite{raghavachari1970theorem}. 
\end{proof}

\vspace{-0.3cm}

\subsection{Theorem 6}

\begin{proof}
Recall the notation $b_{\{\Lambda, \boldsymbol{\Delta}\}, \kappa}(\Lambda, \boldsymbol{\Delta})$ as the Bahadur-Rosenbaum slope under Type-II confounding heterogeneity evaluated at sensitivity parameter values $\Lambda, \Delta_{1},\dots \Delta_{K}$. A necessary condition for $b^{+}_{\kappa}(\Gamma, a) - b_{\{\Gamma\}, \kappa}(\Gamma)$ to be non-decreasing in $\Gamma$ is that the derivative of $b^{+}_{\kappa}(\Gamma, a) - b_{\{\Gamma\}, \kappa}(\Gamma)$ evaluated at $\Gamma = 1$ is non-negative:
\begin{equation*}
\frac{\partial}{\partial \Gamma}\bigg(b^{+}_{\kappa}(\Gamma, a) - b_{\{\Gamma\}, \kappa}(\Gamma)\bigg)\bigg|_{\Gamma \to 1^{+}} \geq 0.
\end{equation*}
First, evaluating $(\partial / \partial \Gamma)b_{\{\Gamma\}, \kappa}(\Gamma)$ at $\Gamma = 1$, from the proof of Lemma~\ref{Lem: Bahadur-Rosenbaum slope convex in Gamma}, note:
\begin{align*}
\frac{\partial}{\partial \Gamma}b_{\{\Gamma\}, \kappa}(\Gamma)\bigg|_{\Gamma = 1} & = 2\int_{0}^{\widetilde{\xi}_{\{\Gamma\}, \kappa}(\Gamma)} -\frac{\partial}{\partial \Gamma}c^{(1)}_{\{\Gamma\}, \kappa}(\Gamma, t)\bigg|_{\Gamma = 1}  dt   = -\mathbb{E}_{\eta}(\tanh(\eta_{\kappa i}\widetilde{\xi}_{\{\Gamma\}, \kappa}(1))).
\end{align*}

Now, define $g = (\Gamma - 1)/(\Gamma + 1)$ so that under the calibration $a = 1$, we have $(\Lambda - 1)(\Lambda + 1) = \sqrt{g}$. Denote $\mathcal{W}_{K} = \{w_{k} \geq 0:\sum_{k=1}^{K}w_{k} = K\}$. Then, under the restriction $\{\Lambda, \Delta, \boldsymbol{\Lambda}, \boldsymbol{\Delta}\} \in \mathcal{B}_{\Gamma, a}$, note
\begin{equation*}
b_{k} = \frac{\Delta_{k} - 1}{\Delta_{k} + 1} = \sqrt{g}w_{k},
\end{equation*}
where $w_{k} \in \mathcal{W}_{K}$. 

Note that $\Delta_{k} = (1 + \sqrt{g}w_{k})/(1-\sqrt{g}w_{k})$. Now, for $g < K^{-2}$ we have $b_{k} = \sqrt{g}w_{k} < 1$, so optimizing over $\mathcal{W}_{K}$ is equivalent to optimizing over $\mathcal{W}_{K}$ with $b_{k} \leq 1$ for all $k$, given $\Gamma$ is sufficiently close to $1$. Denote $h_{ik}(t) = \tanh(t\phi_{\kappa ik})$ with $t \geq 0$. Then, the moment generating function can be rewritten as
\begin{align*}
  M_{i,g,w}(t)& =\prod_{k=1}^K\cosh(t\phi_{\kappa ik})\; \mathbb{E}_{v_i}\left[\prod_{k=1}^K \{1+v_i b_k h_{ik}(t)\}\right] \\
  & = \prod_{k=1}^K\cosh(t\phi_{\kappa ik})\; \sum_{S\subseteq\{1,\ldots,K\}} \mathbb{E}(v_i^{|S|})\prod_{k\in S}b_k h_{ik}(t) \\
  & = \prod_{k=1}^K\cosh(t\phi_{\kappa ik})\; \left[\sum_{\substack{S\subseteq\{1,\ldots,K\}\\|S|\text{ even}}} g^{|S|/2}\prod_{k\in S}w_kh_{ik}(t) + \sum_{\substack{S\subseteq\{1,\ldots,K\}\\|S|\text{ odd}}} g^{(|S|+1)/2}\prod_{k\in S}w_kh_{ik}(t) \right] \\
  & =\prod_{k=1}^K\cosh(t\phi_{\kappa ik}) \left\{1+gA_{i,w}(t)+R_{i,g,w}(t)\right\},
\end{align*}
where $A_{i,w}(t):=\sum_{k=1}^K w_kh_{ik}(t) +\sum_{1\leq k<\ell\leq K} w_kw_\ell h_{ik}(t)h_{i\ell}(t),$ and $R_{i,g,w}(t)$ captures the remaining terms that consider subsets $S\subseteq\{1,\ldots,K\}$ with $|S| \geq 3$. Now, recalling that $0 \leq w_{k} \leq K$ for all $k$ and $0 \leq h_{ik}(t) \leq 1$, for $0 \leq g \leq 1$ we have
\begin{equation*}
  0\leq R_{i,g,w}(t) \leq g^2\sum_{m=3}^K \binom{K}{m}K^m =: C_Kg^2,
\end{equation*}
so $R_{i,g,w}(t) = O(g)$ when $0 \leq g \leq 1$. Then, note that 
\begin{equation*}
  0\leq A_{i,w}(t) \leq \sum_k w_k+\sum_{k<\ell}w_kw_\ell \leq K+\frac12\left(\sum_k w_k\right)^2 =K+\frac{K^2}{2},
\end{equation*}
so $A_{i,w}(t) = O(g)$ as well. Now, taking the logarithm of $M_{i,g,w}(t)$ and applying a Taylor expansion, we have
\begin{equation*}
\log M_{i,g,w}(t) =\sum_{k=1}^K\log\cosh(t\phi_{\kappa ik}) +gA_{i,w}(t)+O(g^2).
\end{equation*}
Denote $c^{(0)}_{g, w}(t) := \mathbb{E}_{F}(\log M_{i,g,w}(t))$, $c^{(0)}_{0}(t) := \mathbb{E}_{F}(\sum_{k=1}^{K}\phi_{\kappa ik}\tanh(t\phi_{\kappa ik}))$, and $a_{w}(t) := \mathbb{E}_{F}(A_{i, w}(t))$. Then, note $c_{g,w}^{(0)}(t) =c_0^{(0)}(t)+g a_w(t)+O(g^2)$. For each $(g,w)$, define $\xi_w(g)$ as solving
\begin{equation*}
  \frac{\partial}{\partial t}c_{g,w}^{(0)}\{\xi_w(g)\}  =\mu_{F,\kappa}.
\end{equation*}
Denote the Bahadur-Rosenbaum slope $b_{w}(\Gamma):=2\left[\mu_{F,\kappa}\xi_w(g)-c_{g,w}^{(0)}\{\xi_w(g)\}\right]$. Define $H_{g,w}(t):=\mu_{F,\kappa}t-c_{g,w}^{(0)}(t)$ so that by definition, $(\partial / \partial t)H_{g,w}(\xi_{w}(g)) = 0$. By convexity of the cumulant-generating function, we have $b_{w}(\Gamma) = 2\sup_{t\geq 0}H_{g,w}(t)$. At $g=0$, the function $H_{0}(t):=\mu_{F, \kappa}t - c^{(0)}_{0}(t)$ has unique maximizer $\xi_{\mathrm{II},0}$, where $\xi_{\mathrm{II},0}$ satisfies $c^{(1)}_{0}(\xi_{\mathrm{II},0}) = \mu_{F, \kappa}$. For any compact interval $J \subset [0, \infty)$, one can show that as $g \to 0$ from the right side, we have
\begin{equation*}
\sup_{w\in\mathcal W_K}\sup_{t\in J} \left|c_{g,w}^{(1)}(t)-c_0^{(1)}(t)\right| \to 0.
\end{equation*}
This implies that as $g \to 0$ from the right side, $\sup_{w\in\mathcal W_K}|\xi_{w}(g) - \xi_{\mathrm{II},0}| \to 0$. Now, note that $\sup_{t\geq 0}H_{g,w}(t) \geq H_{g, w}(\xi_{\mathrm{II},0}) = H_{0}(\xi_{\mathrm{II},0}) - ga_{w}(\xi_{\mathrm{II},0}) + O(g^{2})$. Additionally, since $H_0\{\xi_w(g)\}\leq H_0(\xi_{\mathrm{II},0})$, we have
\begin{align*}
  \sup_{t\geq0}H_{g,w}(t) &=H_{g,w}\{\xi_w(g)\}\\
  &=H_0\{\xi_w(g)\} -g a_w\{\xi_w(g)\}+O(g^2)\\
  &\leq H_0(\xi_{\mathrm{II},0}) -g a_w(\xi_{\mathrm{II},0})+o(g).
\end{align*}
Thus, $b_{w}(\Gamma) = b_{0} - 2g\mathbb{E}_F(A_{i,w}(\xi_{\mathrm{II},0}))+o(g)$ uniformly over $w \in \mathcal{W}_{K}$. Therefore, we have
\begin{equation*}
b_\kappa^+(\Gamma,1) = \sup_{w\in\mathcal W_K}b_w(\Gamma) = b_0-2g\inf_{w\in\mathcal W_K}
       \mathbb{E}_F(A_{i,w}(\xi_{\mathrm{II},0}))+o(g).
\end{equation*}
Now, recalling $h_{ik}(\xi_{\mathrm{II},0}) =\tanh(\phi_{\kappa ik}\xi_{\mathrm{II},0})$ and defining $h_k:=\mathbb{E}_F(h_{ik})$, we have 
\begin{align*}
  \mathbb{E}_F(A_{i,w}(\xi_{\mathrm{II},0}))
  &=\sum_{k=1}^K w_k h_k +\sum_{1\leq k<\ell\leq K} w_kw_\ell \mathbb{E}_F(h_{ik}(\xi_{\mathrm{II},0}) h_{i\ell}(\xi_{\mathrm{II},0})) \geq \sum_{k=1}^K w_k h_k\\
  &\geq \left(\sum_{k=1}^K w_k\right)\min_{1\leq k\leq K}h_k =K\min_{1\leq k\leq K}h_k.
\end{align*}
The first inequality holds because every pair-product term is nonnegative. The second holds because each $h_k$ is at least its minimum and every $w_k$ is nonnegative. This lower bound is attained. Choose an index $k_*$ such that $h_{k_*}=\min_kh_k$ and take the vertex  $w=K e_{k_*}$, where $e_{k_*}$ is the $k_*$th standard basis vector in $\mathbb{R}^K$. This vector belongs to $\mathcal{W}_K$. At this vertex, all pair-product terms $w_kw_\ell$ with $k<\ell$ are zero, and the singleton term equals $Kh_{k_*}$. Therefore, 
\begin{equation}
  \inf_{w\in\mathcal W_K} \mathbb{E}_F(A_{i,w}(\xi_{\mathrm{II},0})) =K\min_{1\leq k\leq K} \mathbb{E}_F(\tanh\{\phi_{\kappa ik} \xi_{\mathrm{II},0}\}).
\end{equation}
Therefore, $b_\kappa^+(\Gamma,1) =b_0-2gK\min_{1\leq k\leq K} \mathbb{E}_F(\tanh\{\phi_{\kappa ik} \xi_{\mathrm{II},0}\})+o(g).$ Since $g/(\Gamma-1)=1/(\Gamma+1)\to1/2$, the right derivative is
\begin{equation*}
\frac{\partial}{\partial \Gamma}b_{\kappa}^{+}(\Gamma, 1)\bigg|_{\Gamma \to 1^{+}} = -K\min_{1\leq k\leq K} \mathbb{E}_F(\tanh\{\phi_{\kappa ik} \xi_{\mathrm{II},0}\}).
\end{equation*}
Therefore, 
\begin{equation*}
\frac{\partial}{\partial \Gamma} b^{+}_{\kappa}(\Gamma, a)\bigg|_{\Gamma \to 1^{+}} \geq \frac{\partial}{\partial \Gamma}b_{\{\Gamma\}, \kappa}(\Gamma)\bigg|_{\Gamma \to 1^{+}}
\end{equation*}
only if the following inequality holds:
\begin{equation*}
\mathbb{E}_{F}\big(\tanh(\eta_{\kappa i}\widetilde{\xi}_{\{\Gamma\}, \kappa}(1))\big) \geq K\min_{1\leq k\leq K} \mathbb{E}_{F}\big(\tanh(\phi_{\kappa ik}\widetilde{\xi}_{\mathrm{II}, 0})\big).
\end{equation*}

\end{proof}

\vspace{-0.5cm}

\section{Additional Simulations and Calculations}

\vspace{-0.3cm}

\subsection{Design Sensitivity}

\subsubsection*{Rank-based Test Statistic}

We consider the same underlying data generation process for $Y_{ik}V_{i}$, where $(Y_{i1}V_{i}, Y_{i2}V_{i}) \iid N(\boldsymbol{\tau}, \Sigma)$, with diagonal entries $\Sigma_{11} = \Sigma_{22} = 1$ and off-diagonal entries $\Sigma_{12} = \Sigma_{21} = \rho$. 
Here we consider a rank-based test statistic, setting $\widetilde{q}_{ik}=F_{|Y_{k}V|}(|Y_{ik}V_{i}|)$ corresponding to the Wilcoxon signed-rank test \citep{woolson2007Wilcoxon} when ranks are scaled from $0$ to $1$. We consider the cases when $(\tau_{1}, \tau_{2})=(0.25, 0.25)$ and $(\tau_{1}, \tau_{2})=(0.1, 0.5)$.

{\renewcommand{\arraystretch}{0.70}
\begin{table}[ht]
\vspace{-0.1cm}
\centering
\begin{tabular}[t]{c|ccccccc}
\multicolumn{8}{c}{$Y_{i1}V_{i} \sim N(0.25, 1), Y_{i2}V_i \sim N(0.25, 1)$, $\widetilde{q}_{ik} = F_{|Y_{k}V|}(|Y_{ik}V_{i}|)$} \\
\toprule
$(\kappa_{1}, \kappa_{2})$ & $\widetilde{\Gamma}_{-0.5}$ & $\widetilde{\Gamma}_{0}$ & $\widetilde{\Gamma}_{0.5}$ & $\widetilde{\Lambda}^{*}$, $\widetilde{\Delta}^{*}$ & $\widetilde{\Lambda}^{*}_{1}$, $\widetilde{\Lambda}^{*}_{2}$ & $\widetilde{\Delta}^{*}_{1}$ & $\widetilde{\Delta}^{*}_{2}$ \\
\hline 
(1, 0) & 1.76 & 1.76 & 1.76 & 1.76 & 1.33 & 1.76 & 1.00 \\
(0.75, 0.25) & 2.35 & 2.08 & 1.89 & 1.76 & 1.33 & 2.17 & 1.00 \\
(0.5, 0.5) & 3.44 & 2.34 & 1.96 & 1.76 & 1.33 & 1.76 & 1.76 \\
(0.25, 0.75) & 2.35 & 2.07 & 1.90 & 1.76 & 1.33 & 1.00 & 2.17 \\
(0, 1) & 1.76 & 1.76 & 1.76 & 1.76 & 1.33 & 1.00 & 1.76 \\
\bottomrule
\end{tabular} 
\\
\vspace{0.5cm}
\vspace{-0.1cm}
\centering
\begin{tabular}[t]{c|ccccccc}
\multicolumn{8}{c}{$Y_{i1}V_{i} \sim N(0.1, 1), Y_{i2}V_i \sim N(0.5, 1)$, $\widetilde{q}_{ik} = F_{|Y_{k}V|}(|Y_{ik}V_{i}|)$} \\
\toprule
$(\kappa_{1}, \kappa_{2})$ & $\widetilde{\Gamma}_{-0.5}$ & $\widetilde{\Gamma}_{0}$ & $\widetilde{\Gamma}_{0.5}$ & $\widetilde{\Lambda}^{*}$, $\widetilde{\Delta}^{*}$ & $\widetilde{\Lambda}^{*}_{1}$, $\widetilde{\Lambda}^{*}_{2}$ & $\widetilde{\Delta}^{*}_{1}$ & $\widetilde{\Delta}^{*}_{2}$ \\
\hline 
(1, 0) & 1.25 & 1.25 & 1.25 & 1.25 & 1.11 & 1.25 & 1.00 \\
(0.75, 0.25) & 1.91 & 1.75 & 1.64 & 1.55 & 1.24 & 1.80 & 1.00 \\
(0.5, 0.5) & 4.27 & 2.72 & 2.22 & 1.92 & 1.39 & 1.93 & 1.93 \\
(0.25, 0.75) & 4.10 & 3.60 & 2.80 & 2.44 & 1.56 & 1.35 & 3.07 \\
(0, 1) & 3.17 & 3.17 & 3.17 & 3.17 & 1.78 & 1.00 & 3.17 \\
\bottomrule
\end{tabular}
\label{tab: design sensitivity example 2}
\caption{Infimum design sensitivity values across the classic, amplification, Type-I confounding heterogeneity and Type-II confounding heterogeneity frameworks when $Y_{i1}V_{i} \sim N(\tau_{1}, 1), Y_{i2}V_i \sim N(\tau_{2}, 1)$ and $\widetilde{q}_{ik}$ is chosen to be $\widetilde{q}_{ik}=F_{|Y_{k}V|}(|Y_{ik}V_{i}|)$, corresponding to the Wilcoxon signed-rank test \citep{woolson2007Wilcoxon}.}
\end{table}}

\newpage
The rank-based test statistic with $\widetilde{q}_{ik} = F_{|Y_{k}V|}(|Y_{ik}V_{i}|)$ has lower design sensitivity than the permutation t-test in both cases when $(\tau_{1}, \tau_{2})=(0.25, 0.25)$ and when $(\tau_{1}, \tau_{2})=(0.1, 0.5)$. The only exception is in the second case under the classic framework when both outcomes are uncorrelated $(\rho = 0)$ and $\kappa = (0.25, 0.75)$.

\vspace{-0.3cm}
\subsubsection*{Heavy-tailed outcomes}

We consider a heavier-tailed outcome distribution, where $Y_{i1}V_{i} \iid \tau_{1} + \sqrt{3/5}\cdot t_{5}$ and $Y_{i2}V_{i} \iid \tau_{2} + \sqrt{3/5}\cdot t_{5}$, where $t_{5}$ is the t-distribution with 5 degrees of freedom. To clarify, the mean of $Y_{i1}V_{i}$ is $\tau_{1}$ and the mean of $Y_{i2}V_{i}$ is $\tau_{2}$, and both $Y_{i1}V_{i}$ and $Y_{i2}V_{i}$ are scaled to have variance $1$. We consider four cases: 1a) $(\tau_{1}, \tau_{2})=(0.25, 0.25), \widetilde{q}_{ik} = |Y_{ik}V_{i}|$, 1b) $(\tau_{1}, \tau_{2})=(0.1, 0.5), \widetilde{q}_{ik} = |Y_{ik}V_{i}|$, 2a) $(\tau_{1}, \tau_{2})=(0.25, 0.25), \widetilde{q}_{ik}=F_{|Y_{k}V|}(|Y_{ik}V_{i}|)$, and 2b) $(\tau_{1}, \tau_{2})=(0.1, 0.5), \widetilde{q}_{ik}=F_{|Y_{k}V|}(|Y_{ik}V_{i}|)$. We estimate each design sensitivity and infimum design sensitivity using Monte Carlo estimation with $I=10{,}000{,}000$ matched pairs.

{\renewcommand{\arraystretch}{0.70}
\begin{table}[h]
\centering
\begin{tabular}[t]{c|ccccccc}
\multicolumn{8}{c}{$Y_{i1}V_{i} \iid 0.25 + \sqrt{3/5}\cdot t_{5}, Y_{i2}V_{i} \iid 0.25 + \sqrt{3/5}\cdot t_{5}$, $\widetilde{q}_{ik} = |Y_{ik}V_{i}|$} \\
\toprule
$(\kappa_{1}, \kappa_{2})$ & $\widetilde{\Gamma}_{-0.5}$ & $\widetilde{\Gamma}_{0}$ & $\widetilde{\Gamma}_{0.5}$ & $\widetilde{\Lambda}^{*}$, $\widetilde{\Delta}^{*}$ & $\widetilde{\Lambda}^{*}_{1}$, $\widetilde{\Lambda}^{*}_{2}$ & $\widetilde{\Delta}^{*}_{1}$ & $\widetilde{\Delta}^{*}_{2}$ \\
\hline 
(1, 0) & 1.97 & 1.97 & 1.97 & 1.97 & 1.40 & 1.97 & 1.00 \\
(0.75, 0.25) & 2.78 & 2.33 & 2.12 & 1.97 & 1.40 & 2.51 & 1.03 \\
(0.5, 0.5) & 3.84 & 2.54 & 2.19 & 1.97 & 1.40 & 1.97 & 1.97 \\
(0.25, 0.75) & 2.78 & 2.33 & 2.12 & 1.97 & 1.40 & 1.03 & 2.51 \\
(0, 1) & 1.97 & 1.97 & 1.97 & 1.97 & 1.40 & 1.00 & 1.97 \\
\bottomrule
\end{tabular} 
\\
\vspace{0.5cm}
\centering
\begin{tabular}[t]{c|ccccccc}
\multicolumn{8}{c}{$Y_{i1}V_{i} \iid 0.1 + \sqrt{3/5}\cdot t_{5}, Y_{i2}V_{i} \iid 0.5 + \sqrt{3/5}\cdot t_{5}$, $\widetilde{q}_{ik} = |Y_{ik}V_{i}|$} \\
\toprule
$(\kappa_{1}, \kappa_{2})$ & $\widetilde{\Gamma}_{-0.5}$ & $\widetilde{\Gamma}_{0}$ & $\widetilde{\Gamma}_{0.5}$ & $\widetilde{\Lambda}^{*}$, $\widetilde{\Delta}^{*}$ & $\widetilde{\Lambda}^{*}_{1}$, $\widetilde{\Lambda}^{*}_{2}$ & $\widetilde{\Delta}^{*}_{1}$ & $\widetilde{\Delta}^{*}_{2}$ \\
\hline 
(1, 0) & 1.31 & 1.31 & 1.31 & 1.31 & 1.15 & 1.31 & 1.00 \\
(0.75, 0.25) & 2.27 & 1.97 & 1.83 & 1.70 & 1.31 & 2.13 & 1.00 \\
(0.5, 0.5) & 4.97 & 3.05 & 2.55 & 2.21 & 1.49 & 2.09 & 2.32 \\
(0.25, 0.75) & 5.04 & 3.85 & 3.31 & 2.88 & 1.70 & 1.49 & 3.63 \\
(0, 1) & 3.83 & 3.83 & 3.83 & 3.83 & 1.96 & 1.00 & 3.83 \\
\bottomrule
\end{tabular}
\label{tab: design sensitivity heavy tail, normal}
\caption{Infimum design sensitivity values across the classic, amplification, Type-I confounding heterogeneity and Type-II confounding heterogeneity frameworks when $Y_{i1}V_{i} \iid \tau_{1} + \sqrt{3/5}\cdot t_{5}, Y_{i2}V_{i} \iid \tau_{2} + \sqrt{3/5}\cdot t_{5}$, and $\widetilde{q}_{ik}$ is chosen to be $\widetilde{q}_{ik}=|Y_{ik}V_{i}|$.}
\end{table}}

{\renewcommand{\arraystretch}{0.70}
\begin{table}[ht]
\centering
\begin{tabular}[t]{c|ccccccc}
\multicolumn{8}{c}{$Y_{i1}V_{i} \iid 0.25 + \sqrt{3/5}\cdot t_{5}, Y_{i2}V_{i} \iid 0.25 + \sqrt{3/5}\cdot t_{5}$, $\widetilde{q}_{ik} = F_{|Y_{k}V|}(|Y_{ik}V_{i}|)$} \\
\toprule
$(\kappa_{1}, \kappa_{2})$ & $\widetilde{\Gamma}_{-0.5}$ & $\widetilde{\Gamma}_{0}$ & $\widetilde{\Gamma}_{0.5}$ & $\widetilde{\Lambda}^{*}$, $\widetilde{\Delta}^{*}$ & $\widetilde{\Lambda}^{*}_{1}$, $\widetilde{\Lambda}^{*}_{2}$ & $\widetilde{\Delta}^{*}_{1}$ & $\widetilde{\Delta}^{*}_{2}$ \\
\hline 
(1, 0) & 1.90 & 1.90 & 1.90 & 1.90 & 1.38 & 1.90 & 1.00 \\
(0.75, 0.25) & 2.63 & 2.29 & 2.07 & 1.90 & 1.38 & 2.42 & 1.00 \\
(0.5, 0.5) & 4.09 & 2.62 & 2.17 & 1.90 & 1.38 & 1.90 & 1.90 \\
(0.25, 0.75) & 2.63 & 2.29 & 2.07 & 1.90 & 1.38 & 1.00 & 2.42 \\
(0, 1) & 1.90 & 1.90 & 1.90 & 1.90 & 1.38 & 1.00 & 1.90 \\
\bottomrule
\end{tabular} 
\\
\vspace{0.5cm}
\centering
\begin{tabular}[t]{c|ccccccc}
\multicolumn{8}{c}{$Y_{i1}V_{i} \iid 0.1 + \sqrt{3/5}\cdot t_{5}, Y_{i2}V_{i} \iid 0.5 + \sqrt{3/5}\cdot t_{5}$, $\widetilde{q}_{ik} = F_{|Y_{k}V|}(|Y_{ik}V_{i}|)$} \\
\toprule
$(\kappa_{1}, \kappa_{2})$ & $\widetilde{\Gamma}_{-0.5}$ & $\widetilde{\Gamma}_{0}$ & $\widetilde{\Gamma}_{0.5}$ & $\widetilde{\Lambda}^{*}$, $\widetilde{\Delta}^{*}$ & $\widetilde{\Lambda}^{*}_{1}$, $\widetilde{\Lambda}^{*}_{2}$ & $\widetilde{\Delta}^{*}_{1}$ & $\widetilde{\Delta}^{*}_{2}$ \\
\hline 
(1, 0) & 1.29 & 1.29 & 1.29 & 1.29 & 1.14 & 1.29 & 1.00 \\
(0.75, 0.25) & 2.04 & 1.86 & 1.74 & 1.63 & 1.27 & 1.93 & 1.00 \\
(0.5, 0.5) & 5.07 & 3.05 & 2.46 & 2.07 & 1.44 & 2.07 & 2.07 \\
(0.25, 0.75) & 4.83 & 3.76 & 3.19 & 2.70 & 1.64 & 1.53 & 3.38 \\
(0, 1) & 3.65 & 3.65 & 3.65 & 3.65 & 1.91 & 1.00 & 3.65 \\
\bottomrule
\end{tabular}
\label{tab: design sensitivity heavy tail, rank}
\caption{Infimum design sensitivity values across the classic, amplification, Type-I confounding heterogeneity and Type-II confounding heterogeneity frameworks when $Y_{i1}V_{i} \iid \tau_{1} + \sqrt{3/5}\cdot t_{5}, Y_{i2}V_{i} \iid \tau_{2} + \sqrt{3/5}\cdot t_{5}$ and $\widetilde{q}_{ik}$ is chosen to be $\widetilde{q}_{ik}=F_{|Y_{k}V|}(|Y_{ik}V_{i}|)$, corresponding to the Wilcoxon signed-rank test \citep{woolson2007Wilcoxon}.}
\end{table}}

The permutation t-test has slightly higher design sensitivity when $Y_{i1}V_{i} \iid \tau_{1} + \sqrt{3/5}\cdot t_{5}$ and $Y_{i2}V_{i} \iid \tau_{2} + \sqrt{3/5}\cdot t_{5}$ in both the case when $(\tau_{1}, \tau_{2})=(0.25, 0.25)$ and when $(\tau_{1}, \tau_{2})=(0.1, 0.5)$, compared to when outcomes are normally distributed. The rank-based test statistic also has improved design sensitivity when $Y_{i1}V_{i} \iid \tau_{1} + \sqrt{3/5}\cdot t_{5}$ and $Y_{i2}V_{i} \iid \tau_{2} + \sqrt{3/5}\cdot t_{5}$ across most cases of $\kappa$ tested. 

\newpage
\subsection{Finite-sample Power}

We show the simulated finite-sample power when $(Y_{i1}V_{i}, Y_{i2}V_{i}) \iid N((0.1, 0.5), \Sigma)$ with diagonal entries $\Sigma_{11} = \Sigma_{22} = 1$ and off-diagonal entries $\Sigma_{12} = \Sigma_{21} = 0$. We consider $q_{ik}=|Y_{ik}V_{i}|$, corresponding to the permutation t-test \citep{welch1937z}. We simulate $1000$ datasets with $I = 300$ matched pairs and use a significance level of $\alpha = 0.05$. Table~\ref{tab: finite-sample power} shows the simulated finite-sample power when $a=1/3, 1, 3$, recalling that for a fixed $\Gamma \geq 1$, the sensitivity parameter values $\{\Lambda, \Delta, \boldsymbol{\Lambda}, \boldsymbol{\Delta}\} \in \mathcal{B}_{\Gamma, a}$, defined in the main text.

{\renewcommand{\arraystretch}{0.75}
\begin{table}
    \centering
    \footnotesize
    \setlength{\tabcolsep}{2pt}
    \begin{tabular}{ccc|ccccccccccc}
    \toprule
        & & & \multicolumn{11}{c}{$\Gamma$} \\
        $a$ & $\kappa$ & Framework & 1 & 1.2 & 1.4 & 1.6 & 1.8 & 2.0 & 2.2 & 2.4 & 2.6 & 2.8 & 3.0 \\
        \toprule
        \multirow{12}{*}{$1/3$} & \multirow{4}{*}{$(0.5, 0.5)$} & classic & 1 & 1 & 0.999 & 0.988 & 0.926 & 0.76 & 0.525 & 0.287 & 0.135 & 0.055 & 0.02 \\
         & & Inf CH & 1 & 1 & 0.976 & 0.741 & 0.279 & 0.046 & 0.008 & 0 & 0 & 0 & 0 \\
         & & amplification & 1 & 1 & 0.98 & 0.801 & 0.372 & 0.077 & 0.015 & 0.001 & 0 & 0 & 0 \\
         & & Sup CH & 1 & 1 & 0.989 & 0.866 & 0.541 & 0.209 & 0.056 & 0.014 & 0.001 & 0 & 0 \\
         \cline{2-14}
         & \multirow{4}{*}{$(0.25, 0.75)$} & classic & 1 & 1 & 1 & 1 & 0.998 & 0.981 & 0.913 & 0.776 & 0.579 & 0.386 & 0.228 \\
         & & Inf CH & 1 & 1 & 0.992 & 0.647 & 0.079 & 0.001 & 0 & 0 & 0 & 0 & 0 \\
         & & amplification & 1 & 1 & 1 & 0.997 & 0.955 & 0.766 & 0.423 & 0.177 & 0.049 & 0.003 & 0.001 \\
         & & Sup CH & 1 & 1 & 1 & 1 & 1 & 1 & 1 & 0.998 & 0.992 & 0.955 & 0.852 \\
         \cline{2-14}
         & \multirow{4}{*}{$(0, 1)$} & classic & 1 & 1 & 1 & 1 & 0.997 & 0.979 & 0.898 & 0.757 & 0.542 & 0.351 & 0.193 \\
         & & Inf CH & 1 & 1 & 0.979 & 0.428 & 0.011 & 0 & 0 & 0 & 0 & 0 & 0 \\
         & & amplification & 1 & 1 & 1 & 1 & 0.997 & 0.979 & 0.898 & 0.757 & 0.542 & 0.351 & 0.193 \\
         & & Sup CH & 1 & 1 & 1 & 1 & 1 & 1 & 1 & 1 & 1 & 1 & 1 \\
         \midrule 
        \multirow{12}{*}{$1$} & \multirow{4}{*}{$(0.5, 0.5)$} & classic & 1 & 1 & 0.999 & 0.988 & 0.926 & 0.76 & 0.525 & 0.287 & 0.135 & 0.055 & 0.02 \\
         & & Inf CH & 1 & 1 & 0.975 & 0.733 & 0.273 & 0.044 & 0.007 & 0 & 0 & 0 & 0 \\
         & & amplification & 1 & 1 & 0.978 & 0.771 & 0.334 & 0.058 & 0.012 & 0.001 & 0 & 0 & 0 \\
         & & Sup CH & 1 & 1 & 0.989 & 0.866 & 0.516 & 0.15 & 0.03 & 0.001 & 0 & 0 & 0 \\
         \cline{2-14}
         & \multirow{4}{*}{$(0.25, 0.75)$} & classic & 1 & 1 & 1 & 1 & 0.998 & 0.981 & 0.913 & 0.776 & 0.579 & 0.386 & 0.228 \\
         & & Inf CH & 1 & 1 & 0.992 & 0.647 & 0.158 & 0.019 & 0.001 & 0 & 0 & 0 & 0 \\
         & & amplification & 1 & 1 & 1 & 0.997 & 0.952 & 0.753 & 0.404 & 0.157 & 0.045 & 0.002 & 0.001 \\
         & & Sup CH & 1 & 1 & 1 & 1 & 1 & 0.999 & 0.989 & 0.94 & 0.794 & 0.535 & 0.318 \\
         \cline{2-14} 
         & \multirow{4}{*}{$(0, 1)$} & classic & 1 & 1 & 1 & 1 & 0.997 & 0.979 & 0.898 & 0.757 & 0.542 & 0.351 & 0.193 \\
         & & Inf CH & 1 & 1 & 0.979 & 0.428 & 0.065 & 0.007 & 0 & 0 & 0 & 0 & 0 \\
         & & amplification & 1 & 1 & 1 & 1 & 0.997 & 0.979 & 0.898 & 0.757 & 0.542 & 0.351 & 0.193 \\
         & & Sup CH & 1 & 1 & 1 & 1 & 1 & 1 & 1 & 1 & 1 & 0.999 & 0.995 \\
         \midrule 
        \multirow{12}{*}{$3$} & \multirow{4}{*}{$(0.5, 0.5)$} & classic & 1 & 1 & 0.999 & 0.988 & 0.926 & 0.76 & 0.525 & 0.287 & 0.135 & 0.055 & 0.02 \\
         & & Inf CH & 1 & 1 & 0.973 & 0.705 & 0.255 & 0.043 & 0.005 & 0 & 0 & 0 & 0 \\
         & & amplification & 1 & 1 & 0.974 & 0.725 & 0.283 & 0.05 & 0.006 & 0.001 & 0 & 0 & 0 \\
         & & Sup CH & 1 & 1 & 0.977 & 0.77 & 0.346 & 0.061 & 0.011 & 0.001 & 0 & 0 & 0 \\
         \cline{2-14} 
         & \multirow{4}{*}{$(0.25, 0.75)$} & classic & 1 & 1 & 1 & 1 & 0.998 & 0.981 & 0.913 & 0.776 & 0.579 & 0.386 & 0.228 \\
         & & Inf CH & 1 & 1 & 0.999 & 0.956 & 0.694 & 0.305 & 0.075 & 0.005 & 0.001 & 0 & 0 \\
         & & amplification & 1 & 1 & 1 & 0.997 & 0.937 & 0.717 & 0.381 & 0.133 & 0.038 & 0.002 & 0.001 \\
         & & Sup CH & 1 & 1 & 1 & 1 & 0.996 & 0.953 & 0.79 & 0.501 & 0.263 & 0.09 & 0.028 \\
         \cline{2-14}  
         & \multirow{4}{*}{$(0, 1)$} & classic & 1 & 1 & 1 & 1 & 0.997 & 0.979 & 0.898 & 0.757 & 0.542 & 0.351 & 0.193 \\
         & & Inf CH & 1 & 1 & 1 & 0.984 & 0.871 & 0.6 & 0.34 & 0.143 & 0.045 & 0.013 & 0.004 \\
         & & amplification & 1 & 1 & 1 & 1 & 0.997 & 0.979 & 0.898 & 0.757 & 0.542 & 0.351 & 0.193 \\
         & & Sup CH & 1 & 1 & 1 & 1 & 1 & 1 & 0.998 & 0.985 & 0.959 & 0.889 & 0.783 \\
        \bottomrule
         
    \end{tabular}
    \caption{The simulated power over $1000$ simulated datasets of $I=300$ matched pairs, where $\kappa = (0.5, 0.5), (0.25, 0.75), (0, 1)$ and $a = 1/3, 1, 3$.}
    \label{tab: finite-sample power}
\end{table}}

\begin{figure}[ht] 
    \centering
    \makebox[\textwidth][c]{%
    \begin{tikzpicture}

        \node[] at (-6, 0) {\includegraphics[width=0.35\textwidth]{figs/Finite-Sample_Power/power_a3_kappa.5.5.png}};
        \node[] at (0, 0) {\includegraphics[width=0.35\textwidth]{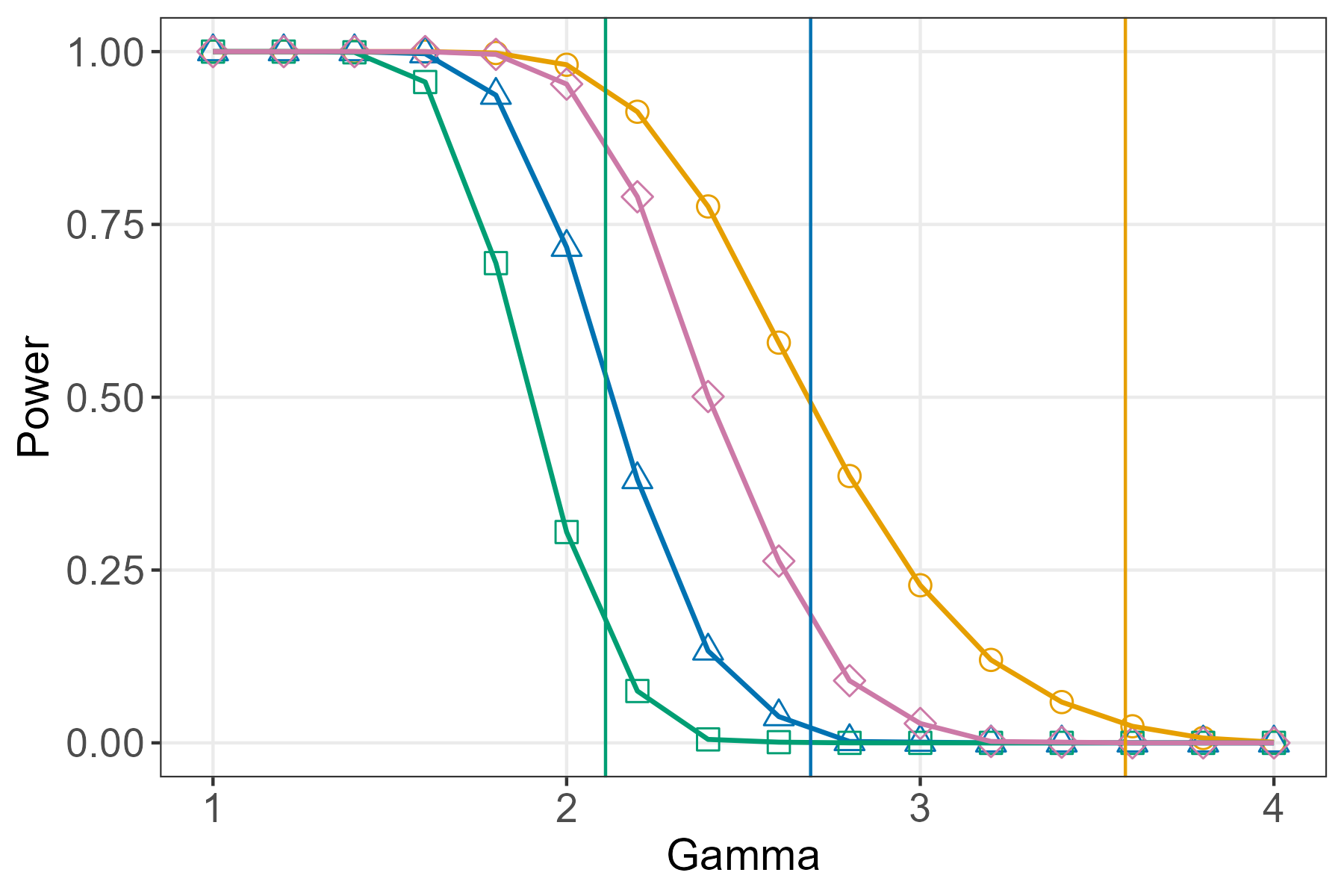}};
        \node[] at (6, 0) {\includegraphics[width=0.35\textwidth]{figs/Finite-Sample_Power/power_a3_kappa01.png}};

                
    \end{tikzpicture}
    }
    \caption{The simulated power over $1000$ simulated datasets of $I=300$ matched pairs, where $\kappa = (0.5, 0.5)$ (left) $\kappa = (0.25, 0.75)$ (middle), and $(0, 1)$ (right) with $a=1$. The vertical line for the classic framework (orange) corresponds to $\widetilde{\Gamma}_{0}$, the vertical line for the amplification framework (blue) corresponds to the value of $\Gamma$ satisfying $\Gamma = \widetilde{\Delta}^{*}$, and the vertical line for the worst-case confounding heterogeneity scenario (green) corresponds to the value of $\Gamma$ satisfying $\Gamma/(\Gamma + 1) = 0.5(\widetilde{\Delta}^{*}_{1}/(\widetilde{\Delta}^{*}_{1} + 1) + \widetilde{\Delta}^{*}_{2}/(\widetilde{\Delta}^{*}_{2} + 1))$. Note that for $\kappa = (0.5, 0.5)$, the vertical lines for the amplification and infimum confounding heterogeneity frameworks overlap, and for $\kappa = (0, 1)$, the vertical lines for the classic and amplification frameworks overlap.}
    \label{fig: finite-sample power, a = 1}
\end{figure}

\begin{figure}[h] 
    \centering
    \makebox[\textwidth][c]{%
    \begin{tikzpicture}

        \node[] at (-3, 0) {\includegraphics[width=0.35\textwidth]{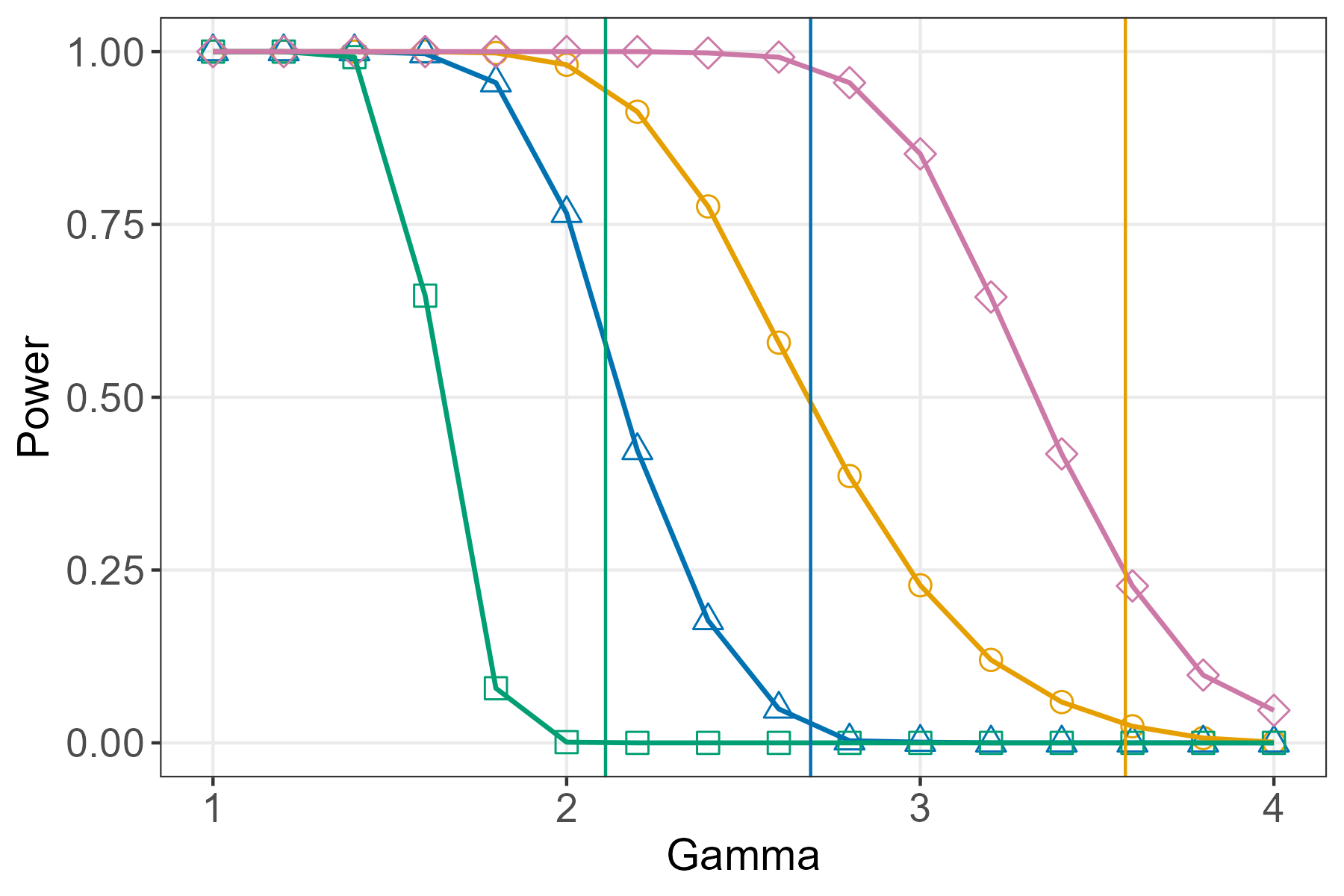}};
        \node[] at (3, 0) {\includegraphics[width=0.35\textwidth]{figs/Finite-Sample_Power/power_a3_kappa.25.75.png}};
                
    \end{tikzpicture}
    }
    \caption{The simulated power over $1000$ simulated datasets of $I=300$ matched pairs, where $\kappa = (0.25, 0.75)$ and $a=1/3$ (left) and $a = 3$ (right). The vertical line for the classic framework (orange) corresponds to $\widetilde{\Gamma}_{0}$, the vertical line for the amplification framework (blue) corresponds to the value of $\Gamma$ satisfying $\Gamma = \widetilde{\Delta}^{*}$, and the vertical line for the worst-case confounding heterogeneity scenario (green) corresponds to the value of $\Gamma$ satisfying $\Gamma/(\Gamma + 1) = 0.5(\widetilde{\Delta}^{*}_{1}/(\widetilde{\Delta}^{*}_{1} + 1) + \widetilde{\Delta}^{*}_{2}/(\widetilde{\Delta}^{*}_{2} + 1))$.}
    \label{fig: finite-sample power, kappa = (0.25 0.75)}
\end{figure}

\newpage

\subsection{Bahadur-Rosenbaum Efficiency}

We show the calculated Bahadur-Rosenbaum efficiency when $(Y_{i1}V_{i}, Y_{i2}V_{i}) \iid N((0.1, 0.5), \Sigma)$ with diagonal entries $\Sigma_{11} = \Sigma_{22} = 1$ and off-diagonal entries $\Sigma_{12} = \Sigma_{21} = 0$. We consider $q_{ik}=|Y_{ik}V_{i}|$, corresponding to the permutation t-test \citep{welch1937z}. Table~\ref{tab: Bahadur-Rosenbaum slope table} shows the relative Bahadur-Rosenbaum efficiency and simulated ratio of the minimum number of matched pairs required to achieve 95\% power at a significance level of $\alpha = 0.01$ with sensitivity parameter values $\{\Lambda, \Delta, \boldsymbol{\Lambda}, \boldsymbol{\Delta}\} \in \mathcal{B}_{\Gamma, a}$ for $a = 1/3, 1, 3$.

{\renewcommand{\arraystretch}{0.75}
\begin{table}
\centering
\footnotesize
\setlength{\tabcolsep}{3pt}
\makebox[\textwidth][c]{%

\begin{tabular}[t]{cc|ccccc|ccccc|ccccc}
\toprule
& $(\kappa_{1},\kappa_{2})$ & & \multicolumn{3}{c}{$(0.5, 0.5)$} &  &  & \multicolumn{3}{c}{$(0.25, 0.75)$} & & & \multicolumn{3}{c}{$(0, 1)$} & \\
a & $\Gamma$ & 1 & 1.1 & 1.3 & 1.5 & 1.7 & 1 & 1.1 & 1.3 & 1.5 & 1.7 & 1 & 1.1 & 1.3 & 1.5 & 1.7 \\
\midrule 
\multirow{6}{*}{$1/3$} & \multirow{2}{*}{Infimum CH} & 1.01 & 0.91 & 0.71 & 0.49 & 0.25 & 1.01 & 0.86 & 0.57 & 0.29 & 0.07 & 1.00 & 0.83 & 0.51 & 0.22 & 0.02 \\
& & (1.02) & (0.93) & (0.72) & (0.48) & (0.25) & (1.02) & (0.84) & (0.55) & (0.27) & (0.06) & (1.00) & (0.82) & (0.49) & (0.21) & (0.02) \\
\cline{2-17}
& \multirow{2}{*}{amplification} & 1.01 & 0.92 & 0.73 & 0.53 & 0.31 & 1.01 & 0.96 & 0.86 & 0.75 & 0.63 & 1.00 & 1.00 & 1.00 & 1.01 & 1.00 \\
& & (1.02) & (0.93) & (0.74) & (0.53) & (0.32) & (1.02) & (0.96) & (0.87) & (0.75) & (0.64) & (1.00) & (1.00) & (0.99) & (0.99) & (1.00) \\
\cline{2-17}
& \multirow{2}{*}{Supremum CH} & 1.01 & 0.94 & 0.80 & 0.64 & 0.45 & 1.01 & 1.07 & 1.24 & 1.46 & 1.74 & 1.00 & 1.19 & 1.66 & 2.33 & 3.24 \\
& & (1.02) & (0.96) & (0.81) & (0.63) & (0.45) & (1.02) & (1.09) & (1.28) & (1.51) & (1.88) & (1.00) & (1.17) & (1.64) & (2.33) & (3.33) \\
\midrule
\multirow{6}{*}{$1$} & \multirow{2}{*}{Infimum CH} & 1.01 & 0.89 & 0.68 & 0.47 & 0.25 & 1.01 & 0.86 & 0.56 & 0.29 & 0.08 & 1.00 & 0.83 & 0.51 & 0.21 & 0.03 \\
& & (1.02) & (0.89) & (0.68) & (0.46) & (0.25) & (1.02) & (0.84) & (0.55) & (0.27) & (0.07) & (1.00) & (0.82) & (0.49) & (0.20) & (0.03) \\
\cline{2-17}
& \multirow{2}{*}{amplification} & 1.01 & 0.90 & 0.69 & 0.49 & 0.29 & 1.01 & 0.95 & 0.83 & 0.71 & 0.60 & 1.00 & 1.00 & 1.00 & 1.00 & 1.01 \\
& & (1.02) & (0.89) & (0.69) & (0.48) & (0.29) & (1.02) & (0.94) & (0.83) & (0.71) & (0.60) & (1.00) & (0.98) & (0.99) & (0.98) & (1.00) \\
\cline{2-17}
& \multirow{2}{*}{Supremum CH} & 1.01 & 0.94 & 0.80 & 0.64 & 0.44 & 1.01 & 1.08 & 1.24 & 1.45 & 1.69 & 1.00 & 1.19 & 1.66 & 2.31 & 3.16 \\
& & (1.02) & (0.94) & (0.80) & (0.64) & (0.44) & (1.02) & (1.09) & (1.26) & (1.52) & (1.80) & (1.00) & (1.17) & (1.68) & (2.31) & (3.26) \\
\midrule 
\multirow{6}{*}{$3$} & \multirow{2}{*}{Infimum CH} & 0.87 & 0.77 & 0.59 & 0.41 & 0.23 & 0.92 & 0.83 & 0.62 & 0.46 & 0.31 & 1.00 & 0.86 & 0.70 & 0.58 & 0.48 \\
& & (0.85) & (0.76) & (0.57) & (0.40) & (0.22) & (0.91) & (0.81) & (0.61) & (0.44) & (0.30) & (1.00) & (0.86) & (0.69) & (0.58) & (0.47) \\
\cline{2-17}
& \multirow{2}{*}{amplification} & 0.87 & 0.77 & 0.59 & 0.42 & 0.24 & 0.91 & 0.86 & 0.75 & 0.65 & 0.54 & 1.00 & 1.00 & 1.00 & 1.00 & 1.00 \\
& & (0.85) & (0.76) & (0.57) & (0.41) & (0.24) & (0.91) & (0.84) & (0.74) & (0.64) & (0.54) & (1.00) & (1.00) & (1.00) & (1.00) & (1.00) \\
\cline{2-17}
& \multirow{2}{*}{Supremum CH} & 0.87 & 0.87 & 0.66 & 0.47 & 0.28 & 0.92 & 1.00 & 0.97 & 0.92 & 0.85 & 1.00 & 1.14 & 1.35 & 1.52 & 1.70 \\
& & (0.85) & (0.87) & (0.64) & (0.46) & (0.28) & (0.91) & (1.00) & (0.97) & (0.92) & (0.86) & (1.00) & (1.14) & (1.37) & (1.55) & (1.72) \\
\bottomrule
\end{tabular}}
\caption{The relative Bahadur-Rosenbaum efficiency of the amplification and confounding heterogeneity frameworks compared to the classic framework. The infimum confounding heterogeneity relative efficiency is the infimum Bahadur-Rosenbaum slope divided by the Bahadur-Rosenbaum slope for the classic framework, and the supremum confounding heterogeneity relative efficiency is the supremum Bahadur-Rosenbaum slope divided by the Bahadur-Rosenbaum slope for the classic framework. Values in parentheses are simulated ratios of the minimum sample size required under the classic framework to achieve $95\%$ power to the corresponding minimum sample sizes required under the amplification and confounding heterogeneity frameworks at a significance level of $\alpha = 0.01$. Outcomes are simulated under case 2: $Y_{i1}V_{i} \sim N(0.1, 1), Y_{i2}V_i \sim N(0.5, 1)$, $q_{ik} = |Y_{ik}V_{i}|$.}
\label{tab: Bahadur-Rosenbaum slope table}
\end{table}}

\begin{figure}[ht] 
    \centering
    \makebox[\textwidth][c]{%
    \begin{tikzpicture}

        \node[] at (-6, 0) {\includegraphics[width=0.35\textwidth]{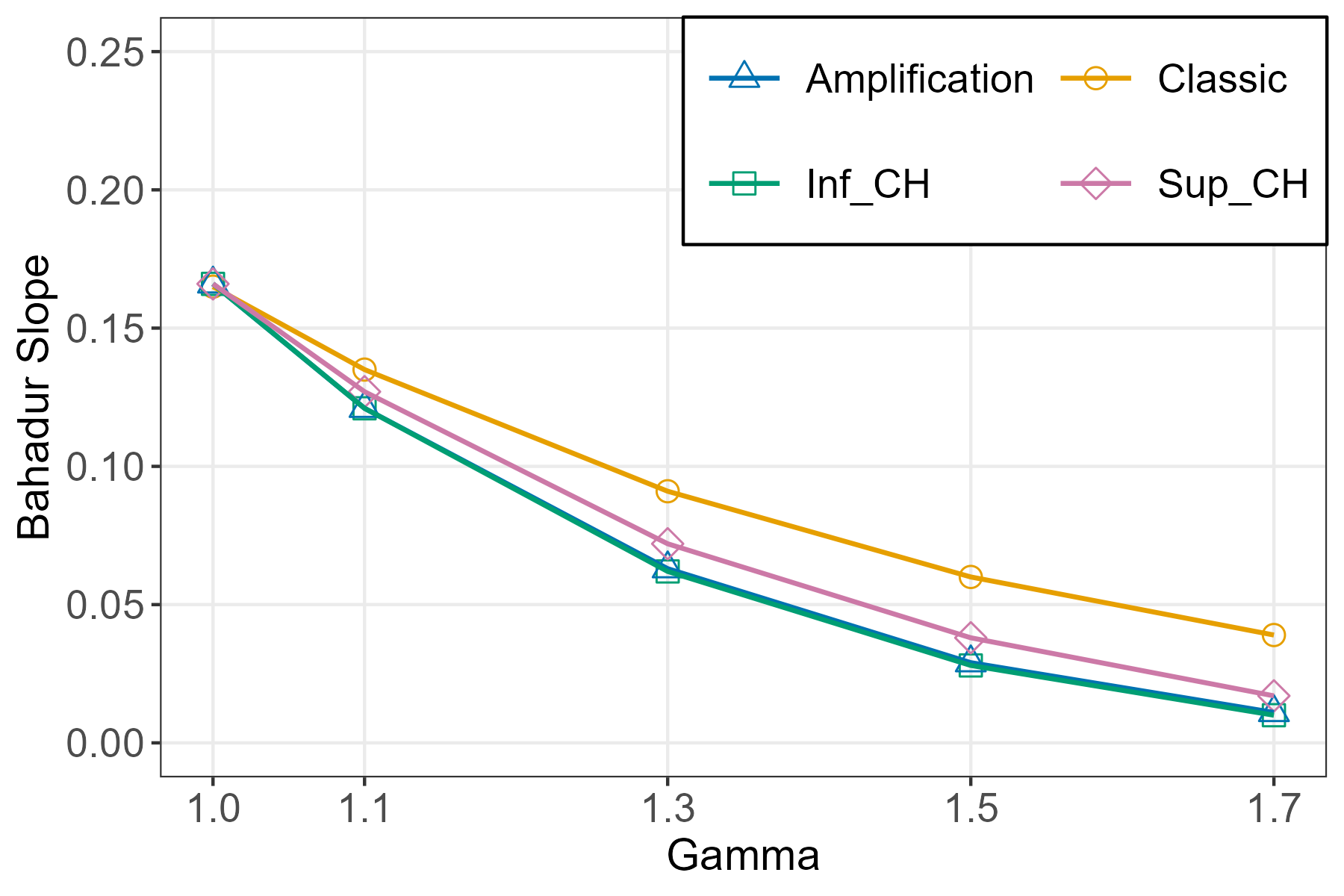}};
        \node[] at (0, 0) {\includegraphics[width=0.35\textwidth]{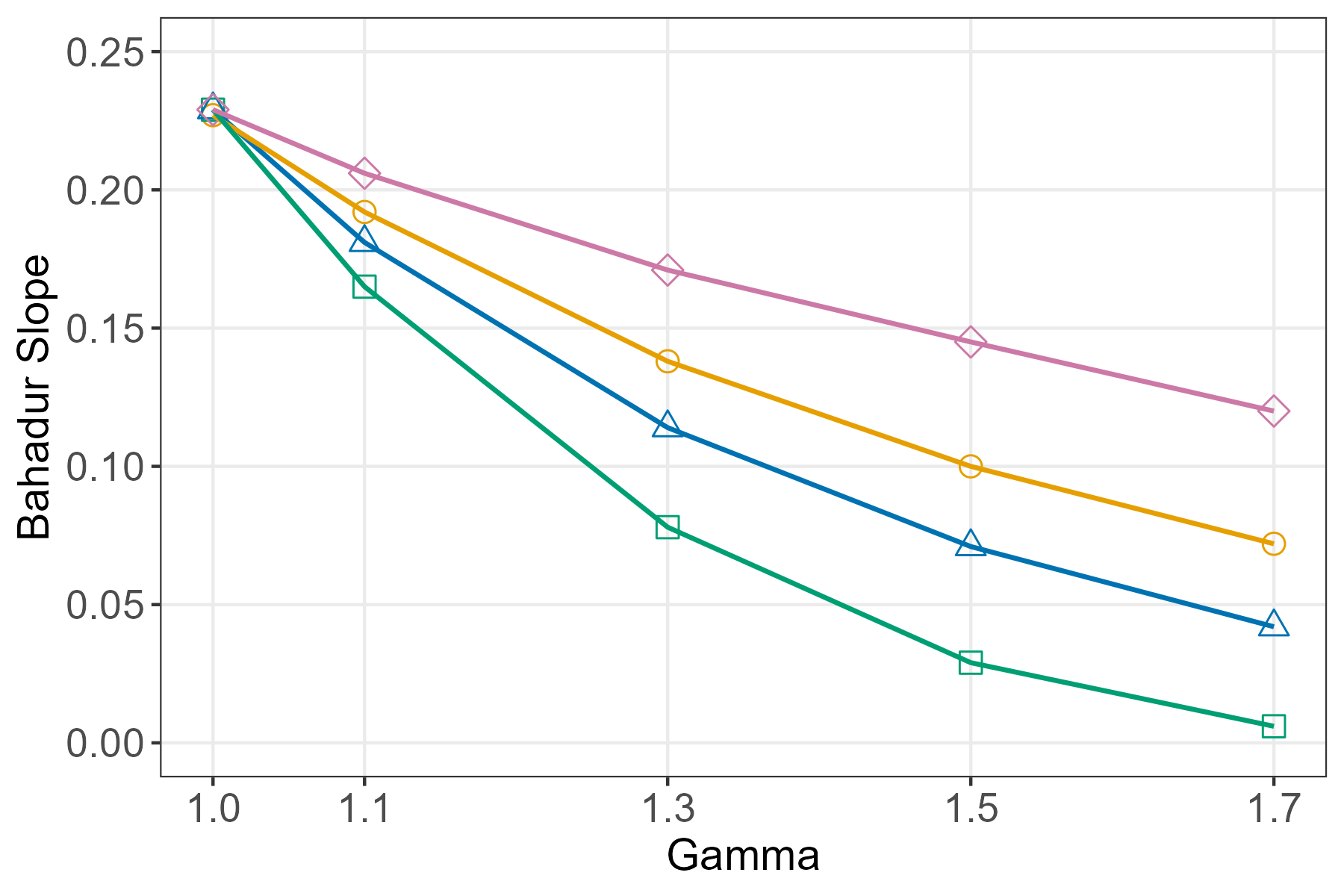}};
        \node[] at (6, 0) {\includegraphics[width=0.35\textwidth]{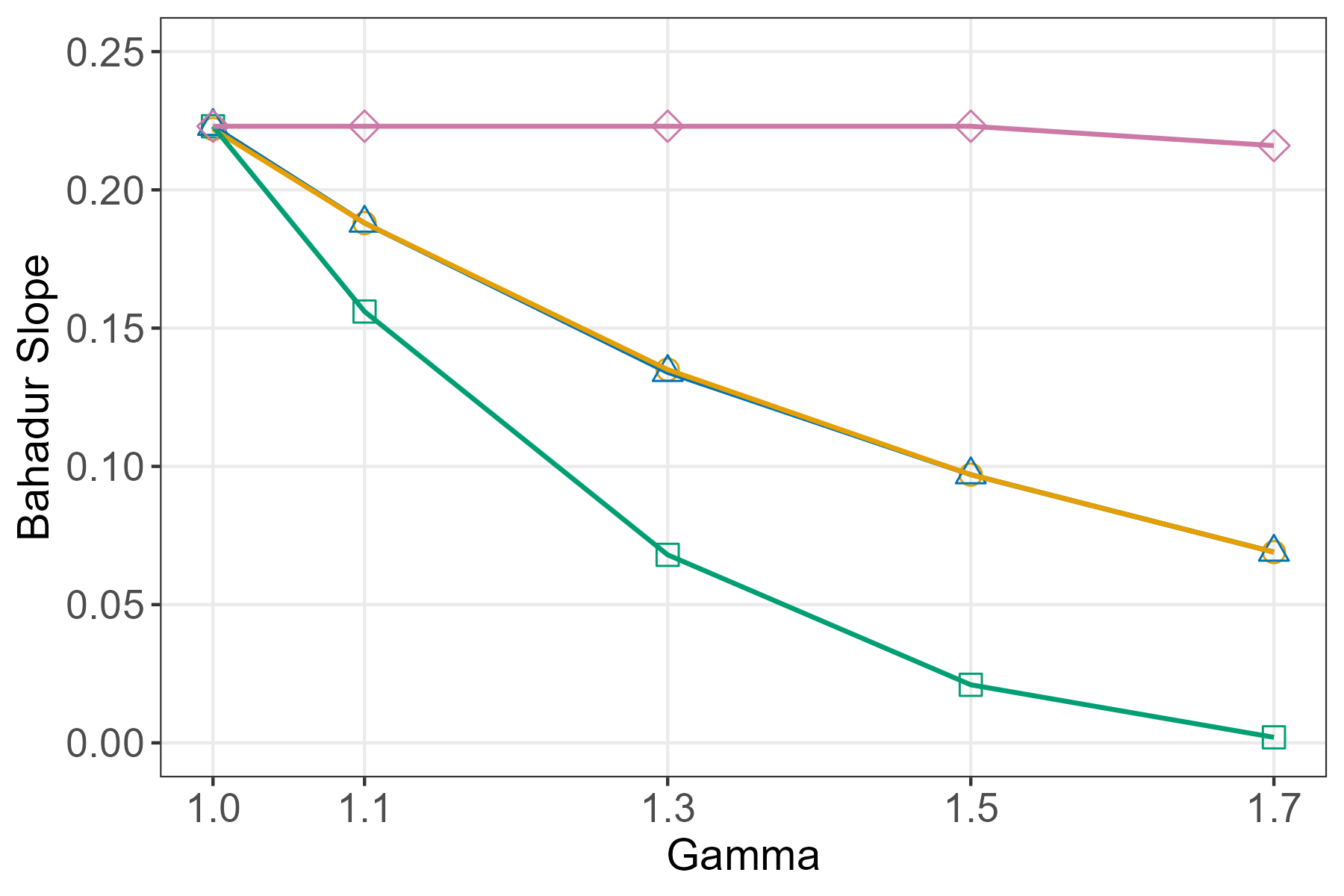}};

                
    \end{tikzpicture}
    }
    \caption{Visualizing the Bahadur-Rosenbaum slope when $Y_{i1}V_{i} \sim N(0.1, 1), Y_{i2}V_i \sim N(0.5, 1)$, $q_{ik} = |Y_{ik}V_{i}|$, $a = 1$, and $\kappa = (0.5, 0.5)$ (left), $\kappa = (0.25, 0.75)$ (middle) and $\kappa = (0, 1)$ (right).}
    \label{fig: Bahadur-Rosenbaum slope figure, a = 1}
\end{figure}

\begin{figure}[ht] 
    \centering
    \makebox[\textwidth][c]{%
    \begin{tikzpicture}

        \node[] at (-3, 0) {\includegraphics[width=0.35\textwidth]{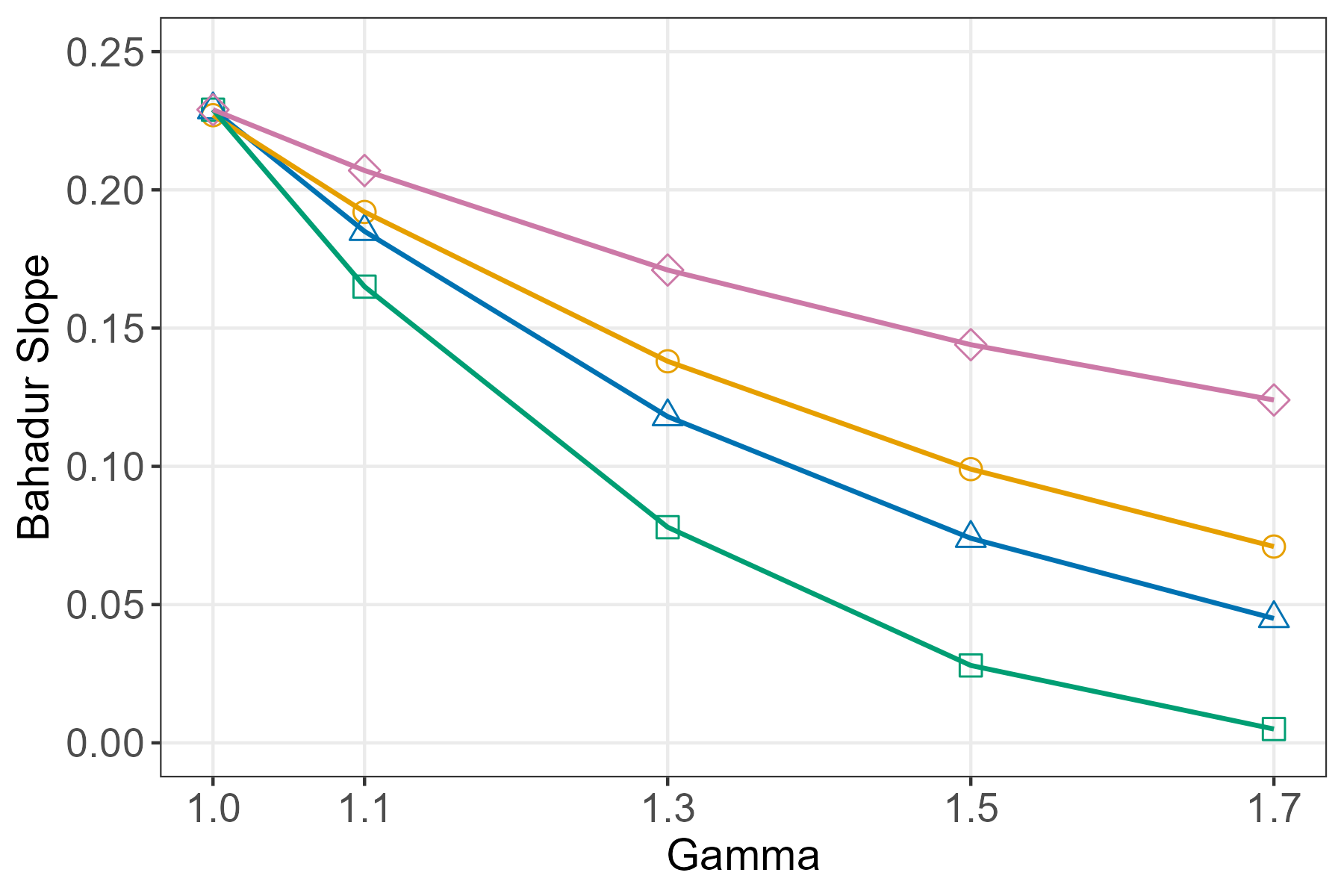}};
        \node[] at (3, 0) {\includegraphics[width=0.35\textwidth]{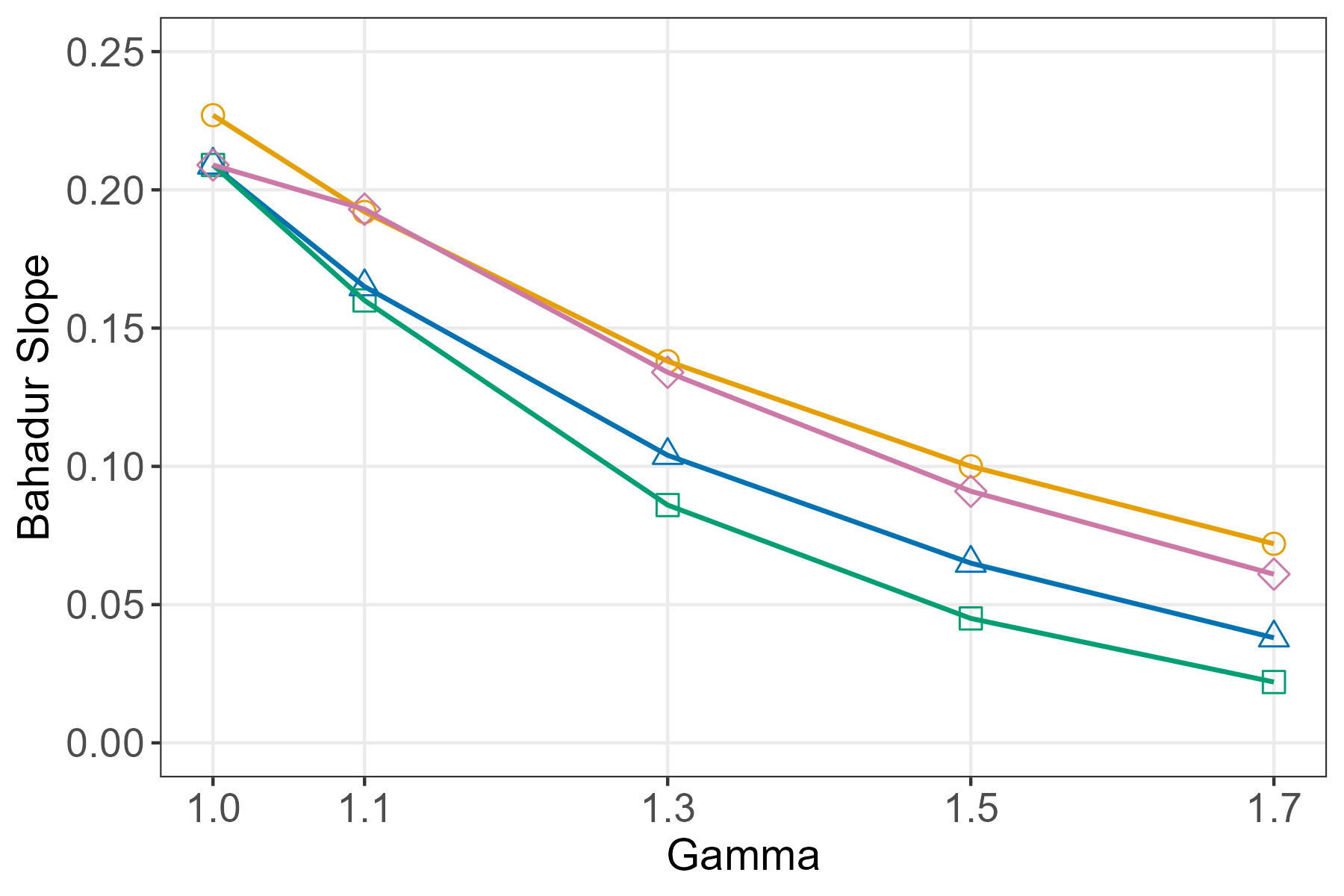}};
                
    \end{tikzpicture}
    }
    \caption{Visualizing the Bahadur-Rosenbaum slope when $Y_{i1}V_{i} \sim N(0.1, 1), Y_{i2}V_i \sim N(0.5, 1)$, $q_{ik} = |Y_{ik}V_{i}|$, $\kappa = (0.25, 0.75)$, and $a = 1/3$ (left) and $a = 3$ (right).}
    \label{fig: Bahadur-Rosenbaum slope figure, kappa = (0.25 0.75)}
\end{figure}

\newpage
\section{Additional Analysis of Smoking and Periodontal Disease Data}

As noted by \cite{rosenbaum2016using}, \cite{fogarty2016sensitivity}, \cite{cohen2018multivariate} and \cite{bekerman2026powerful}, the minimax inequality implies that if unmeasured confounding acts before the practitioner chooses the optimal linear combination of test statistics, the resulting worst-case deviate will be no less than in the case where the practitioner chooses the optimal linear combination of test statistics first, then unmeasured confounding responds to such choice:
\begin{equation*}
\min _{\mathbf{u} \in \mathcal{U}_{\mathcal{P}}} \max _{\kappa \in \mathcal{K}} \frac{\left|\kappa^{T}\left(\mathbf{T}-\boldsymbol{\mu}_{\mathcal{P}, \mathbf{u}}\right)\right|}{\sqrt{\kappa^{T} \boldsymbol{\Sigma}_{\mathcal{P}, \mathbf{u}}\kappa}} \geq \max _{\kappa \in \mathcal{K}} \min _{\mathbf{u} \in \mathcal{U}_{\mathcal{P}}} \frac{\left|\kappa^{T}\left(\mathbf{T}-\boldsymbol{\mu}_{\mathcal{P}, \mathbf{u}}\right)\right|}{\sqrt{\kappa^{T} \boldsymbol{\Sigma}_{\mathcal{P}, \mathbf{u}}\kappa}}.
\end{equation*}
As part of our reanalysis, we compare the effects of confounding heterogeneity across two strategies for calculating the worst-case deviate: 1) ``Confounder acts first'', corresponding to calculating the worst-case deviate as $\min_{\mathbf{u} \in \mathcal{U}_{\mathcal{P}}} \max _{\kappa \in \mathcal{K}} \left|\kappa^{T}\left(\mathbf{T}-\boldsymbol{\mu}_{\mathcal{P}, \mathbf{u}}\right)\right|/\sqrt{\kappa^{T} \boldsymbol{\Sigma}_{\mathcal{P}, \mathbf{u}}\kappa}$, and 2) ``Practitioner acts first'', corresponding to calculating the worst-case deviate as $\max _{\kappa \in \mathcal{K}} \min _{\mathbf{u} \in \mathcal{U}_{\mathcal{P}}} \left|\kappa^{T}\left(\mathbf{T}-\boldsymbol{\mu}_{\mathcal{P}, \mathbf{u}}\right)\right|/\sqrt{\kappa^{T} \boldsymbol{\Sigma}_{\mathcal{P}, \mathbf{u}}\kappa}$. In the ``Practitioner acts first'' case, the linear combination of test statistics chosen is $\kappa = (0.21, 0.79)$, where the upper teeth scores are weighted much more heavily than the scores from the lower teeth. In Table~\ref{tab: sensitivity values, practitioner first}, considering $a = 1/3, 1, 3$, we report the empirical sensitivity values with significance level $\alpha = 0.05$ for the classic and amplification frameworks, as well as the sensitivity value corresponding to the worst-case allocation of confounding heterogeneity (``Infimum CH'') and the sensitivity value corresponding to the best-case allocation of confounding heterogeneity (``Supremum CH'') for the ``practitioner acts first'' case. 

First, the sensitivity values from the ``confounder acts first'' case all dominate the sensitivity values from the ``practitioner acts first'' case even when taking confounding heterogeneity into account, confirming that allowing unmeasured confounding to act before choosing the optimal linear combination of test statistics is a powerful strategy, even in the presence of confounding heterogeneity. Next, since $\kappa = (0.21, 0.79)$, the practitioner heavily weights the upper teeth outcome, so if unmeasured confounding acts after the linear combination of test statistics is chosen, the worst-case allocation of unmeasured confounding strength will allocate as much confounding strength as possible to the second outcome, resulting in a much smaller $\Gamma$ sensitivity value. Additionally, the best-case allocation of unmeasured confounding strength will allocate as much confounding strength as possible to the first outcome.

{\renewcommand{\arraystretch}{0.75}
\begin{table}[t]
    \centering
    \begin{tabular}{cc|ccc|ccc|ccc}
        \multicolumn{11}{c}{Practitioner acts first} \\
        \toprule
         & classic & \multicolumn{3}{c}{amplification} & \multicolumn{3}{c}{Infimum CH} & \multicolumn{3}{c}{Supremum CH} \\
         a & - & 1/3 & 1 & 3 & 1/3 & 1 & 3 & 1/3 & 1 & 3 \\
         \midrule
        $\Gamma$ & 2.60 & 2.51 & 2.50 & 2.46 & 1.72 & 1.76 & 2.12 & 4.10 & 3.45 & 2.87 \\
        $\Lambda$ & -- & 9.79 & 4.79 & 3.18 & 6.41 & 3.21 & 2.70 & 16.21 & 6.75 & 3.73 \\
        $\Delta_{1}$ & -- & 3.26 & 4.79 & 9.53 & 1 & 1.10 & 3.56 & $\infty$ & $\infty$ & $\infty$ \\
        $\Delta_{2}$ & -- & 3.26 & 4.79 & 9.53 & 6.27 & $\infty$ & $\infty$ & 2.20 & 2.88 & 5.11 \\
        \hline 
    \end{tabular}
    \caption{Sensitivity values across sensitivity analysis frameworks.}
    \label{tab: sensitivity values, practitioner first}
\end{table}}

\vspace{-0.5cm}
\bibliographystyle{apalike}
\bibliography{confounding}

\end{document}